\documentclass[useAMS,usenatbib]{mn2e}

\usepackage{amssymb,amsmath}
\usepackage{lscape}
\usepackage{graphicx}
\usepackage[colorlinks=true,citecolor=blue]{hyperref}

\newcommand{\Sp}{_{\mathrm{Sp}}}

\newcommand{\KHI}{_{\mathrm{KHI}}}

\newcommand{\Gal}{_{\mathrm{gal}}}
\newcommand{\Gas}{_{\mathrm{gas}}}
\newcommand{\CF}{_{\mathrm{CF}}}

\newcommand{\Crit}{_{\mathrm{crit}}}
\newcommand{\Max}{_{\mathrm{max}}}
\newcommand{\Min}{_{\mathrm{min}}}
\newcommand{\Up}{_{\mathrm{up}}}
\newcommand{\Down}{_{\mathrm{down}}}

\newcommand{\KHvisc}{_{\mathrm{KHvisc}}}
\newcommand{\KHinvisc}{_{\mathrm{KHinvisc}}}
\newcommand{\Mfp}{_{\mathrm{mfp}}}
\newcommand{\ICM}{_{\mathrm{ICM}}}

\newcommand{\Hot}{_{\mathrm{hot}}}
\newcommand{\Cold}{_{\mathrm{cold}}}

\newcommand{\Erf}{\textrm{erf}}
\newcommand{\Reyn}{\textrm{Re}}
\newcommand{\Mach}{\textrm{Ma}}

\newcommand{\KeV}{\,\textrm{keV}}
\newcommand{\Kpc}{\,\textrm{kpc}}

\newcommand{\Sec}{\,\textrm{s}}
\newcommand{\Myr}{\,\textrm{Myr}}

\newcommand{\Kms}{\,\textrm{km}\,\textrm{s}^{-1}}

\newcommand{\ccm}{\,\textrm{cm}^{-3}}
\newcommand{\gccm}{\,\textrm{g}\,\textrm{cm}^{-3}}

\newcommand{\de}{\partial}

\title[Viscous KHI]{Viscous Kelvin-Helmholtz instabilities in highly ionised plasmas}
\author[Roediger et al.]{
E. Roediger$^{1,3,2}$\thanks{E-mail:
eroediger@hs.uni-hamburg.de (ER)}, 
R. P. Kraft$^{3}$,
P. Nulsen$^{3}$,
E.~Churazov$^{4}$,
W.~Forman$^{3}$,
\newauthor
M.~Br\"uggen$^{1}$,
R.~Kokotanekova$^{5,2,3}$
\\
$^{1}$Hamburger Sternwarte, Universit\"at Hamburg, Gojenbergsweg 112, D-21029 Hamburg, Germany\\
$^{2}$Jacobs University Bremen, PO Box 750 561, 28725 Bremen, Germany\\
$^{3}$ Harvard/Smithsonian Center for Astrophysics, 60 Garden Street, MS-4, Cambridge, MA 02138, USA\\
$^{4}$ Max-Planck-Institut f\"ur Astrophysik, Karl-Schwarzschild-Strasse 1, 85741 Garching, Germany\\ 
$^{5}$ AstroMundus Master Programme, University of Innsbruck, Technikerstr. 25/8, 6020 Innsbruck, Austria \\}
\begin{document}

\date{Accepted 1988 December 15. Received 1988 December 14; in original form 1988 October 11}

\pagerange{\pageref{firstpage}--\pageref{lastpage}} \pubyear{2013}

\maketitle

\label{firstpage}

\begin{abstract}
Transport coefficients in highly ionised plasmas like the intra-cluster medium (ICM) are still ill-constrained. They influence various processes, among them the mixing at shear flow interfaces due to the Kelvin-Helmholtz instability (KHI). The observed structure of potential mixing layers can be used to infer the transport coefficients, but the data interpretation requires a detailed knowledge of the long-term evolution of the KHI under different conditions. 
{
Here we present the first systematic numerical study of the effect of constant and 
temperature-dependent isotropic viscosity over the full range of
possible values.
} 
  We show that moderate viscosities slow down the growth of the  KHI and reduce the height of the KHI rolls and their rolling-up. Viscosities above a critical value suppress the KHI. The effect can be quantified in terms of the Reynolds number $\Reyn=U\lambda/\nu$, where $U$ is the shear velocity, $\lambda$ the perturbation length, and $\nu$ the kinematic viscosity. We derive the critical Re for constant and temperature dependent, Spitzer-like viscosities,  an empirical relation for the viscous KHI growth time as a function of $\Reyn$ and density contrast, and describe special behaviours for Spitzer-like viscosities and high density contrasts. Finally, we briefly discuss several astrophysical situations where the viscous KHI could play a role, i.e., sloshing cold fronts, gas stripping from galaxies, buoyant cavities, ICM turbulence, and high velocity clouds. 
\end{abstract}


\section{Intro}  \label{sec:intro}

The classic Kelvin-Helmholtz instability (KHI) arises due to a shear flow parallel to the interface between two inviscid incompressible fluids. The fluids are in pressure equilibrium, but the shear velocity and possibly the density change discontinuously at the interface. Small perturbations at the interface grow exponentially (\citealt{Chandrasekhar1961,Lamb,Drazin}), and the distorted interface rolls up into the classic Kelvin-Helmholtz (KH) rolls or cat-eye patterns (see, e.g., Fig.~\ref{fig:rolls_nu_Re1000_D1_M05} later-on). As shear flows are ubiquitous in astrophysical fluids, the KHI is a major agent for turbulence generation and mixing. 

Several special conditions or fluid properties can fully or partially suppress the KHI, among them gravity, magnetic fields, and viscosity. Thus, if the dynamical conditions at a shear layer are known, the presence or absence of the KHI can in principle be used to constrain the properties of the fluid. 

This prospect is particularly interesting for the intra-cluster medium (ICM) in galaxy clusters and hot halos of galaxies. At temperatures of up to a few keV, this gas is a highly ionised plasma, and it is weakly magnetised (e.g.~\citealt{Ferrari2008,Bonafede2010}). Its {effective} transport properties, i.e., thermal conduction and viscosity, are still ill-constrained. If the ICM was not magnetised, the transport coefficients due to Coulomb collisions would be large (\citealt{Spitzer1956}). For example, the importance of viscosity can be expressed by the Reynolds number $\Reyn=UL/\nu$, where $U$ is a characteristic velocity of the gas flow, $L$ a characteristic length scale, and $\nu$ the kinematic viscosity. The kinematic viscosity in an unmagnetised plasma scales as $\nu \propto T^{5/2} n^{-1}$ (\citealt{Spitzer1956,Sarazin1988}), and for typical values of ICM temperature $T\ICM$ and electron particle density $n_e$ the Reynolds number becomes
\begin{eqnarray}
\Reyn &=&  10  \; f_{\mu}^{-1}\; 
\left(\frac{U}{400\Kms}\right)\; 
\left(\frac{L}{10\Kpc}\right)  \left(\frac{n_e}{10^{-3}\ccm}\right)  \nonumber\\
&&  \times \left( \frac{kT\ICM}{2.4\KeV}  \right)^{-5/2}. \label{eq:ReICM}
\label{eq:Re}
\end{eqnarray}
We allowed for a reduced viscosity in Eqn.~\ref{eq:Re} by including the viscosity suppression factor $f_{\mu}\le 1$.  {In the presence of magnetic fields the particle mean free path perpendicular to the field lines is reduced dramatically, leading to anisotropic transport coefficients. However, tangled magnetic fields could lead to a reduced isotropic \textit{effective} mean free path and transport coefficients on macroscopic scales.
Microscale instabilities might have even more dramatic
effects on transport in the intracluster plasma (\citealt{Rosin2011}).}
Collisionality in the ICM could be mediated by interactions between magneto-hydrodynamic waves and the particles. The resulting \textit{effective} transport processes in the ICM could range from reduced isotropic conduction and viscosity (\citealt{Narayan2001}) to anisotropic transport coefficients (\citealt{Kunz2012}), 
{
and the effective viscosity and thermal conduction may even experience different suppression factors. 
Isotropic and anisotropic viscosities could affect the KHI differently, e.g., in the latter case the orientation of the magnetic field with respect to the interface may play a role. However, so far  the nature of the \textit{effective} viscosity is still unconstrained observationally.  Therefore, in this paper we focus on the  effect of an isotropic viscosity.}

In the ICM, observable shear flows occur in different dynamical contexts.  Prominent examples are sloshing cold fronts, upstream edges and tails of gas stripped galaxies or merger cores, or the surfaces of buoyantly rising cavities that have been inflated by active galactic nuclei (AGN). The quality of the observational data has become sufficient  to show the structure of the shear layers. For example, the cluster Abell 496 (\citealt{Dupke2007,Roediger2012a496}) has boxy cold fronts with kinks and doublets. We identified distorted fronts also in the merging groups around NGC 7619 and UGC 12491 (\citealt{Roediger2012n7618}). Several gas-stripped elliptical galaxies falling into their host clusters have been observed deeply. They all show an upstream contact discontinuity and a tail of stripped gas, but their detailed structures differ. The tail of M86 starts in a plume, bends, and bifurcates (\citealt{Randall2008}). In M89 (\citealt{Machacek2006a}) and M49 (\citealt{Kraft2011}) the upstream edges have a ragged appearance with horns and kinks. In contrast, the upstream edge of NGC 1404 (\citealt{Machacek2005}) appears to be smooth. Numerous AGN inflated cavities have been observed, despite the fact that in purely hydrodynamical simulations they are disrupted by Rayleigh-Taylor and KHIs. The presence or absence of substructure at these shear layers indicates the presence or suppression of KHIs. 

Several groups started investigating the impact of ICM properties on such shear layers. 
\citet{Reynolds2005} and \citet{Guo2012b} demonstrated that buoyantly rising or even currently inflating cavities can be stabilised by viscosity. {\citet{Dong2009} pointed out that in the case of anisotropic viscosity the evolution of the bubbles depends on the magnetic field orientation because preferentially instabilities parallel to the field lines are suppressed.}   \citet{Lyutikov2006} and \citet{Dursi2008} showed that  magnetic draping can stabilise the cavities, although \citet{Ruszkowski2007} stressed out that this requires magnetic fields with coherence lengths larger than the cavity size. Magnetic draping can also suppress instabilities at gas stripped galaxies (\citealt{Dursi2008,Ruszkowski2012}).Viscosity, however, could have a similar effect (\citealt{Roediger2008visc}). Sloshing cold fronts in hydrodynamical simulations are distorted by KHIs, which can be reduced or suppressed by sufficiently aligned magnetic fields (\citealt{ZuHone2011}) or viscosity (\citealt{Roediger2013virgovisc}).

In all of these situations the KHI occurs in a complex dynamical context. Shear velocities can vary in time and space, the shear layers are curved and experience gravity. In order to provide a solid basis for studies that include these complex dynamical contexts, here we focus on the impact of {an isotropic} viscosity on the KHI in idealised setups. 
{Already the linear stability analysis of the viscous KHI is complicated (see Sect.~\ref{sec:khi_visc}), and, to our knowledge, no previous work investigated the long-term evolution of the viscous KHI. Here we do so with a systematic numerical study.}
 In particular, we 
\begin{itemize}
\item show analytically that viscosity suppresses the KHI below a critical Reynolds number.
\item investigate not only the onset of the KHI, but its long-term evolution over several of linear growth times by means of hydrodynamical simulations. This is important because the dynamical timescales of the processes where we can observe the shear layers operate on much longer timescales than the linear growth time of the KHI.
\item derive an empirical relation for the viscous KHI growth time as a function of Reynolds number and density contrast.
\item show that also a strongly temperature-dependent Spitzer-like viscosity can suppress the KHI.
\end{itemize}

The paper is organised as follows: In Section~\ref{sec:khi_visc} we summarise relevant previous results on the KHI and analytically derive a critical Reynolds number below which the KHI is suppressed. Section~\ref{sec:method} describes the simulation setup and method. Section~\ref{sec:results} reports the results of the simulations, and Sect.~\ref{sec:discussion} discusses  implications of our results. We summarise our findings in Sect.~\ref{sec:summary}. 

\section{KHI in viscous fluids -- analytic considerations} \label{sec:khi_visc}
%
The KHI arises due to a shear flow parallel to the interface between two fluids. In the most simple case, two incompressible inviscid fluids of densities $\rho\Hot$ and $\rho\Cold$ but equal pressures are separated by a planar interface. While the temperature is still irrelevant at this point we choose this notation for consistency with later discussions in the paper.  We consider the 2D case and place the interface at $y=0$.  Without loss of generality, the hot and cold fluid move with velocities $U/2$ and $-U/2$ parallel to the interface, i.e., they are subject to a mutual shear flow. At the interface, both the velocity and density change discontinuously (the densities can also be identical). If a perturbation of length scale $\lambda$ is introduced at the interface, the perturbation grows exponentially with a growth time of
\begin{eqnarray}
\tau\KHinvisc
&=& \frac{\sqrt{\Delta}}{2\pi}\frac{\lambda}{U} \nonumber \\
&=& 3.9\Myr \sqrt{\Delta}\frac{\lambda}{10 \Kpc}\left(\frac{U}{400\Kms}\right)^{-1}  \label{eq:tau_invisc}\\
\textrm{with}\;\Delta&=& \frac{(\rho\Cold+\rho\Hot)^2}{\rho\Cold\rho\Hot} = D_{\rho}(1+1/D_{\rho})^2 \\
\textrm{and}\; D_{\rho} &=& \frac{\rho\Cold}{\rho\Hot}. \nonumber
 \label{eq:Delta}
\end{eqnarray}
This standard scenario and many variations, e.g., the presence of compressibility (see App.~\ref{app:compressibility}),  gravity, surface tension and magnetic fields have been discussed in textbooks, e.g.~\citet{Chandrasekhar1961,Lamb,Drazin}.  

The case of viscous KHI is missing from the extensive discussions in the textbooks because the background flow is not steady. The standard approach of linear perturbation analysis assumes a background state for the spatial distribution of fluid density $\rho$, pressure $p$ and velocity $U$. In the case of the KHI this is the shear flow described in the beginning of this section. Then small perturbations $u\ll U$, and equivalents for other quantities, are added to the background flow. Quantities like $U+u$ are  inserted into the hydrodynamical equations, and these are linearised. Following this approach for the KHI leads to a system of equations for the perturbed quantities like in Eqns.~6 to 10 in \citet{Junk2010} or Eqns.~A1 to A6 in \citet{Kaiser2005}. However, both of these works  made the additional assumption that the background flow is  steady, i.e.,  $\de U/\de t=0$. This is true \textit{only} if the viscosity is small. A non-negligible viscosity, however, smoothes out the shear velocity gradient across the interface, i.e., $\de U/\de t\ne 0$ even without any perturbation. If this aspect is taken into account, the term $\delta\rho\frac{\de U}{\de t}$ needs to be added to the left-hand side of Eqn.~A1 in \citet{Kaiser2005} and Eqn.~6 in \citet{Junk2010}, where $\delta\rho$ is the density perturbation. This term  breaks the symmetry of the linearised equations and makes the calculation of a dispersion relation cumbersome. 

Nonetheless, we can estimate the behaviour of the KHI at low and high viscosities. 
Naturally, at low viscosities the KHI approaches the inviscid case described above. A high viscosity must suppress the KHI below certain length scales for the following reason: The effect of viscosity is to smooth out the velocity gradient between both fluid layers by momentum diffusion (see also Sect.~\ref{eq:smoothshearjump}). If, however, in the classic KHI setup the discontinuity in the shear velocity is smoothed over a length scale $\pm d$ above and below the interface, the KHI is inhibited for wavelengths smaller than $\sim 10 d$ (\citealt{Chandrasekhar1961}, Section 102). Consequently, a given viscosity must suppress the KHI  below a certain length scale. We can estimate this limit quantitatively as follows. 

In the inviscid case with a discontinuous shear velocity, a perturbation of wavelength $\lambda$ grows on the timescale of  $\tau\KHinvisc$ (Eqn.~\ref{eq:tau_invisc}).  During one $\tau\KHinvisc$, a given viscosity $\nu$ widens the jump in  shear velocity to the diffusion length $l_D (t=\tau\KHinvisc) = \pm 2\sqrt{\nu \tau\KHinvisc}$ around the interface. Consequently,  all perturbations with wavelengths $< 10 l_D(\tau\KHinvisc)$ cannot become KH unstable. Thus, the growth of the perturbation of wavelength $\lambda$ will be suppressed if
\begin{eqnarray}
\lambda &<& 10 l_D(\tau\KHinvisc) = 20 \sqrt{\nu \tau\KHinvisc}  \;\;\textrm{or} \\
\nu &>&\nu\Crit = \frac{\pi}{200}\frac{\lambda U}{\sqrt{\Delta}}  \;\;\textrm{or}\\
\Reyn = \frac{\lambda U}{\nu} &<&  \Reyn\Crit \approx 64 \sqrt{\Delta}. \label{eq:viscKHI}
\end{eqnarray}
Here and in the following, we define the Reynolds number for the KHI as
\begin{equation}
\Reyn = \frac{\lambda U}{\nu},  \label{eq:ourRe}
\end{equation}
where $U$ is the shear velocity, i.e., the difference in velocity between both layers, $\lambda$ the perturbation length, and $\nu$ the kinematic viscosity. Note that in other work the length scale used in the Reynolds number may refer to an initial width of the interface instead to the perturbation length. Equation~\ref{eq:viscKHI} shows that the KHI should be suppressed not only for Reynolds numbers around 1, but already for Reynolds numbers around 130, with a moderate dependence on density contrast. 

Viscosity continuously broadens the jump in shear flow, and the choice of comparing the perturbation length scale $\lambda$ to the diffusion length $l_D$ at $t=\tau\KHinvisc$ is somewhat arbitrary. However, two different considerations arrive at similar estimates. The viscous broadening of the interface should dominate and suppress the KHI if the viscous dissipation timescale $\tau_{\nu}=L^2/\nu$ is shorter than the KHI growth time. For the former, $L$ is the length scale of velocity gradients that need to be dissipated. As noted above, the KHI at length scale $\lambda$ is suppressed if the shear flow jump is smoothed over $\sim \lambda/10$; hence we use $L=\lambda/10$. This leads to 
\begin{eqnarray}
\frac{\sqrt{\Delta}}{2\pi} \frac{\lambda}{U} = \tau\KHinvisc &>& \tau_{\nu}=\frac{L^2}{\nu} = \frac{(\lambda/10)^2}{\nu}  \;\;\textrm{or} \\
\Reyn &<&  \Reyn\Crit \approx 16 \sqrt{\Delta}.
\end{eqnarray}
as the condition to suppress the KHI. 

A third estimate comes from demanding that the height of the unsuppressed KH rolls would lag behind the viscous broadening of the shear flow jump up to the time the final height is reached. The inviscid KHI saturates after $\sim 4\tau\KHinvisc$, and the rolls reach a height of typically $\lesssim \lambda/2$, with small perturbations only $\lambda/4$.  Consequently, the condition for a suppressed KHI is 
\begin{eqnarray}
\lambda/2 &<& l_D(4\tau\KHinvisc) = 2 \sqrt{\nu 4 \tau\KHinvisc}    \;\;\textrm{or} \\
\Reyn &<&  \Reyn\Crit \approx 10 \sqrt{\Delta}.
\end{eqnarray}
This third estimate is rather conservative. For example, assuming $\lambda/4$ as the height of the KH rolls leads to a 4 times higher $\Reyn\Crit$, which is close to our first estimate. Thus, our estimates agree within order of magnitude that viscosity should certainly suppress the KHI for $\Reyn<$ several 10s. Our simulations confirm this general result. However, they show that the dependence on density contrast derived here is wrong because none of our estimates takes into account that the viscous broadening of the shear flow jump proceeds asymmetrically between two layers with different densities.

Earlier work on viscous shear flows reaches similar conclusions (e.g., \citealt{Esch1957,Amsden1964,Gerwin1968}).  \citet{Villermaux1998} investigates the stability of a viscous shear flow layer where the jump in velocity is already smoothed over a certain width $2d$. A given smoothing of the velocity jump does not only prevent the KHI at small wavelengths, but also reduces the maximum growth rate. The author stresses that, especially for small initial widths, the viscous spreading can occur faster than the KHI, and even if the KHI should still grow formally, it would not show up because it is not the fastest process. By comparing the viscous spreading rate to the KHI growth rate given the evolving viscously spreading interface, \citet{Villermaux1998} estimates the critical wavelength $\lambda\Crit$ below which the KHI is suppressed as a function of viscosity. In the limit of small initial widths, his Eqn.~13 can be rewritten as
\begin{equation}
\frac{\lambda\Crit U}{\nu} = \frac{2\pi}{b} 10 a,
\end{equation}
expressing his result in terms of our Reynolds number. The parameters $a$ and $b$ are factors of order unity, depending on the exact shape of the smoothed velocity jump. This result agrees well with our simpler order-of-magnitude estimates.

\section{Numerical simulations -- method}  \label{sec:method}

\subsection{Model setup} \label{sec:modelsetup}
We set up a 2D simulation box with a cool gas below $y=0$ and a hot gas above.   The shear flow is set as constant velocities $U/2$ and $-U/2$ in $x$-direction, i.e.~parallel to the interface, in the hot and cold fluid, respectively. Our standard choice is a shear velocity $U$ corresponding to Mach 0.5 in the hot layer. We introduce a sinusoidal perturbation in the velocity component $v_y$, i.e., perpendicular to the interface, of amplitude $v_0$ and wave length $\lambda$. The perturbation is restricted to the layer around the interface and decreases  away from it as in \citet{Junk2010}:
\begin{equation}
v_y=v_0\sin\left(\frac{2\pi x}{\lambda}\right) \exp\left( -\left[\frac{y}{\sigma_y}\right]^2 \right).
\end{equation}
The scale parameter for the width of the perturbation layer is $\sigma_y=0.3\lambda$, the perturbation amplitude is $v_0=0.1 U$. We  verified that the suppression of KHI by viscosity does not arise due to an insufficient perturbation (see Sect.~\ref{sec:testperturb} and Fig.~\ref{fig:testperturb} in appendix). Our simulation box has the size of $(-\lambda,\lambda)$ and $(-2\lambda,2\lambda)$  $x$-direction and $y$-direction, respectively. 
{We verified that the simulation box is sufficiently large in $y$-direction to ensure an unimpeded growth of the KHI by re-simulating some cases with particularly tall KHI rolls with a simulation box twice as high.}
 The simulation box is periodic in $x$-direction and has open boundaries in $y$-direction. We chose this setup over an all-periodic setup  because it is a better representation of real situations where the shear flow arises, e.g., at cold fronts. For subsonic shear velocities relative to the hot layer the open boundaries lead to a mass and energy loss from the simulation volume of less than 1 or 2\% over the course of the simulations, which is irrelevant. We arrive at very similar results when using reflecting boundaries in $y$-direction, but  sound and shock waves originating at the interface are reflected at the boundaries and make the evolution noisier. For supersonic shear flows mass losses are still below 4\%, but shocks leaving the simulation box lead to a loss of total energy of up to 10\% at high viscosities. Shock waves carrying away energy from the shear layer is, however, a realistic effect.
 
The setup described so far employs discontinuities in density and shear velocity at the interface. In low-viscosity simulations this setup is prone to secondary, unintended KHI modes at length scales $<\lambda$ that are seeded by the numerical discretisation. However, the viscosity included in our simulations prevents these secondary modes for most cases, and we use the straightforward setup with the discontinuous interface. This also ensures that the KHI has the best-possible chance to grow and is not slowed down by an initially smoothed interface. Only at  high $\Reyn$ do the secondary modes appear. There we follow the suggestion of \citet{McNally2012khi} and slightly smooth the density and shear velocity jump according to
\begin{eqnarray}
X(y) &=& \left\{
  \begin{array}{l l}
    X\Cold - X_m\exp(y/w) & \quad \text{if $y\le 0$}\\
    X\Hot  + X_m\exp(-y/w) & \quad \text{if $y> 0$}
  \end{array} \right. \label{eq:smooth} \\
  \textrm{with} \; X_m &=& (X\Cold-X\Hot)/2 \\
  \textrm{and} \; X &\in& \{ \rho,v_x  \} \nonumber
\end{eqnarray}
The smoothing scale length is 1 to 2\% of $\lambda$, which does not affect the growth of the intended mode. Secondly, we apply this smoothing in simulations with high density contrasts and high viscosity, where viscous dissipation would lead to excessive heating at a discontinuous interface.

\subsection{Code}

We use the FLASH code (version 3.3, \citealt{Dubey2009}).  FLASH is a modular block-structured adaptive mesh refinement code, parallelised using the Message Passing Interface (MPI) library. It solves the Riemann problem on a Cartesian grid using the Piecewise-Parabolic Method (PPM). We solve the viscous hydrodynamic equations (e.g., \citealt{BatchelorHydro,Landau_hydro}):
\begin{eqnarray}
&&\frac{\partial\rho}{\partial t} + \nabla \cdot (\rho \mathbf v) = 0 \;\,\textrm{(continuity eqn.)} \\
&&\frac{\partial\rho\mathbf{v}}{\partial t} + \nabla\cdot (\rho \mathbf{vv})  + \nabla\cdot \mathbf\Pi= 0 \;\textrm{(momentum eqn.)} \\
&&\frac{\partial\rho E}{\partial t} + \nabla\cdot (\rho E \mathbf{v})  + \nabla\cdot (\mathbf{\Pi}\cdot \mathbf{v})= 0 \;\textrm{(energy eqn.)},
\end{eqnarray}
where $\rho$ is the mass density, $E$ the specific total energy, $\mathbf{v}$ the gas velocity. We assume an ideal equation of state with $\gamma=5/3$. The full pressure tensor $\mathbf\Pi$  includes the viscous stresses:
\begin{eqnarray}
\Pi_{ik} &=& \delta_{ik} P - \pi_{ik} \;\;\textrm{with} \nonumber \\
 \pi_{ik}&=& \mu \left(  \frac{\partial v_i}{\partial x_k} +  \frac{\partial v_k}{\partial x_i} - 2/3\; \delta_{ik} \nabla\cdot\mathbf{v} \right).
\end{eqnarray}
Here, $\mu$ is the dynamic shear viscosity, and we neglect the second viscosity. In the inviscid case with $\mu=0$ we recover the Euler equations, where only the isotropic pressure $P$ appears in the momentum and energy equations. Viscous fluxes for momenta and energy are computed explicitly. We verified the accuracy of the viscosity module on two setups with analytic solutions, the viscous flow between two stationary plates, and the viscous spreading of a shear flow discontinuity (see Appendix~\ref{sec:visctests}).

\subsection{Types of viscosity}

The nature of the viscosity in the ICM is unknown, it may be constant or  strongly temperature dependent. Thus, we investigate the impact of a constant kinematic viscosity $\nu$ and a Spitzer-like kinematic viscosity  $\nu\Sp = A_{\mu} T^{5/2}/n$. The viscosity amplitude $A_{\mu}$ is varied to achieve Reynolds numbers between $\sim 10$ and $10^4$, thus sampling the range of possible Re in the ICM (see Eqn.~\ref{eq:ReICM} and Sect.~\ref{sec:sensibleRe}).  Due to its temperature and density dependence, a Spitzer-like viscosity leads to strong dependence of the ratio of Reynolds numbers in the hot and the cold layer on density contrast $D_{\rho}=\frac{\rho\Cold}{\rho\Hot}$:
\begin{equation}
D_{\Reyn} = \frac{\Reyn\Cold}{\Reyn\Hot} = D_{\rho}^{7/2},
\end{equation}
In case of a constant kinematic viscosity, the Reynolds number is the same in both layers.

We discuss limits on plausible Reynolds numbers in the ICM due to the related length of the mean free path and saturation of momentum transport in Sect.~\ref{sec:sensibleRe}.

\begin{figure*}
\hspace{1cm}$t=2\tau\KHinvisc$ \hfill $4\tau\KHinvisc$ \hfill $10\tau\KHinvisc$ \hfill $20\tau\KHinvisc$ \hfill \phantom{x}\newline
\rotatebox{90}{tracer $F$\phantom{$v_y^1$}}
\includegraphics[trim=0     150 250 160,clip,height=1.83cm]{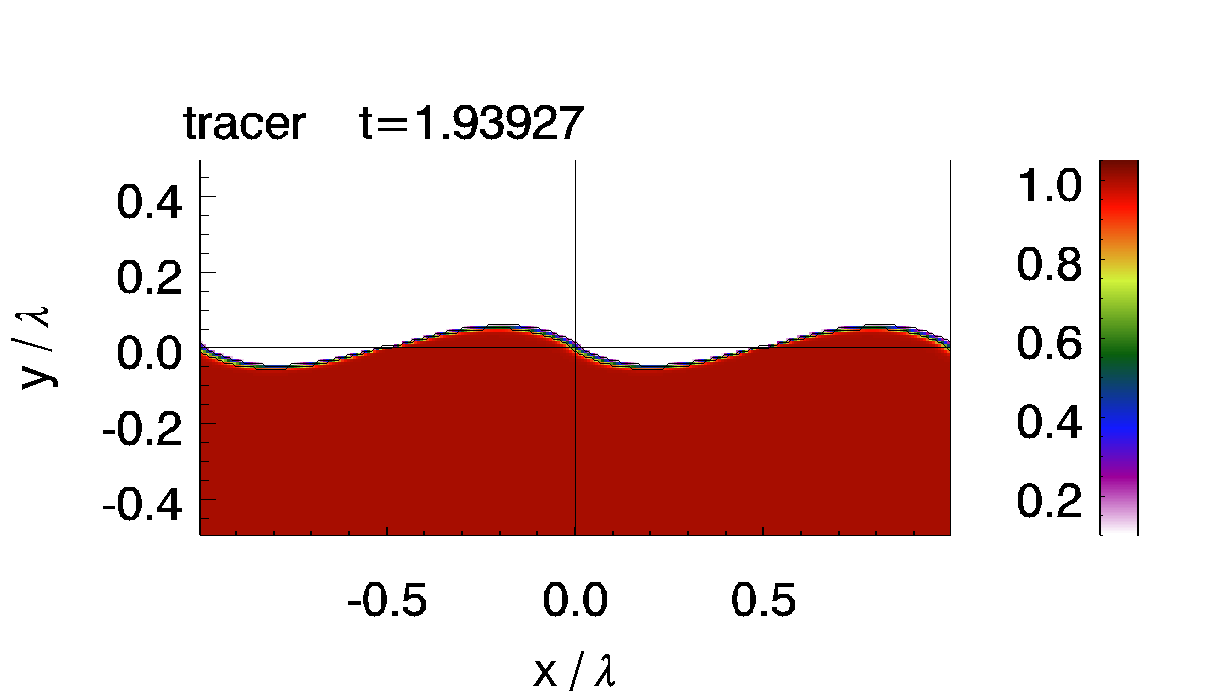}
\includegraphics[trim=190   150 250 160,clip,height=1.83cm]{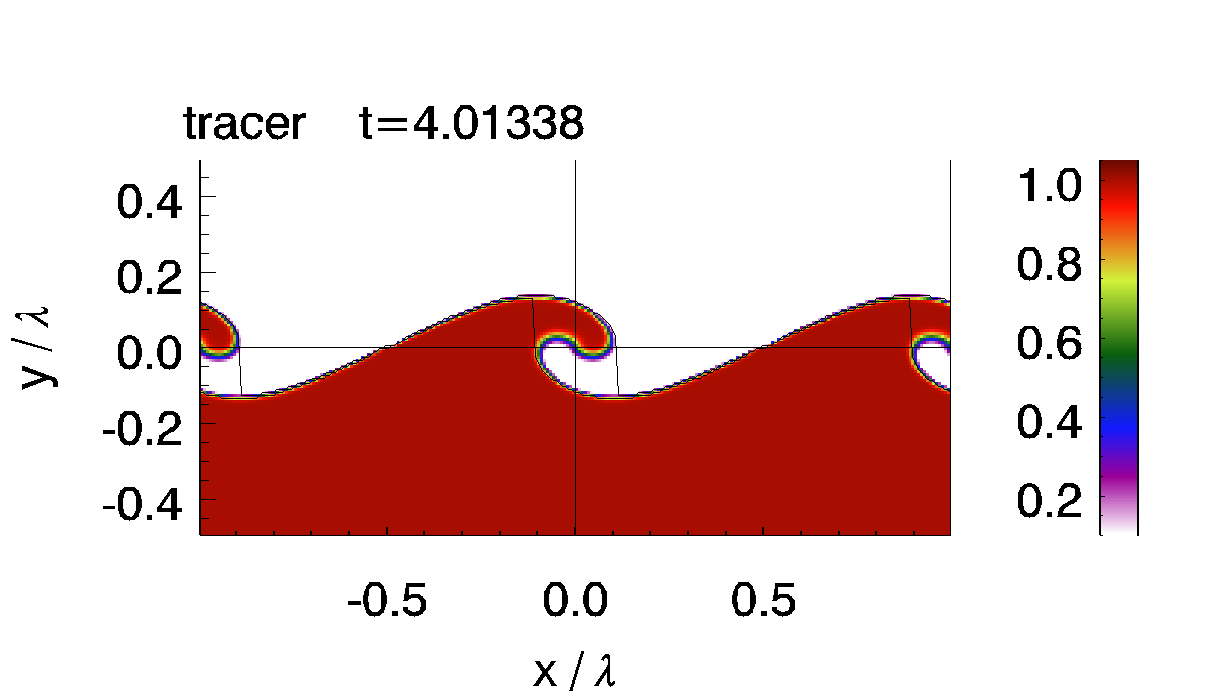}
\includegraphics[trim=190   150 250 160,clip,height=1.83cm]{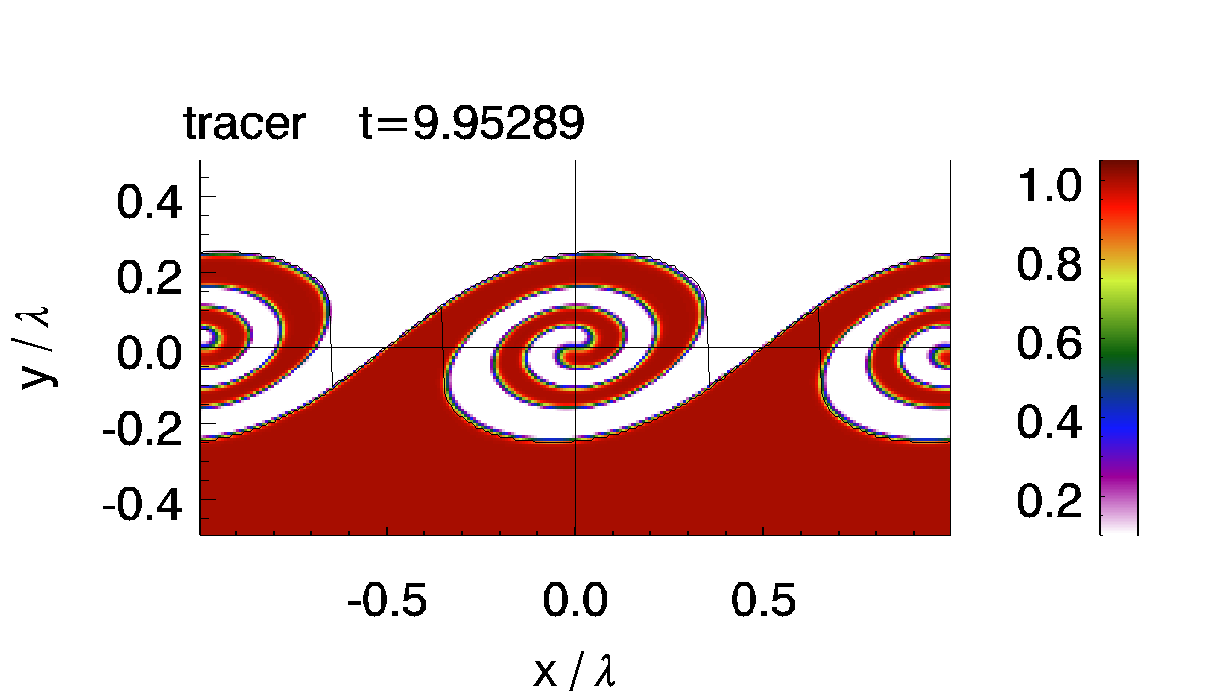}
\includegraphics[trim=190   150     0 160,clip,height=1.83cm]{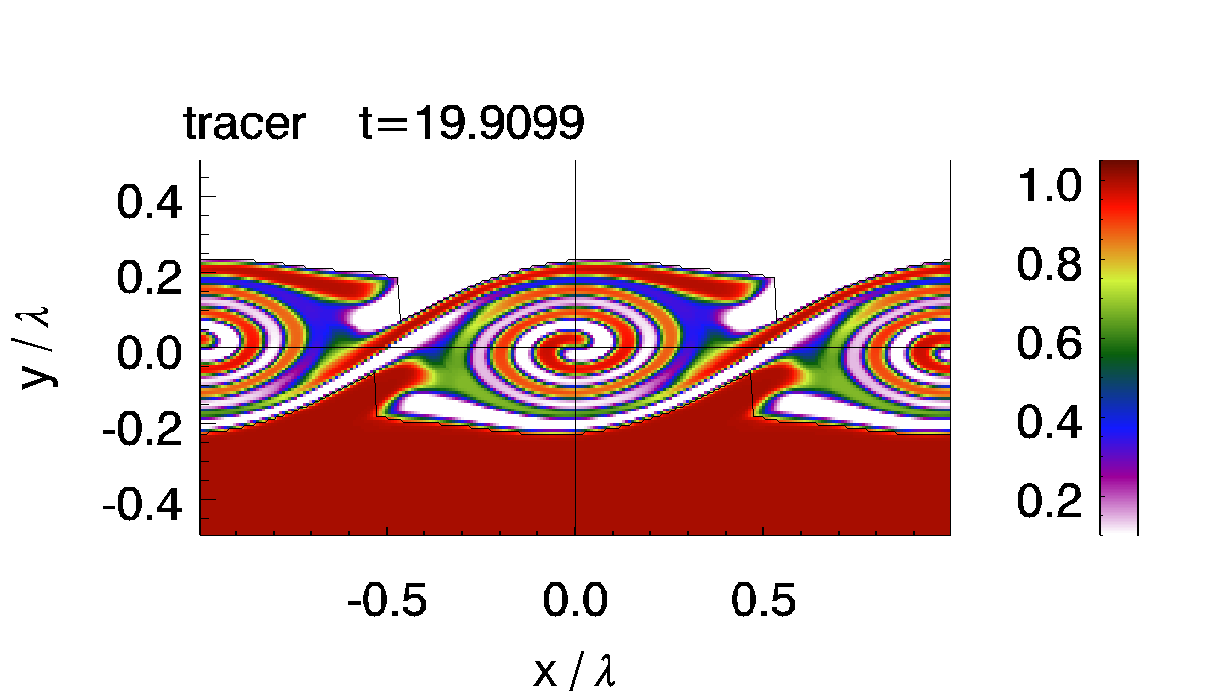}
\newline
\rotatebox{90}{\phantom{xxxxxxx}$v_y/c_s$}
\includegraphics[trim=  0 0 250 160,clip,height=2.55cm]{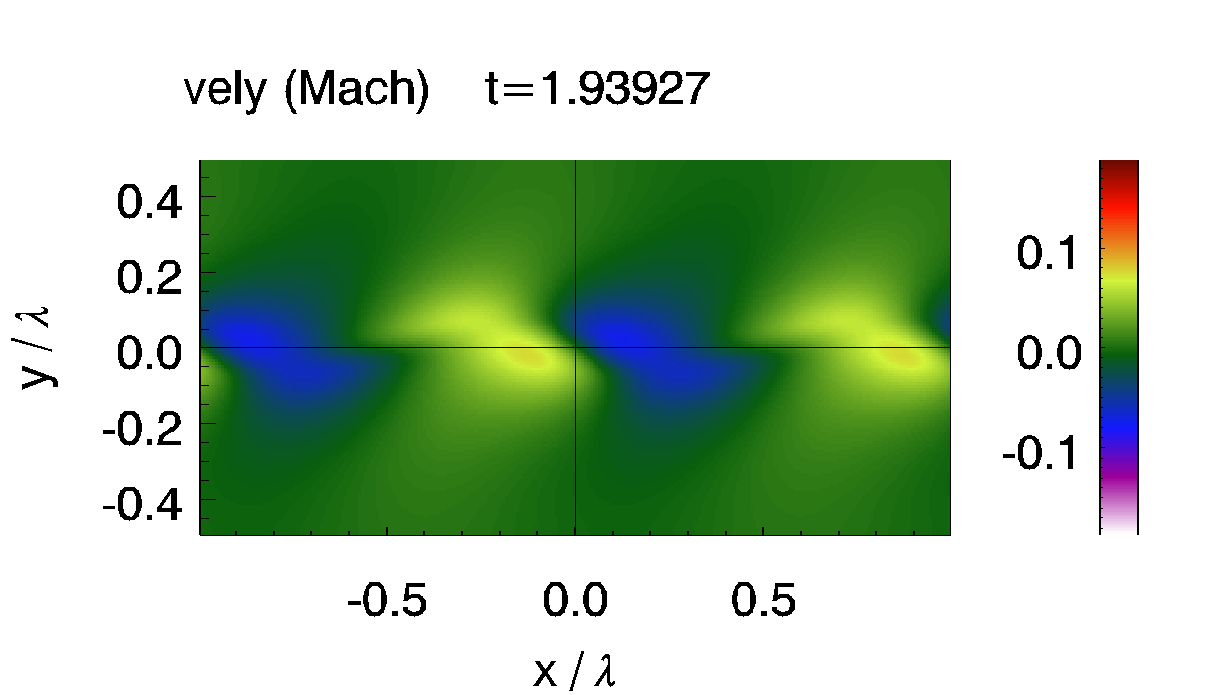}
\includegraphics[trim=190 0 250 160,clip,height=2.55cm]{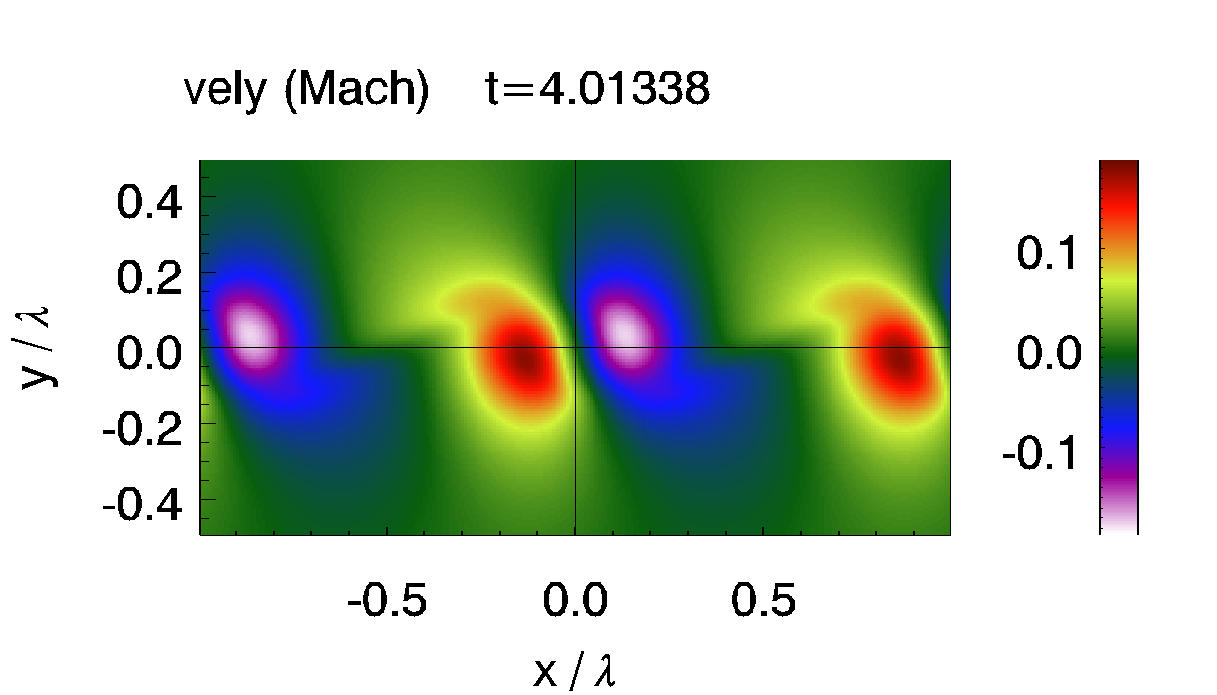}
\includegraphics[trim=190 0 250 160,clip,height=2.55cm]{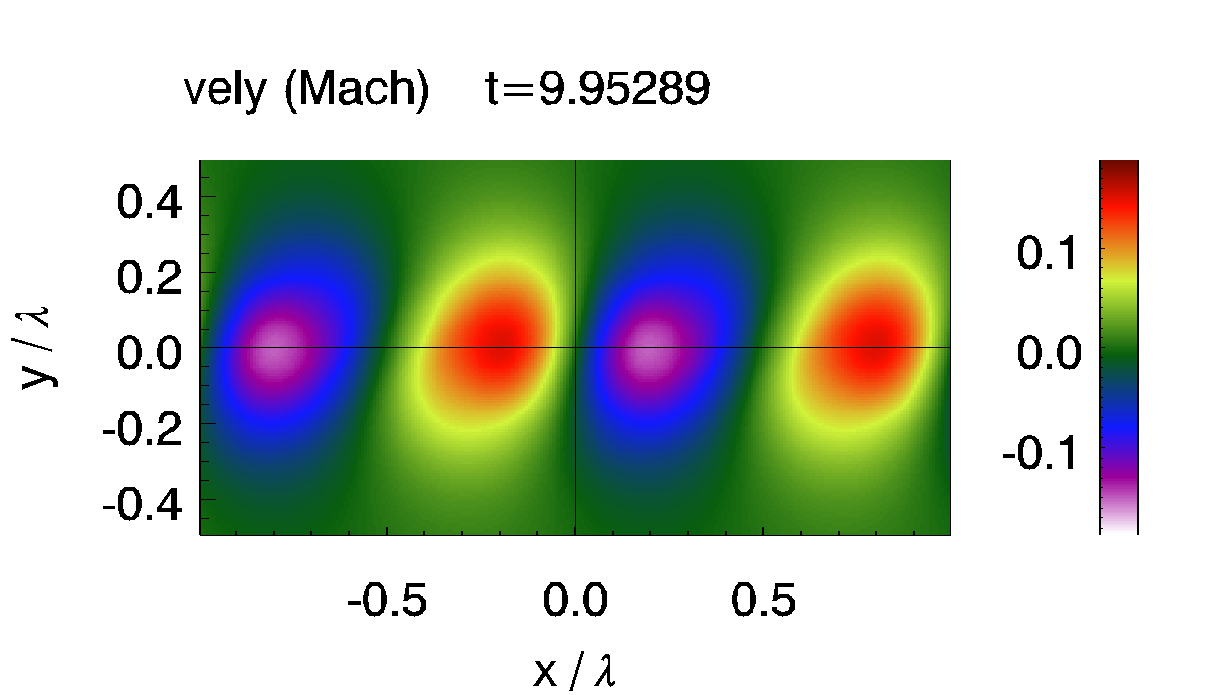}
\includegraphics[trim=190 0     0 160,clip,height=2.55cm]{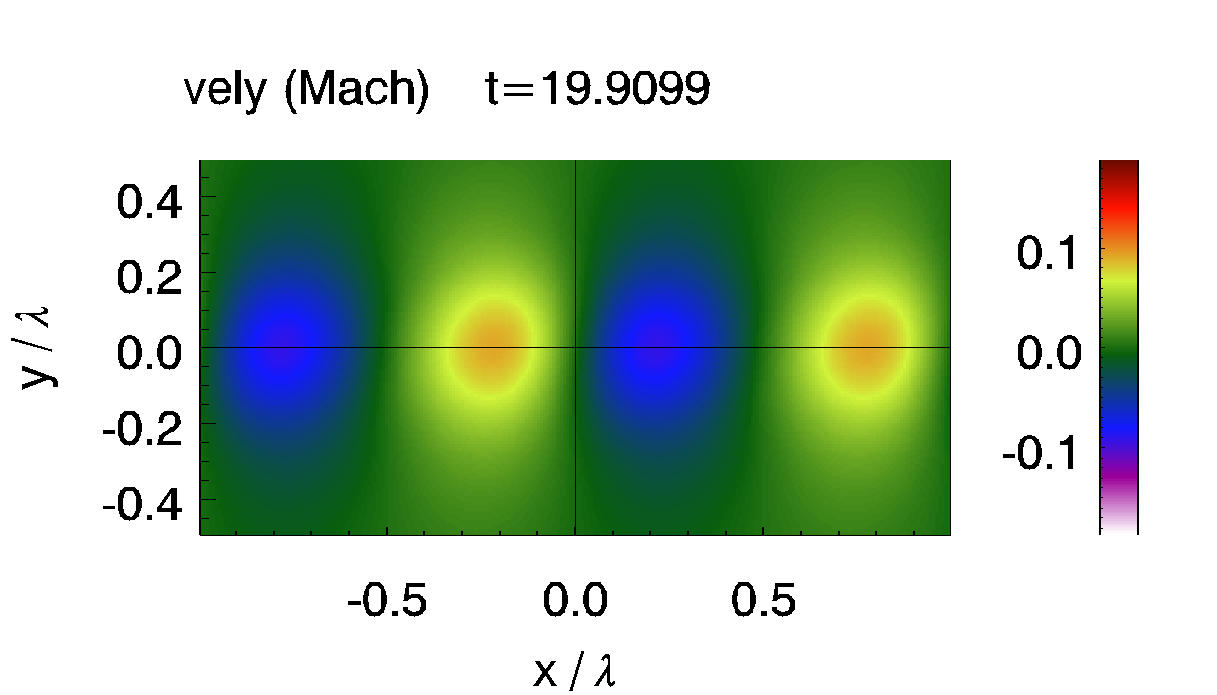}
\caption{The top row shows snapshots of the tracer $F$ for a fiducial Kelvin-Helmholtz test. Initially, the tracer was $F=1$ for $y<0$ and $F=0$ for $y\ge 0$.  Timesteps from left to right are labelled above each column.  
This simulation used a spatially constant kinematic viscosity $\nu$, a density ratio $D_{\rho}=1$ between both layers,  a shear flow of Mach 0.5, and a Reynolds number of $\Reyn=1000$ (see Eqn.~\ref{eq:ourRe}). Snapshots for other $\Reyn$ are shown in Fig.~\ref{fig:rolls_nu_D1_M05}. The thin black lines mark the  ``edges" of the upper and lower fluid, used to measure the height of the KH rolls, see Sect.~\ref{sec:appx_rollheight}. The growth of the KHI rolls is shown in the top panel of Fig.~\ref{fig:thick_vely_nu_M05_D1}. The bottom row colour-codes the vertical velocity $v_y$ in units of the sound speed $c_s$ at corresponding times, highlighting the vortices due to the KH rolls. The temporal evolution of the maximum $v_y$ is shown in the bottom panel of Fig.~\ref{fig:thick_vely_nu_M05_D1}. {Note that the simulation box extends in $y$-direction from $-2\lambda$ to $2\lambda$, i.e., beyond the size of the snapshots.}
}
\label{fig:rolls_nu_Re1000_D1_M05}
\end{figure*}

\subsection{Resolution}
We use a uniform grid and a standard resolution of 128 cells per perturbation length. In Appendix~\ref{sec:resolution} we demonstrate convergence of our results. Furthermore, in Appendix~\ref{sec:resolution} we show that the FLASH code captures the presence and size of KH rolls down to a resolution 16 grid cells per perturbation length. The internal structure of the KH rolls, e.g., regarding mixing and velocity structure, requires a higher resolution that ensures that the width, or thickness, of the KHI mixing layer is resolved beyond the numerical diffusion length of 2 to 3 grid cells.

\section{Simulation results} \label{sec:results}

\subsection{Fiducial KHI -- density ratio 1, constant kinematic viscosity, Reynolds number 1000}
\label{sec:fiducial}

Figure~\ref{fig:rolls_nu_Re1000_D1_M05} displays the typical evolution of the KHI for the most simple case with equal densities in both layers. A low viscosity is applied to achieve $\Reyn=1000$. At this Reynolds number, the KHI at the intended perturbation length is slightly slowed down by viscosity but unintended secondary modes are absent, making this case instructive.    Due to the uniform density throughout the simulation box we use a tracer to visualise the two fluid layers. Initially,  the tracer was set to $F=1$ for $y<0$ and $F=0$ for $y\ge 0$.  The evolution of the tracer fluid is displayed in the top row of Fig.~\ref{fig:rolls_nu_Re1000_D1_M05}. The bottom row displays the distribution of $v_y$, i.e., the velocity component perpendicular to the initial interface.  

 The instability evolves in two phases,  a growth phase followed by a saturation phase. During the first $\sim 3$ to $4\tau\KHinvisc$ the initial perturbation leads to a wave-like distortion of the interface. Simultaneously,  $v_y{}\Max$, the maximum of the vertical velocity, increases to up to 4 times the initial perturbation amplitude. At the end of this growth phase the interface starts to roll up, leading to the classic KH roll or cat-eye pattern over the next few $\tau\KHinvisc$. In the following saturation phase, these rolls continue to spin over many $\tau\KHinvisc$. The low viscosity  slowly dissipates the vortices, decreasing $v_y{}\Max$. The height of the rolls, or the thickness of the mixing layer, increases until about $8\tau\KHinvisc$ and then saturates as well. Given that we introduced only a single perturbation mode with a length scale of half the box width, the simulation box contains only 2 identical KHI rolls. In a more realistic perturbation spectrum, small perturbations start growing first, and with time larger and larger perturbations dominate the appearance. 
 

\subsection{Constant kinematic viscosity}

Next we investigate the effect of varying the viscosity. In this section we apply a spatially constant kinematic viscosity, which leads to equal Reynolds numbers in both layers regardless of density contrast.

\subsubsection{Density ratio 1}  \label{sec:Drho1}
%
\begin{figure*}
\begin{center}
\hspace{0.5cm} $\Reyn=100$ \hfill $\Reyn=200$ \hfill $\Reyn=300$ \hfill $\Reyn=1000$ \hfill $\Reyn=10^4$ \hfill\phantom{x}\newline
\rotatebox{90}{\phantom{xx}$4\tau\KHinvisc$}\hfill%
\includegraphics[trim=  0 150 250 160,clip,height=1.5cm]{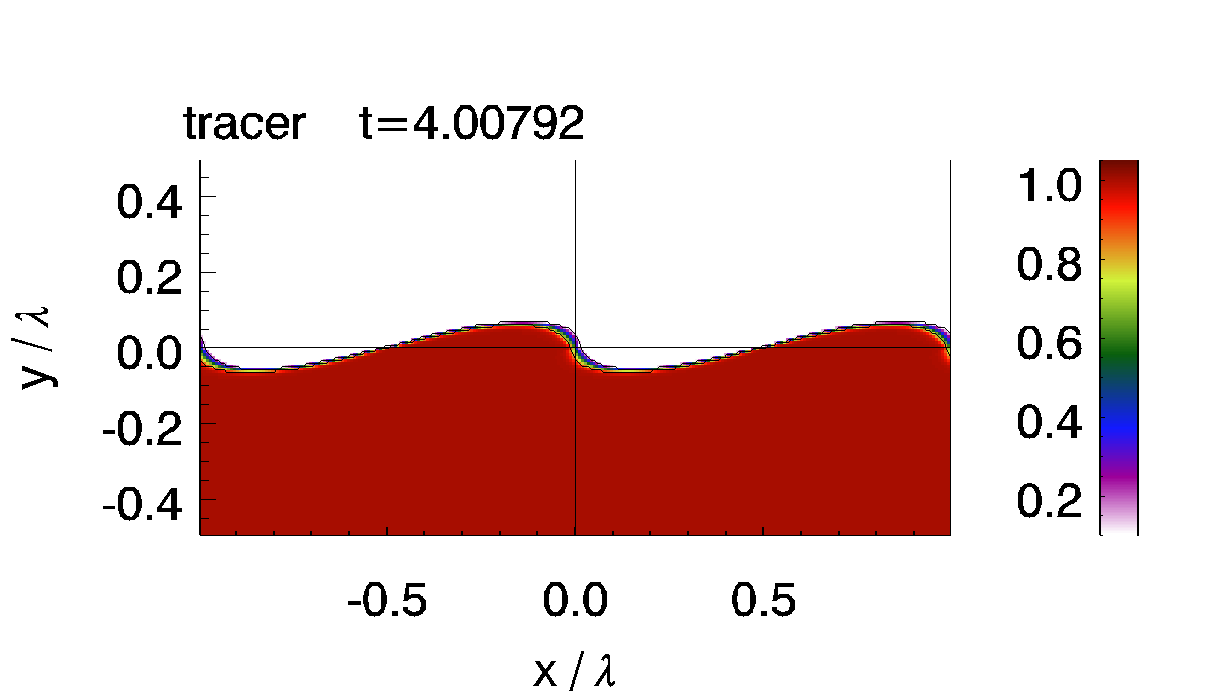}
\includegraphics[trim=190 150 250 160,clip,height=1.5cm]{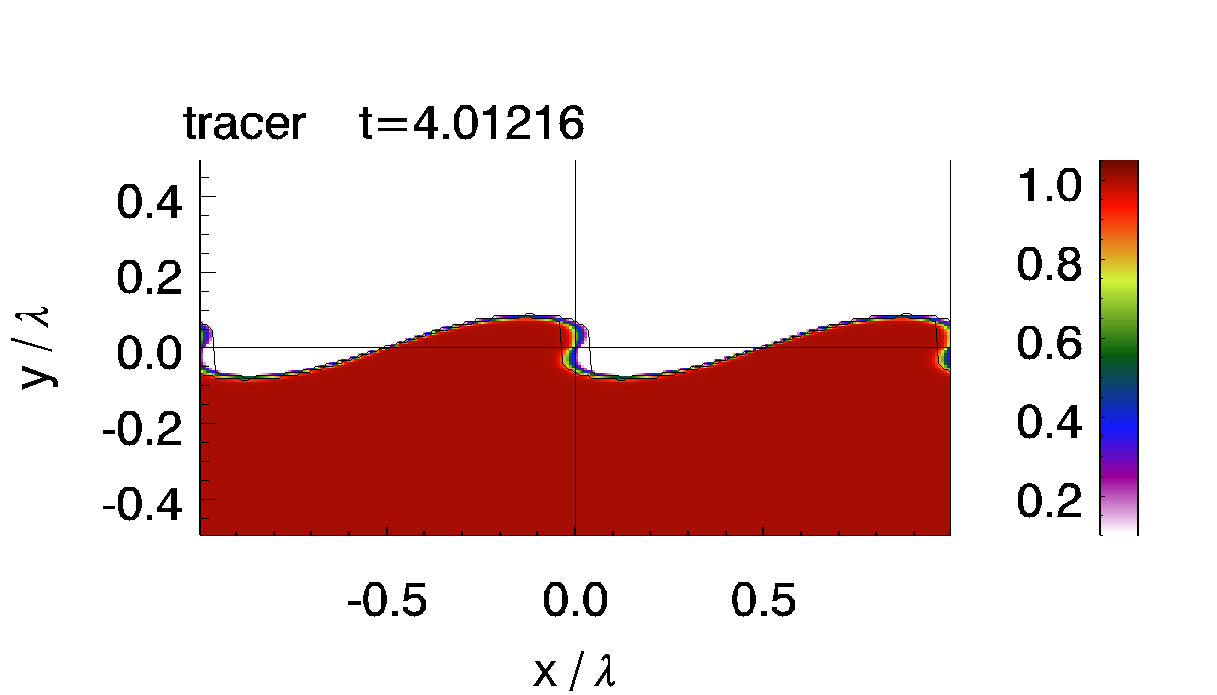}
\includegraphics[trim=190 150 250 160,clip,height=1.5cm]{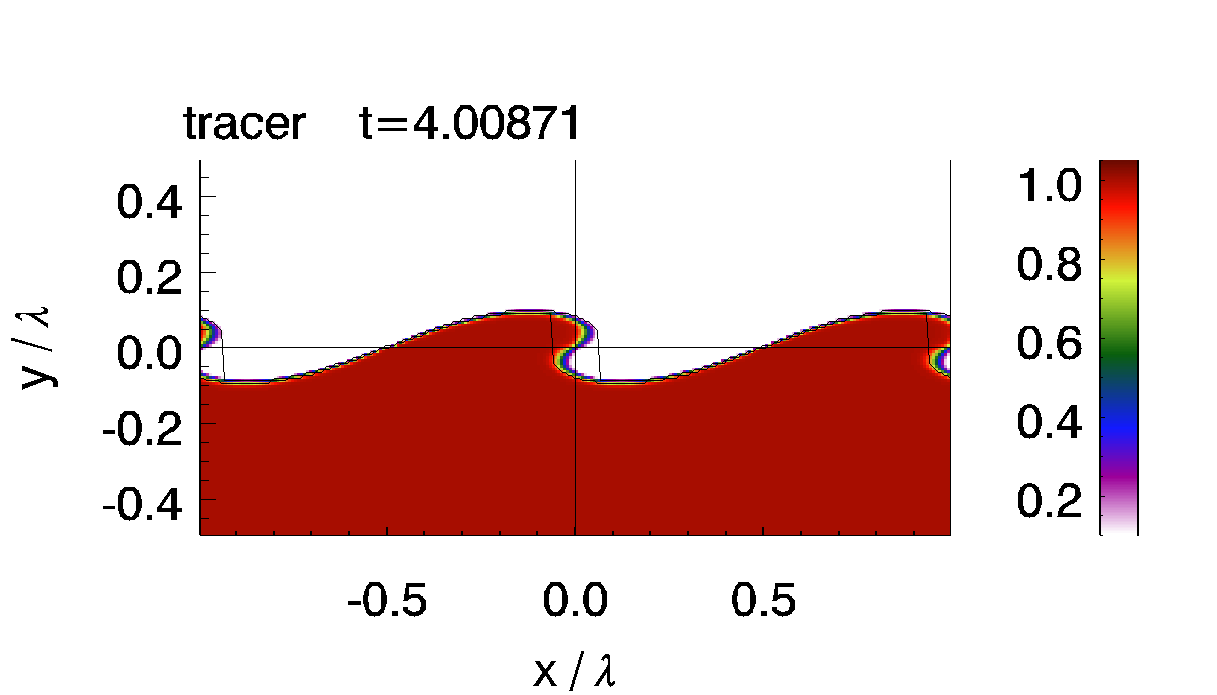}
\includegraphics[trim=190 150 250 160,clip,height=1.5cm]{Figs/Nu_D1/metf_Re1000_0031}
\includegraphics[trim=190 150    0 160,clip,height=1.5cm]{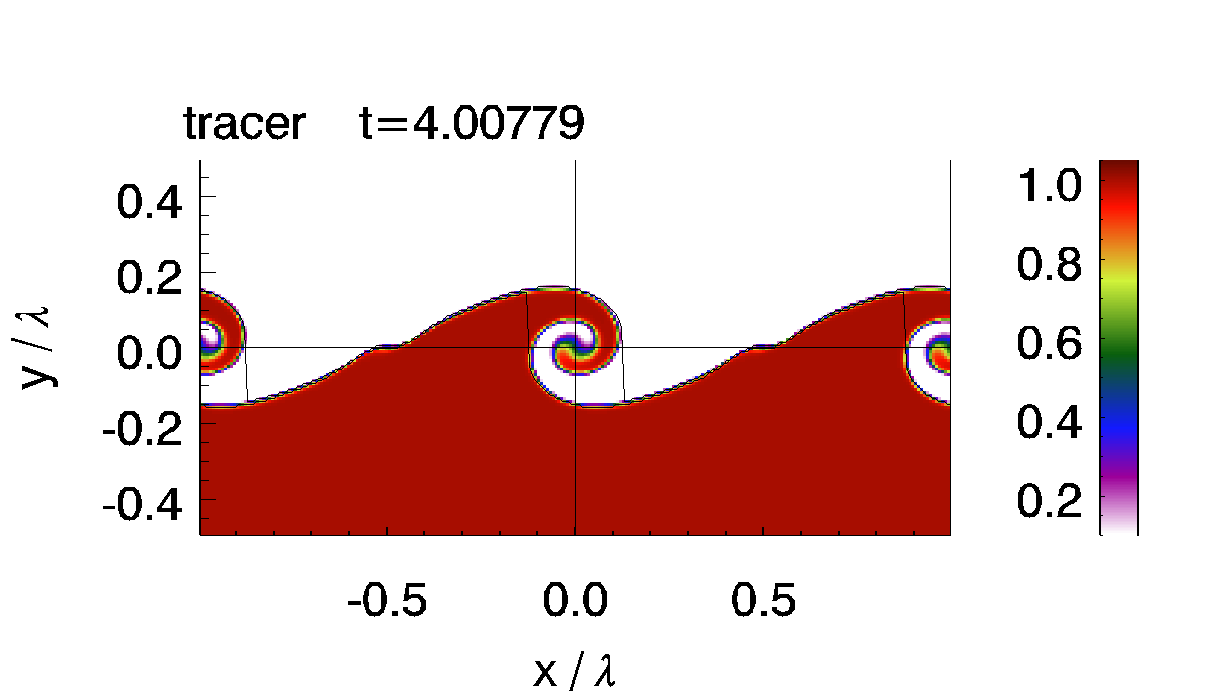}
\newline
\rotatebox{90}{\phantom{xx}$10\tau\KHinvisc$}\hfill%
\includegraphics[trim=  0 150 250 160,clip,height=1.5cm]{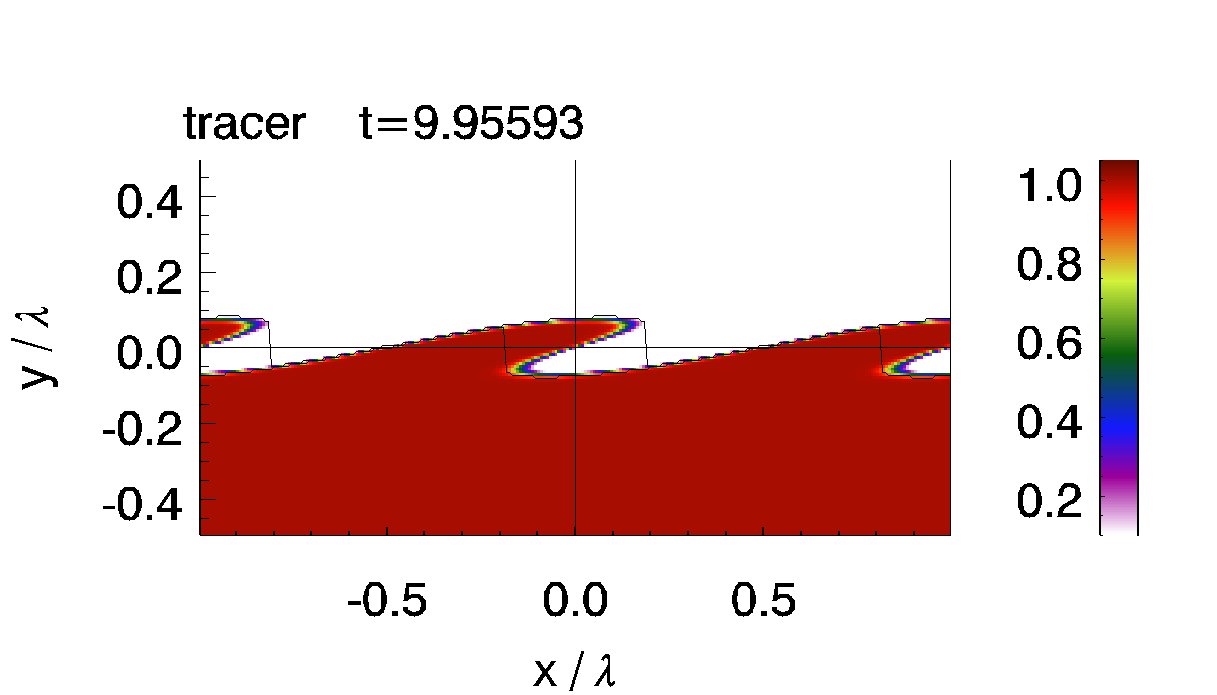}
\includegraphics[trim=190 150 250 160,clip,height=1.5cm]{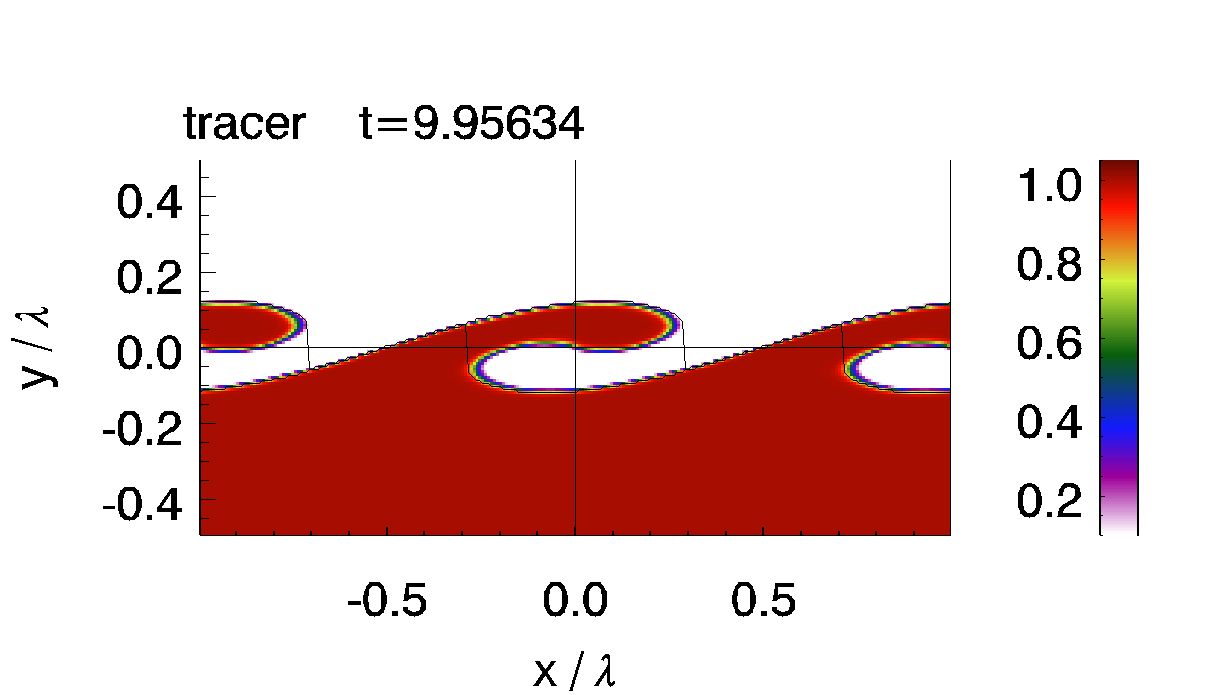}
\includegraphics[trim=190 150 250 160,clip,height=1.5cm]{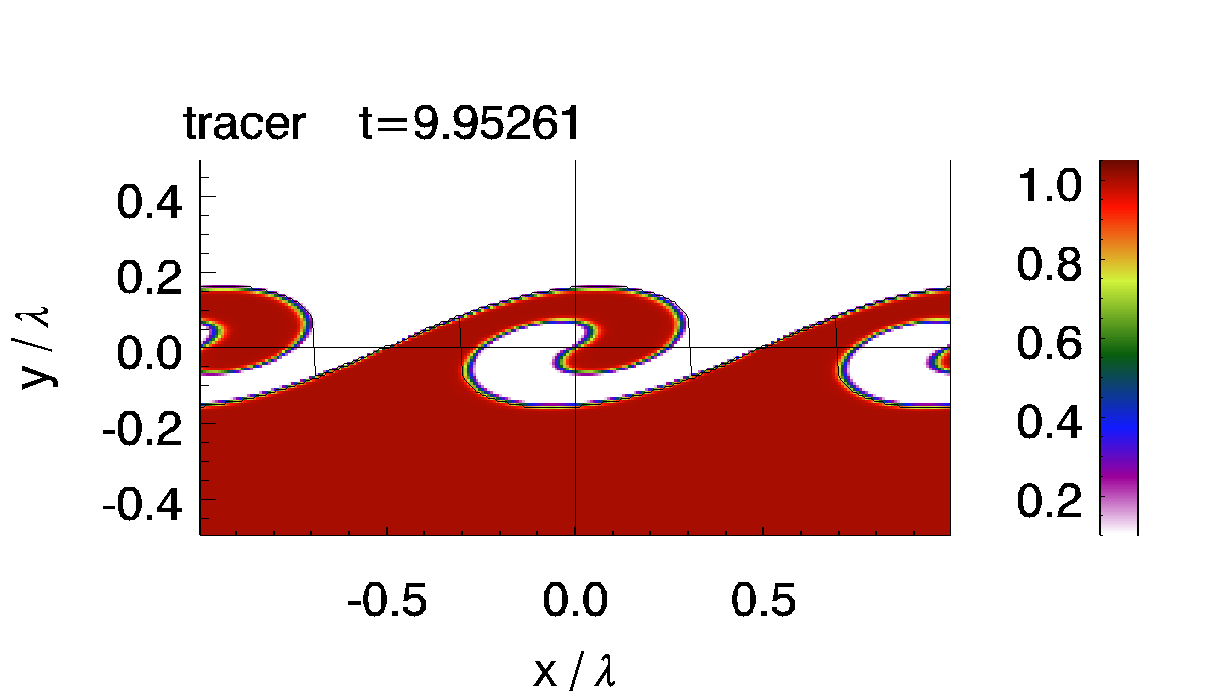}
\includegraphics[trim=190 150 250 160,clip,height=1.5cm]{Figs/Nu_D1/metf_Re1000_0077}
\includegraphics[trim=190 150    0 160,clip,height=1.5cm]{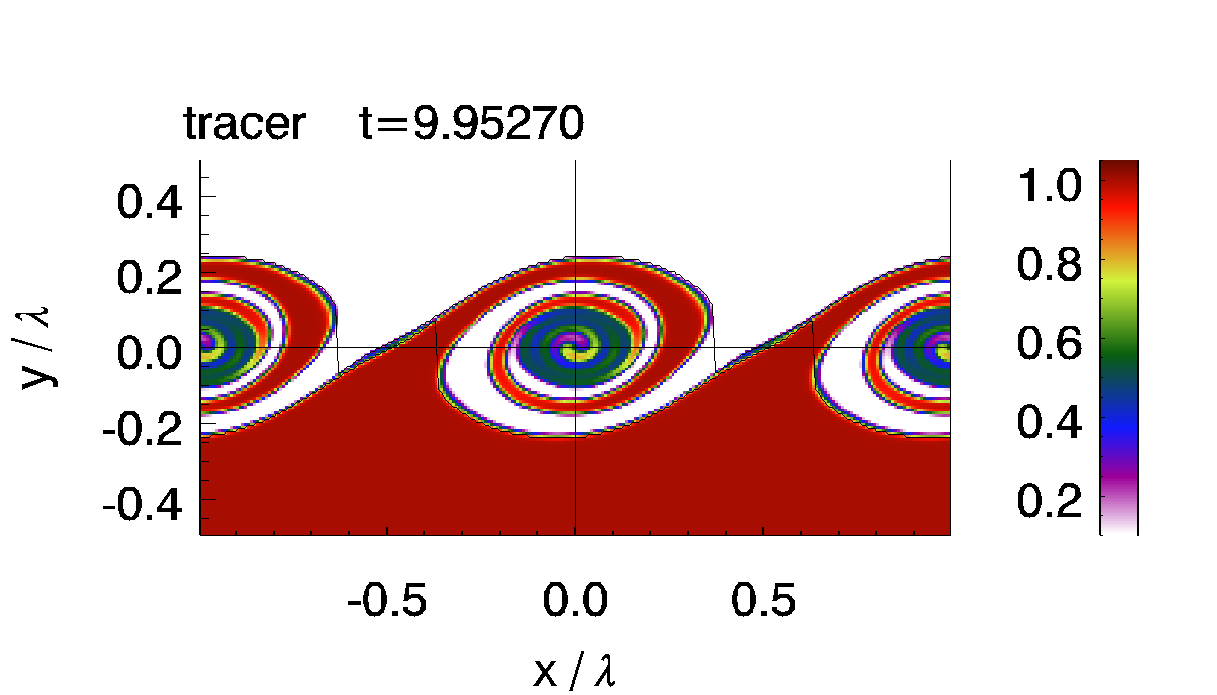}
\newline
\rotatebox{90}{\phantom{xxx}$20\tau\KHinvisc$}\hfill%
\includegraphics[trim=  0 0 250 160,clip,height=2.09cm]{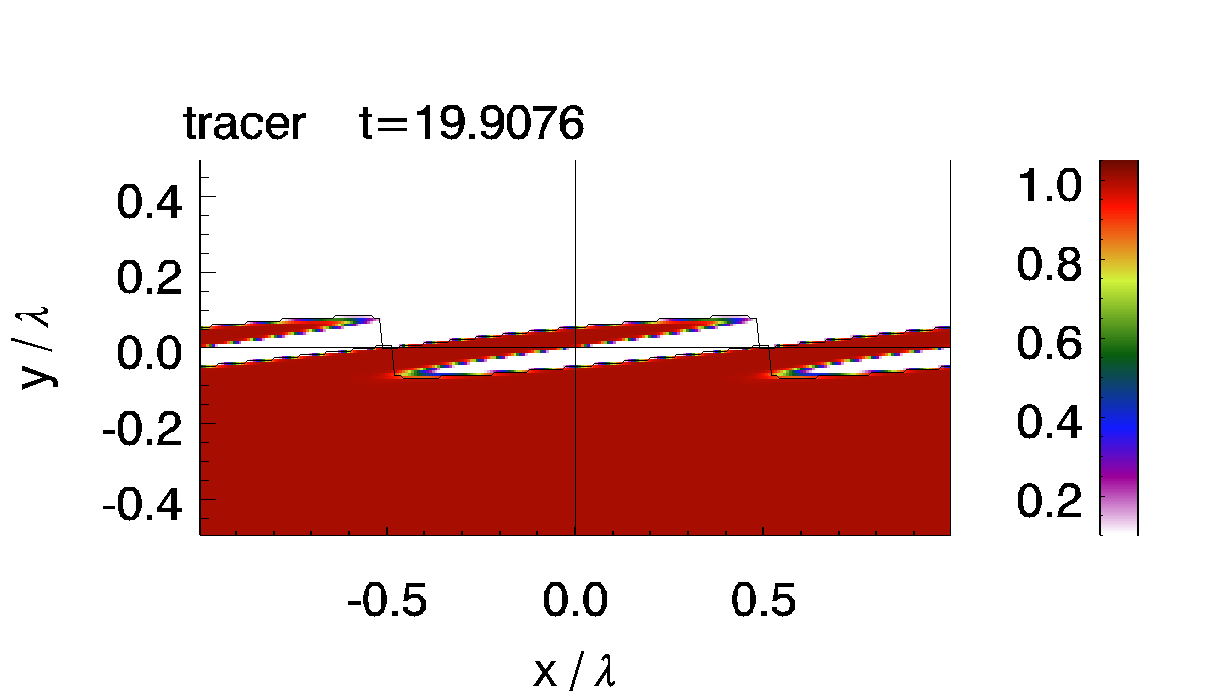}
\includegraphics[trim=190 0 250 160,clip,height=2.09cm]{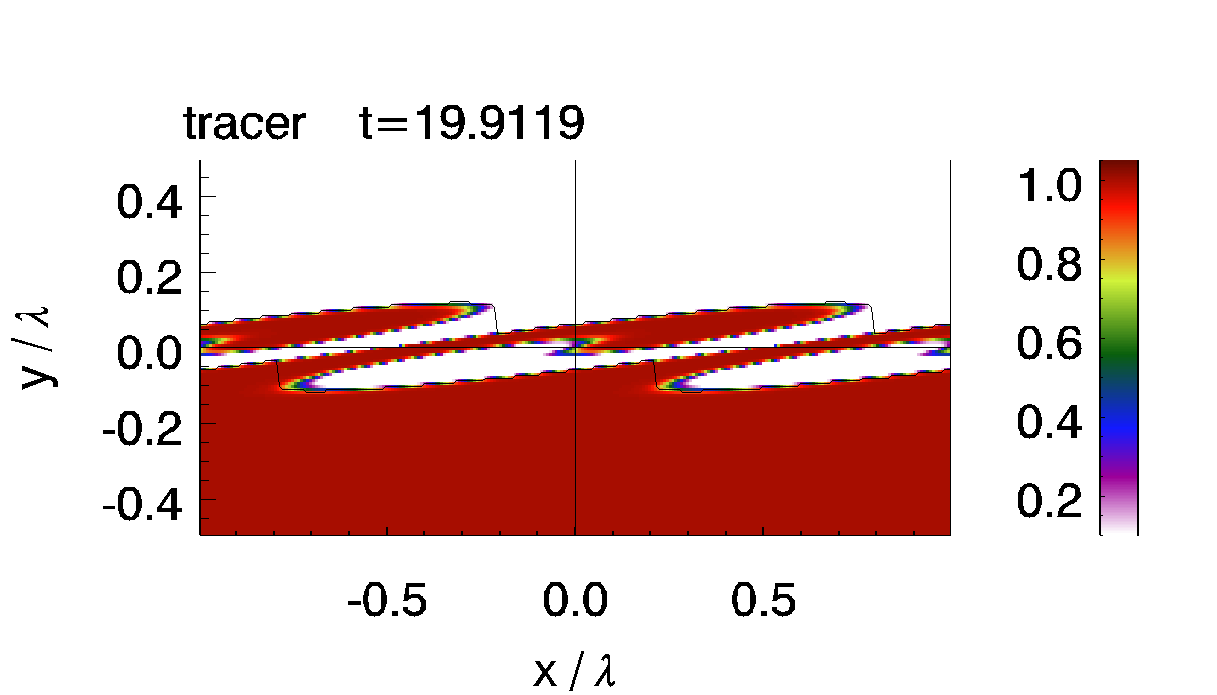}
\includegraphics[trim=190 0 250 160,clip,height=2.09cm]{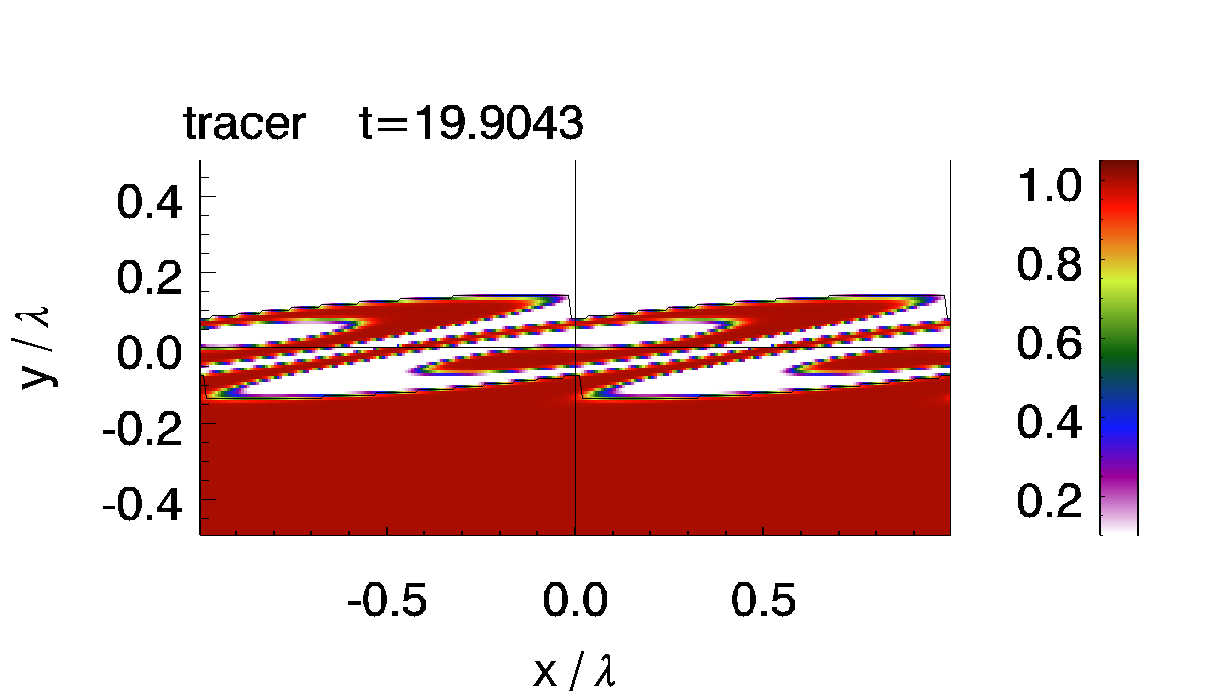}
\includegraphics[trim=190 0 250 160,clip,height=2.09cm]{Figs/Nu_D1/metf_Re1000_0154}
\includegraphics[trim=190 0    0 160,clip,height=2.09cm]{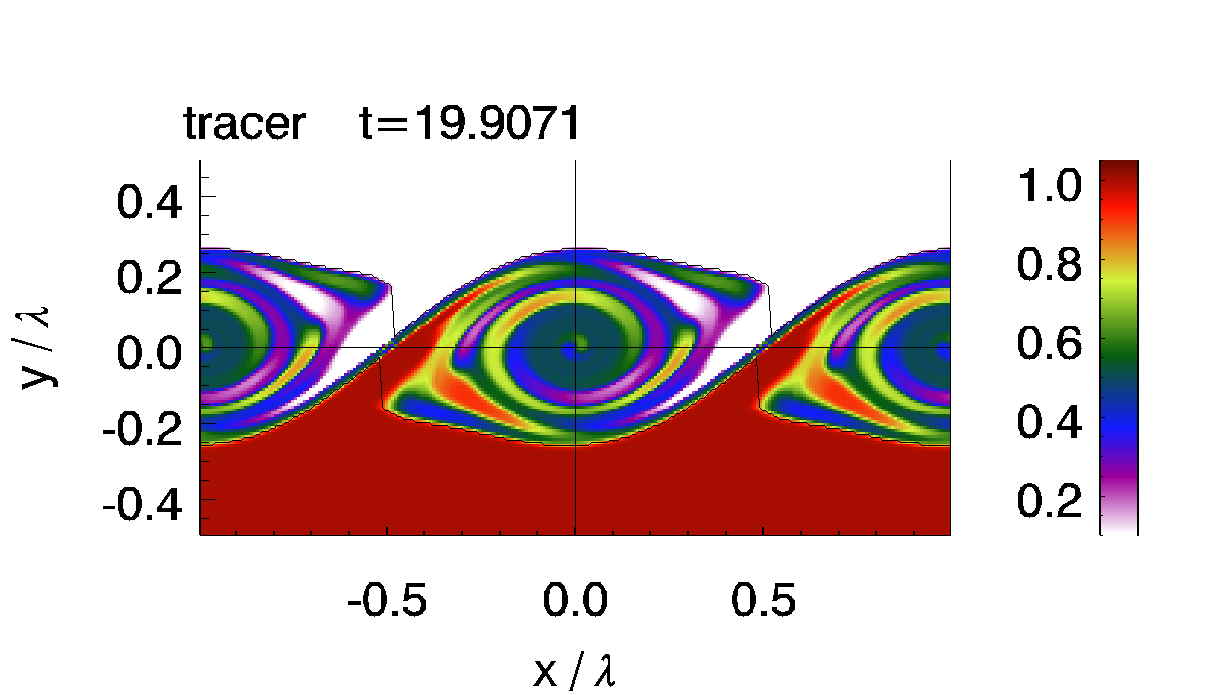}
\caption{Snapshots showing the tracer $F$ for various Reynolds numbers (labels at top of each column) and 3 timesteps (labels on the left). These simulations used a constant kinematic viscosity $\nu$, a density contrast $D_{\rho}=1$,  and a shear flow of Mach  0.5. The KHI is significantly slowed down below $\Reyn \le 300$. For $\Reyn \le 200$, the  deformation of the interface is solely due to the initial perturbation, i.e., the KHI itself is suppressed.  Figure~\ref{fig:thick_vely_nu_M05_D1} compares the height of the KH rolls and the evolution of the maximal vertical velocity. For $\Reyn=10^4$ we smoothed the initial interface over 1\% of the perturbation length scale to suppress secondary instabilities (see Eqn.~\ref{eq:smooth}).}
\label{fig:rolls_nu_D1_M05}
\end{center}
\end{figure*}

\begin{figure}
\centering\includegraphics[trim=0 0 0 0,clip,width=0.42\textwidth]{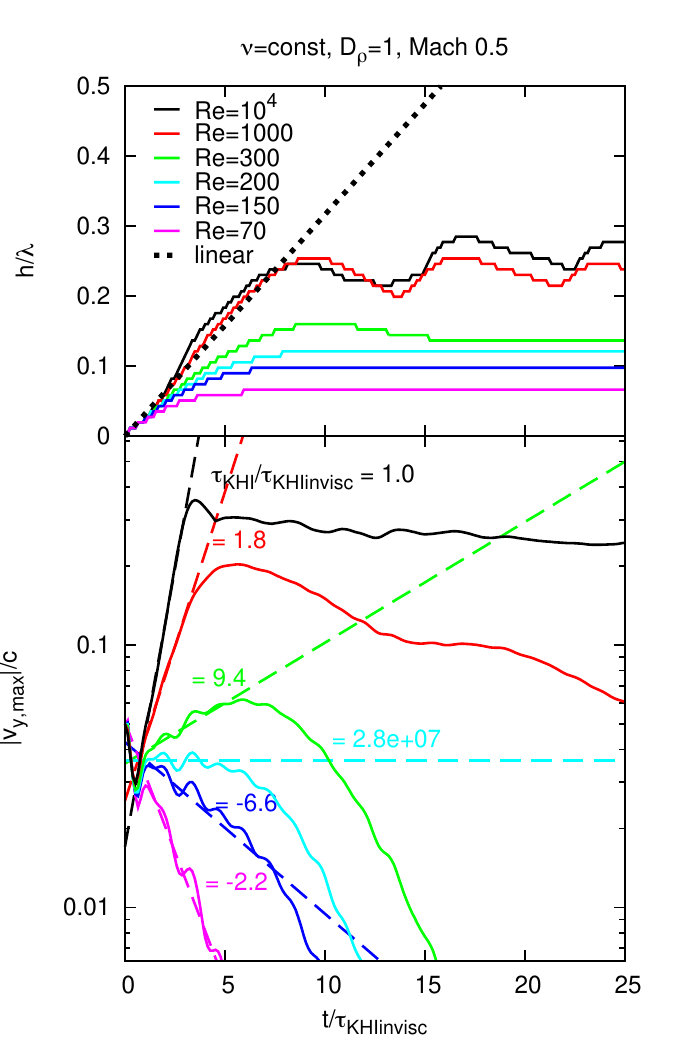}
\caption{Temporal evolution of the height of the KH rolls (top) and of the maximum velocity in $y$ direction, $|v_y|{}\Max$ (bottom). Different line colours code different Reynolds numbers, see legend. Figs.~\ref{fig:rolls_nu_Re1000_D1_M05} and \ref{fig:rolls_nu_D1_M05} display corresponding snapshots. All runs are for a constant kinematic viscosity, $D_{\rho}=1$, and shear flow = Mach  0.5.  In the top panel, the expected linear increase of the interface distortion as it would occur purely due to the initial perturbation is shown as the black dotted line labelled "linear". The smaller height of the KH rolls for $\Reyn \le 300$ is due to viscosity. The maximum $v_y$ shows an initial exponential growth for $\Reyn \gtrsim 200$. After $\sim 5\tau\KHinvisc$, $v_y{}\Max$ decreases approximately exponentially due to viscous dissipation, i.e.,  faster for lower Reynolds numbers. We fit  the initial increase  with  exponential functions that are shown as thin dashed lines of matching colour. The resulting growth times are stated in the plot in units of $\tau\KHinvisc$ in corresponding colours. The initial exponential growth time is the viscous KHI growth time (see also Sect.~\ref{sec:Drho1}, Fig.~\ref{fig:tau_of_Re}, and Appendix~\ref{sec:appx_analysis}). Note that we ignore the initial fluctuations of $v_y{}\Max$ during the first $\tau\KHinvisc$ because they reflect sound waves from the initialisation and not the KHI growth.
}
\label{fig:thick_vely_nu_M05_D1}
\end{figure}

We start with the equal density case. Fig.~\ref{fig:rolls_nu_D1_M05} displays the evolution of the instability for Reynolds numbers between 100 and $10^4$. Figure~\ref{fig:thick_vely_nu_M05_D1} shows the evolution of the width of the KHI-induced mixing layer (top), and of $v_y{}\Max$ (bottom).  Section~\ref{sec:appx_analysis} explains how both quantities are measured, and how the viscous KHI growth time is measured from $v_y{}\Max(t)$.  

As indicated above, the evolution of the viscous KHI is governed by the competition of the actual instability and viscosity. The former increases the height of the KH rolls and leads to an exponential growth of velocities in $y$-direction. Indeed, for $\Reyn=10^4$, the exponential growth time of $v_y{}\Max$ recovers the analytic estimate for the inviscid KHI growth time $\tau\KHinvisc$. Viscosity, on the other hand, dissipates velocity gradients. A low viscosity cannot slow down the widening of the mixing interface, but the rolling up of the KH rolls. Already at $\Reyn=1000$ the rolls curl up slower than at $\Reyn=10^4$ (Fig.~\ref{fig:rolls_nu_D1_M05}). At $\Reyn=300$ only rudimentary rolls are formed, and at lower $\Reyn$ the interface does not roll up at all. For $\Reyn \lesssim 300$ also the width of the mixing layer, or height of the KH rolls, is reduced. This visual impression from Fig.~\ref{fig:rolls_nu_D1_M05} is confirmed in the upper panel of Fig.~\ref{fig:thick_vely_nu_M05_D1}. The dissipation of velocity shear becomes clear in the bottom panel of Fig.~\ref{fig:thick_vely_nu_M05_D1}. Instead of a saturation of $v_y{}\Max$ at late times like in the $\Reyn=10^4$ case, higher viscosities leads to an approximately exponential decrease. The decay time is shorter for lower Reynolds numbers, which translates into the reduced curling-up of the KH rolls. Furthermore, the peak value of $v_y{}\Max$ is reduced with increasing viscosity, and  the initial exponential growth of $v_y{}\Max$ is slowed down with decreasing $\Reyn$. Below $\Reyn=200$ there is no initial growth anymore,  the instability is suppressed. This agrees with the drastic reduction of the mixing layer width with decreasing $\Reyn$.

\begin{figure}
\centering\includegraphics[trim=0 0 0 0,clip,width=0.35\textwidth]{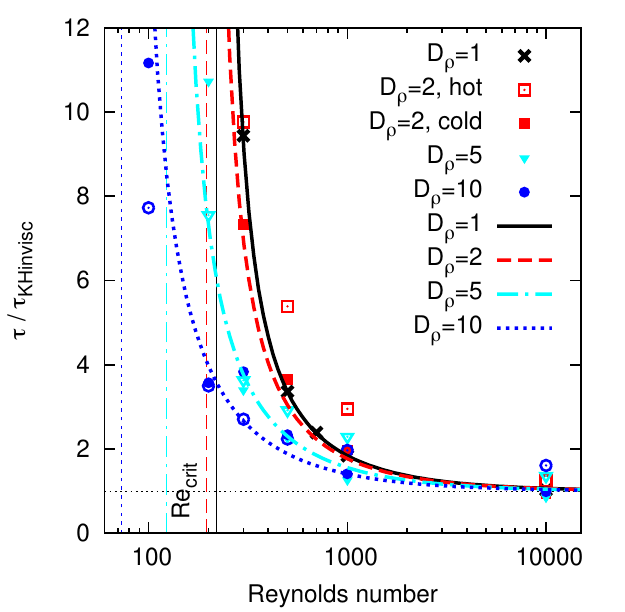}
\caption{Viscous KHI growth time as a function of Reynolds number and density contrast, for the case of a spatially constant kinematic viscosity. Colours code the density contrast, see legend. The symbols show the growth times derived from the maximum and minimum $v_y$. For density contrasts $>1$, open symbols show the growth time derived from the maximum $v_y$, i.e., velocities towards the hot layer, solid symbols show the minimum $v_y$, i.e., derived from the fastest velocity to towards the cold layer. For equal densities, the KHI evolves symmetrically, and only one symbol is shown.  Lines show the empirical relation given Eqn.~\ref{eq:tau_Re_fit} combined with Eqns.~\ref{eq:ReCritFit} and \ref{eq:Re0Fit}. We mark the critical Reynolds number $\Reyn\Crit$, below which the KHI does not grow at all, with vertical lines of matching colour.}
\label{fig:tau_of_Re}
\end{figure}

We summarise the influence of viscosity in Fig.~\ref{fig:tau_of_Re} by plotting the measured viscous KH growth times as a function of Reynolds number. The black crosses are for the equal density case. The viscous growth time increases strongly from $\Reyn=1000$ down to $\Reyn=300$. Below $\Reyn\sim 200$ growth times are negative, i.e, the instability is suppressed. Thus, the numerically derived critical Reynolds number is even about 50\% higher than the analytic estimate in Eqn.~\ref{eq:viscKHI}. 

The dependence of the numerically derived growth times on Reynolds number can be approximated by the empirical relation
\begin{equation}
\tau\KHvisc = \tau\KHinvisc \left( 1 + \frac{\Reyn_0}{\Reyn - \Reyn\Crit}  \right)\;\;\textrm{for}\;\Reyn>\Reyn\Crit,    \label{eq:tau_Re_fit}
\end{equation}
where $\Reyn\Crit$ is the critical Reynolds number below which the KHI is suppressed, and $\Reyn_0$ scales the decrease of $\tau\KHvisc$ with $\Reyn$. For the case of equal densities the numeric values of $\Reyn\Crit=220$ and $\Reyn_0=660$ fit the simulation results as shown by the black line in Fig.~\ref{fig:tau_of_Re}.  The dependence of  $\Reyn\Crit$ and $\Reyn_0$ on density contrast is given in Eqns.~\ref{eq:ReCritFit} and \ref{eq:Re0Fit} below.

\clearpage

\subsubsection{Density contrasts up to 10}

\begin{figure*}
\begin{center}
\hspace{0.5cm} $\Reyn=30$ \hfill $\Reyn=100$ \hfill $\Reyn=300$ \hfill $\Reyn=1000$ \hfill $\Reyn=10^4$ \hfill\phantom{x}\newline
\rotatebox{90}{\phantom{xx}$4\tau\KHinvisc$}\hfill%
\includegraphics[trim=0     150 250 160,clip,height=1.5cm]{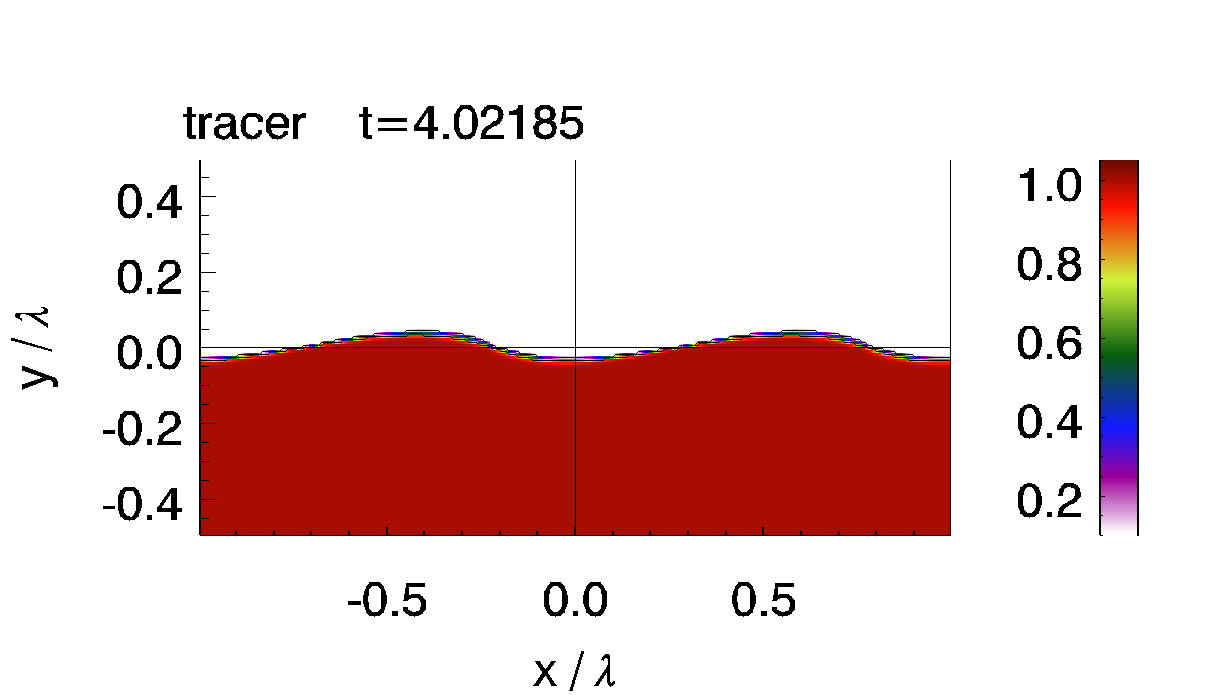}
\includegraphics[trim=190 150 250 160,clip,height=1.5cm]{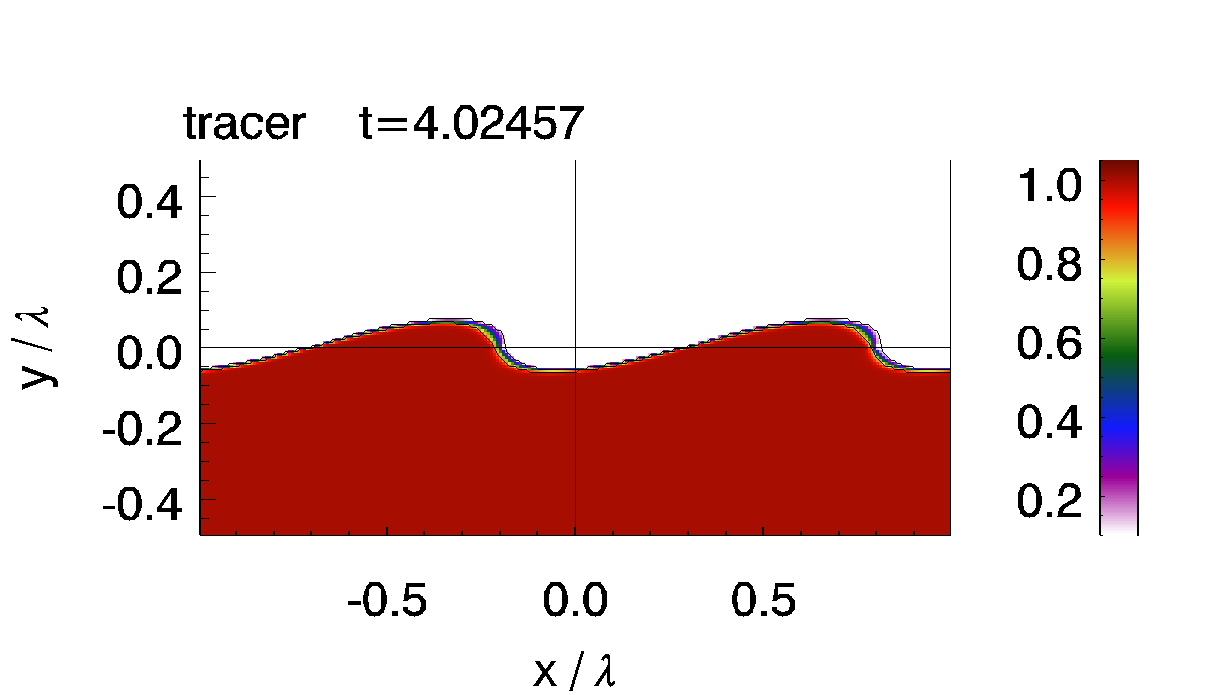}
\includegraphics[trim=190 150 250 160,clip,height=1.5cm]{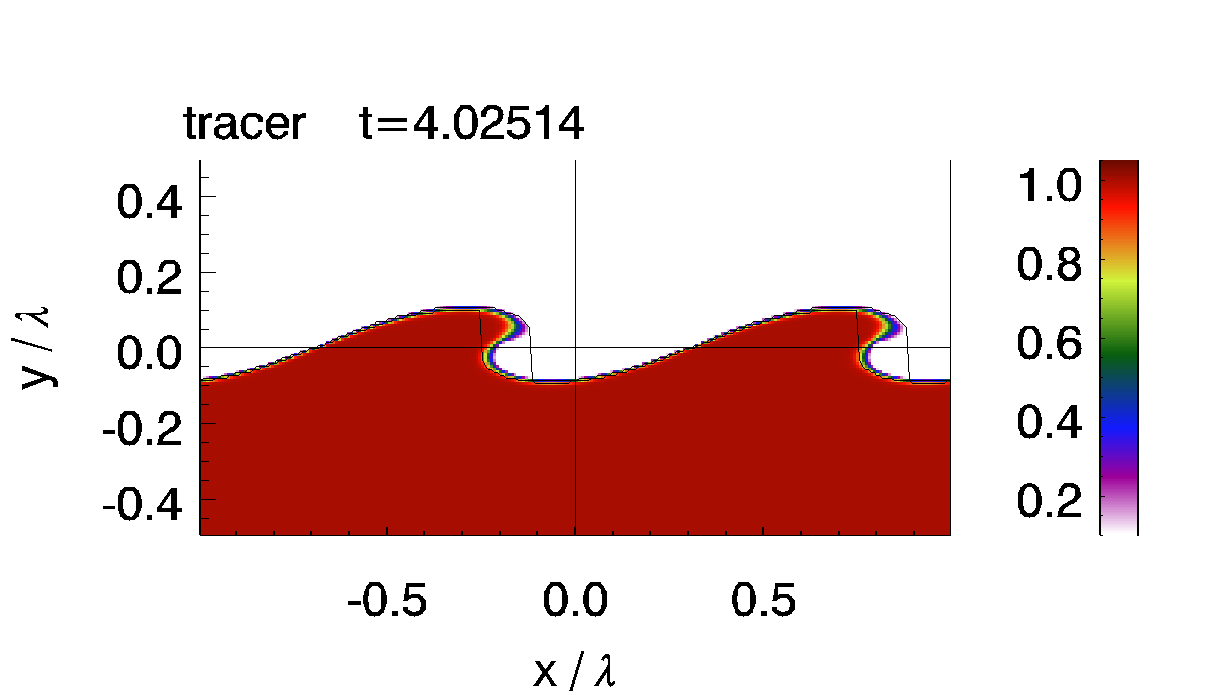}
\includegraphics[trim=190 150 250 160,clip,height=1.5cm]{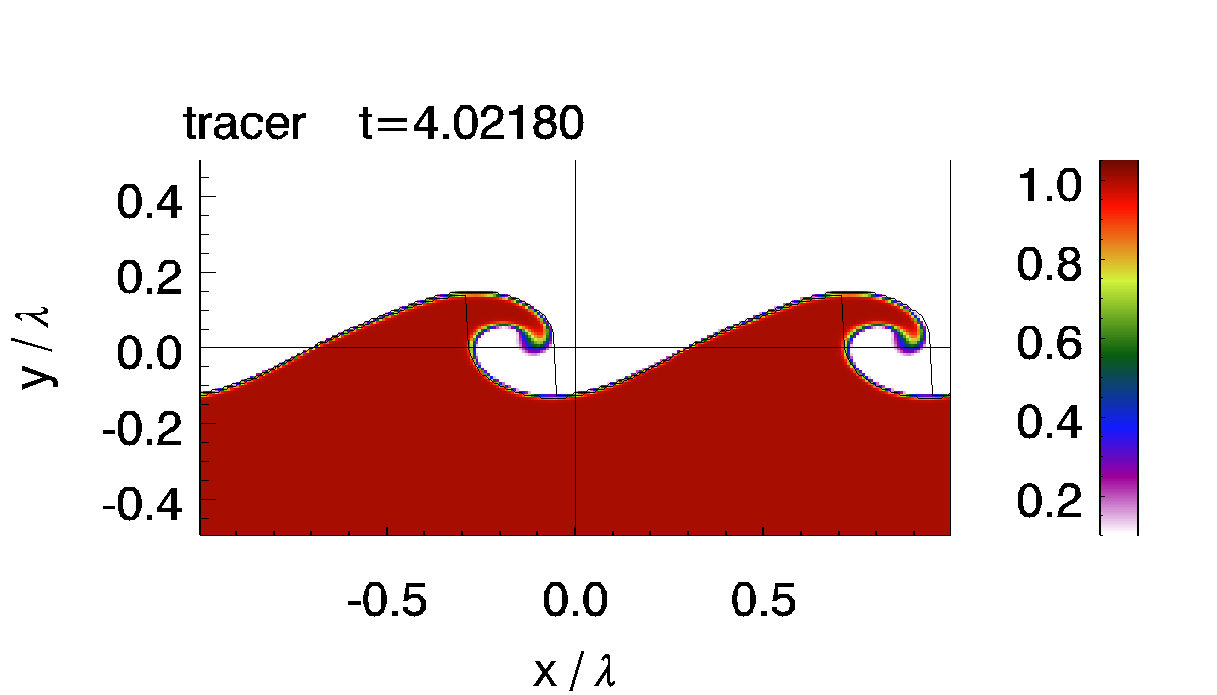}
\includegraphics[trim=190 150     0 160,clip,height=1.5cm]{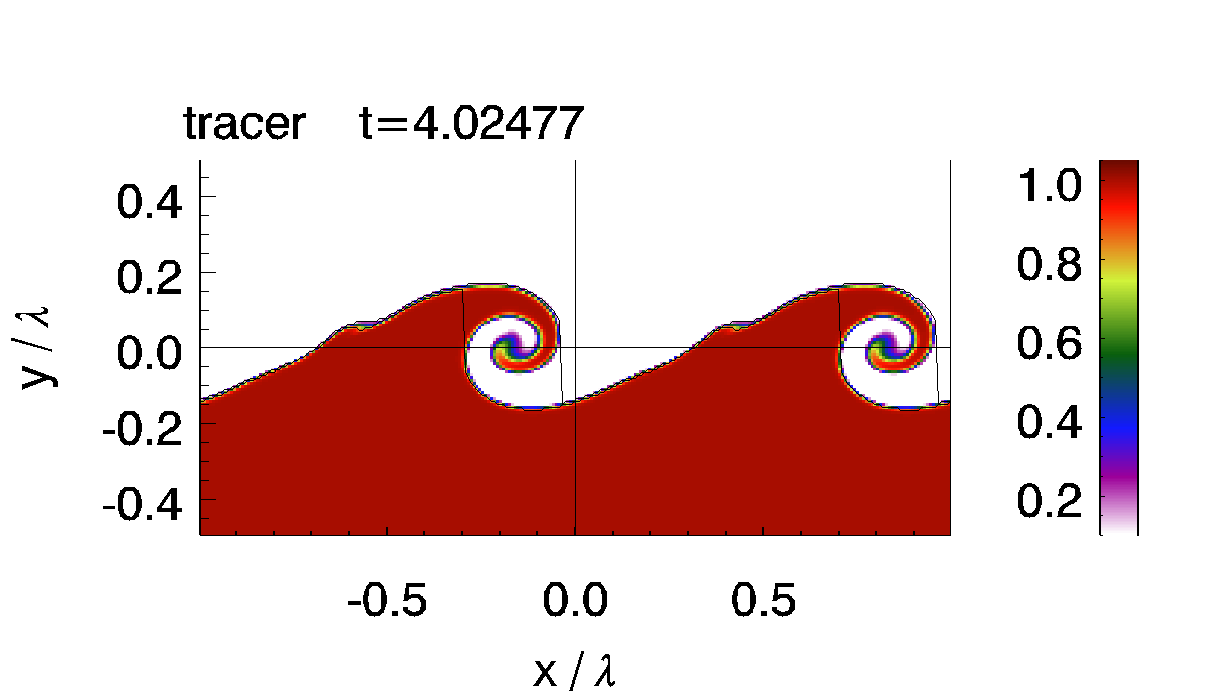}
\newline
\rotatebox{90}{\phantom{xx}$10\tau\KHinvisc$}\hfill%
\includegraphics[trim=0     150 250 160,clip,height=1.5cm]{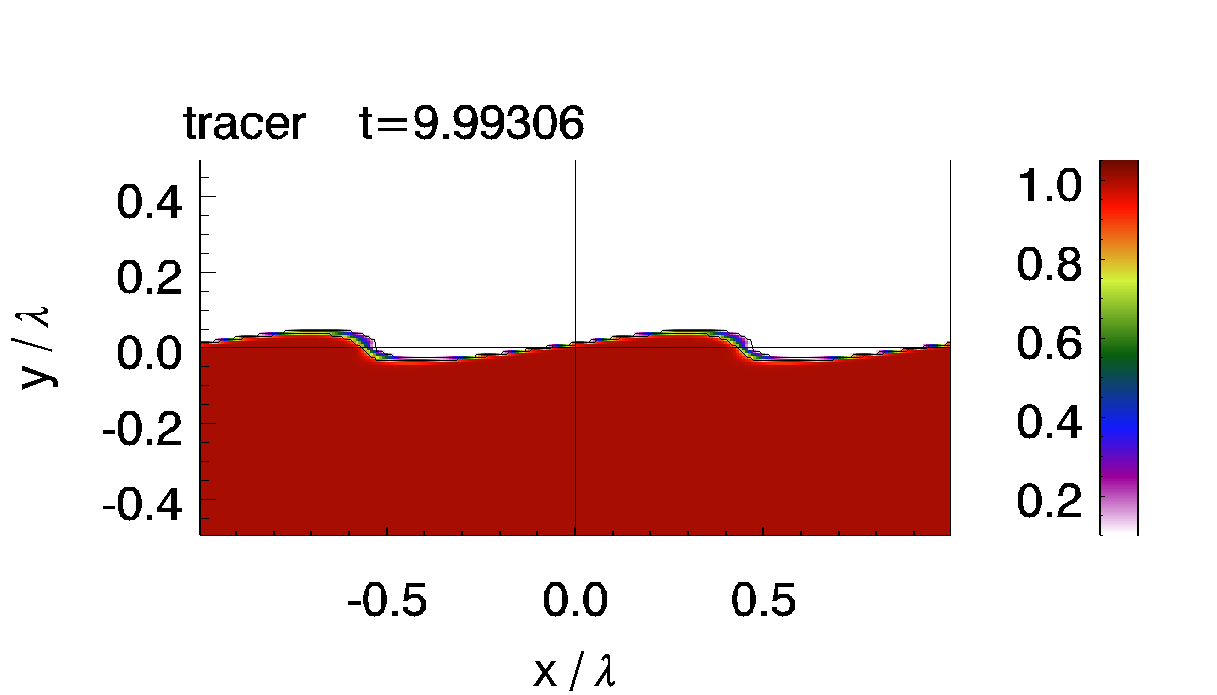}
\includegraphics[trim=190 150 250 160,clip,height=1.5cm]{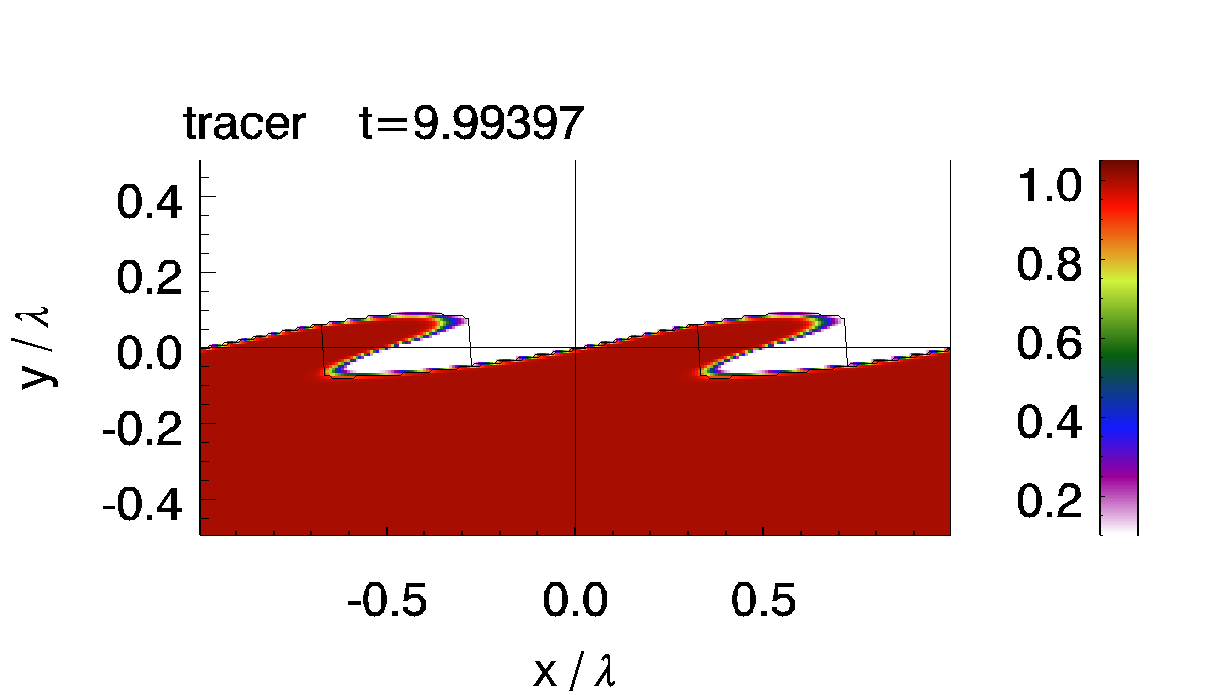}
\includegraphics[trim=190 150 250 160,clip,height=1.5cm]{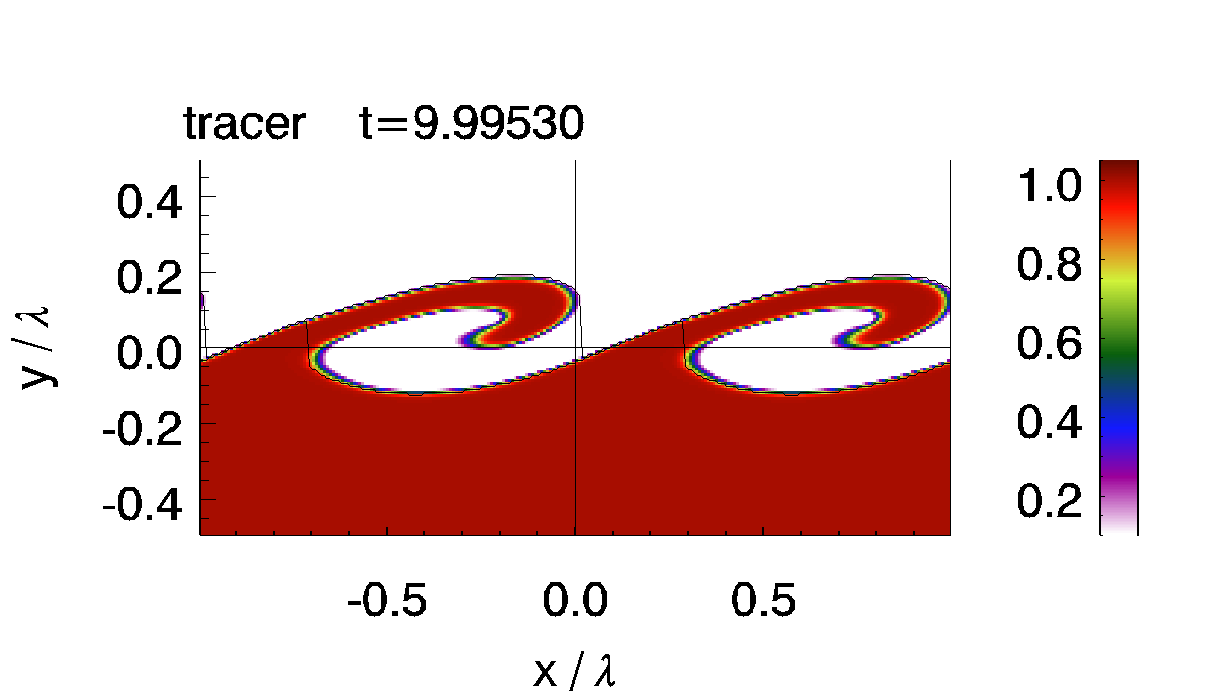}
\includegraphics[trim=190 150 250 160,clip,height=1.5cm]{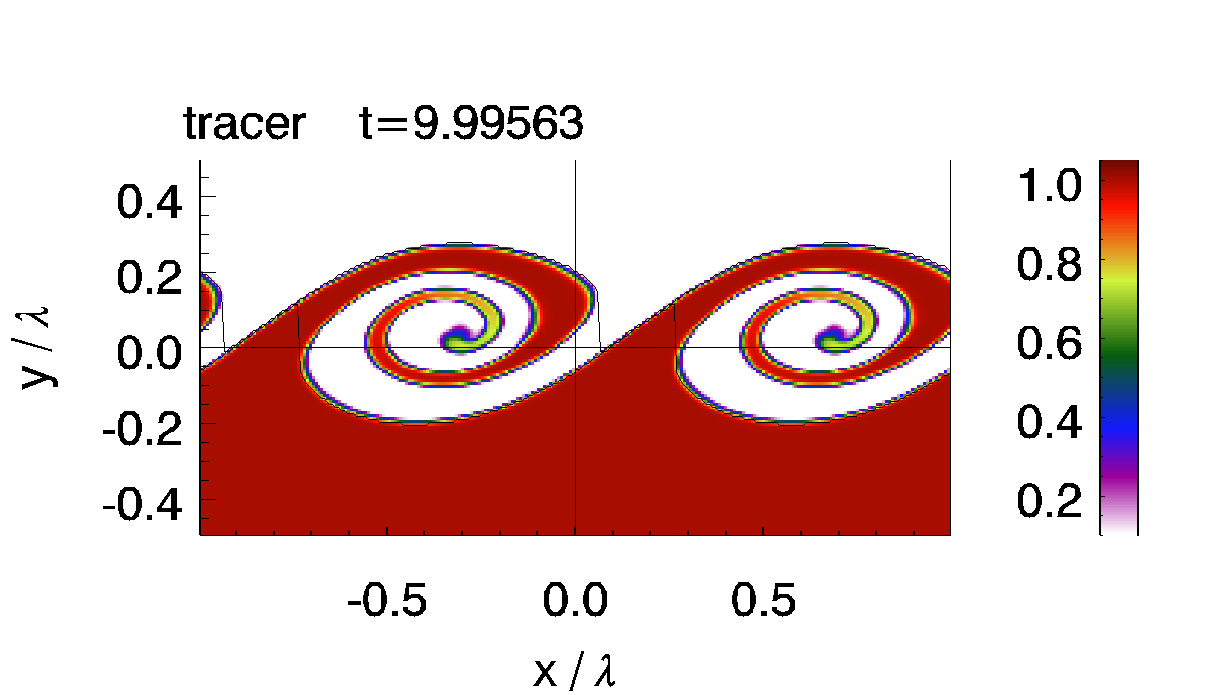}
\includegraphics[trim=190 150     0 160,clip,height=1.5cm]{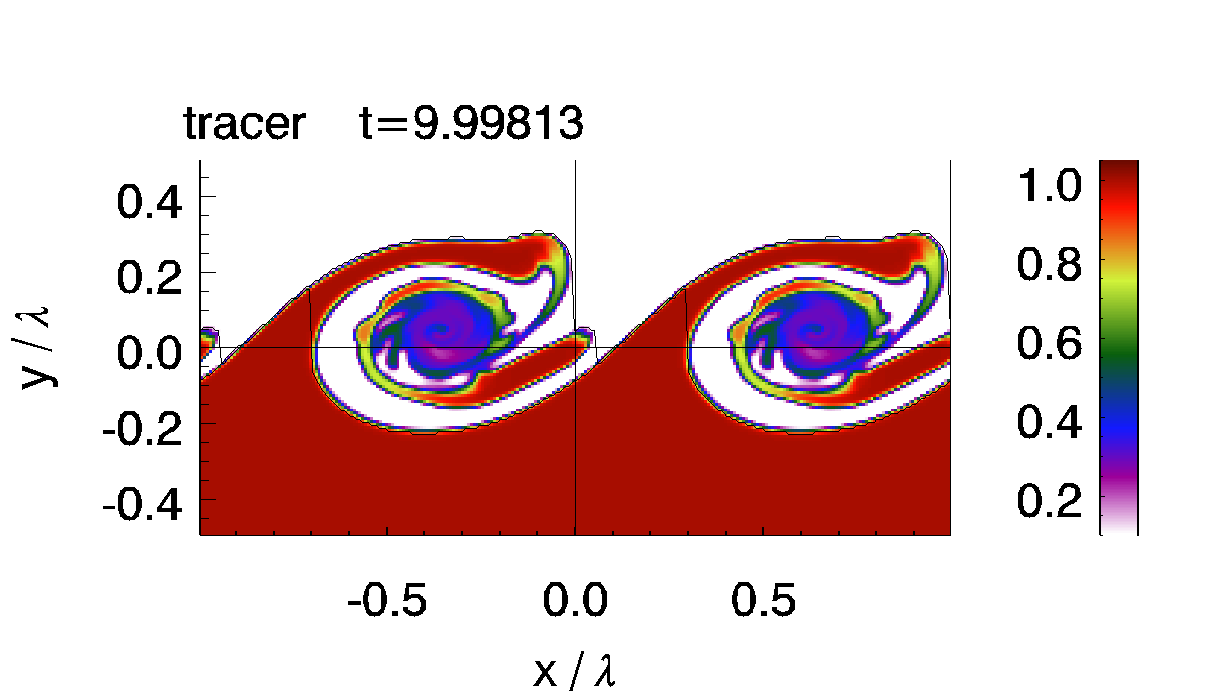}
\newline
\rotatebox{90}{\phantom{xx}$20\tau\KHinvisc$}\hfill%
\includegraphics[trim=0 0     250 160,clip,height=2.09cm]{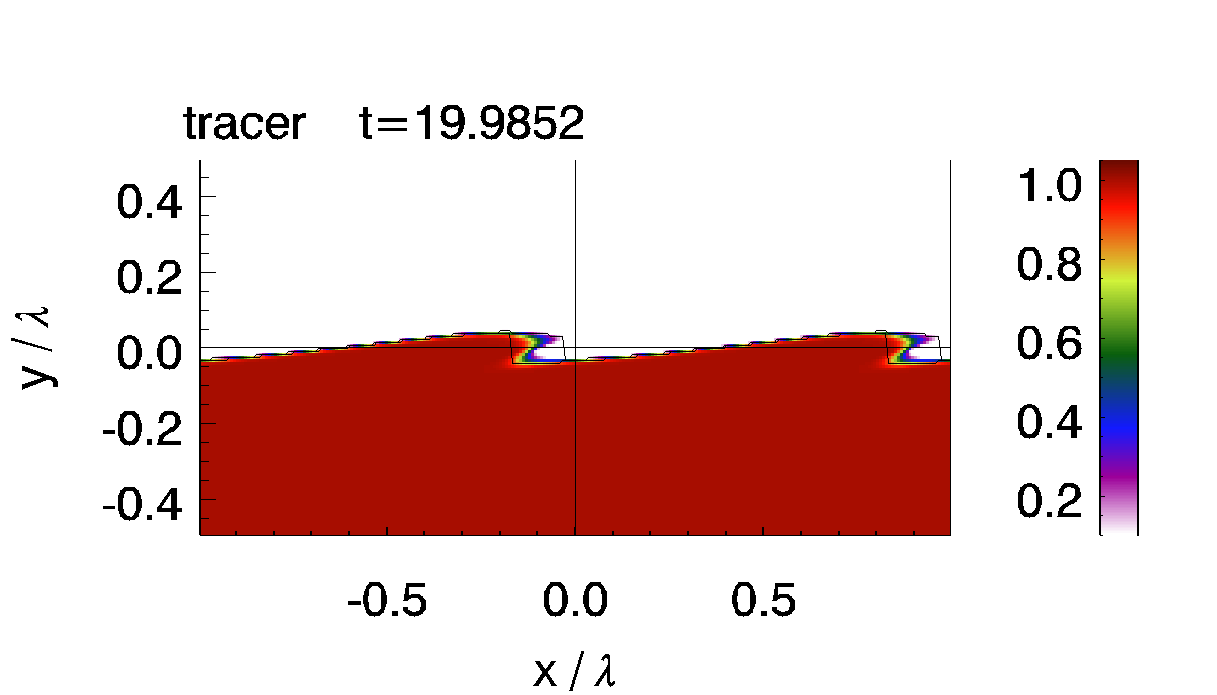}
\includegraphics[trim=190 0 250 160,clip,height=2.09cm]{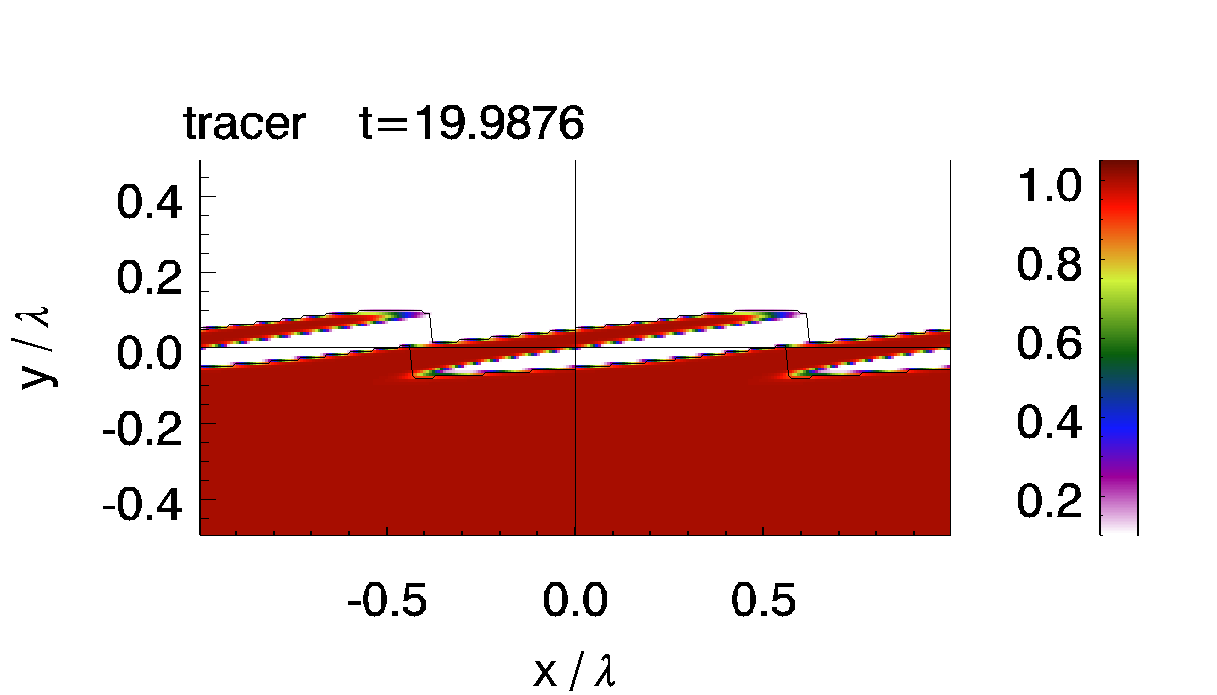}
\includegraphics[trim=190 0 250 160,clip,height=2.09cm]{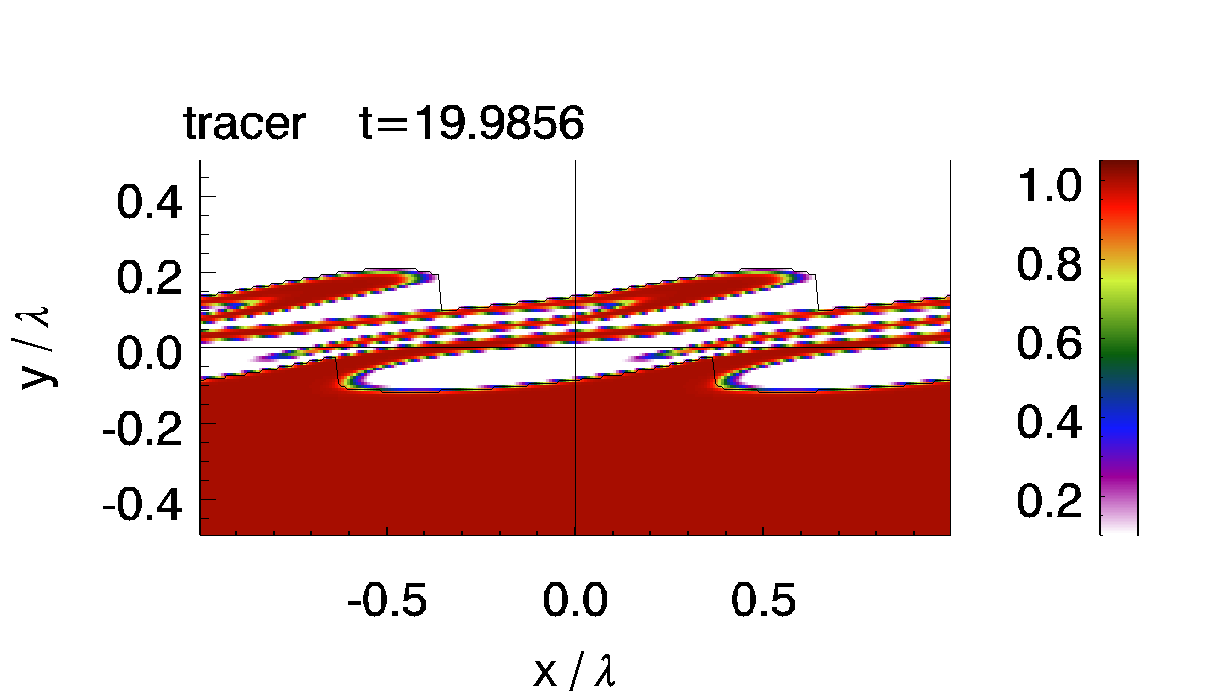}
\includegraphics[trim=190 0 250 160,clip,height=2.09cm]{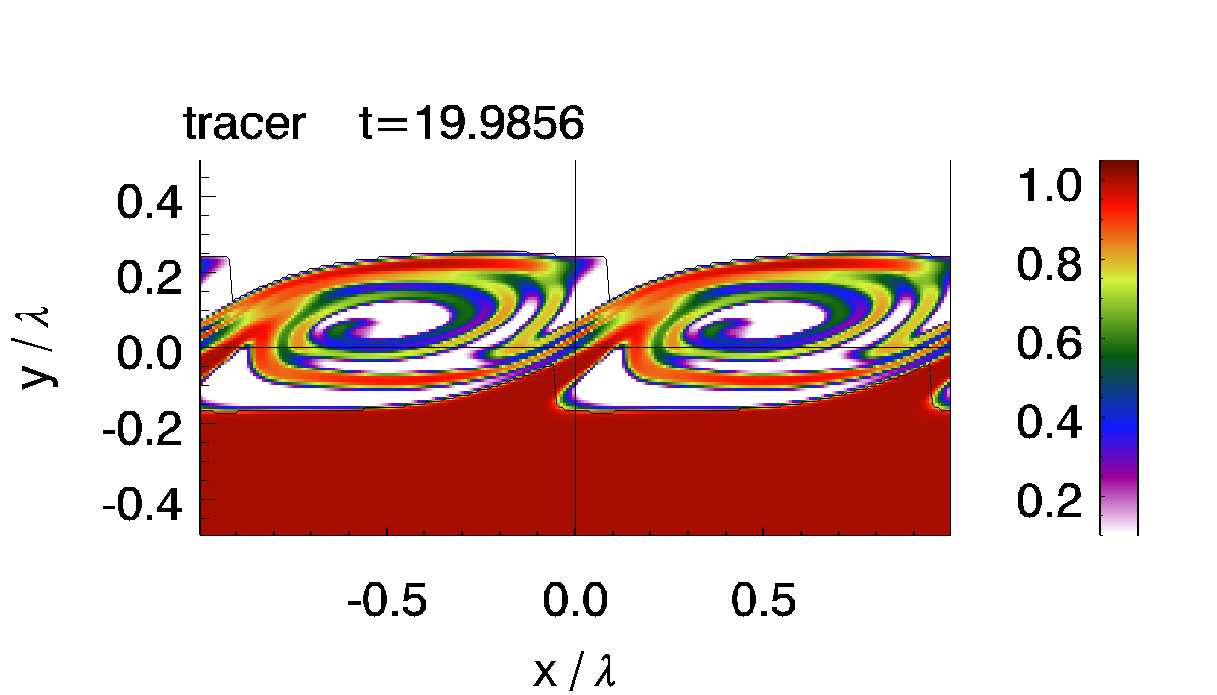}
\includegraphics[trim=190 0 0     160,clip,height=2.09cm]{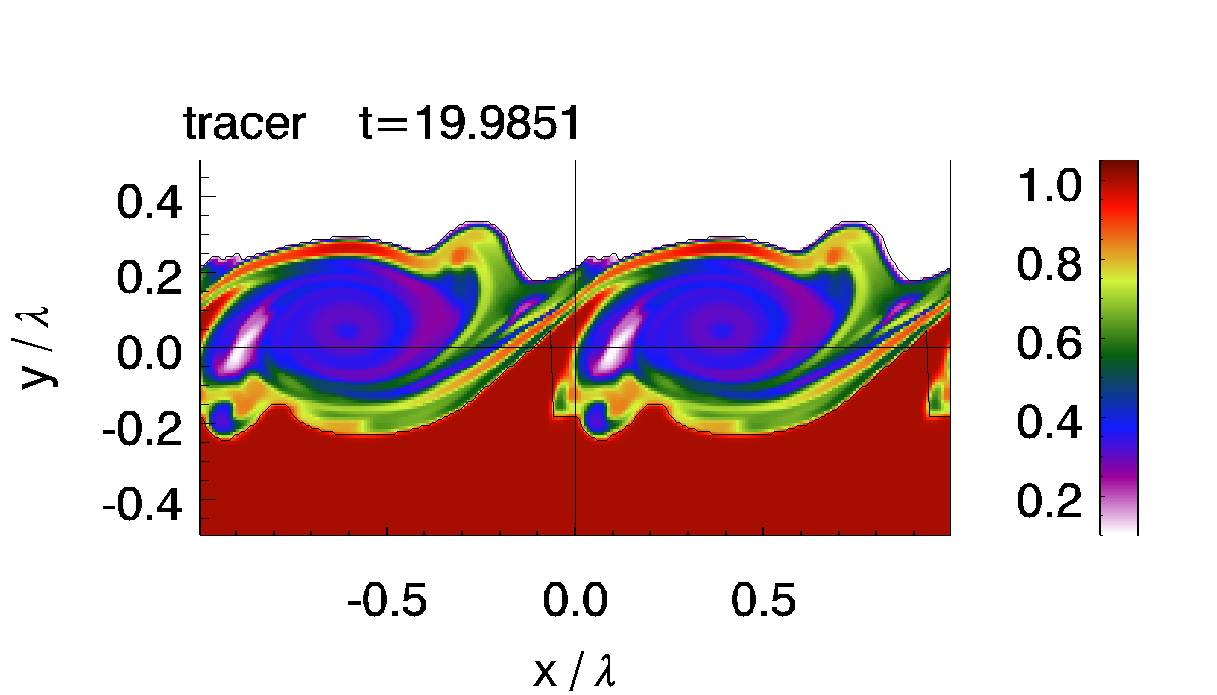}
\caption{Tracer slices like in Fig.~\ref{fig:rolls_nu_D1_M05}, but for constant kinematic viscosity $\nu$, density ratio 2, and shear flow of Mach number 0.5. For $\Reyn=10^4$ we smoothed the initial interface over 1\% of the perturbation length scale to suppress secondary instabilities (see Eqn.~\ref{eq:smooth}).}
\label{fig:rolls_nu_D2_M05}
\end{center}
\end{figure*}
%

\begin{figure*}
\begin{center}
\hspace{0.5cm} $\Reyn=30$ \hfill $\Reyn=100$ \hfill $\Reyn=300$ \hfill $\Reyn=1000$ \hfill $\Reyn=10^4$ \hfill\phantom{x}\newline
\rotatebox{90}{\phantom{xx}$4\tau\KHinvisc$}\hfill%
\includegraphics[trim=0     150 250 160,clip,height=1.5cm]{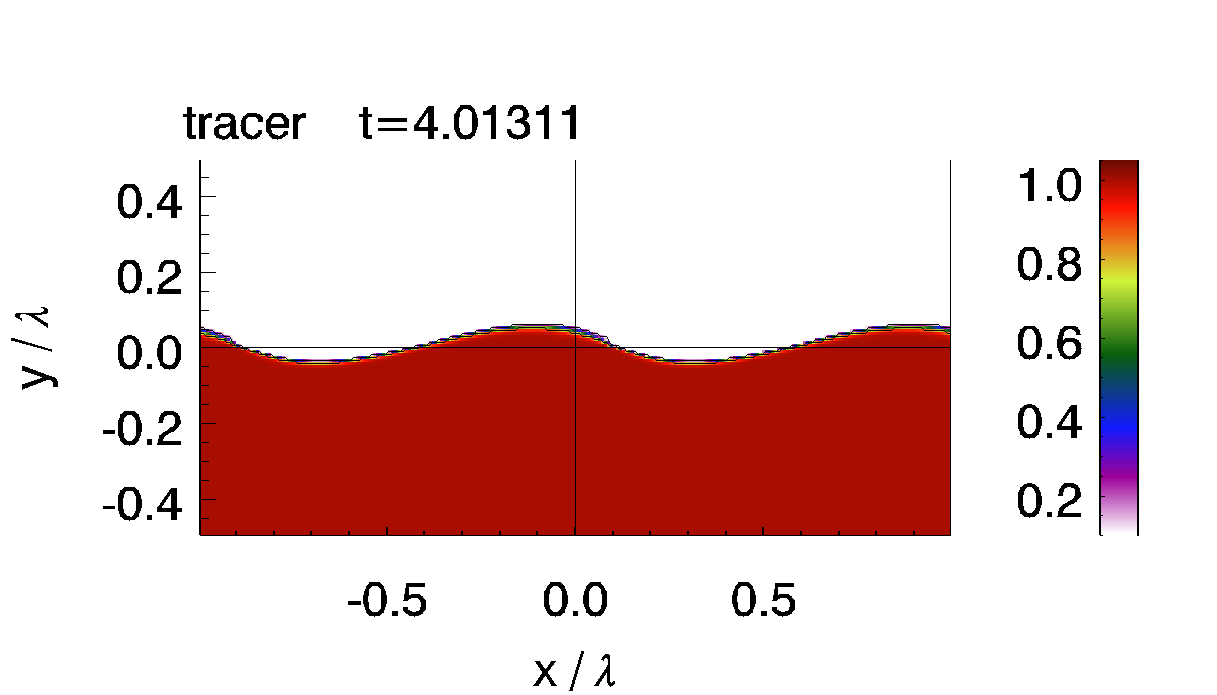}
\includegraphics[trim=190 150 250 160,clip,height=1.5cm]{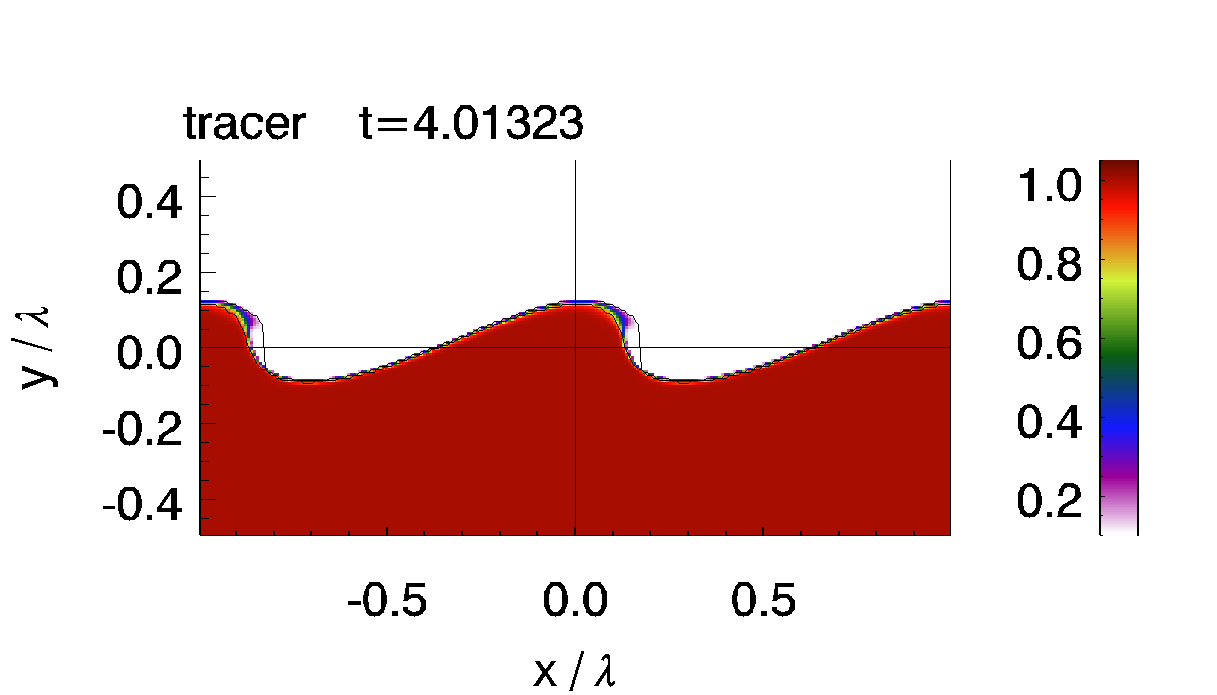}
\includegraphics[trim=190 150 250 160,clip,height=1.5cm]{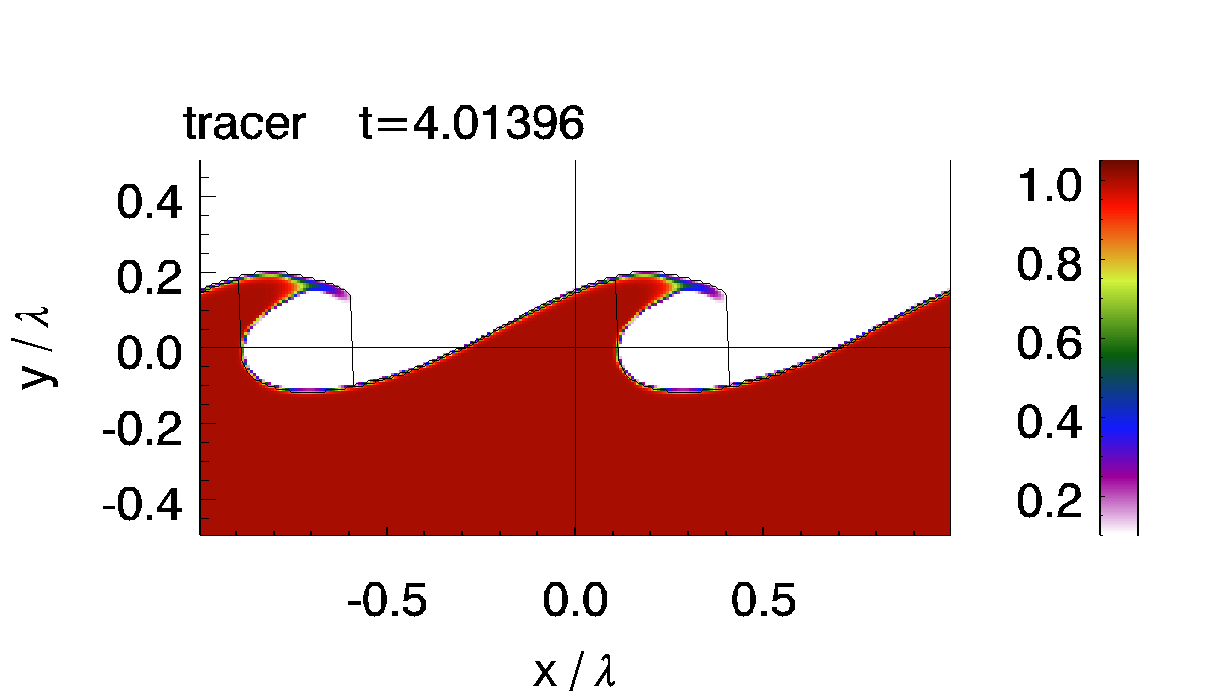}
\includegraphics[trim=190 150 250 160,clip,height=1.5cm]{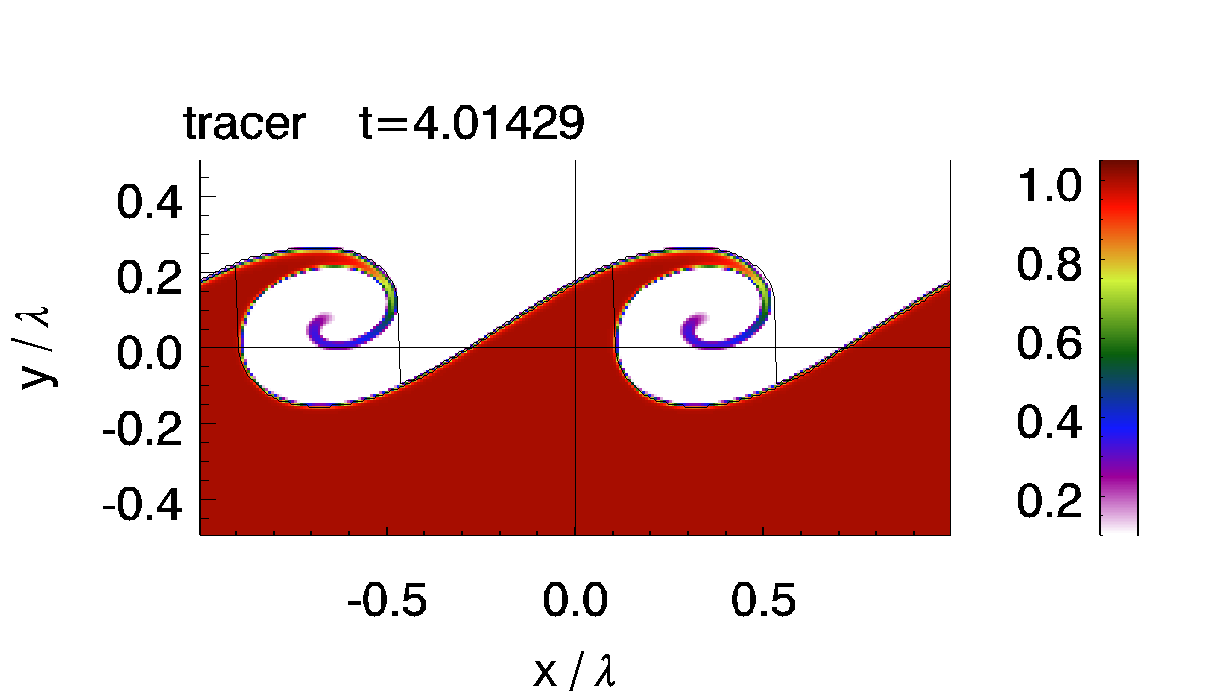}
\includegraphics[trim=190 150    0  160,clip,height=1.5cm]{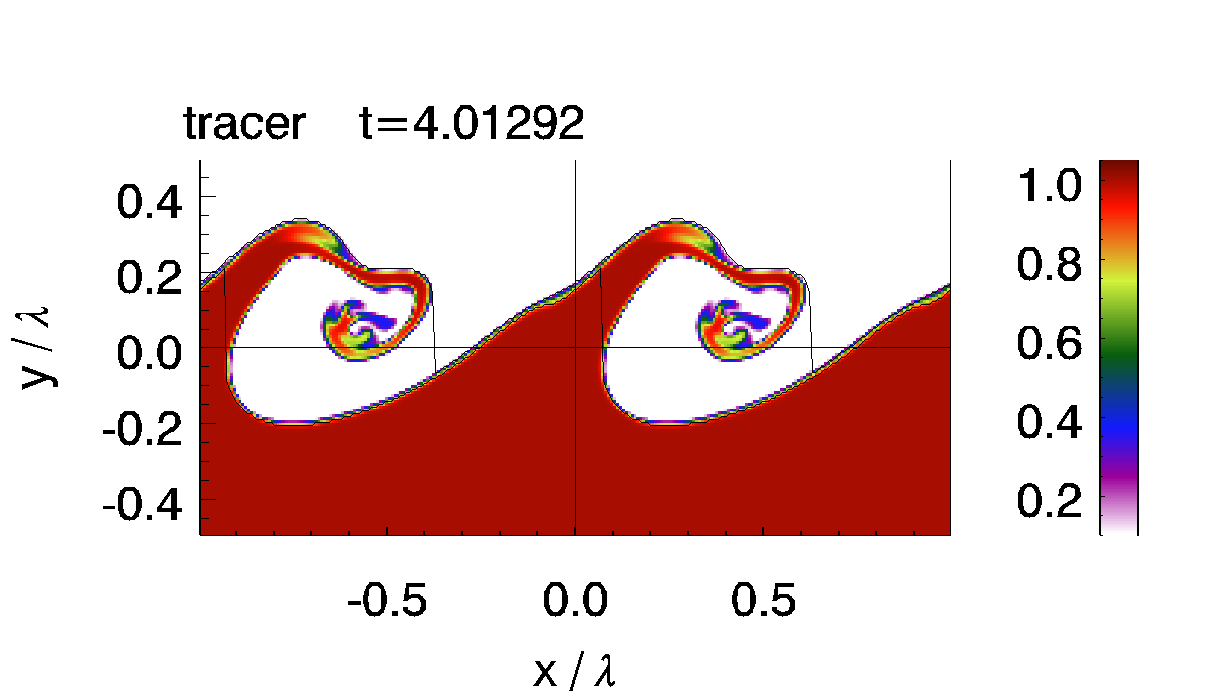}
\newline
\rotatebox{90}{\phantom{xx}$10\tau\KHinvisc$}\hfill%
\includegraphics[trim=0     150 250 160,clip,height=1.5cm]{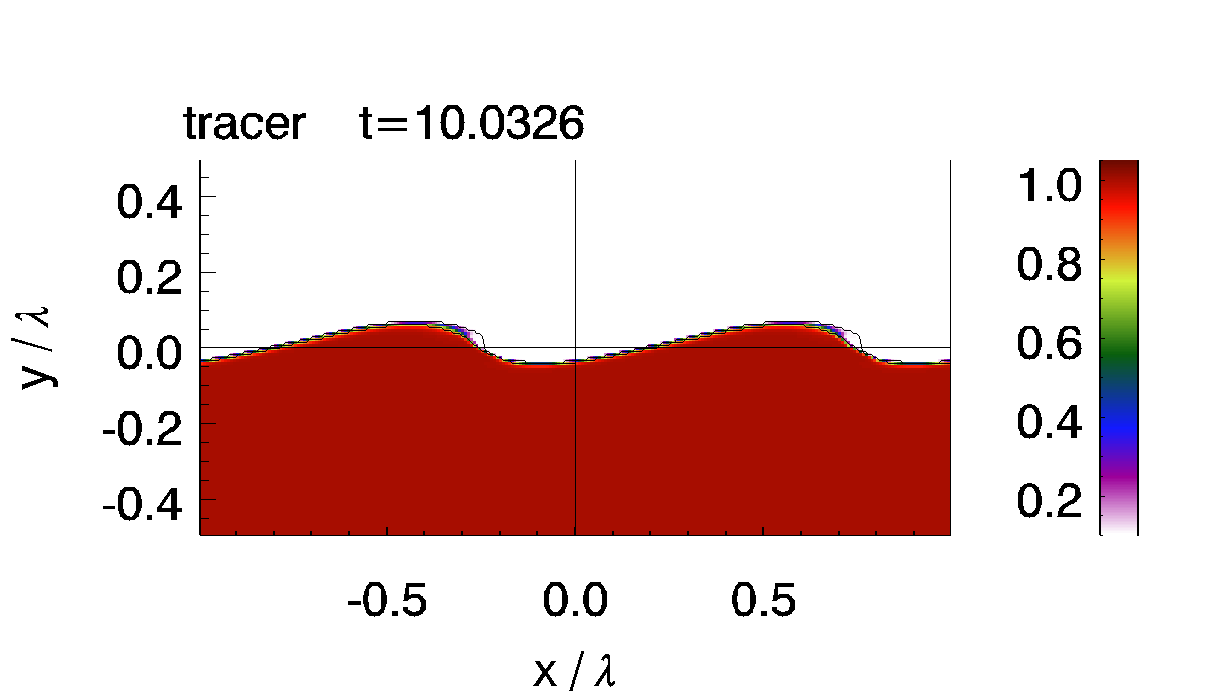}
\includegraphics[trim=190 150 250 160,clip,height=1.5cm]{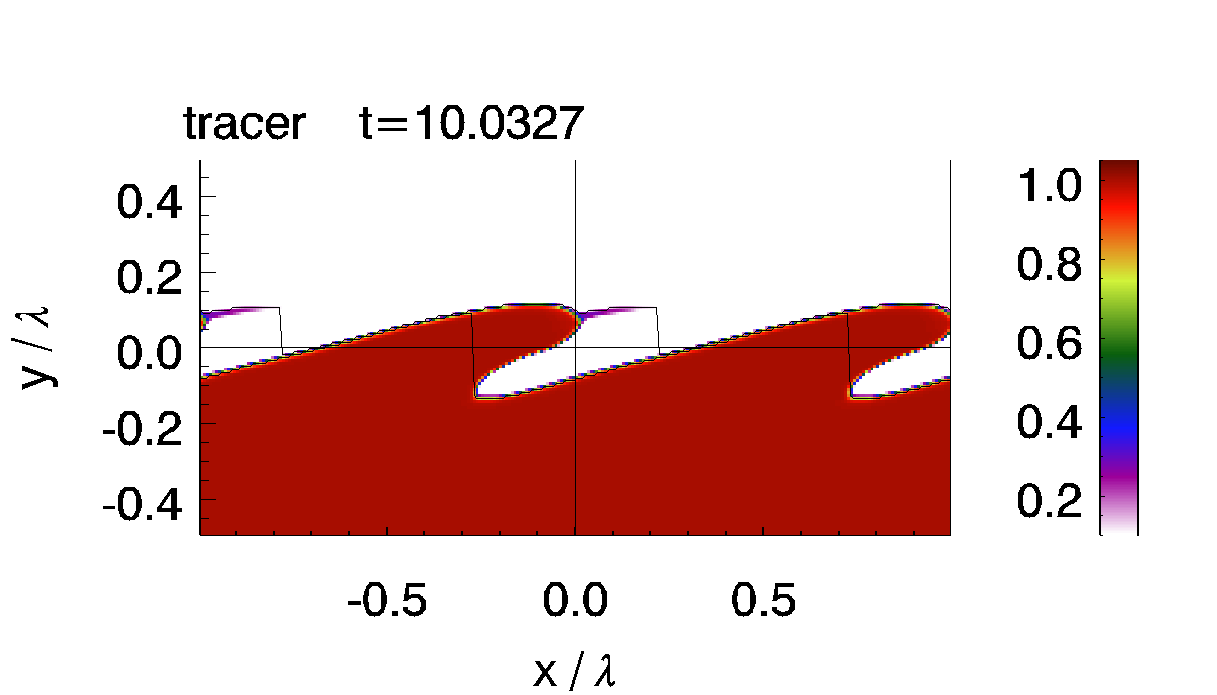}
\includegraphics[trim=190 150 250 160,clip,height=1.5cm]{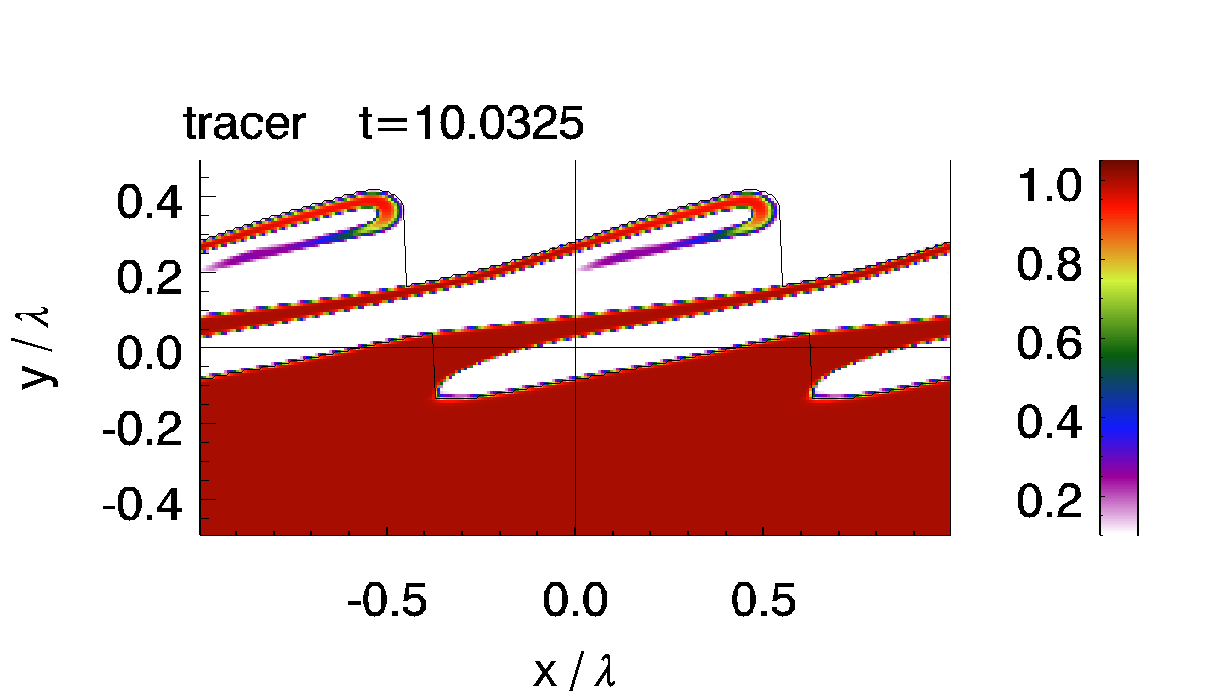}
\includegraphics[trim=190 150 250 160,clip,height=1.5cm]{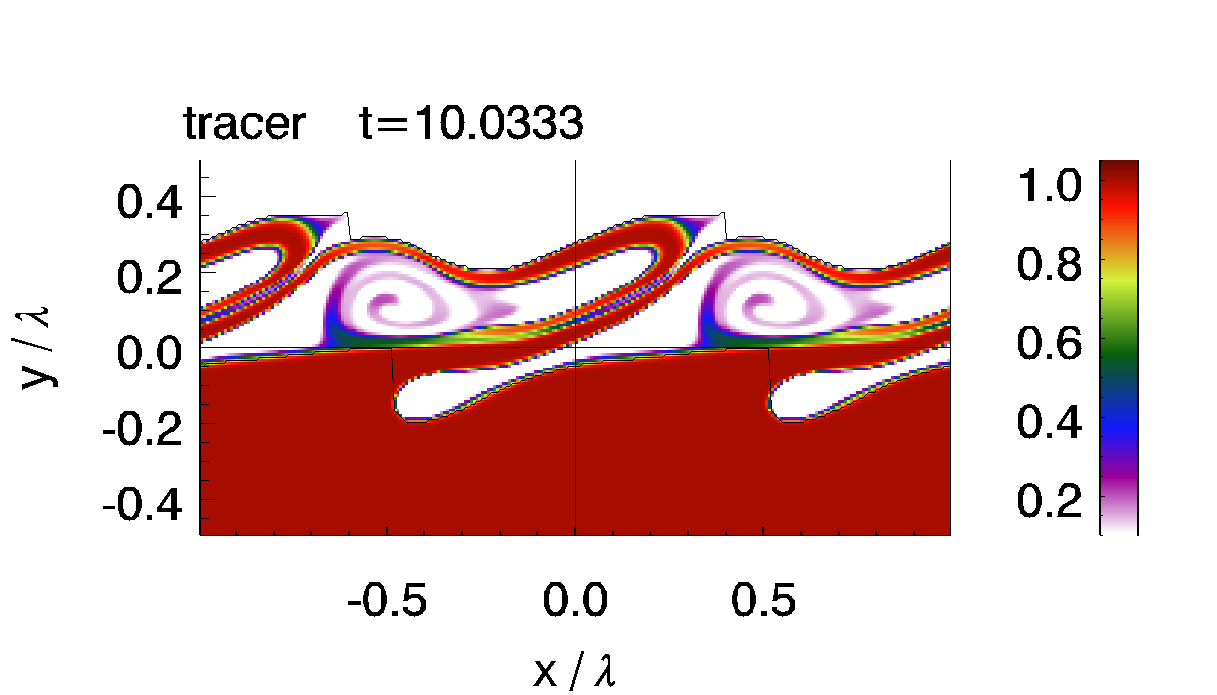}
\includegraphics[trim=190 150     0 160,clip,height=1.5cm]{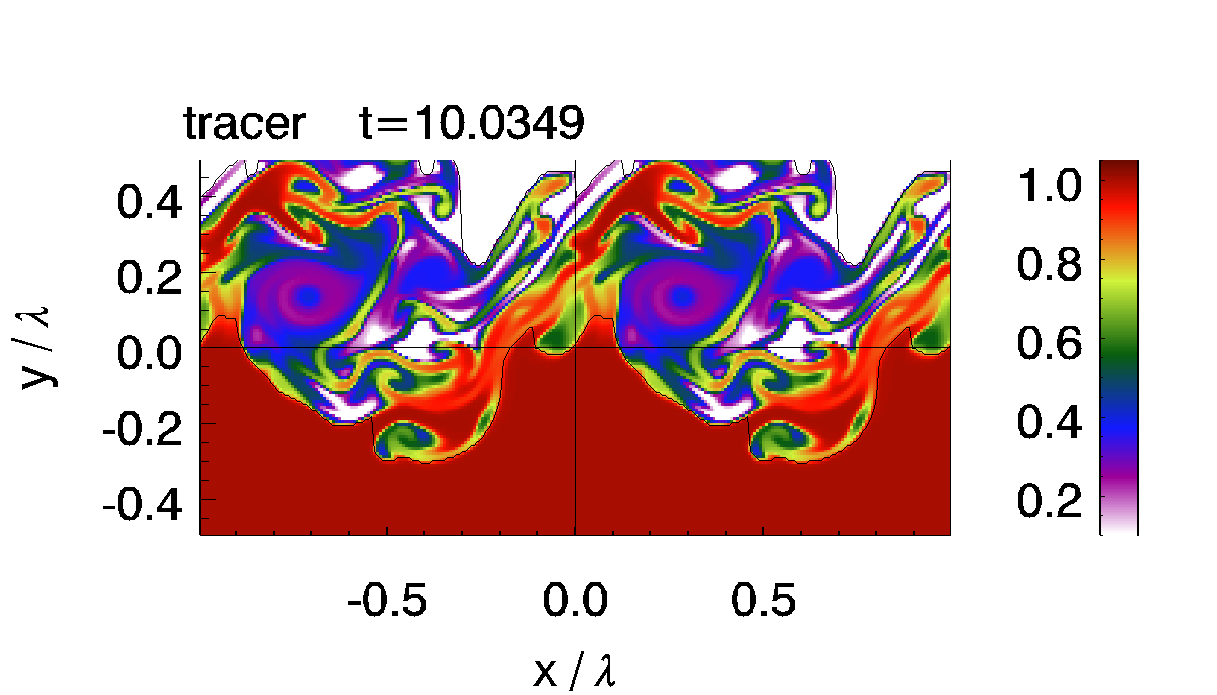}
\newline
\rotatebox{90}{\phantom{xx}$20\tau\KHinvisc$}\hfill%
\includegraphics[trim=0     0 250 160,clip,height=2.09cm]{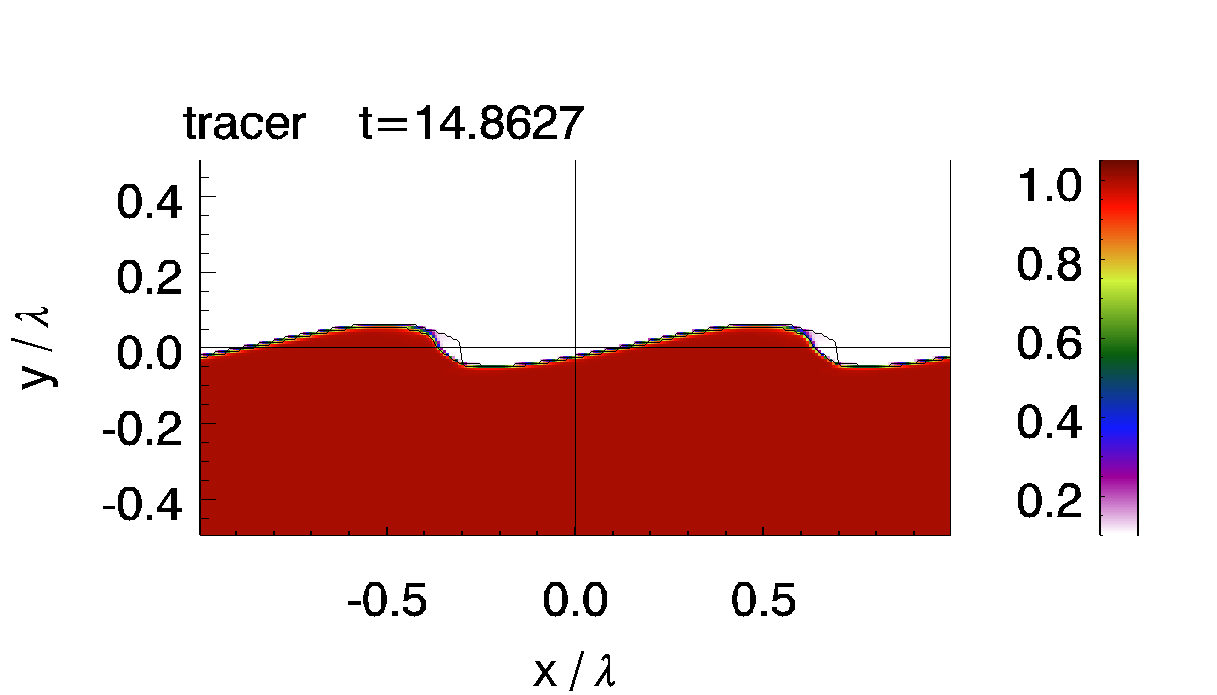}
\includegraphics[trim=190 0 250 160,clip,height=2.09cm]{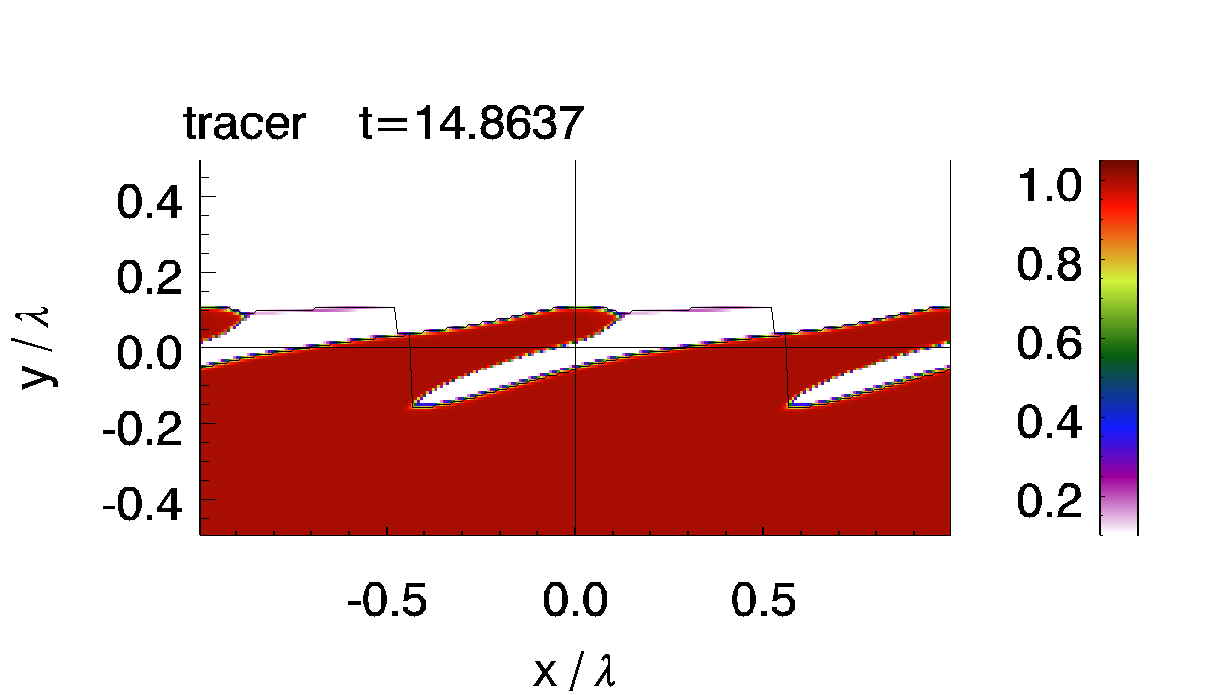}
\includegraphics[trim=190 0 250 160,clip,height=2.09cm]{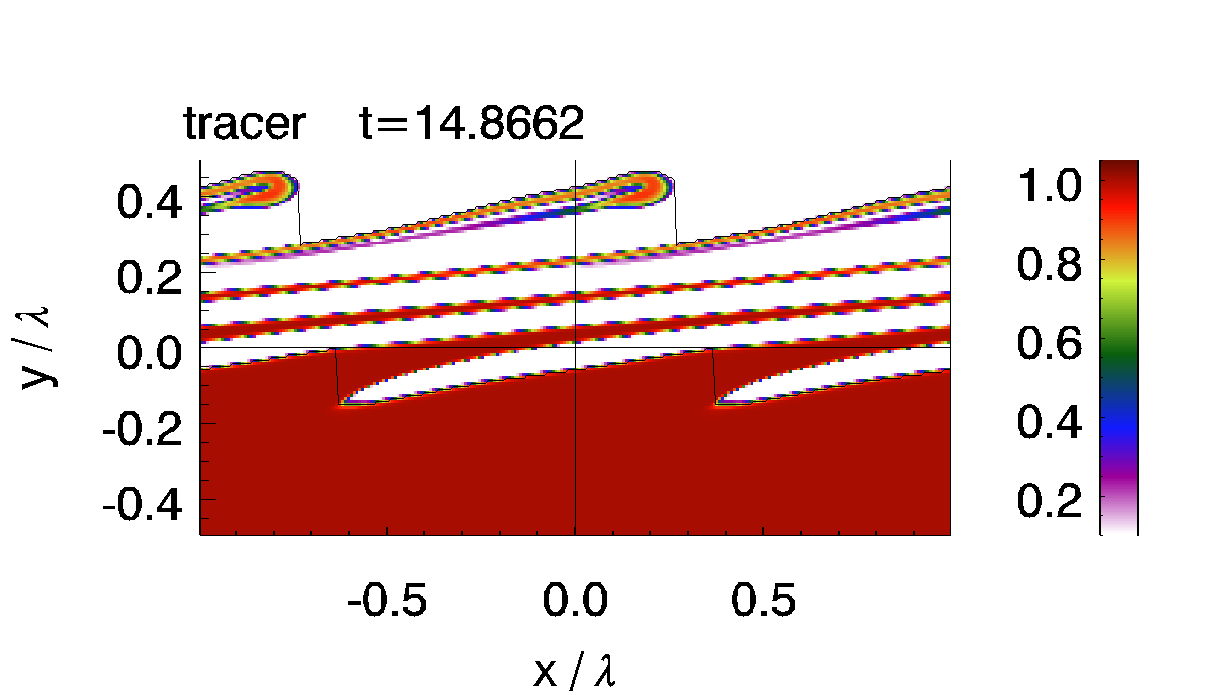}
\includegraphics[trim=190 0 250 160,clip,height=2.09cm]{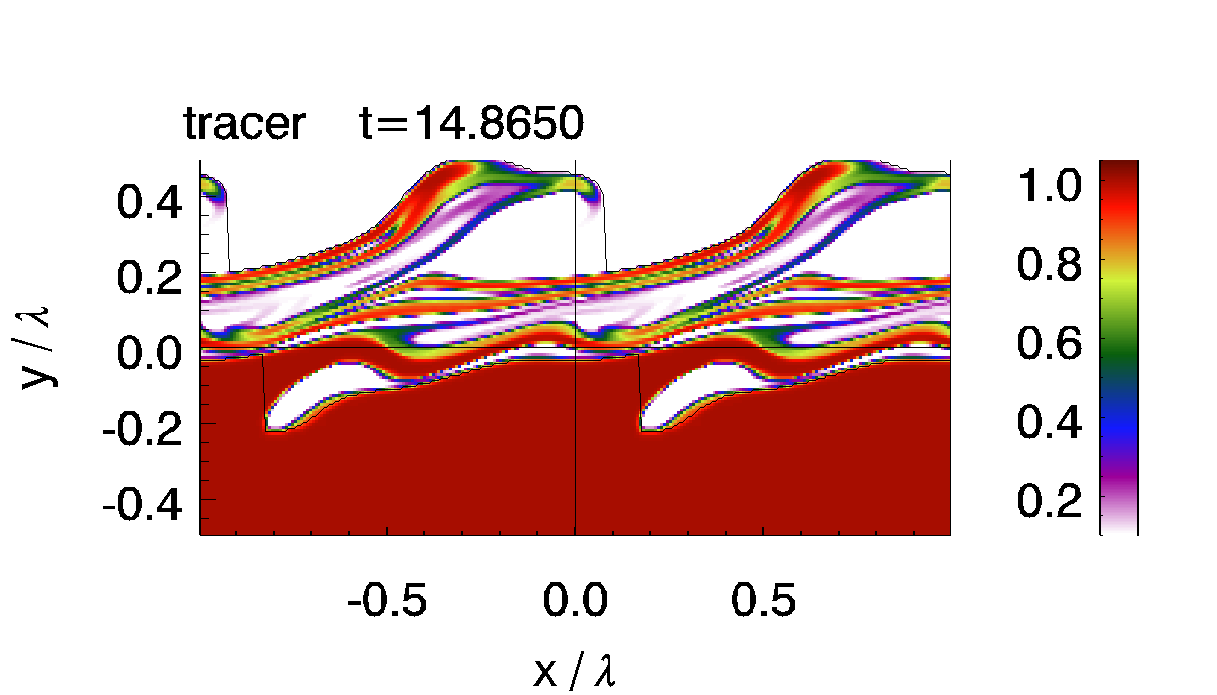}
\includegraphics[trim=190 0 0     160,clip,height=2.09cm]{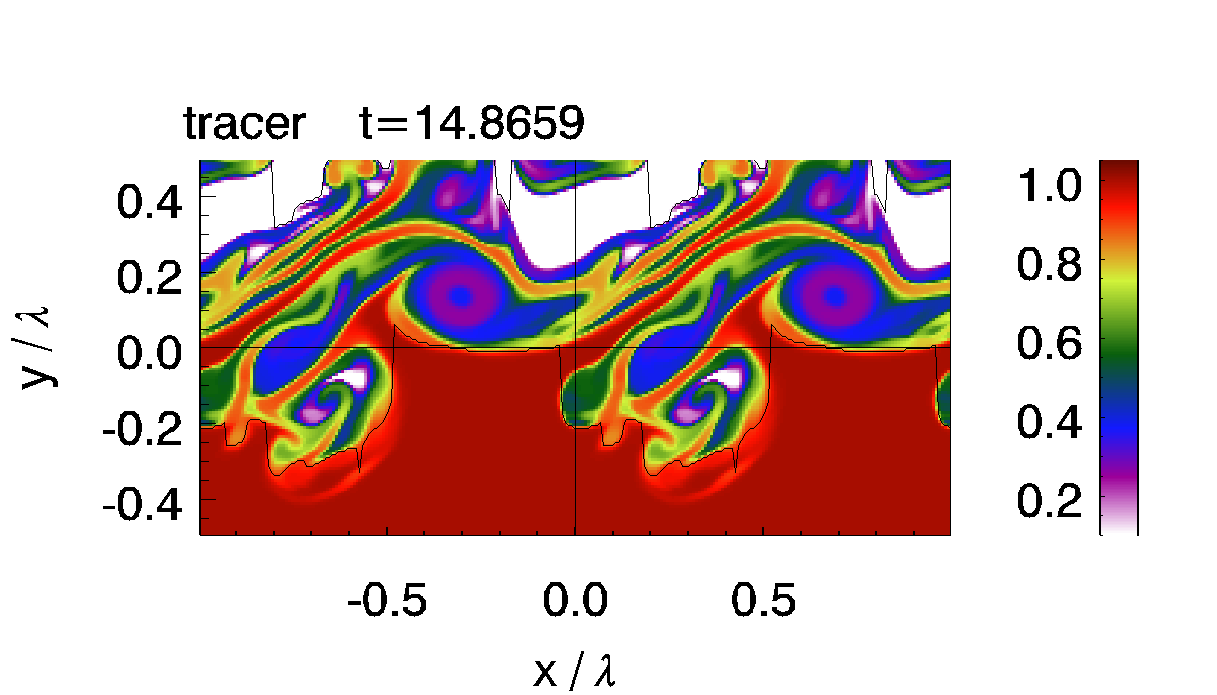}
\caption{Tracer slices like in Fig.~\ref{fig:rolls_nu_D1_M05}, but for constant kinematic viscosity $\nu$, density ratio 10, and shear flow of Mach number 0.5.  For $\Reyn=10^4$ we smoothed the initial interface over 1\% of the perturbation length scale to suppress secondary instabilities (see Eqn.~\ref{eq:smooth}).}
\label{fig:rolls_nu_D10_M05}
\end{center}
\end{figure*}

\begin{figure*}
\includegraphics[trim=0 0 13 0,clip,width=0.45\textwidth]{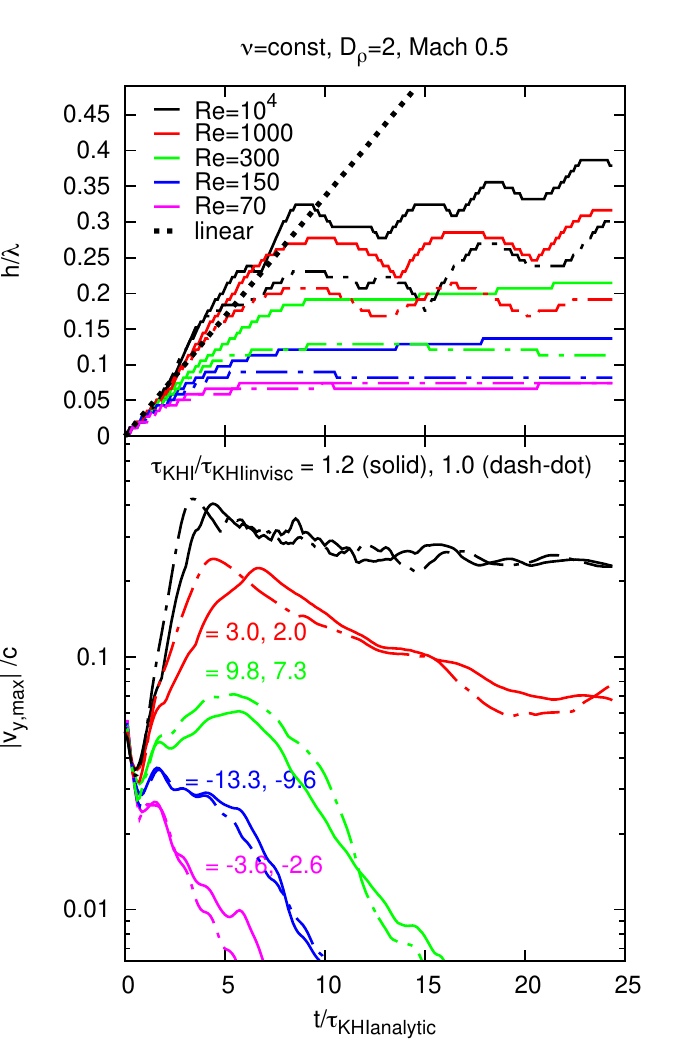}
\includegraphics[trim=0 0 13 0,clip,width=0.45\textwidth]{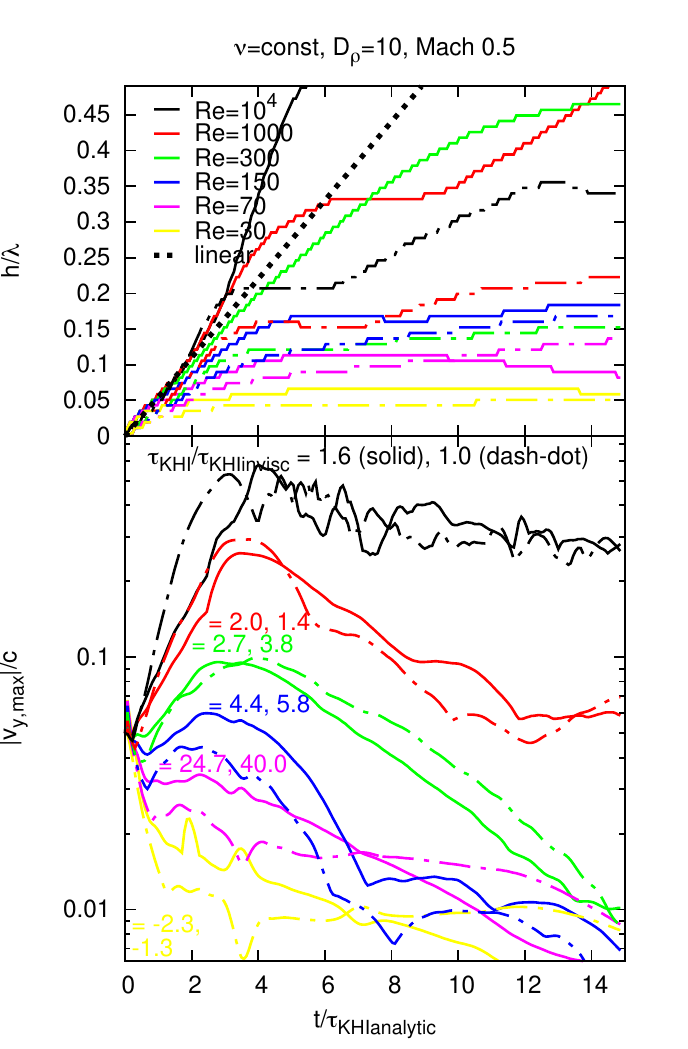}
\caption{Same as Fig.~\ref{fig:thick_vely_nu_M05_D1}, but for different density contrasts, see title of each column. Due to the unequal densities in the two layers, the KHI evolves asymmetrical, and we plot two lines for each simulation. The solid lines show the evolution towards the hot layer (height of KH rolls above initial interface, maximum of $v_y$), the dash-dotted lines show the height of KH rolls below initial interface and the minimum of $v_y$. Exponential growth times for the velocity extrema are given in the bottom panels with the matching font colour. The first number refers to the solid line, the second to the dash-dotted. Qualitatively, the behaviour is very similar to the constant density case, but for a density contrast of 10 a somewhat higher viscosity is needed to fully suppress the KHI.
}
\label{fig:thick_vely_nu_M05}
\end{figure*}

Figures~\ref{fig:rolls_nu_D2_M05} and \ref{fig:rolls_nu_D10_M05} display tracer slices for the viscous KHI for density ratios $D_{\rho}=2$ and 10 between the two layers, respectively. The unequal densities between both layers result in an asymmetric evolution of the KHI. The fingers or filaments of cool gas are thinner than the hot ones because the denser gas has more inertia and is thus more difficult to displace. The KHI rolls extend further into the hot layer than into the cold layer, and the growth times of $v_y{}\Max$ and $v_y{}\Min$ differ slightly.  We  took this into account in Fig.~\ref{fig:thick_vely_nu_M05} by measuring the height of the KH rolls above and below the initial interface, and by distinguishing between $v_y{}\Max$ and $v_y{}\Min$ instead of using only $v_y{}\Max$. At the higher density contrast of 10 and high Re, the late evolution of the KHI takes on more complex dynamics. 

The impact of viscosity is very similar as in the $D_{\rho}=1$ case. The dissipation of velocity shears again leads to reduced rolling up of the interface, slower growth times, decreasing heights of the KHI rolls, and finally suppression of the KHI at low Reynolds numbers. For $D_{\rho}=2$, the limiting Re is similar to the $D_{\rho}=1$ case. For $D_{\rho}=10$ the growth rates derived from $v_y{}\Max(t)$ and $v_y{}\Min(t)$ imply that the KHI still grows initially for $\Reyn=100$. However, the height of the distortions of the interface remains well behind the $\Reyn\ge 300$ case, and the interface does not roll up. Therefore we consider the KHI suppressed in this case as well. Nonetheless, at the higher density contrast, viscosity seems to be less able to slow down the KHI. For example, at $\Reyn=300$, the derived growth time is only slowed down by a factor of 3 compared to the inviscid case, whereas at lower density contrast it was slowed down by a factor of almost 10. A closer look at the dynamics reveals that viscosity does not only need to work against the KHI, but also against the increased amount of momentum in the denser layer. We revisit this point  in Sect.~\ref{sec:highcontrast}.  

We again summarise the influence of viscosity in Fig.~\ref{fig:tau_of_Re} by plotting the measured viscous KH growth times as a function of Reynolds number. Symbols of different colours code different $D_{\rho}$. Again, we can approximate $\tau\KHvisc(\Reyn)$ by Eqn.~\ref{eq:tau_Re_fit} but parameters $\Reyn\Crit$ and $\Reyn_0$ depending on $D_{\rho}$. The variation of $\Reyn\Crit$ and $\Reyn_0$ with density contrast can be approximated by
\begin{eqnarray}
\Reyn\Crit &=& 880 / \Delta, \label{eq:ReCritFit}  \\ 
\Reyn_0 &=& 1320 / \sqrt{\Delta}, \label{eq:Re0Fit}
\end{eqnarray}
where $\Delta$ depends on the density contrast $D_{\rho}$ as in Eqn.~\ref{eq:Delta}. Thus, Eqn.~\ref{eq:tau_Re_fit} combined with Eqns.~\ref{eq:ReCritFit} and \ref{eq:Re0Fit} provides an empirical relation for the viscous KHI growth time as a function of $\Reyn$ up to density contrasts of 10. This empirical relation is shown by lines of matching colour in Fig.~\ref{fig:tau_of_Re}. We note that due to being based solely on the evolution of the $v_y$ extrema, this empirical relation slightly underestimates the ability of the viscosity to suppress the KHI at high density contrasts. For $D_{\rho}=10$ it states $\Reyn\Crit=73$, but we discussed above that the KHI is already suppressed for $\Reyn=100$. If viewed in detail, the suppression of the KHI by viscosity is a gradual process, and the difference in $\Reyn\Crit$ arrises due to a different definition of when suppression of the  KHI is reached. Thus, the empirical relation is useful within a factor of 1.5. 

Our simulations find that $\Reyn\Crit$ slightly decreases with density contrast, which is opposite of what was expected from all analytical estimates in Sect.~\ref{sec:khi_visc}. The reason is that none of those estimates takes into account that the diffusion of momentum into the denser, cold layer will be slower due to the higher momentum in the dense layer. The slower diffusion of momentum into the cold layer leads to a slower widening of the shear flow discontinuity into the cold layer, and hence only smaller instabilities can be suppressed -- or, stated differently, a lower $\Reyn$ is required to suppress a given perturbation.

\subsection{Spitzer viscosity} \label{sec:spitzer}

The density dependence and strong temperature dependence of the Spitzer-like viscosity introduces a strong difference of Reynolds numbers between the hot and the cold layer. Density contrasts of 2 or 10 result in a ratios of Reynolds numbers of  11 and 3160, respectively. Consequently, the cold layer always will be more turbulent, and we can expect that a higher viscosity is needed to suppress the KHI. The Reynolds number stated in the following refers to the Reynolds number in the hot layer, as this is the crucial one. 

\begin{figure*}
\begin{center}
\hspace{0.5cm} $\Reyn=10$ \hfill $\Reyn=30$ \hfill $\Reyn=100$ \hfill $\Reyn=1000$ \hfill $\Reyn=10^4$ \hfill\phantom{x}\newline
\rotatebox{90}{\phantom{xx}$4\tau\KHinvisc$}\hfill%
\includegraphics[trim=0     150 250 160,clip,height=1.5cm]{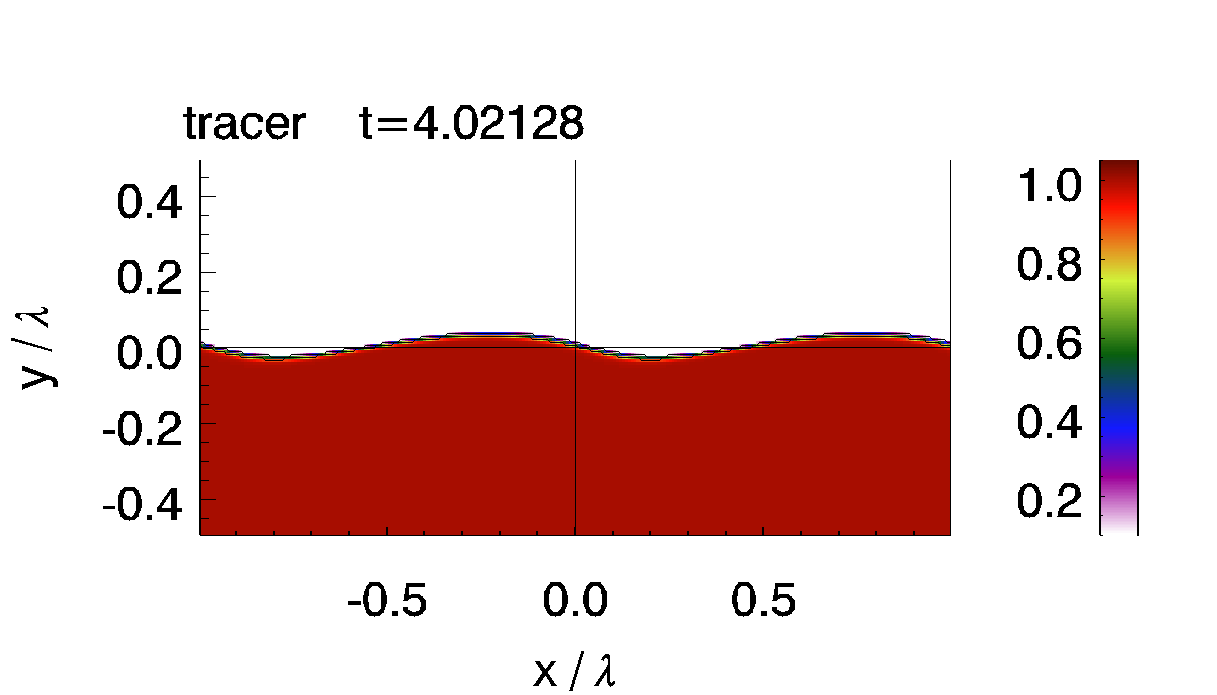}
\includegraphics[trim=190 150 250 160,clip,height=1.5cm]{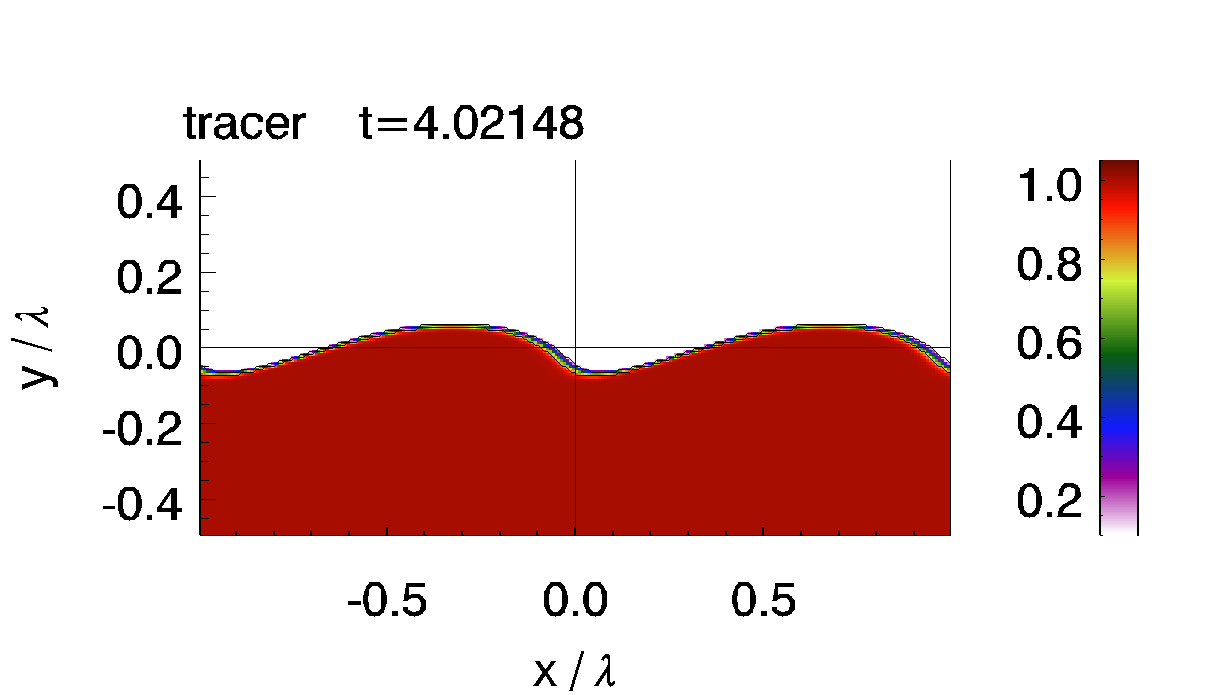}
\includegraphics[trim=190 150 250 160,clip,height=1.5cm]{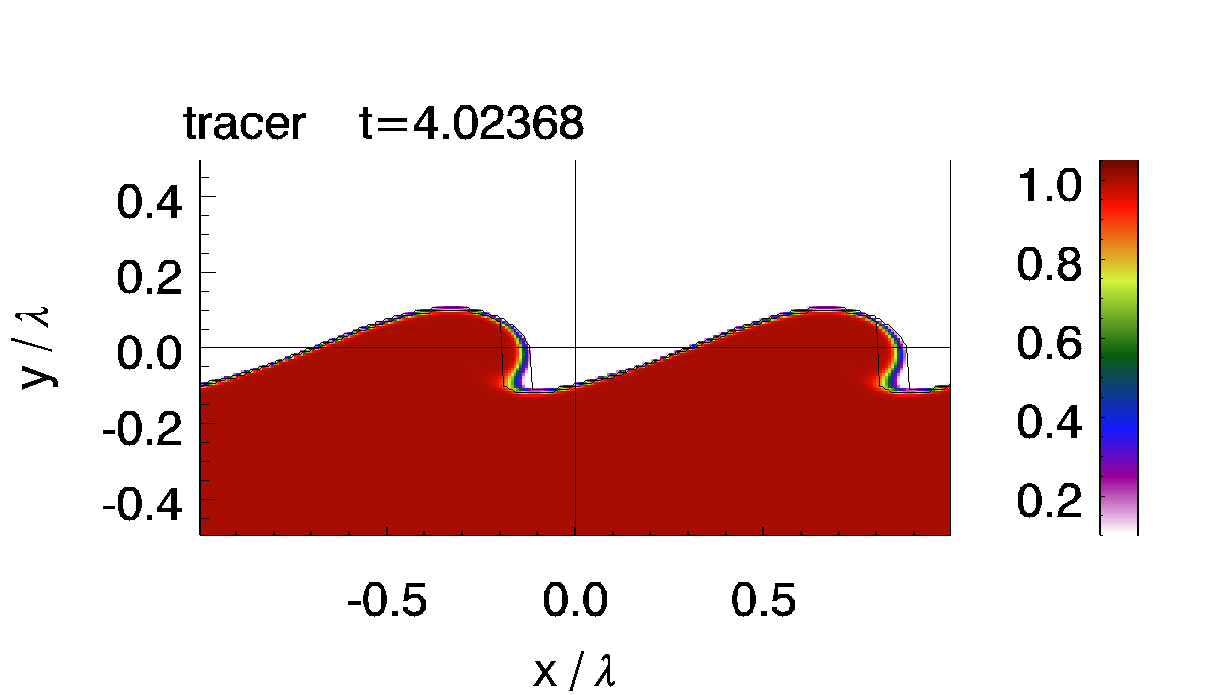}
\includegraphics[trim=190 150 250 160,clip,height=1.5cm]{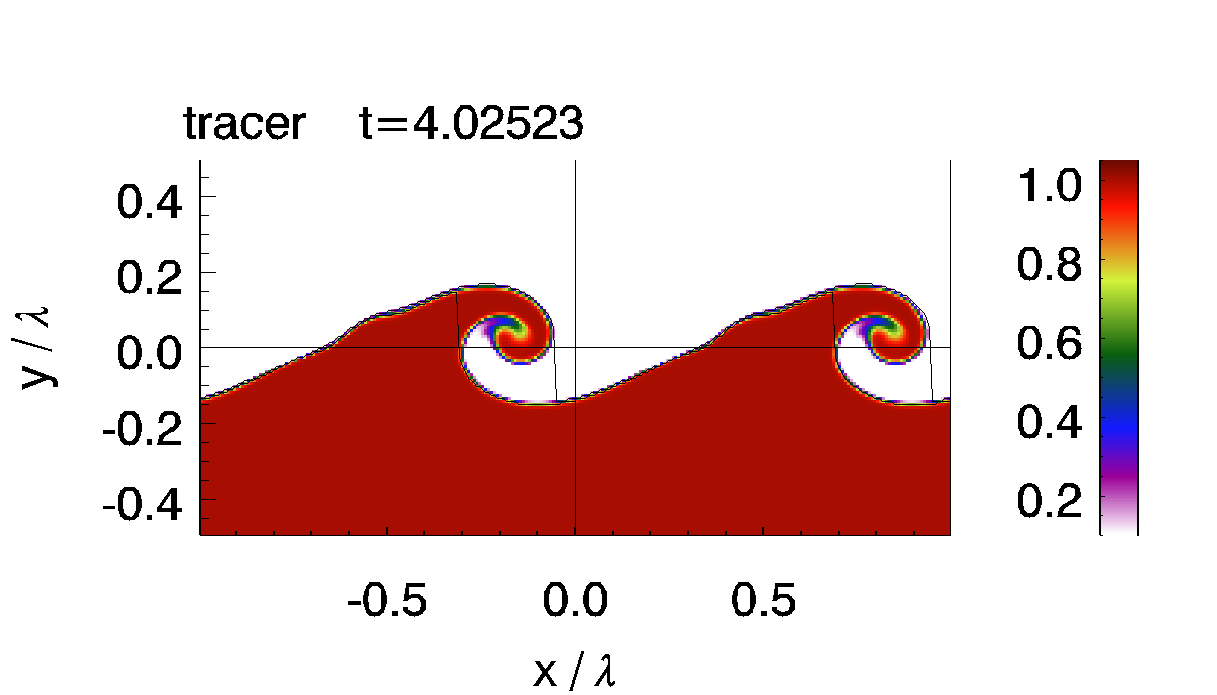}
\includegraphics[trim=190 150     0 160,clip,height=1.5cm]{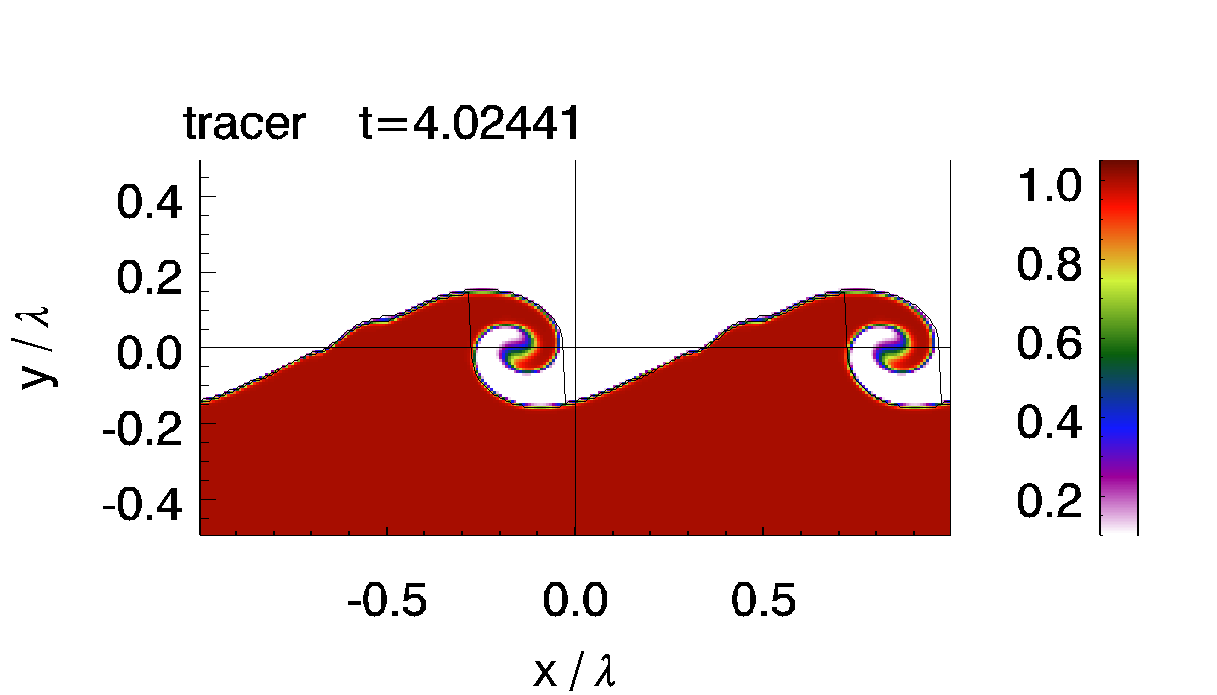}
\newline
\rotatebox{90}{\phantom{xx}$10\tau\KHinvisc$}\hfill%
\includegraphics[trim=0     150 250 160,clip,height=1.5cm]{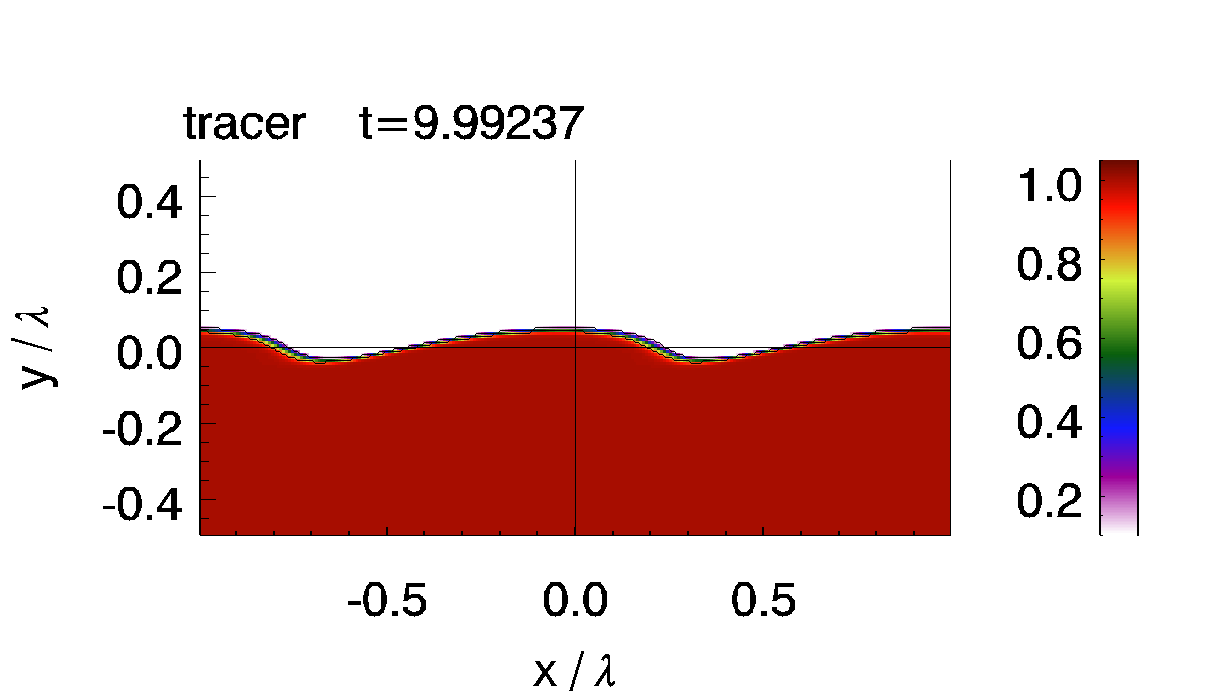}
\includegraphics[trim=190 150 250 160,clip,height=1.5cm]{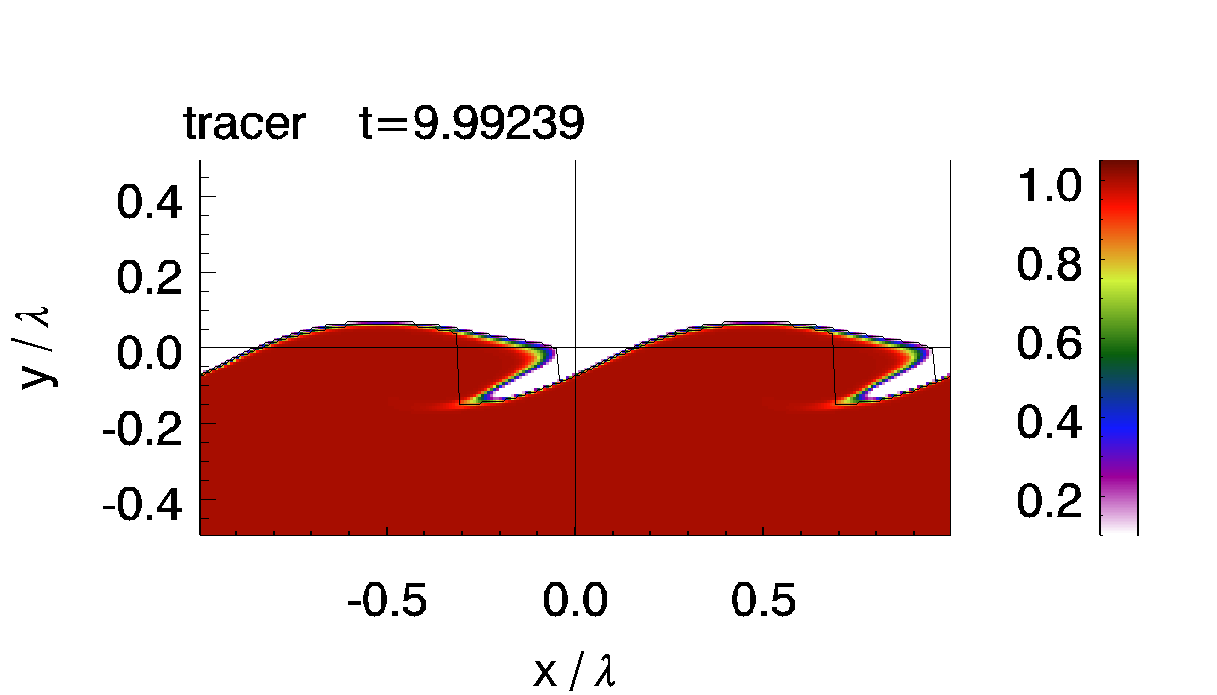}
\includegraphics[trim=190 150 250 160,clip,height=1.5cm]{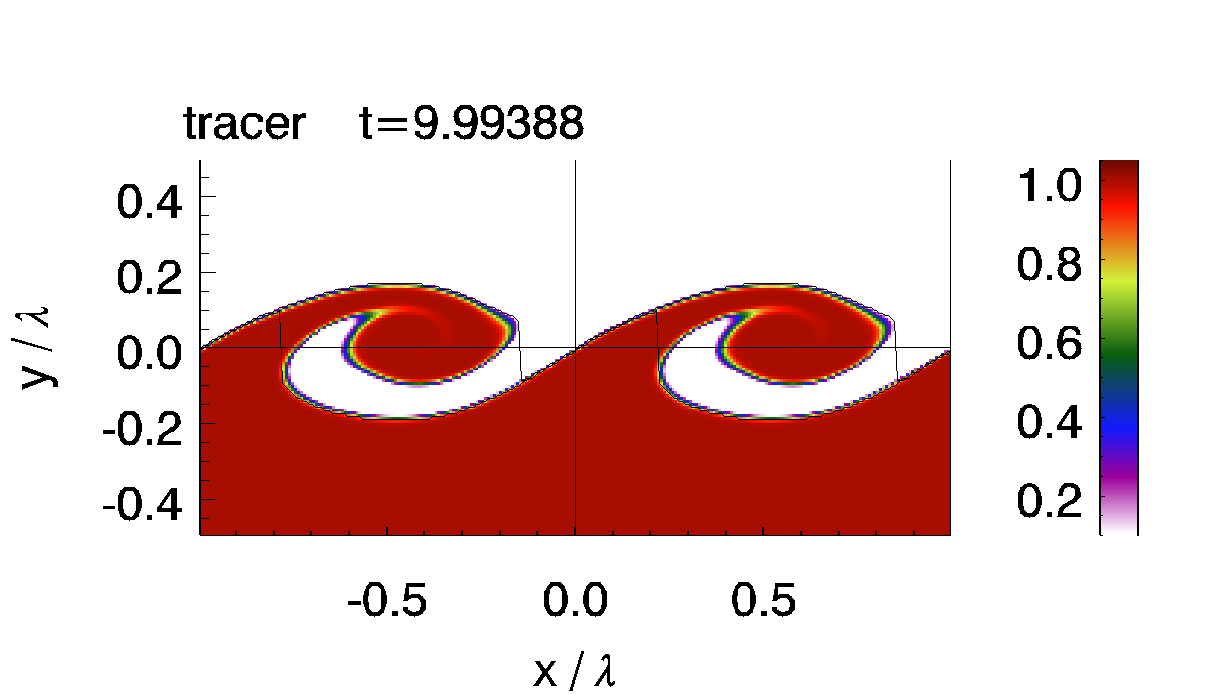}
\includegraphics[trim=190 150 250 160,clip,height=1.5cm]{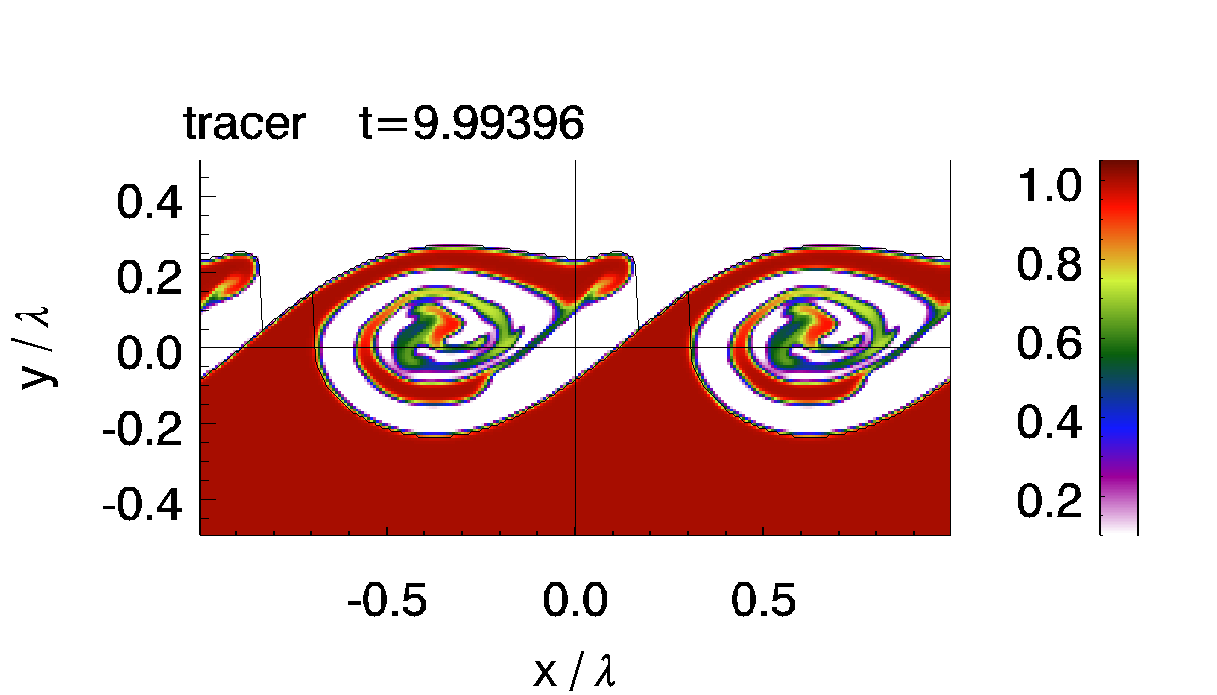}
\includegraphics[trim=190 150     0 160,clip,height=1.5cm]{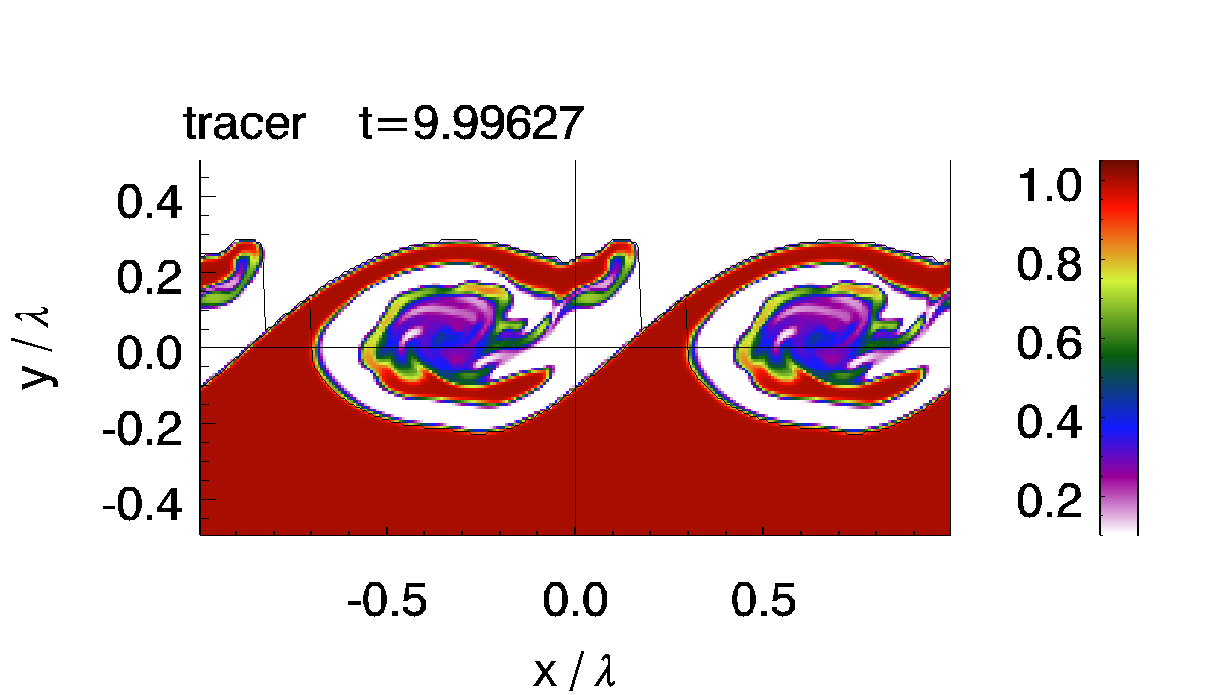}
\newline
\rotatebox{90}{\phantom{xx}$20\tau\KHinvisc$}\hfill%
\includegraphics[trim=0     0 250 160,clip,height=2.09cm]{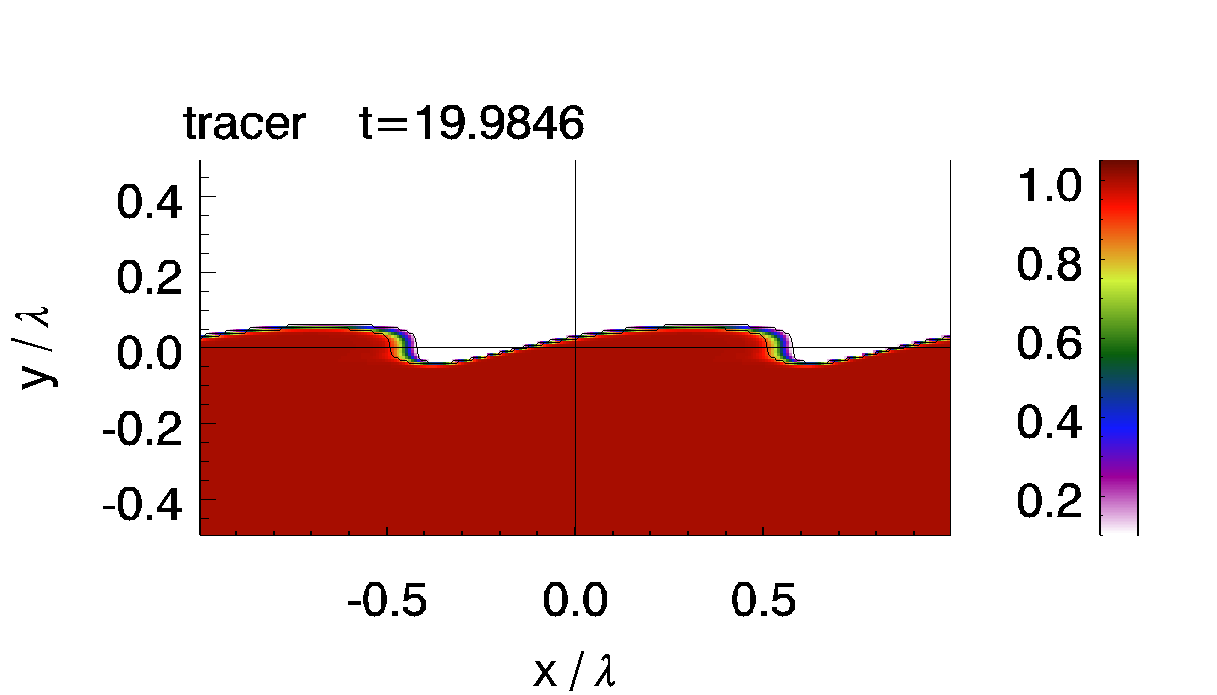}
\includegraphics[trim=190 0 250 160,clip,height=2.09cm]{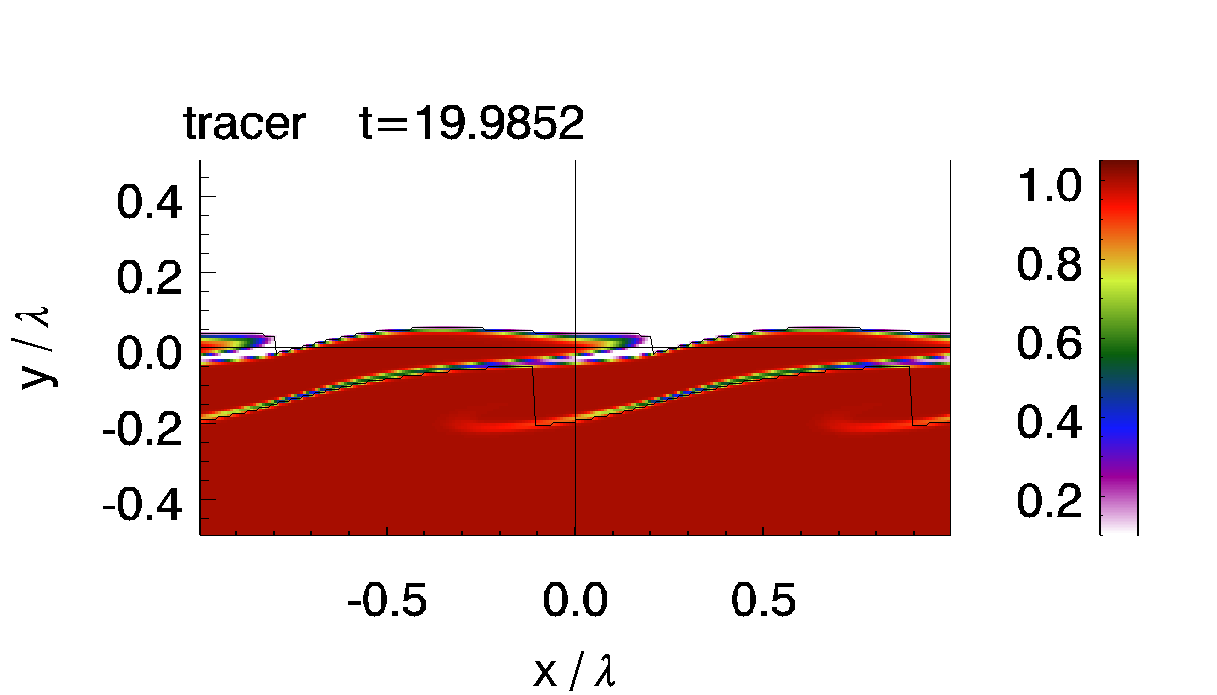}
\includegraphics[trim=190 0 250 160,clip,height=2.09cm]{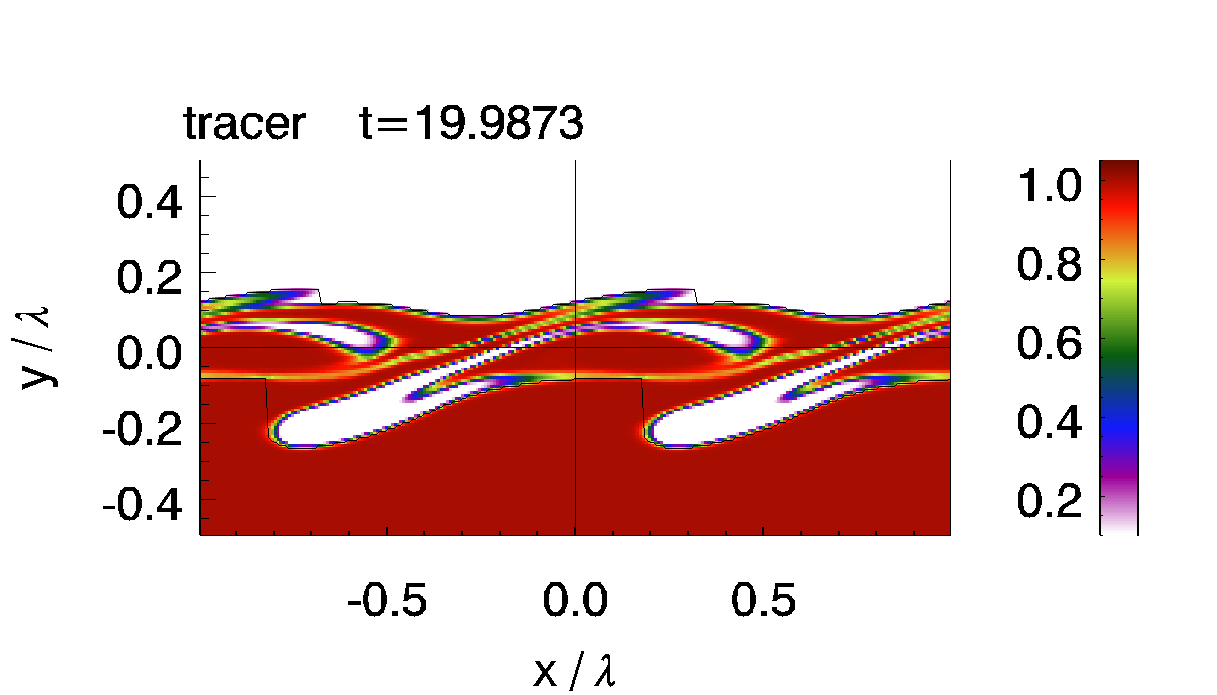}
\includegraphics[trim=190 0 250 160,clip,height=2.09cm]{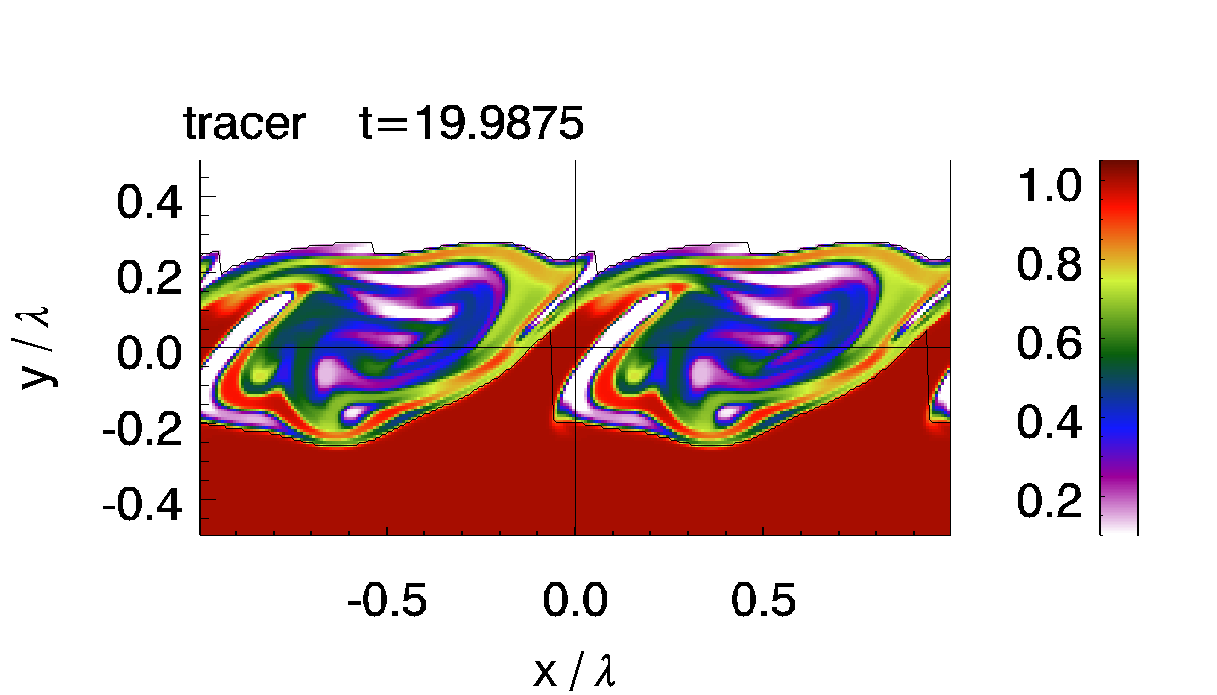}
\includegraphics[trim=190 0     0 160,clip,height=2.09cm]{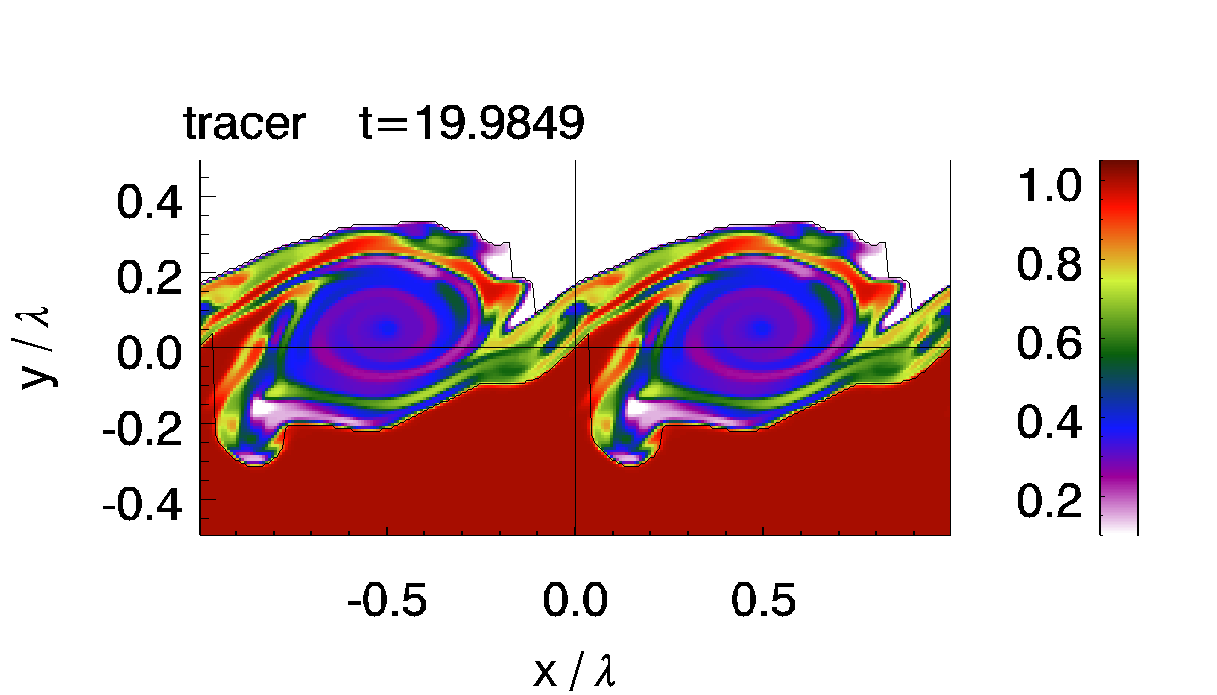}
\caption{Tracer slices for simulations with Spitzer viscosity, density ratio 2, and shear flow of Mach number 0.5. Columns are for different $\Reyn$ (see top of each column), where the Reynolds number is given for the hot (upper) layer. Re is 11 times higher in the cool (bottom) layer. For $\Reyn=1000$ and $\Reyn=10^4$ we smoothed the initial interface over 1\% and 2\% of the perturbation length scale, respectively, to suppress secondary instabilities (see Eqn.~\ref{eq:smooth}).}
\label{fig:rolls_Sp_D2_M05}
\end{center}
\end{figure*}
%

\begin{figure*}
\begin{center}
\hspace{0.5cm} $\Reyn=3$ \hfill $\Reyn=10$ \hfill $\Reyn=30$ \hfill $\Reyn=100$ \hfill $\Reyn=1000$ \hfill\phantom{x}\newline
\rotatebox{90}{\phantom{xx}$4\tau\KHinvisc$}\hfill%
\includegraphics[trim=0     150 250 160,clip,height=1.5cm]{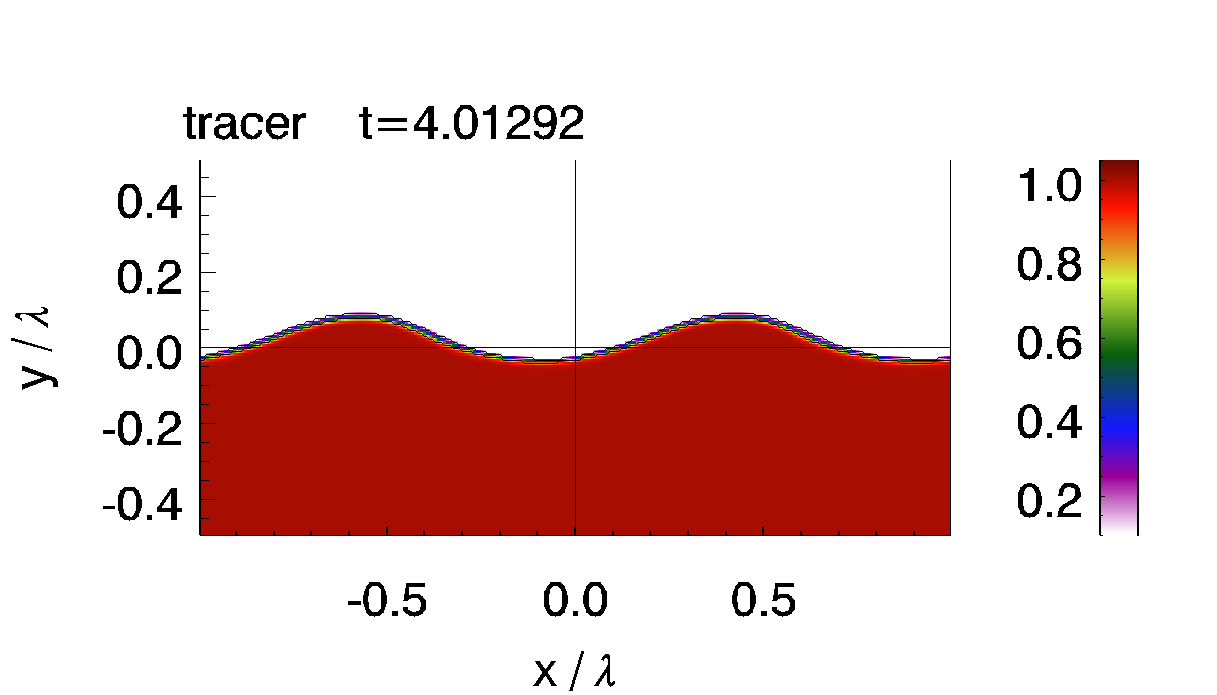}
\includegraphics[trim=190 150 250 160,clip,height=1.5cm]{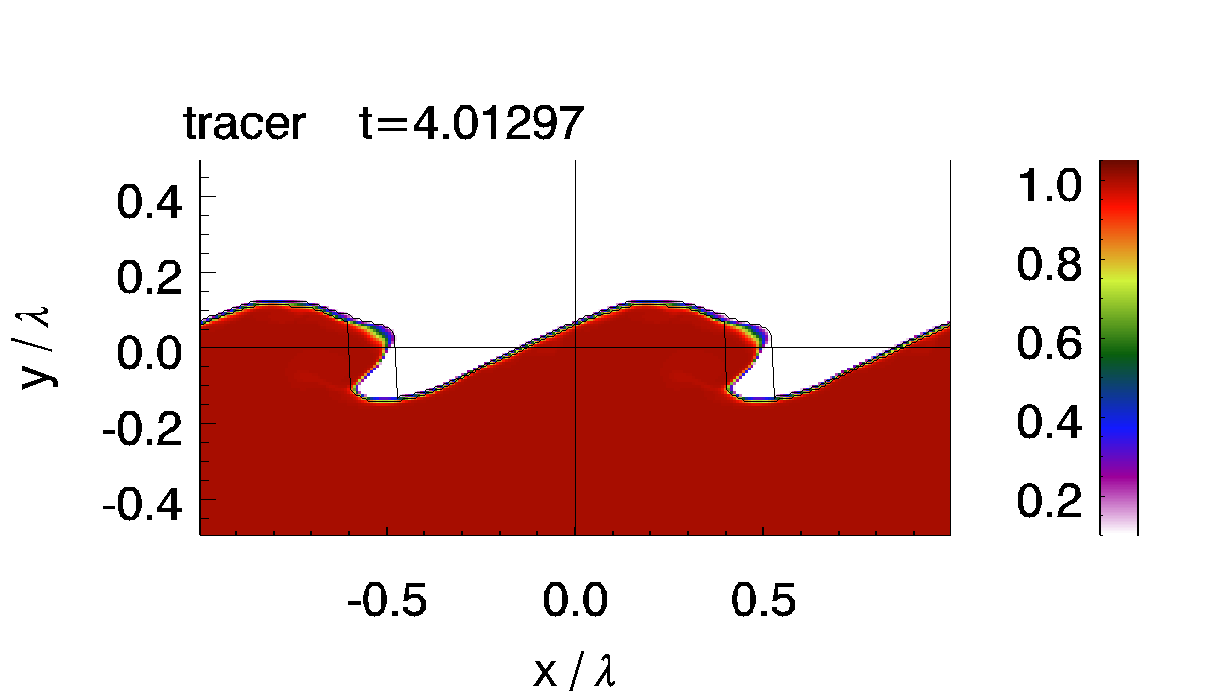}
\includegraphics[trim=190 150 250 160,clip,height=1.5cm]{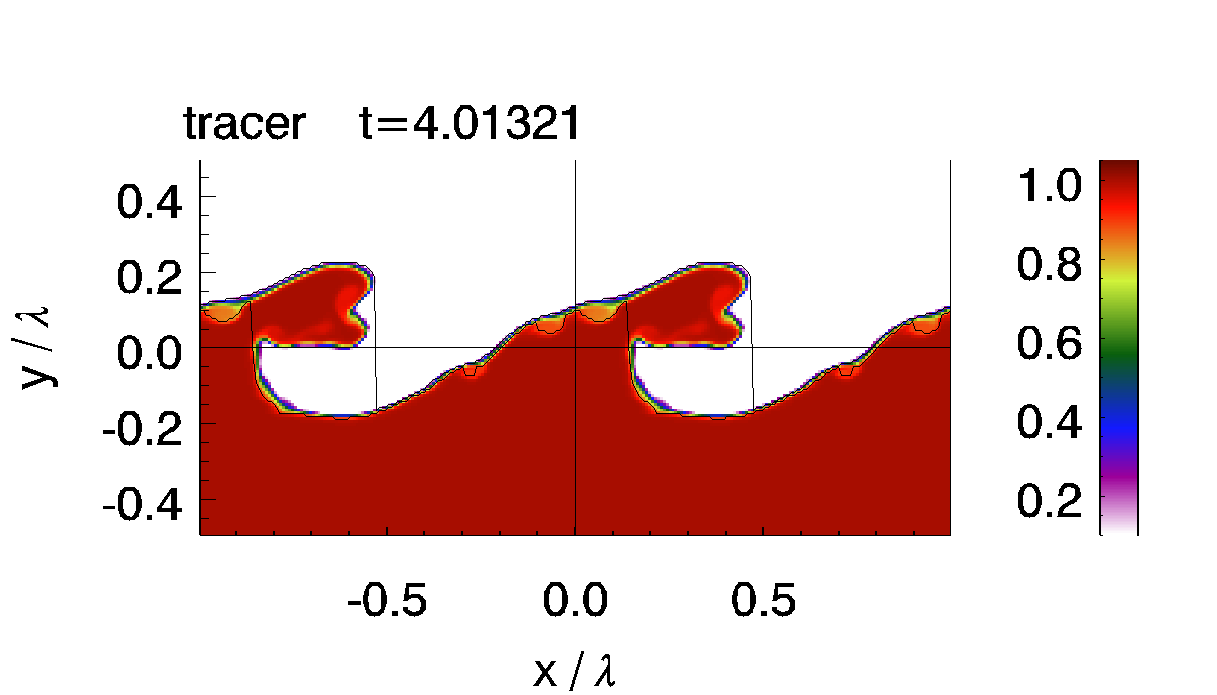}
\includegraphics[trim=190 150 250 260,clip,height=1.5cm]{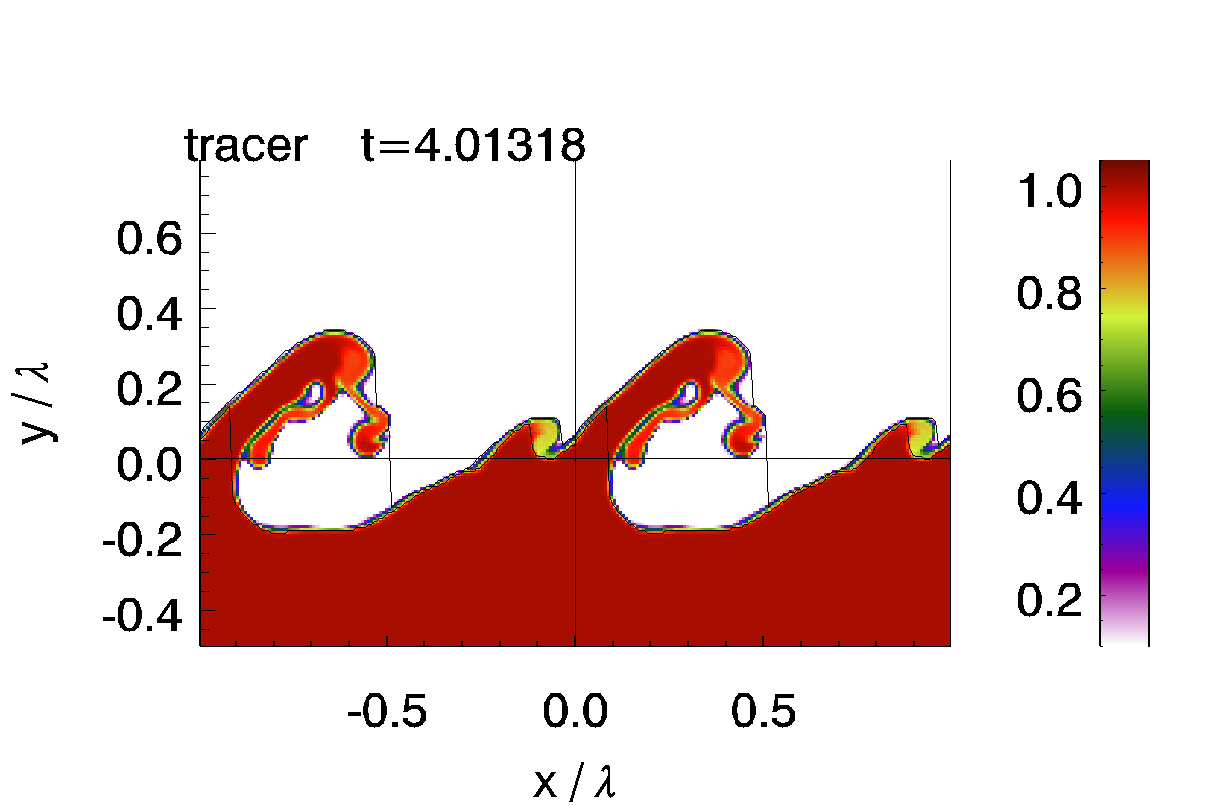}
\includegraphics[trim=190 150     0 260,clip,height=1.5cm]{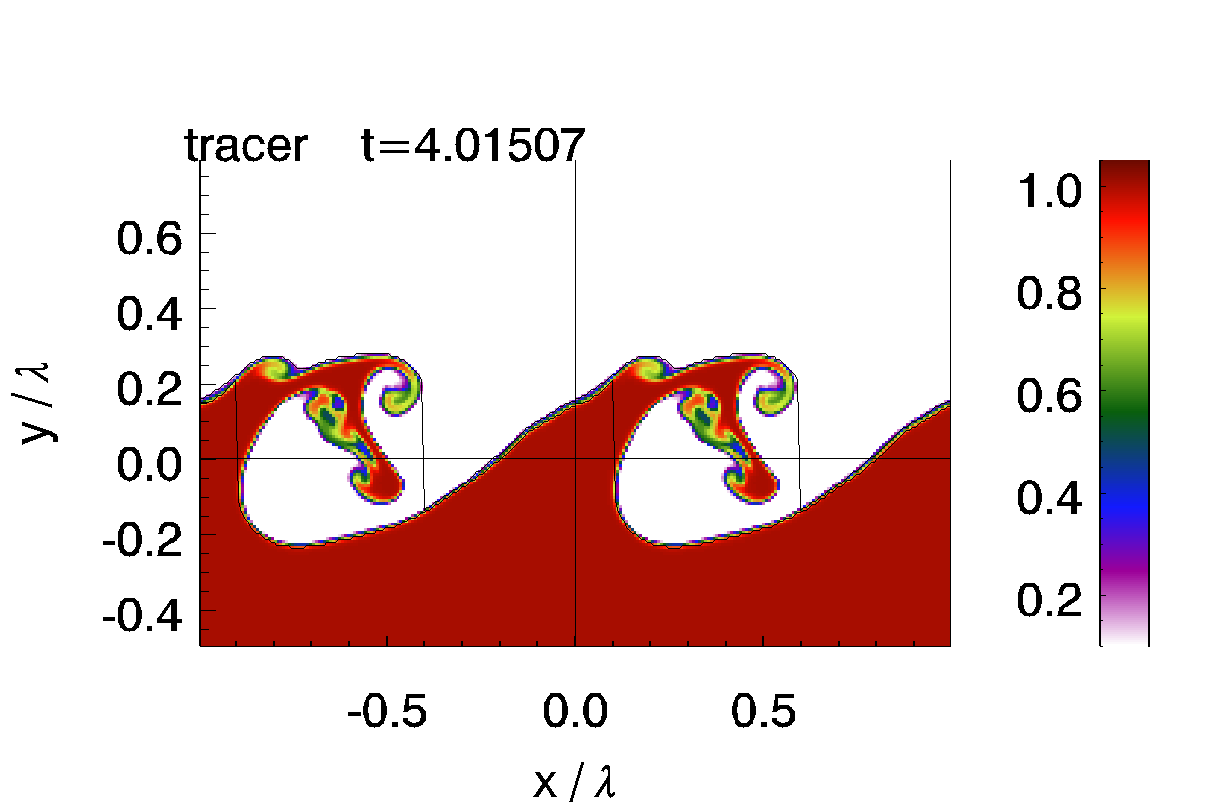}
\newline
\rotatebox{90}{\phantom{xx}$10\tau\KHinvisc$}\hfill%
\includegraphics[trim=0     150 250 160,clip,height=1.5cm]{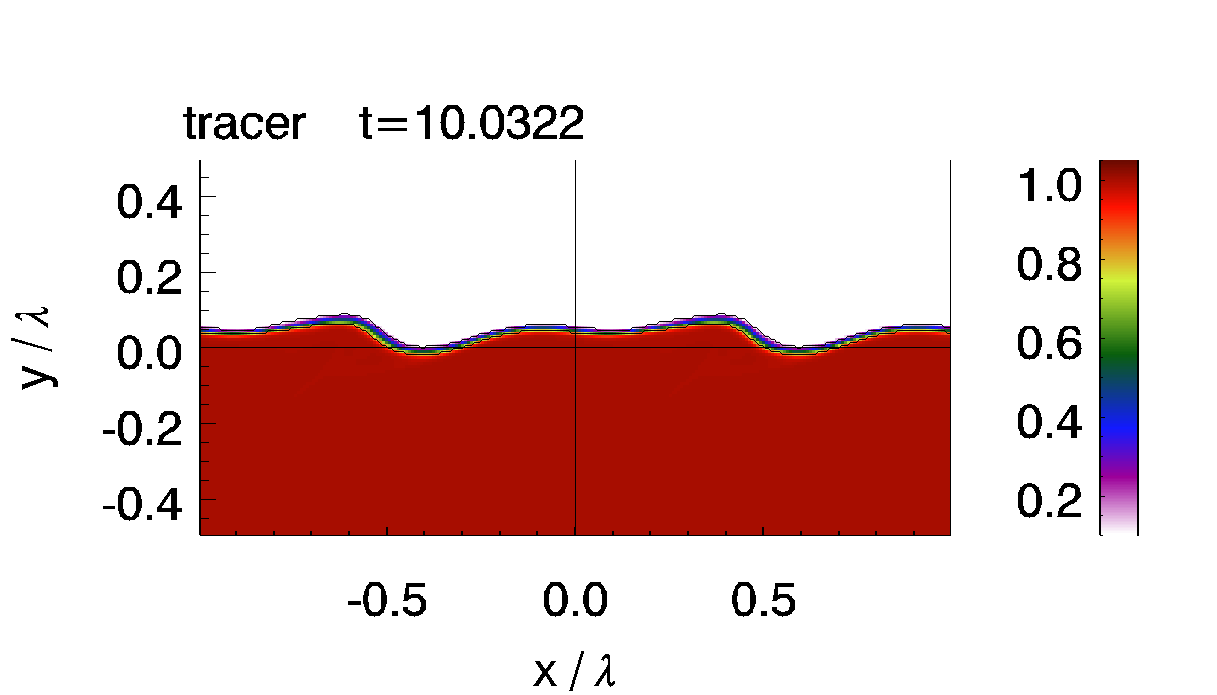}
\includegraphics[trim=190 150 250 160,clip,height=1.5cm]{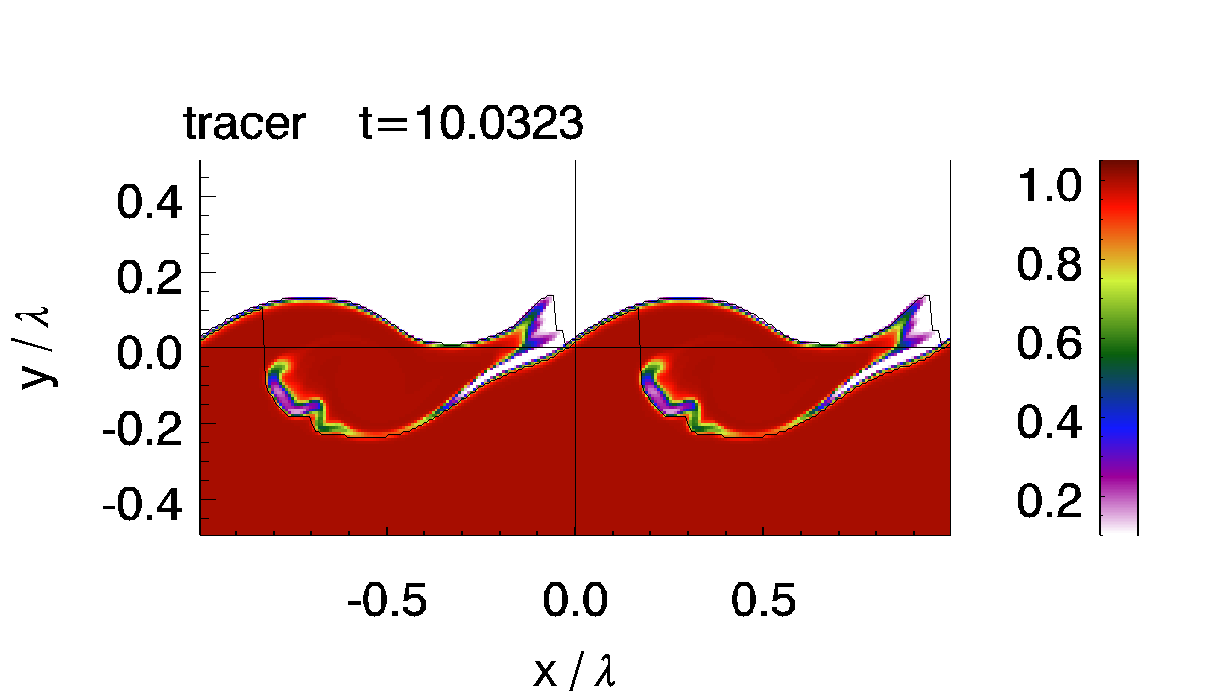}
\includegraphics[trim=190 150 250 160,clip,height=1.5cm]{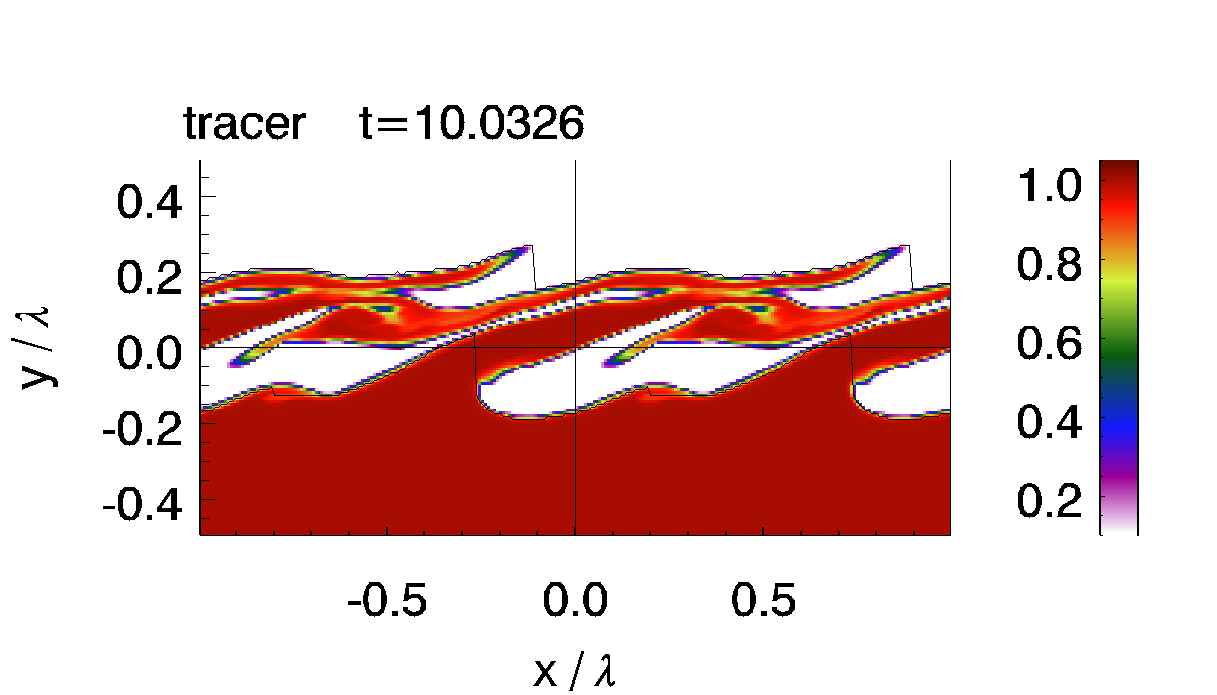}
\includegraphics[trim=190 150 250 170,clip,height=1.82cm]{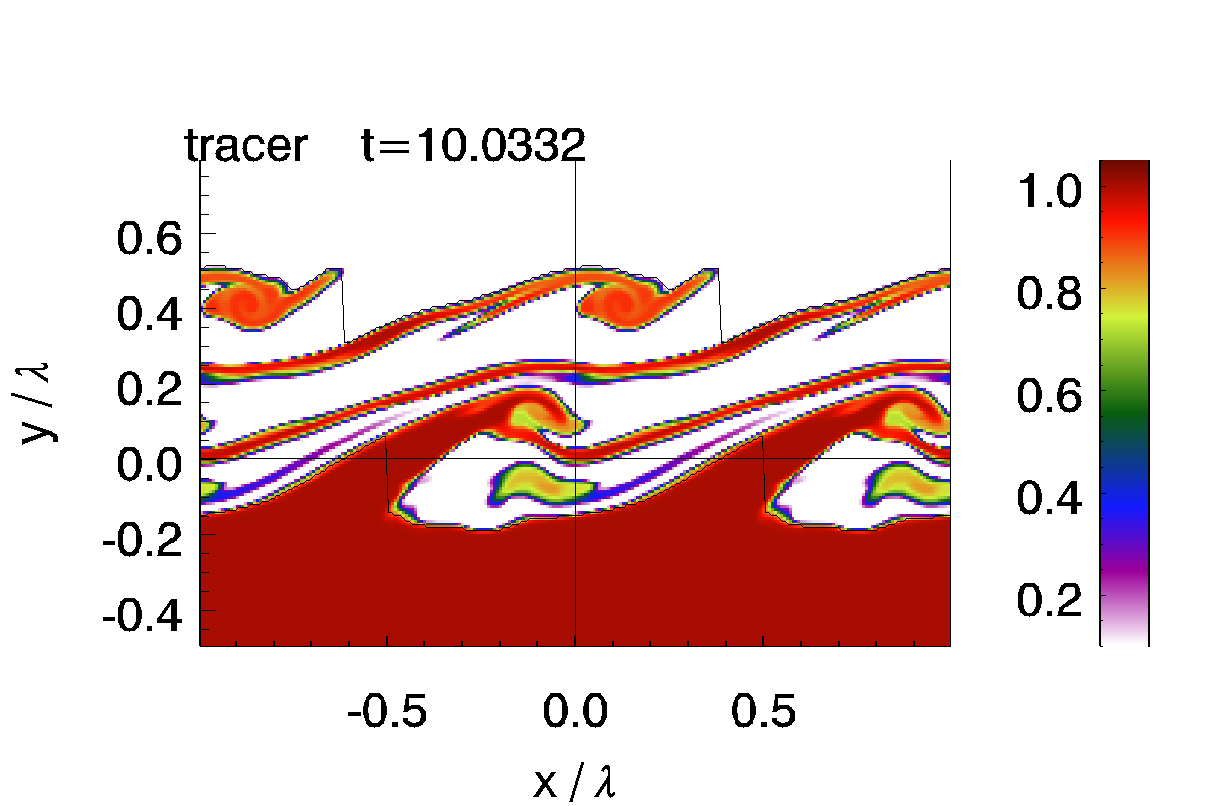}
\includegraphics[trim=190 150     0 170,clip,height=1.82cm]{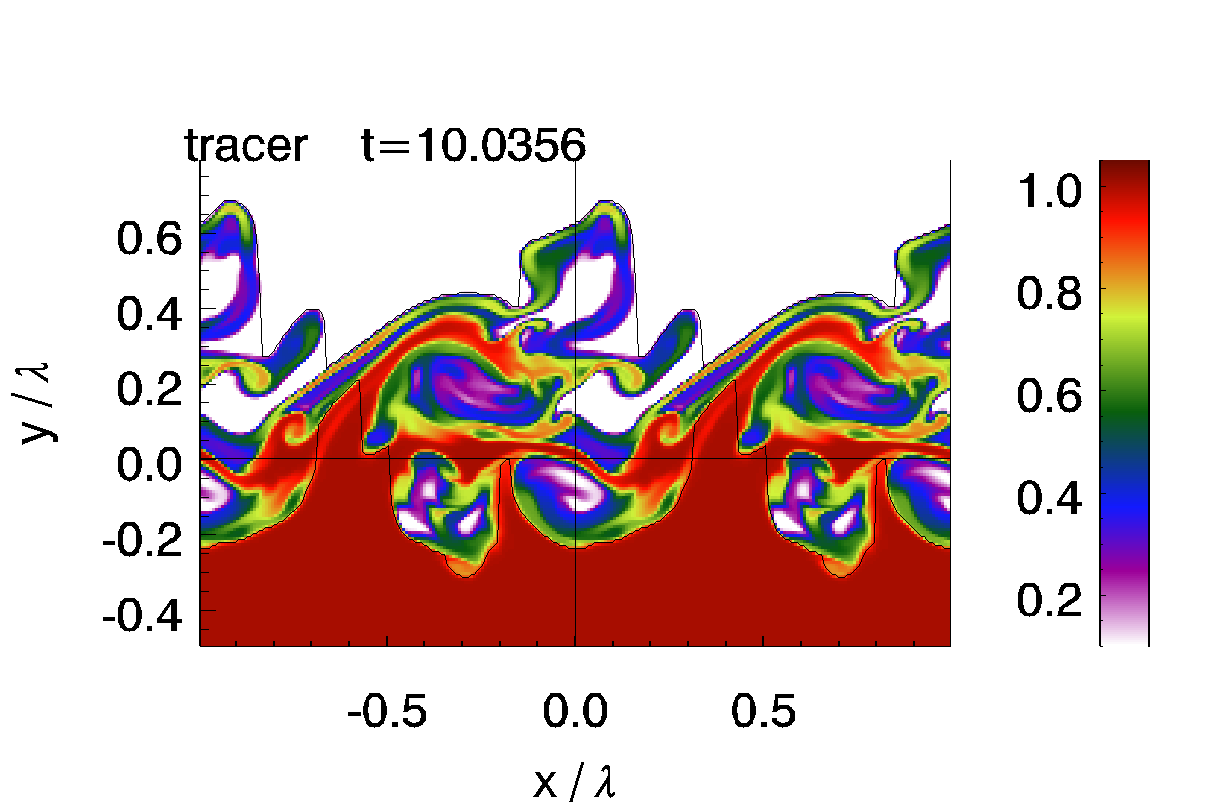}
\newline
\rotatebox{90}{\phantom{xx}$15\tau\KHinvisc$}\hfill%
\includegraphics[trim=0     0 250 160,clip,height=2.1cm]{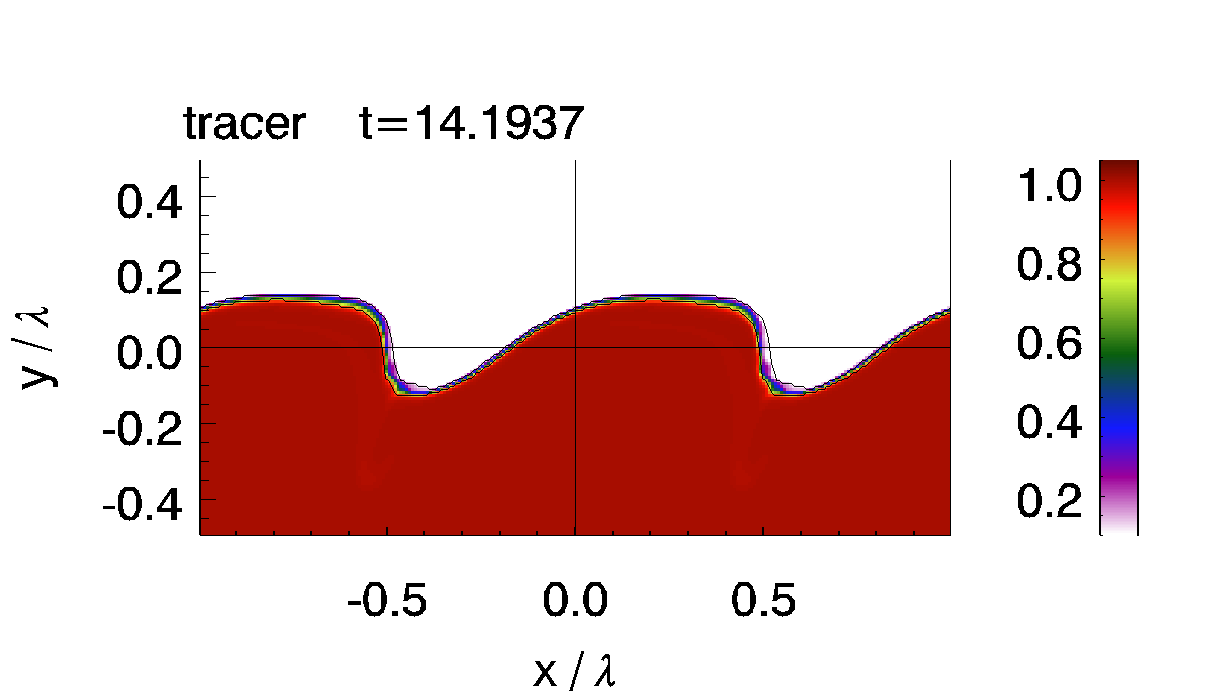}
\includegraphics[trim=190 0 250 160,clip,height=2.1cm]{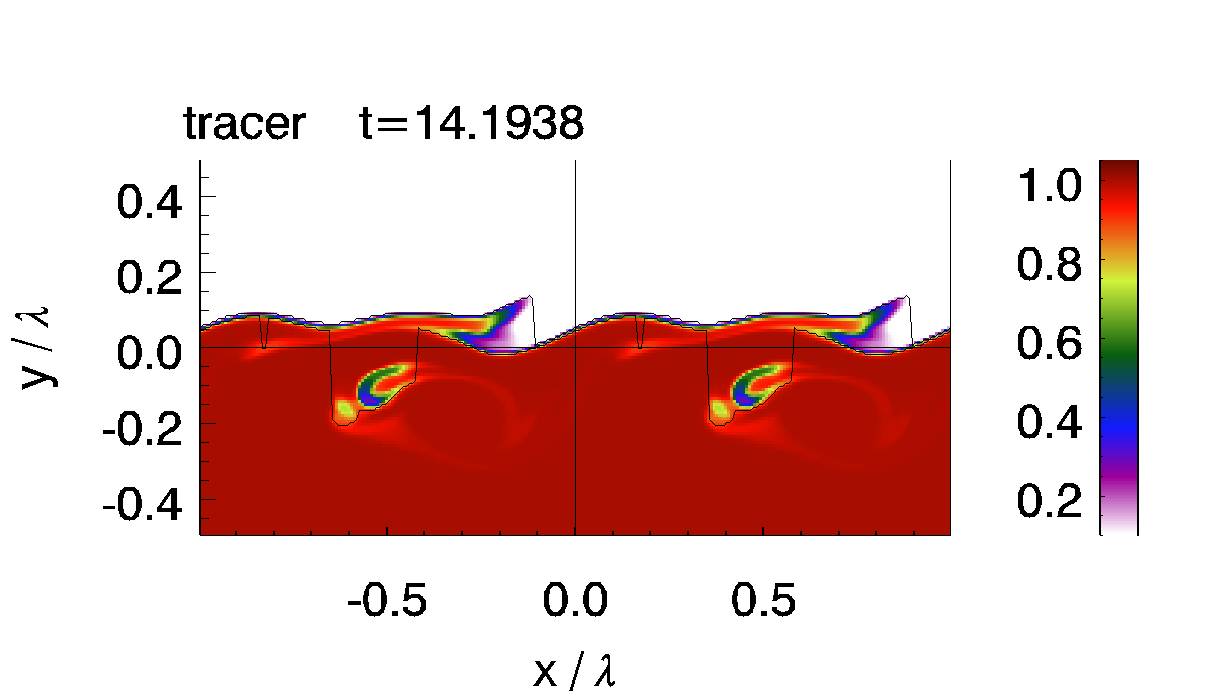}
\includegraphics[trim=190 0 250 160,clip,height=2.1cm]{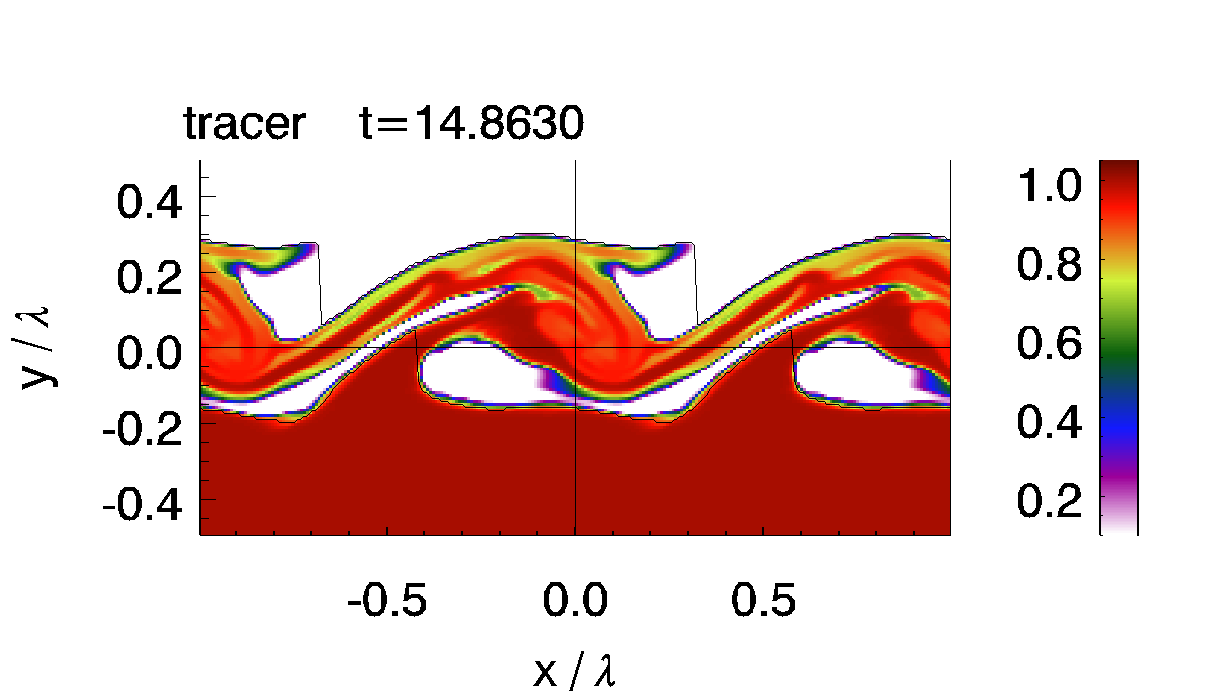}
\includegraphics[trim=190 0 250 170,clip,height=2.3cm]{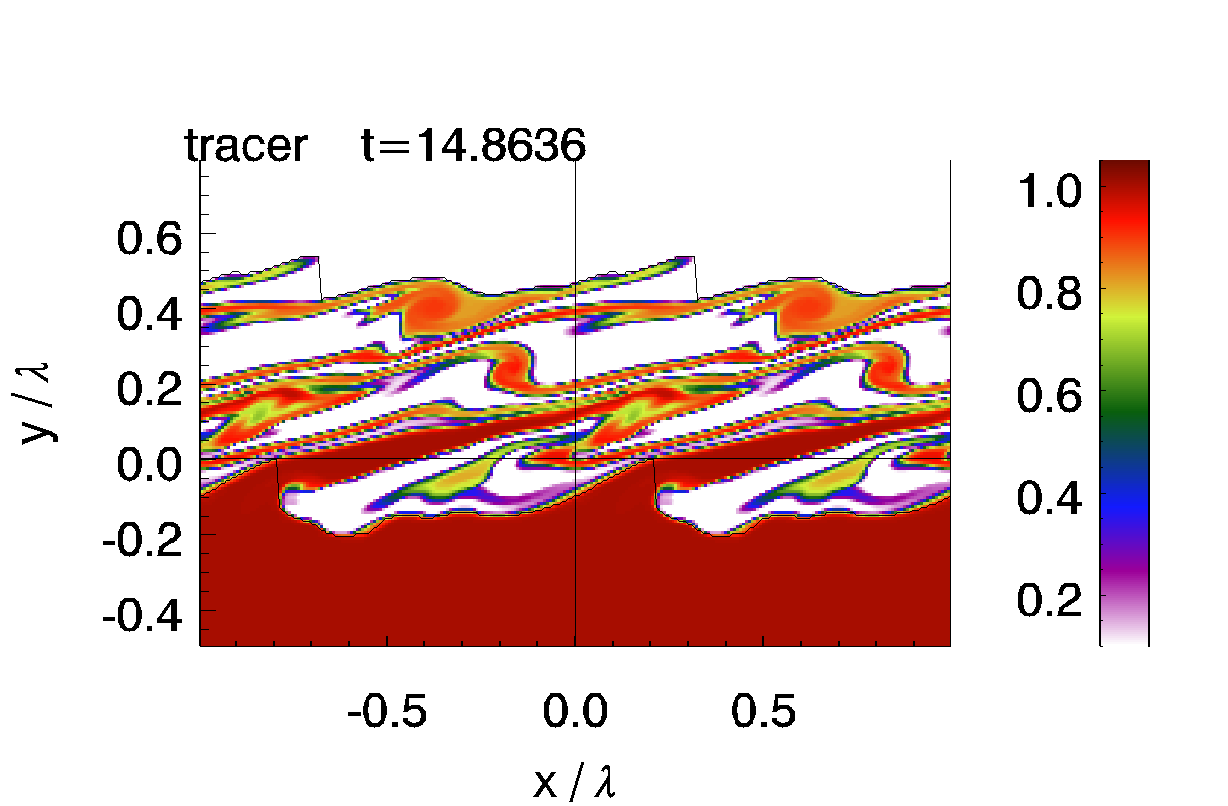}
\includegraphics[trim=190 0 0     170,clip,height=2.3cm]{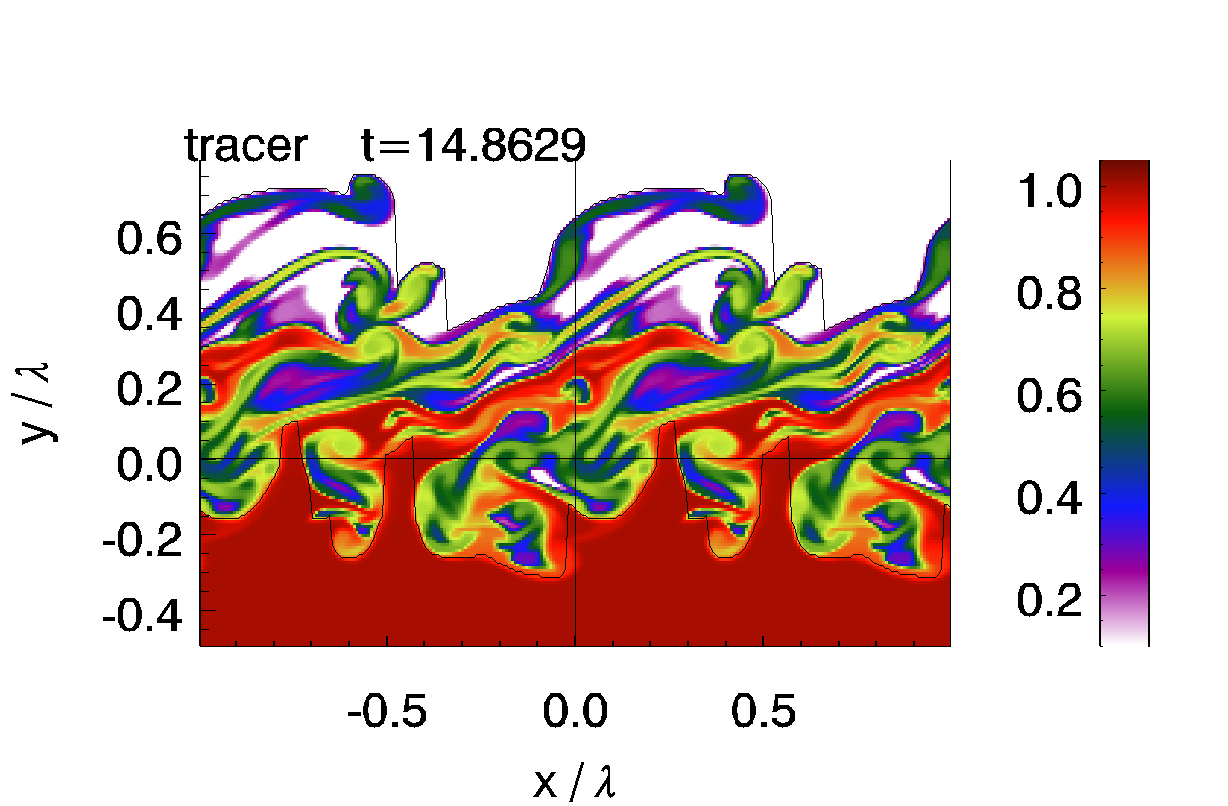}
\caption{Tracer slices for simulations with Spitzer viscosity, density ratio 10, and shear flow of Mach number 0.5. Columns are for different $\Reyn$ (see top of each column), where the Reynolds number is given for the hot (upper) layer. Re is formally 3000 times higher in the cool (bottom) layer.  Fig.~\ref{fig:Tempflow_Sp_D10_Re10} shows a zoom-in on the flow patterns for the $\Reyn=10$ case. For $\Reyn \ge 30$ we smoothed the initial interface over 1\% of the perturbation length and for $\Reyn = 1000$ over 2\% of the perturbation length to suppress secondary instabilities (see Eqn.~\ref{eq:smooth}).}
\label{fig:rolls_Sp_D10_M05}
\end{center}
\end{figure*}

\begin{figure*}
\includegraphics[trim=0 0 10 0,clip,width=0.45\textwidth]{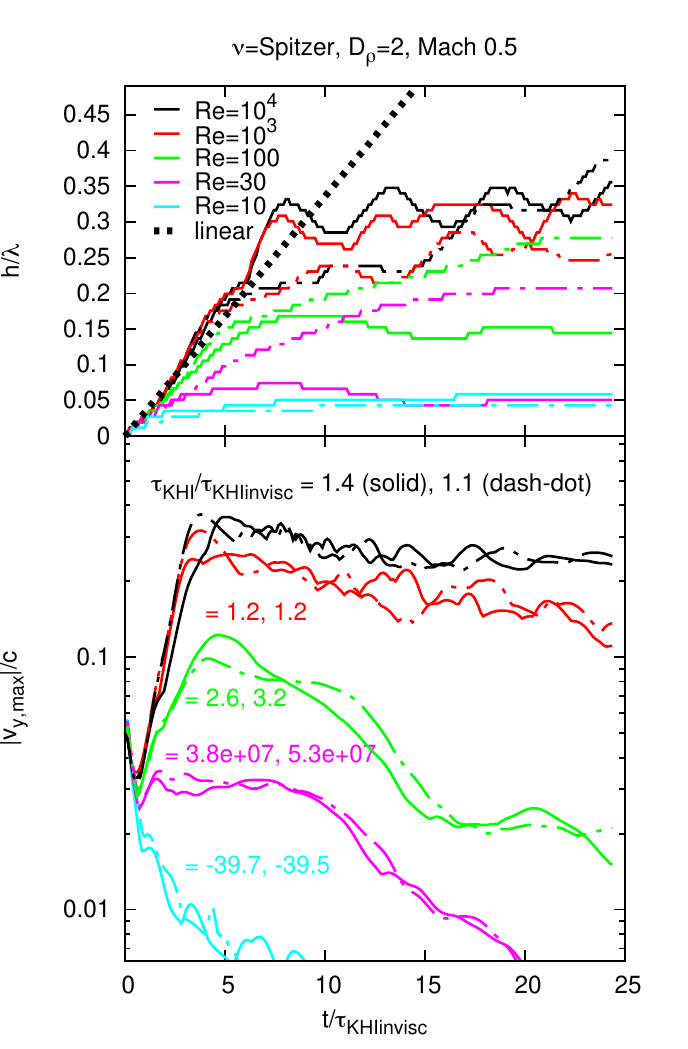}
\includegraphics[trim=0 0 10 0,clip,width=0.45\textwidth]{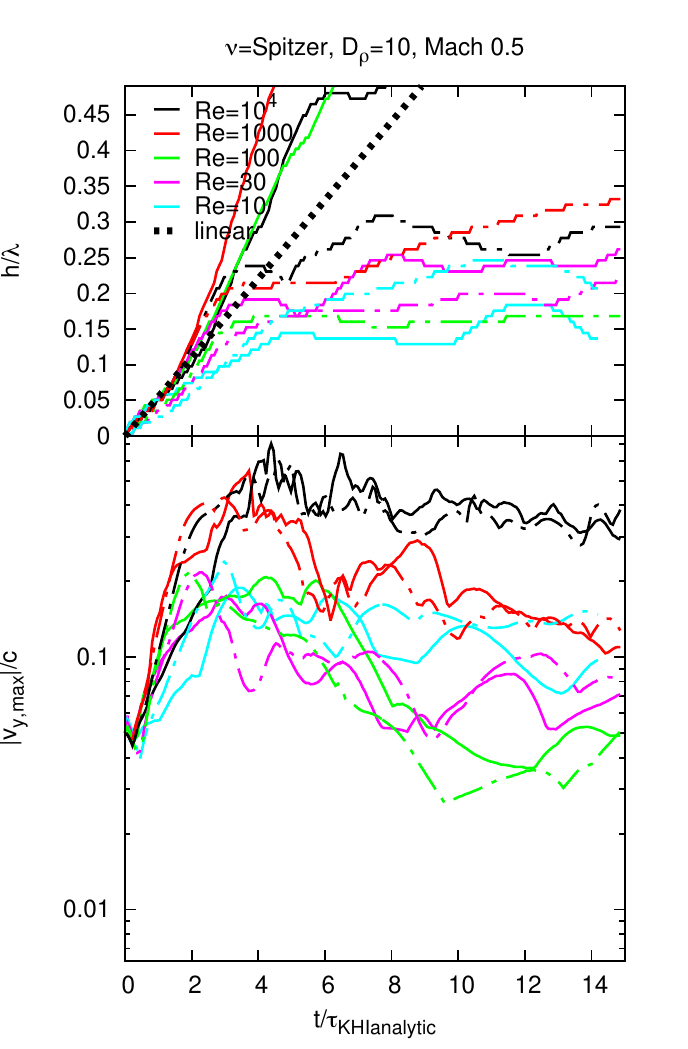}
\caption{Temporal evolution of the thickness of the KHI mixing layer (top) and of the $v_y$ velocity extrema. All runs are for a Spitzer viscosity and shear flow = Mach  0.5. The two columns are for density contrasts 2 and 10, see label at top. Different line colours code different Reynolds numbers in the hot layer, see legend. Exponential growth times for the velocity extrema are given in the bottom panels with the matching font colour. See Fig.~\ref{fig:thick_vely_nu_M05} and Sect.~\ref{sec:spitzer} for details.}
\label{fig:thick_vely_Sp_M05}
\end{figure*}

Figures \ref{fig:rolls_Sp_D2_M05} and \ref{fig:rolls_Sp_D10_M05} display tracer slices for the Spitzer-viscous KHI. In Fig.~\ref{fig:thick_vely_Sp_M05} we plot the evolution of the height of the KH rolls and of  $v_y{}\Min$ and $v_y{}\Max$ for different Reynolds numbers. Qualitatively, the same trends as before apply. Increasing viscosity slows down the rolling up of the interface, its widening, and finally suppresses the instability. 

For $D_{\rho}=2$ the Spitzer-viscous KHI evolves similarly to the constant $\nu$ case except for minor differences. The critical Reynolds number is reduced to 30 compared to 200 in the equal kinematic viscosity case. The KHI is first suppressed on the hot side, i.e., at $\Reyn$ somewhat larger than $\Reyn\Crit$ no cool fingers are drawn upwards, but hot fingers can be drawn downwards.

\begin{figure*}
colour-coded temperature in keV with velocity vectors \hfill\hfill vertical velocity $v_y/c_s$  \hfill \phantom{x}\newline
\includegraphics[trim=  0 113 0 150,clip,width=0.49\textwidth]{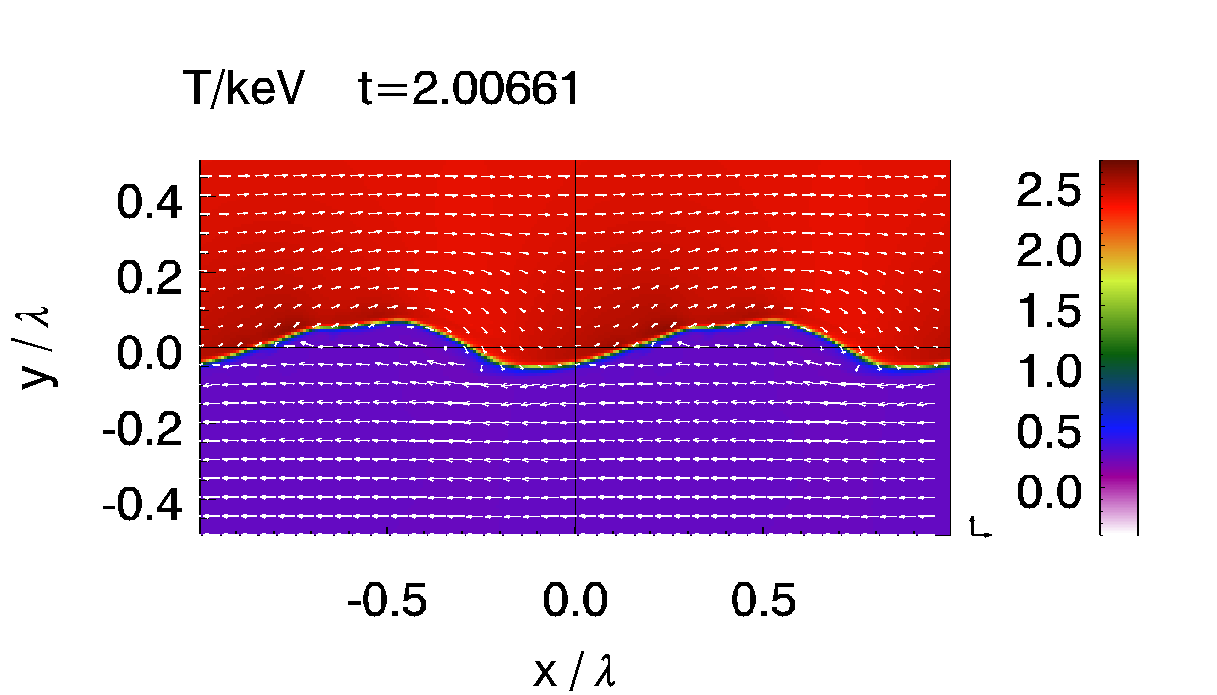}
\includegraphics[trim=90 113 0 150,clip,width=0.46\textwidth]{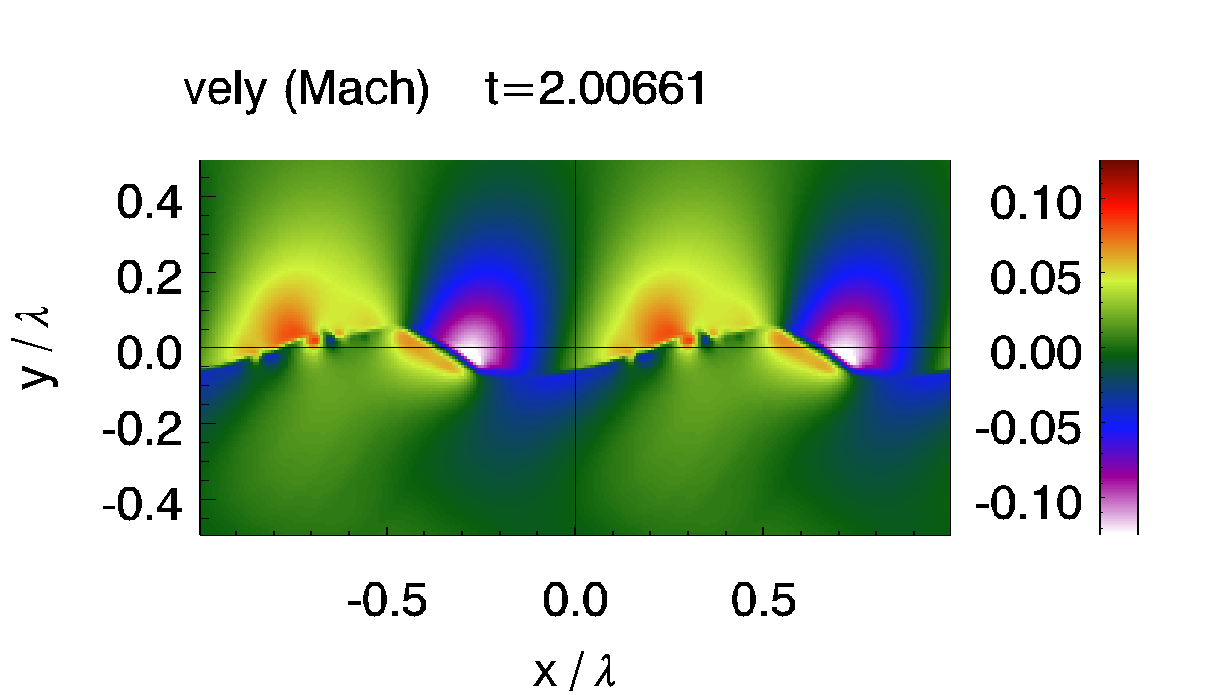}
\includegraphics[trim=  0 113 0 150,clip,width=0.49\textwidth]{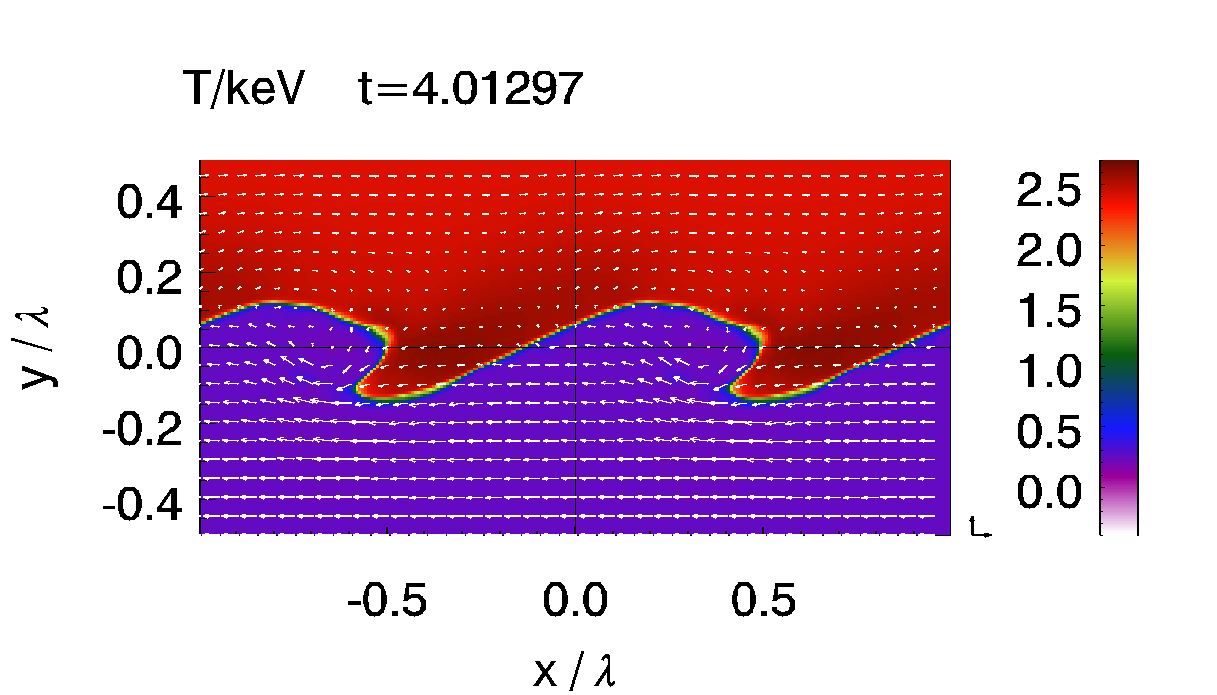}
\includegraphics[trim=90 113 0 150,clip,width=0.46\textwidth]{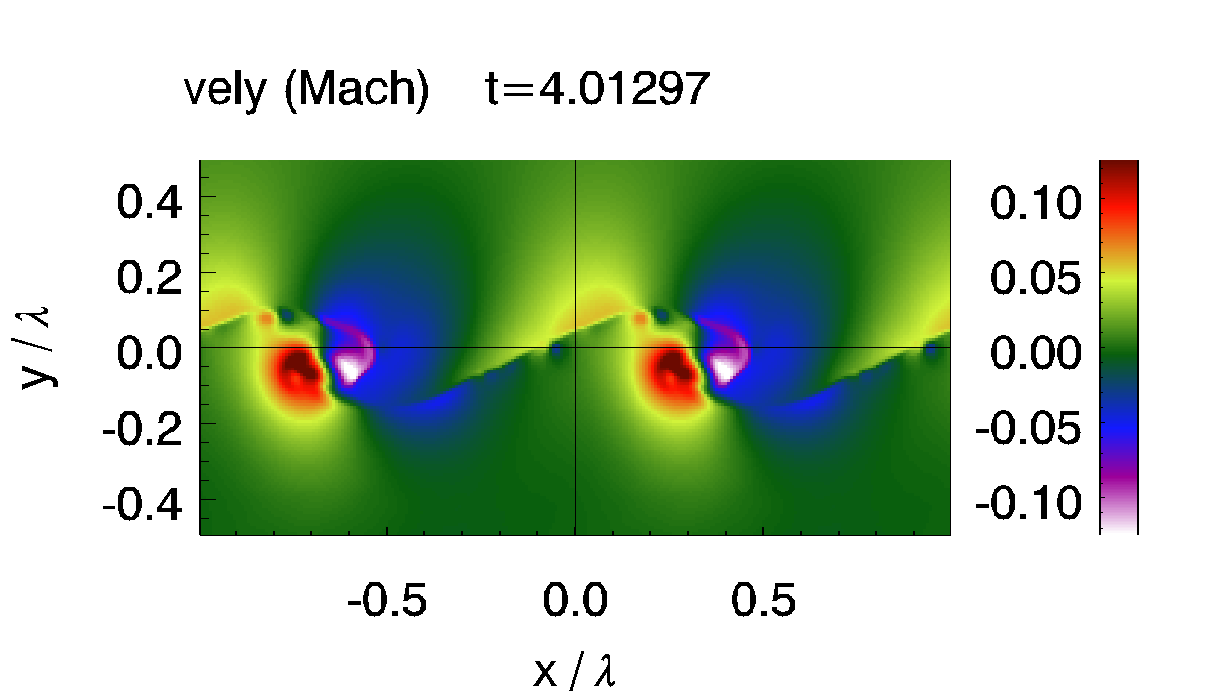}
\includegraphics[trim=  0 113 0 150,clip,width=0.49\textwidth]{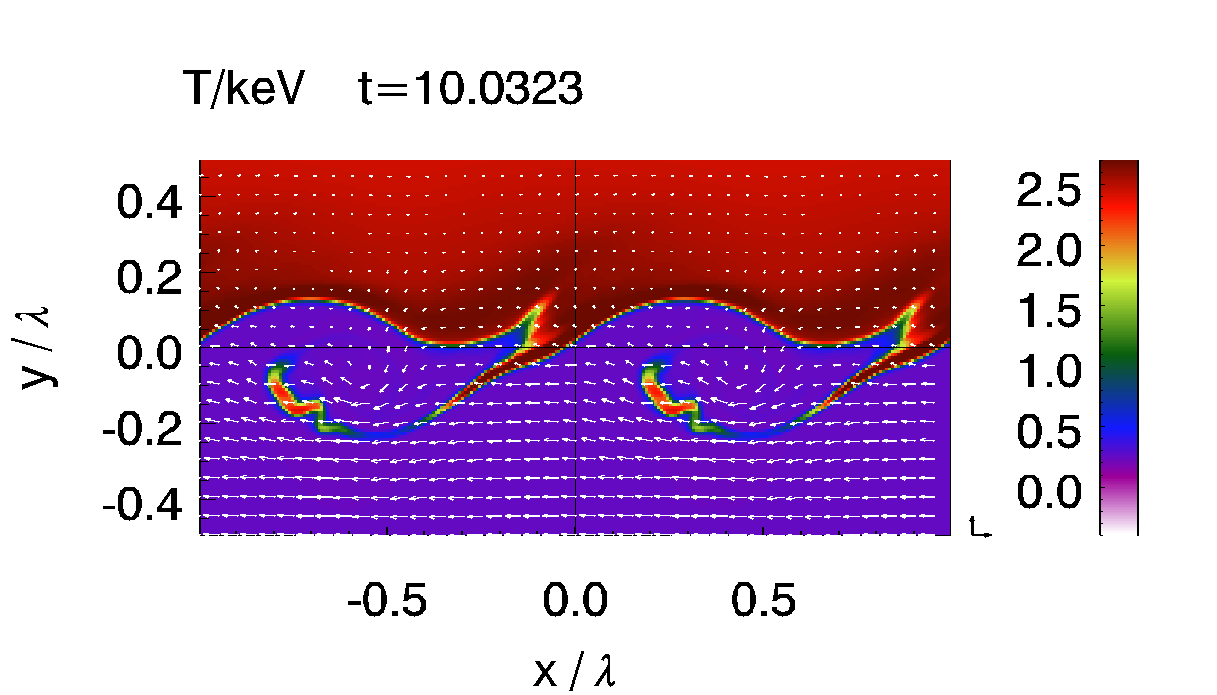}
\includegraphics[trim=90 113 0 150,clip,width=0.46\textwidth]{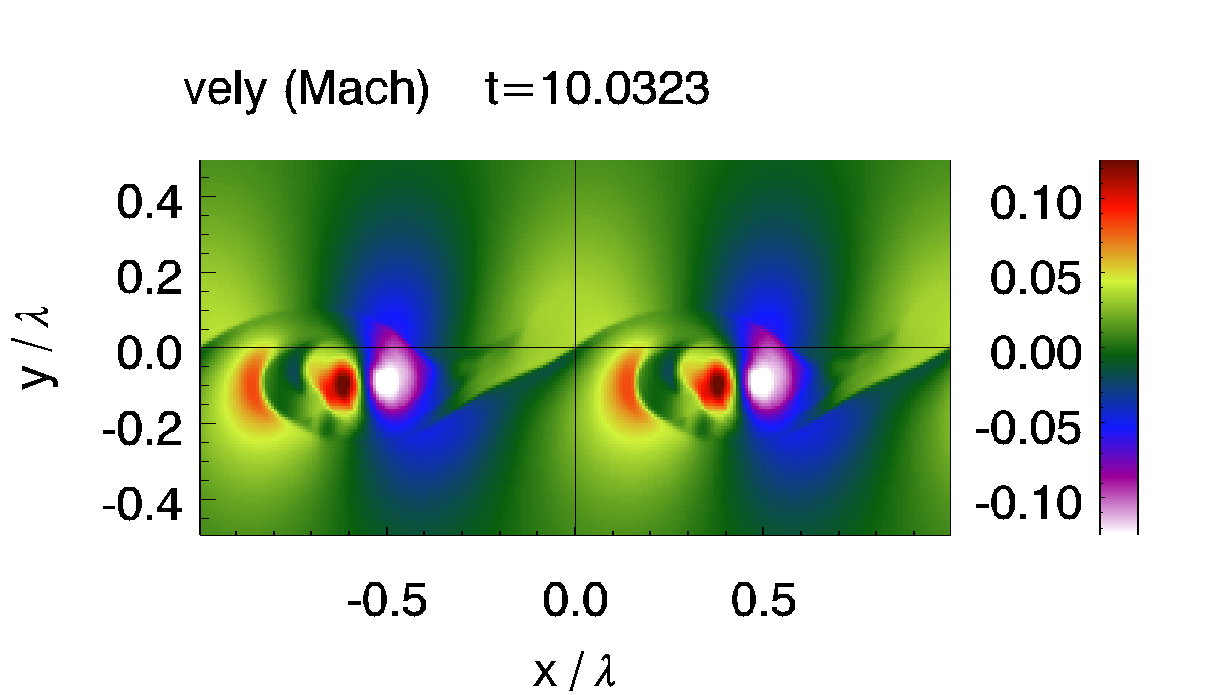}
\includegraphics[trim=  0 0 0 150,clip,width=0.49\textwidth]{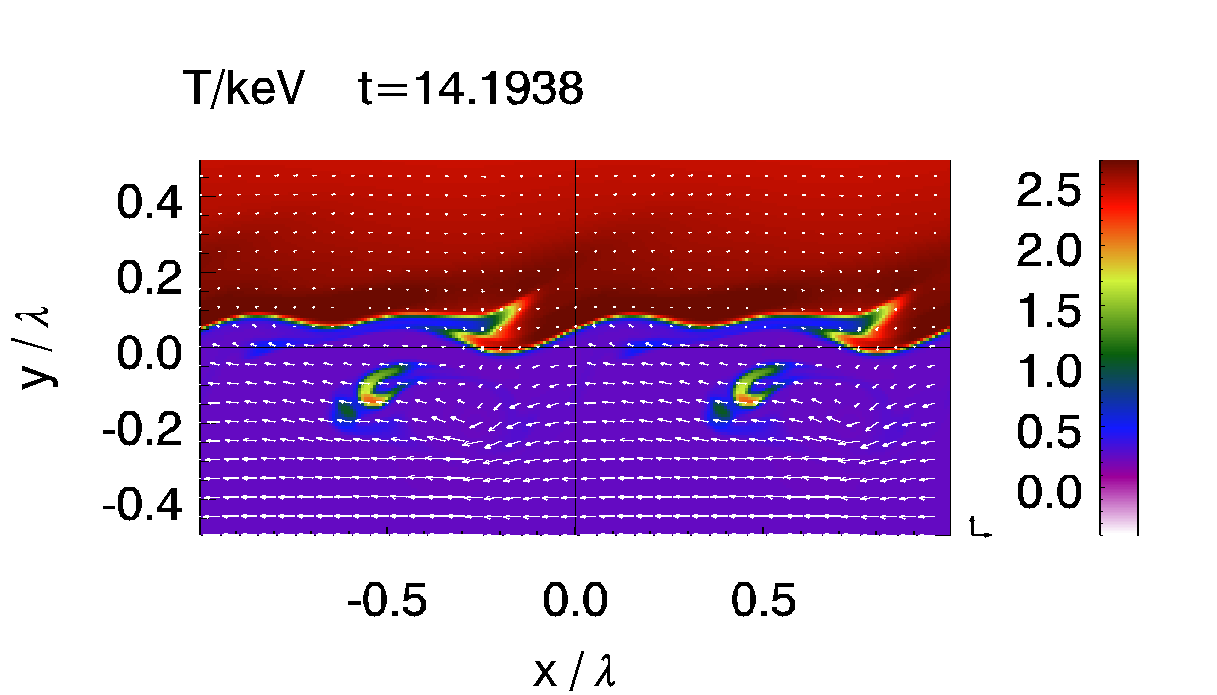}
\includegraphics[trim=90 0 0 150,clip,width=0.46\textwidth]{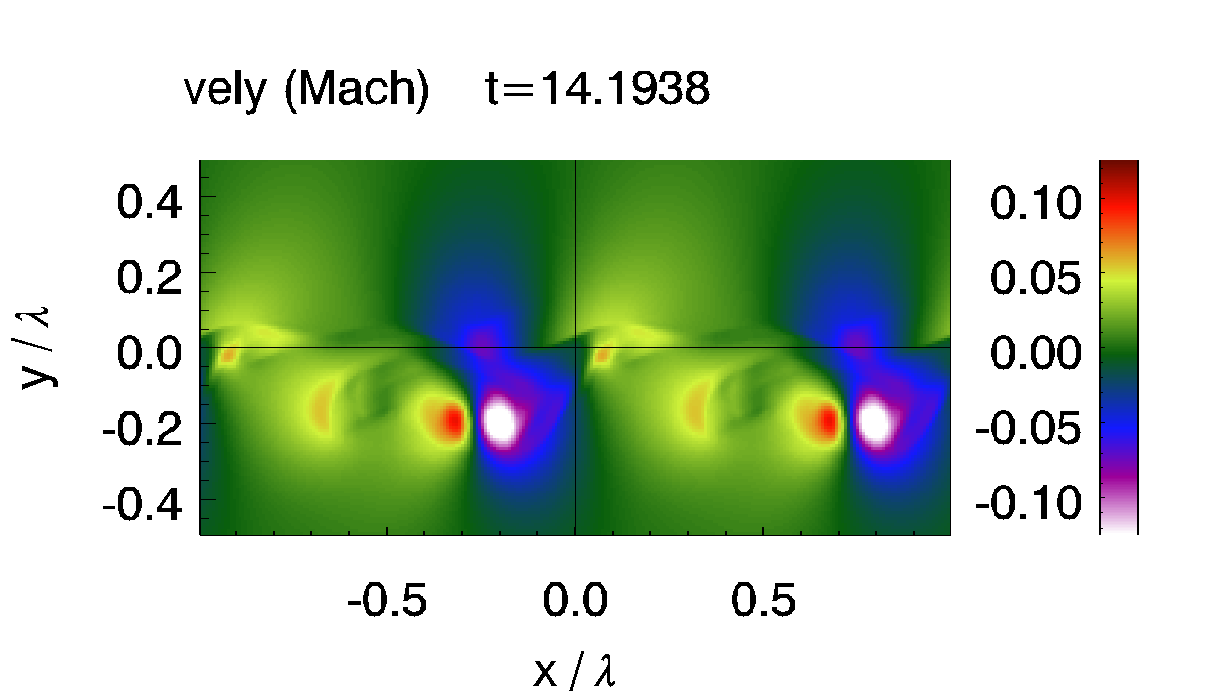}
\caption{Details of the flow patterns for Spitzer viscosity, density contrast 10, shear flow Mach 0.5, Reynolds numbers 10 and $3160$ in hot and cold layer, respectively. The left column shows the colour-coded temperature  in keV with velocity vectors overplotted.  The right column colour-codes $v_y/c_s$. The snapshots are at  timesteps 2, 4, 10, 15 $\tau\KHinvisc$ from top to bottom. Note that all turbulence occurs in the cold gas.}
\label{fig:Tempflow_Sp_D10_Re10}
\end{figure*}

At $D_{\rho}=10$, the high density contrast and the highly asymmetric viscosity lead to an  untypical and very irregular morphology of the KH rolls also at high Re. At $D_{\rho}=2$, the viscosity on the cool side can still add to the suppression of the KHI, whereas at $D_{\rho}=10$ the cool side is always turbulent ($\Reyn > 1000$). This leads to complex flow patterns in the cool layer even if the Reynolds number is low. The initial instability  induces vortices in the high-Reynolds number cold gas, which can remain there for a long time. This effect is shown for $\Reyn=10$ (in the hot layer) in Fig.~\ref{fig:Tempflow_Sp_D10_Re10}. Consequently, the maximum and minimum $v_y$ are not a good tracer of an instability anymore, because they mainly trace vortices in the cool gas. Thus, using $v_y{}\Max$ and $v_y{}\Min$ alone as a diagnostic for the growth of the KHI leads to the impression that the instability is not suppressed at all (right panel of Fig.~\ref{fig:thick_vely_Sp_M05}). The evolution of the height of the KH rolls and the snapshots, however, show that the KHI is suppressed for $\Reyn\lesssim 30$. Thus, at high density contrasts, a Spitzer-like viscosity puts the KHI in a hybrid state, where it is able to induce turbulence in the cold layer, but does not mix both fluids.
%
%


\subsection{High density contrast 100} \label{sec:highcontrast}

\subsubsection{Low viscosity}

\begin{figure}
\hspace{0.5cm} discontinuous interface \hfill smoothed interface \hfill\phantom{x}\newline
\includegraphics[trim=0     120 250 160,clip,height=2.8cm]{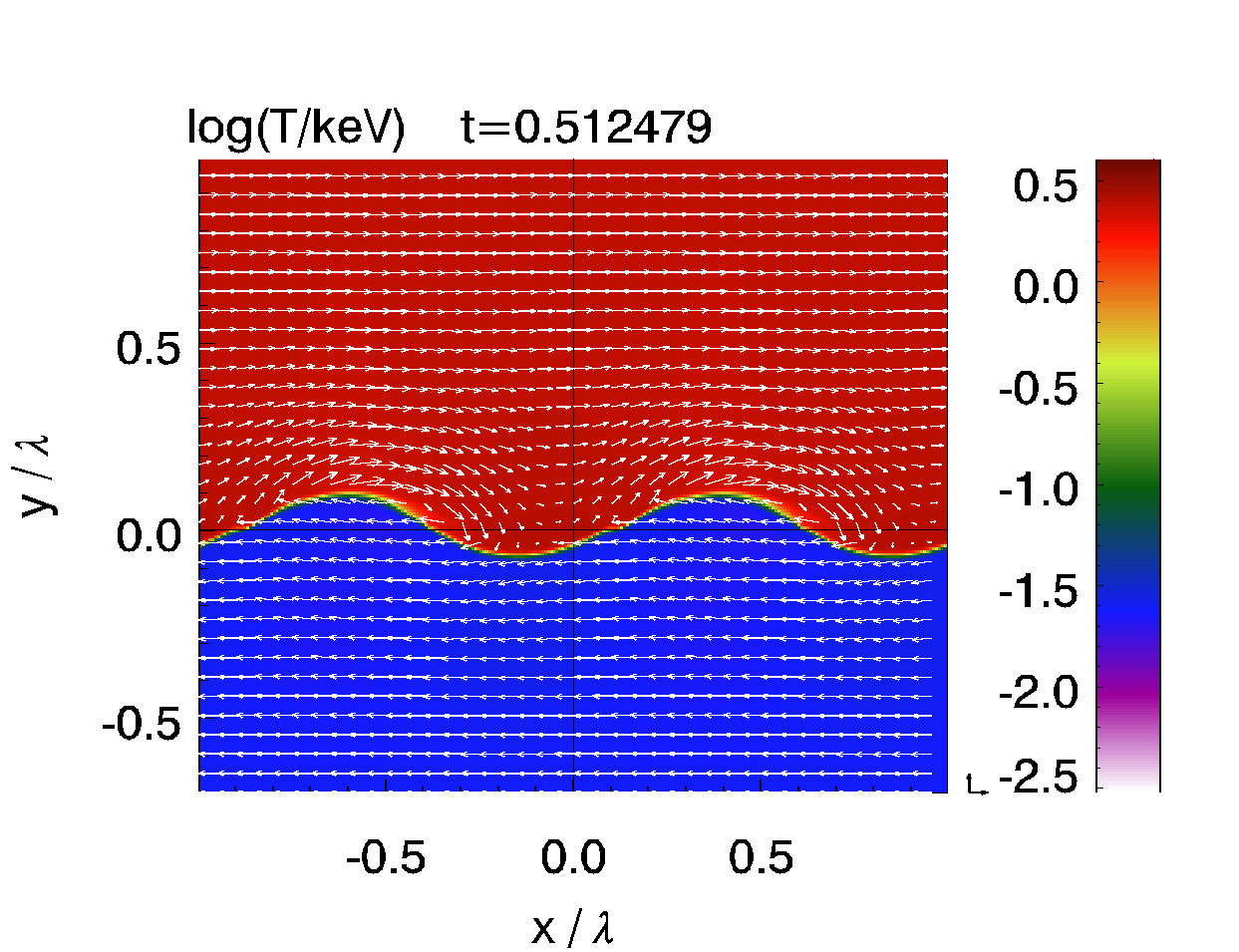}
\includegraphics[trim=200 120     0 160,clip,height=2.8cm]{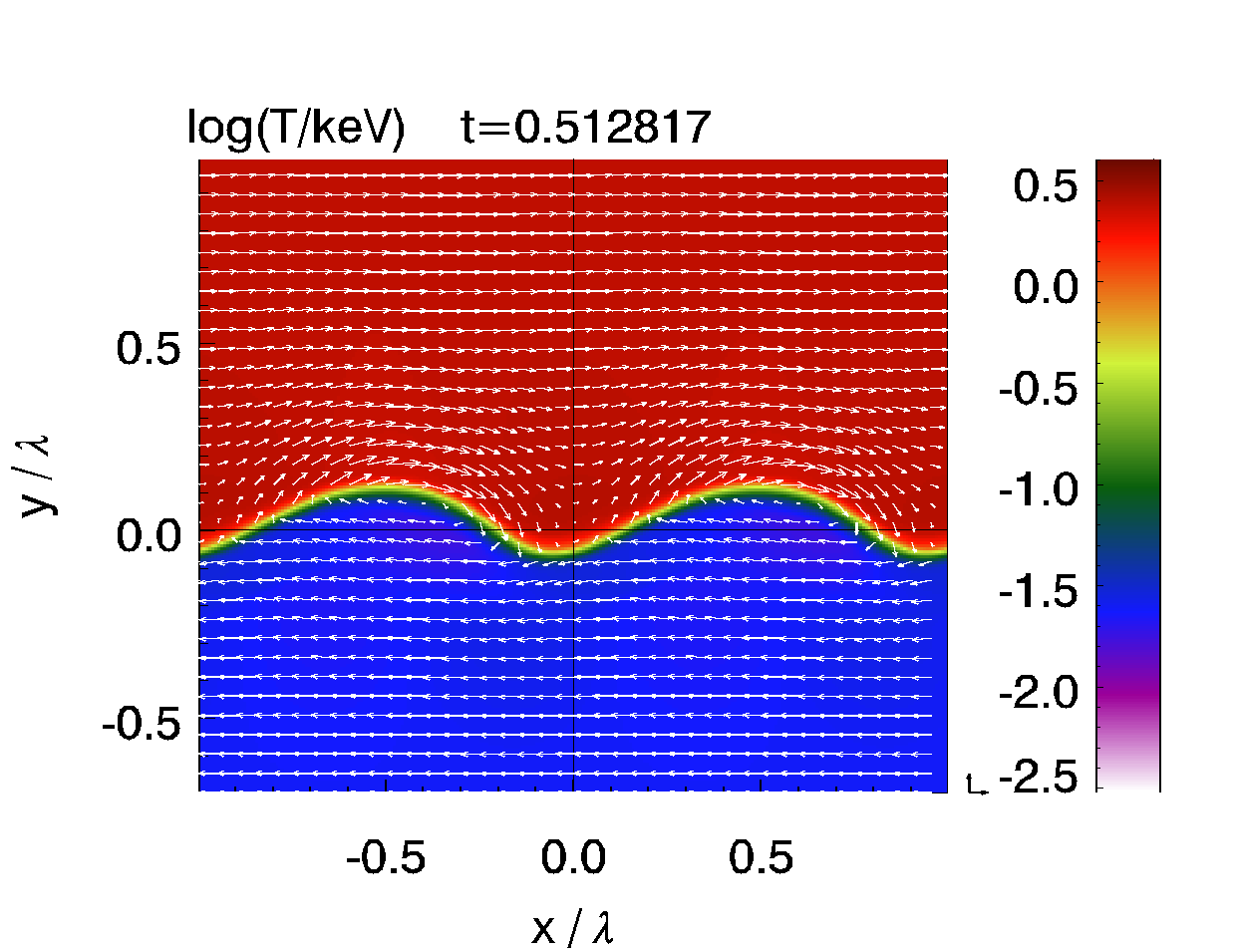}
\newline
\includegraphics[trim=0     120 250 160,clip,height=2.8cm]{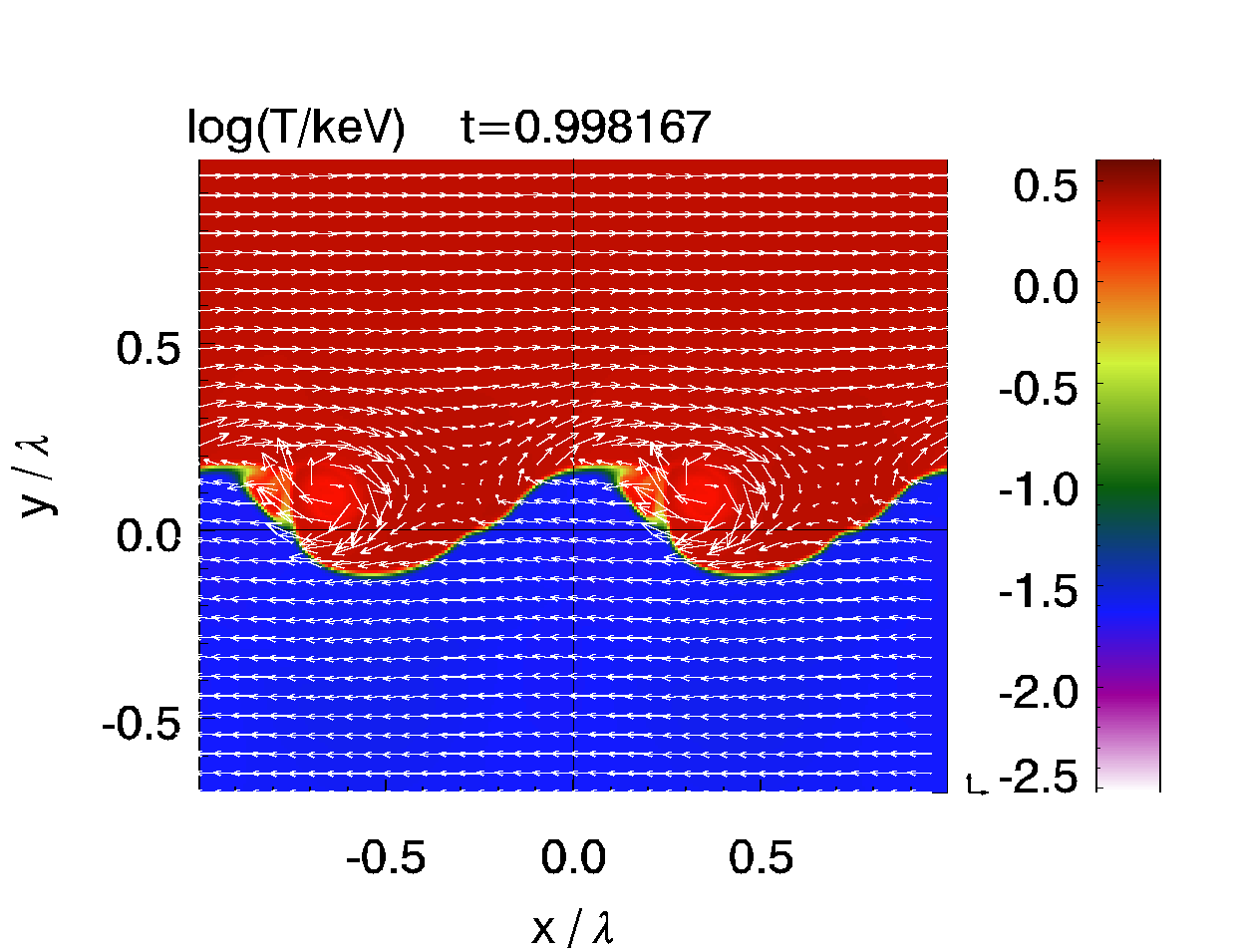}
\includegraphics[trim=200 120     0 160,clip,height=2.8cm]{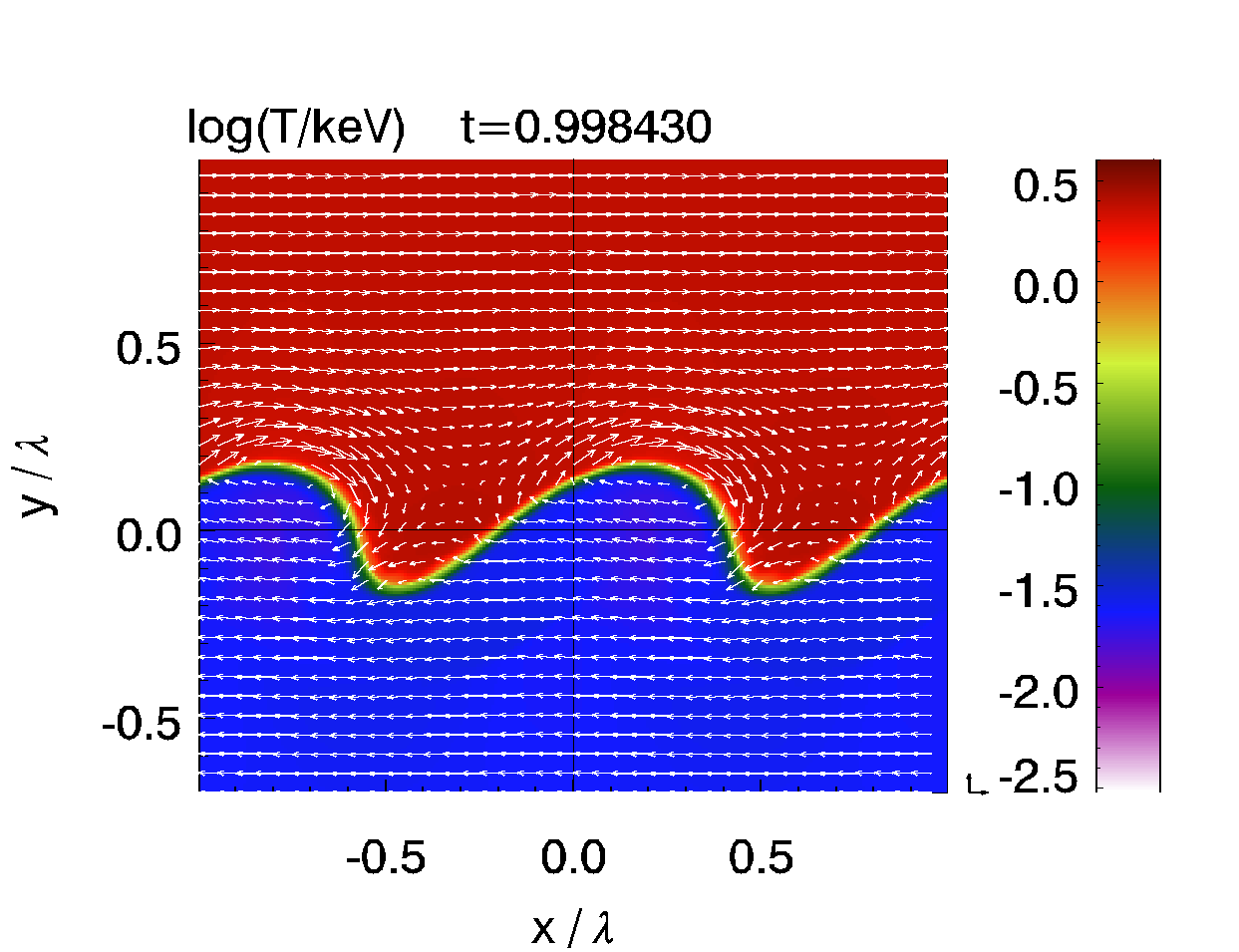}
\newline
\includegraphics[trim=0     120 250 160,clip,height=2.8cm]{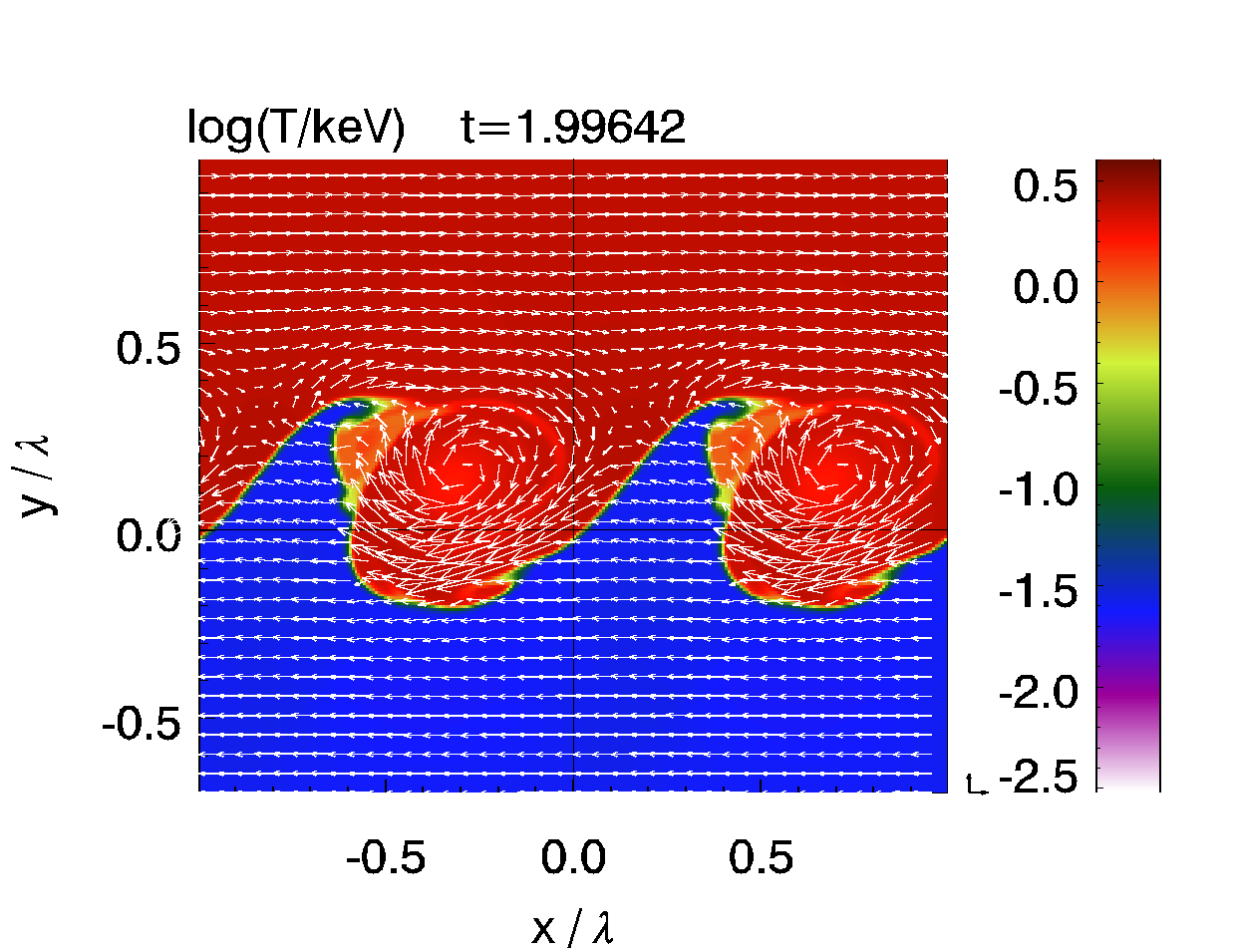}
\includegraphics[trim=200 120     0 160,clip,height=2.8cm]{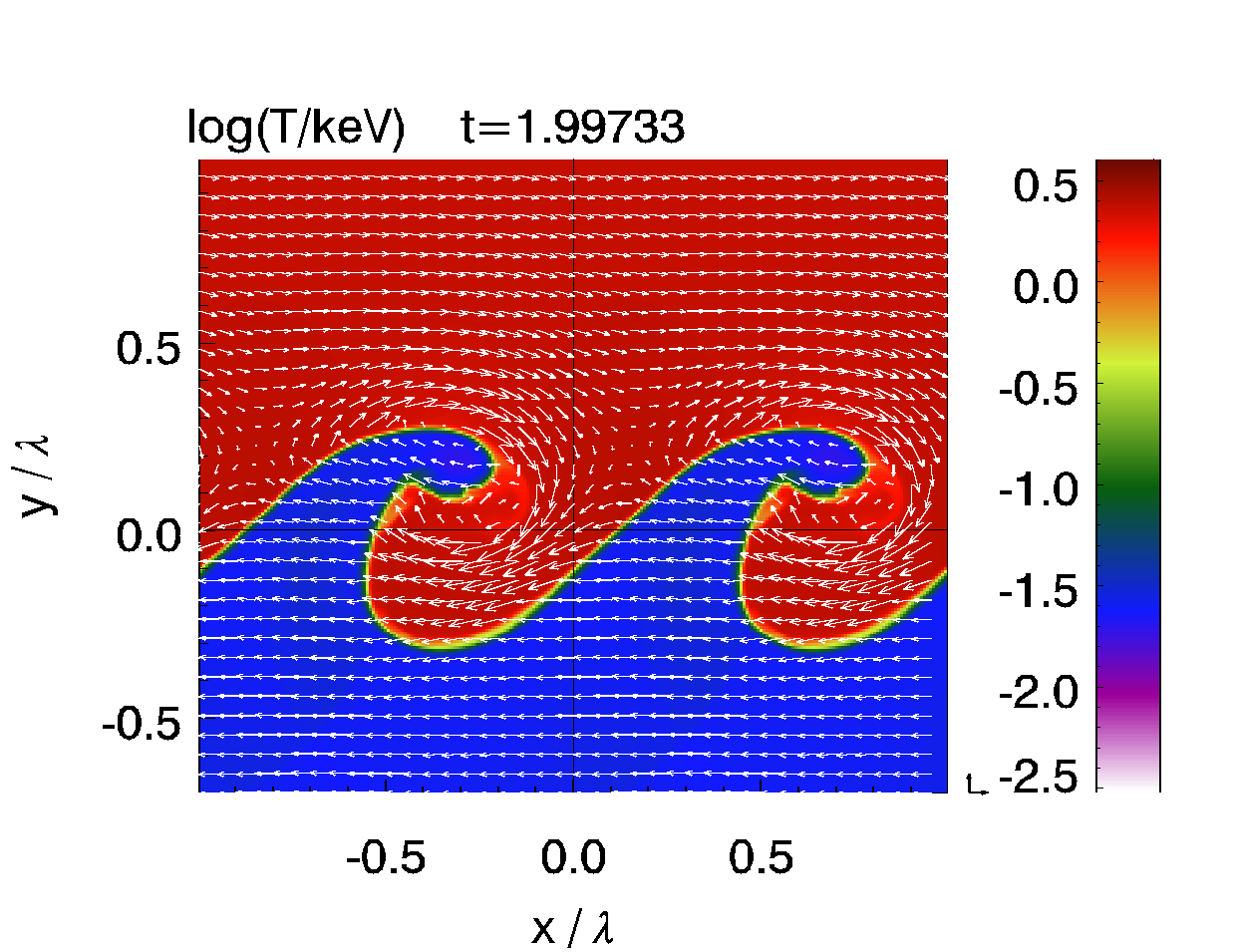}
\newline
\includegraphics[trim=0     120 250 160,clip,height=2.8cm]{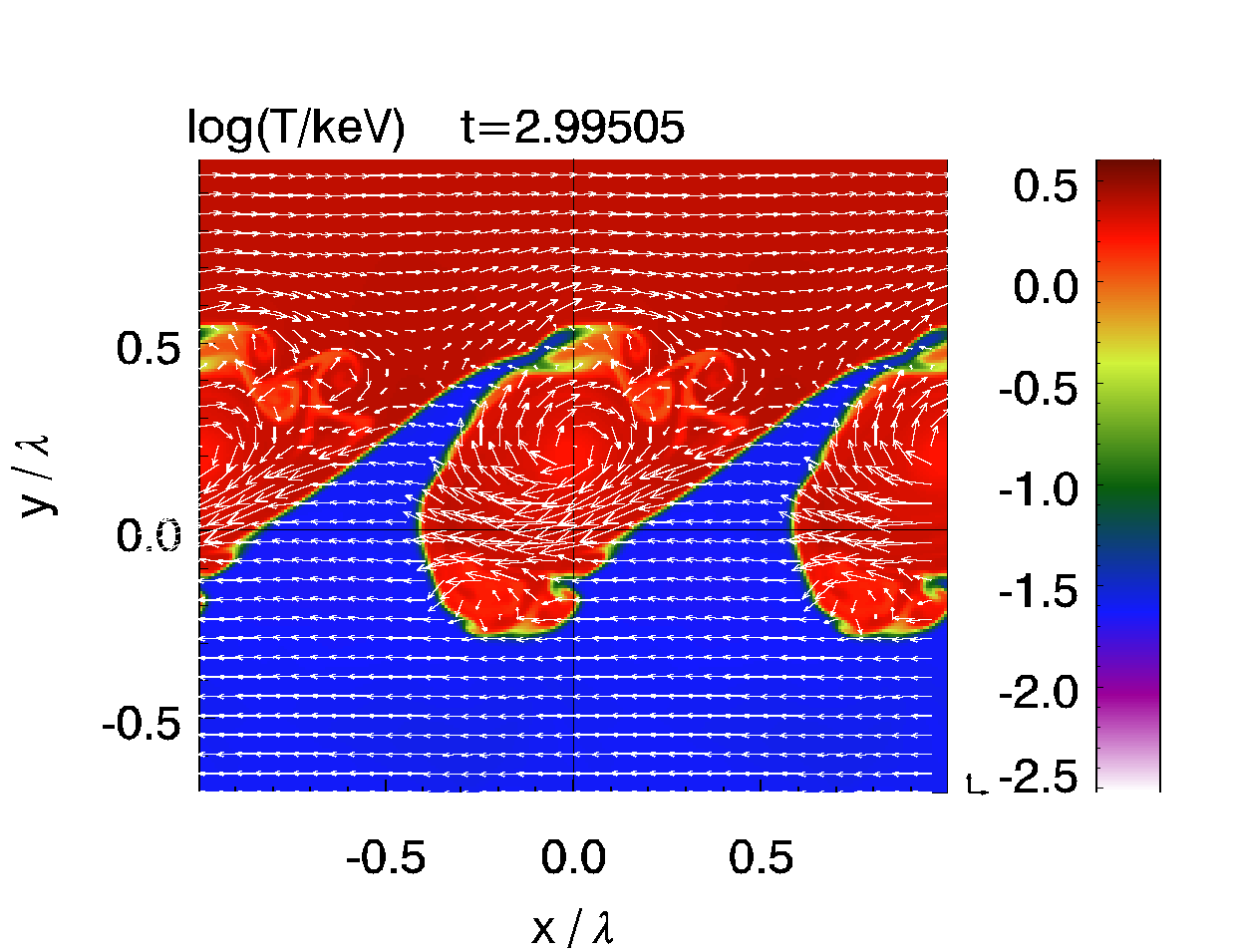}
\includegraphics[trim=200 120     0 160,clip,height=2.8cm]{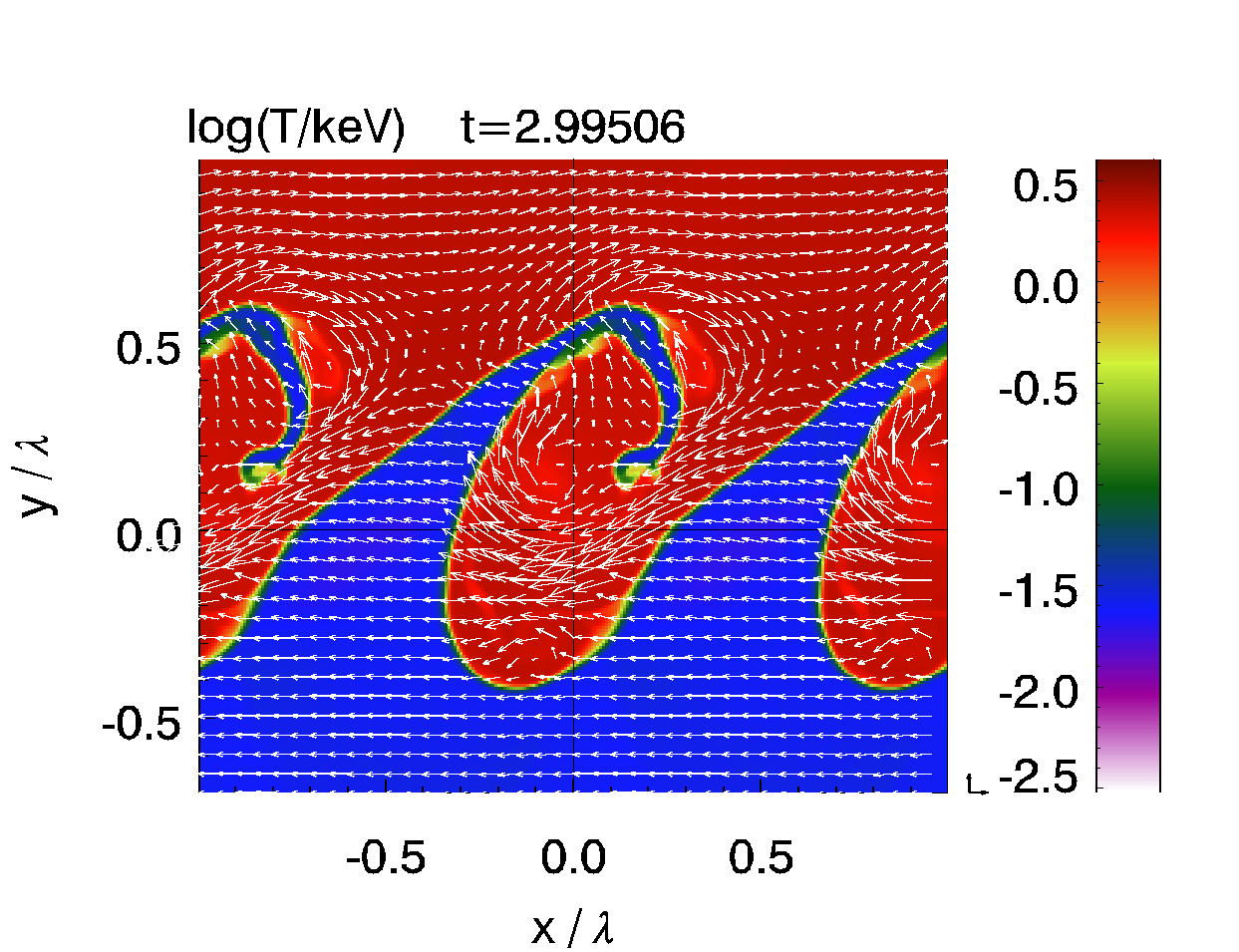}
\newline
\includegraphics[trim=0     120 250 160,clip,height=2.8cm]{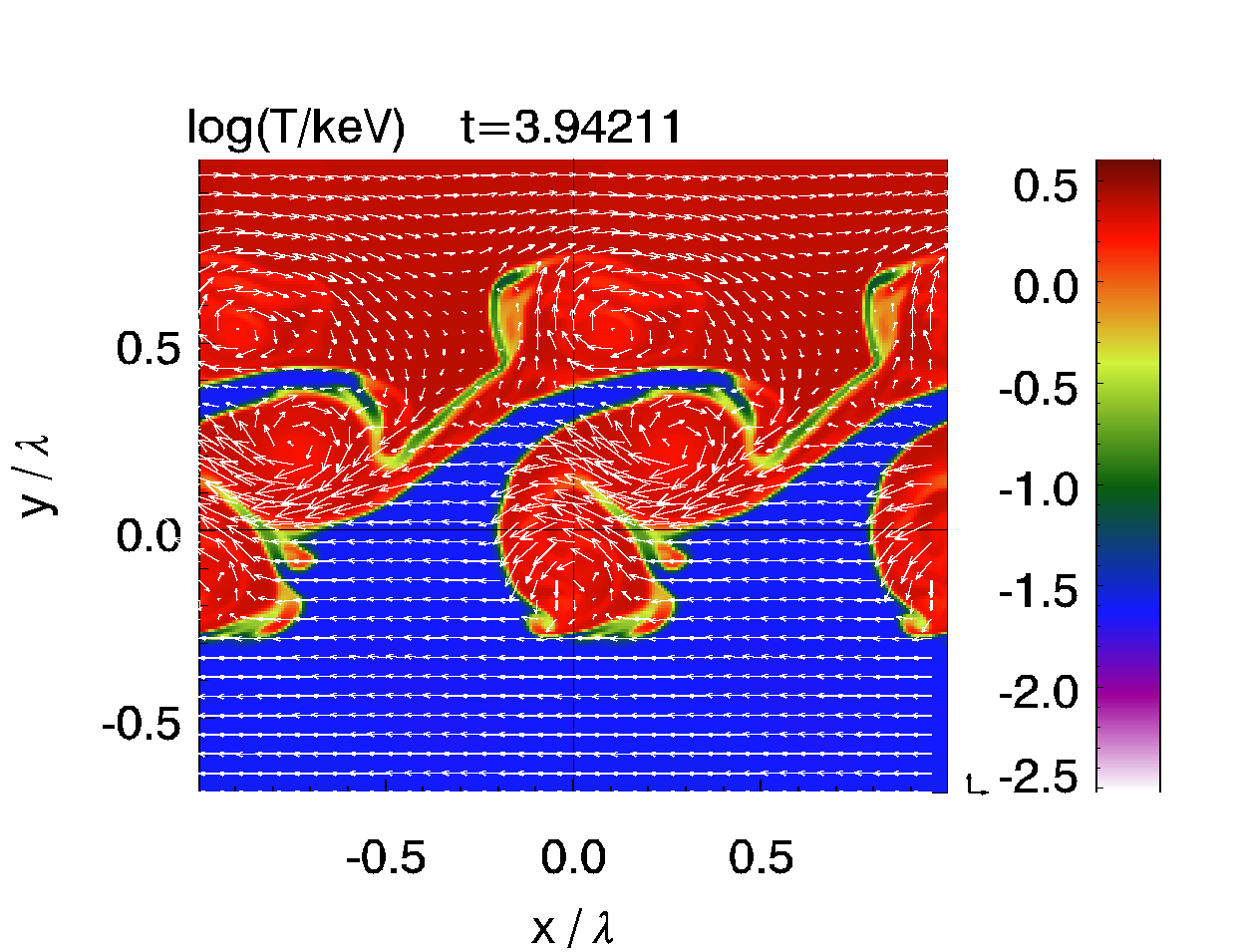}
\includegraphics[trim=200 120     0 160,clip,height=2.8cm]{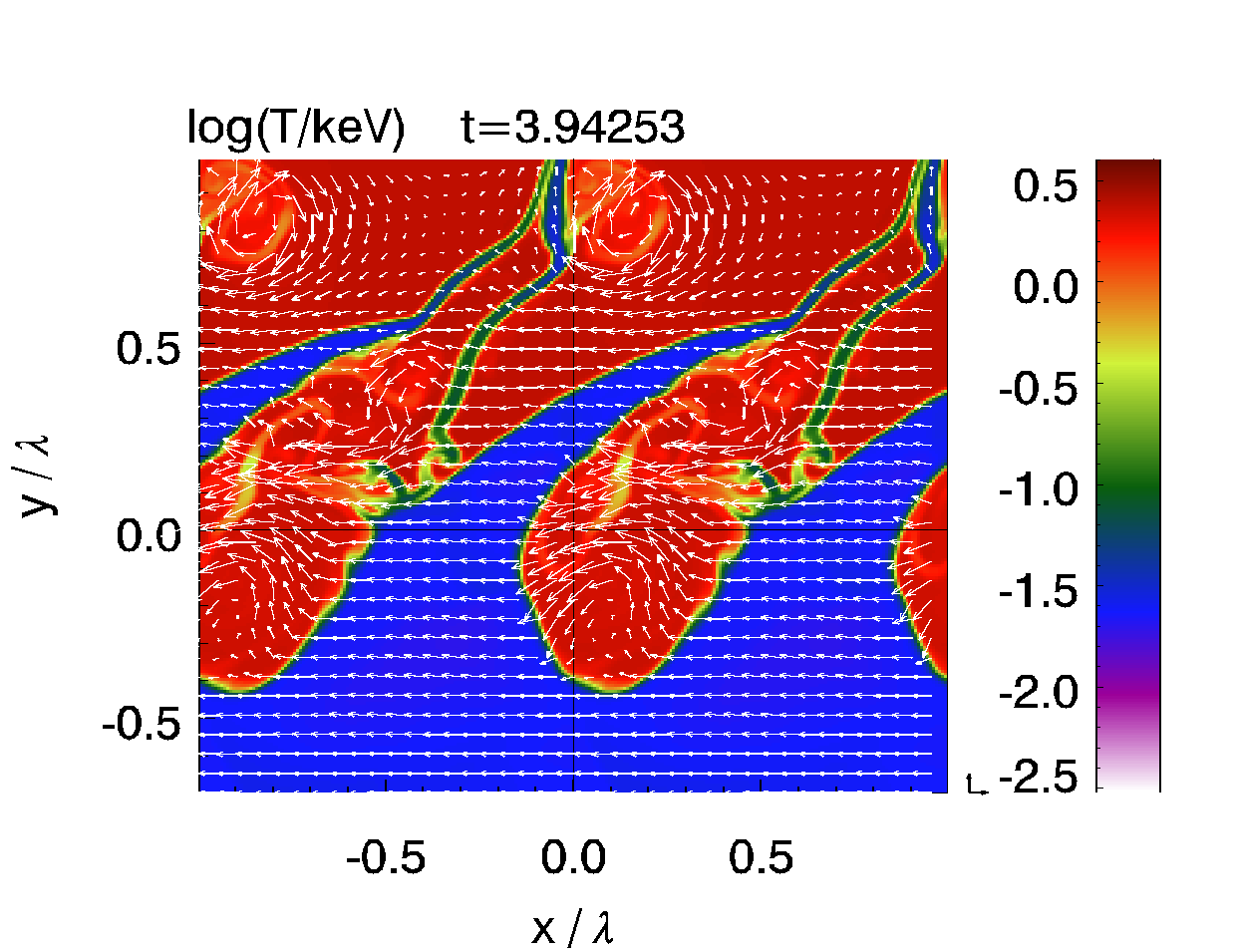}
\newline
\includegraphics[trim=0     0 250 160,clip,height=3.3cm]{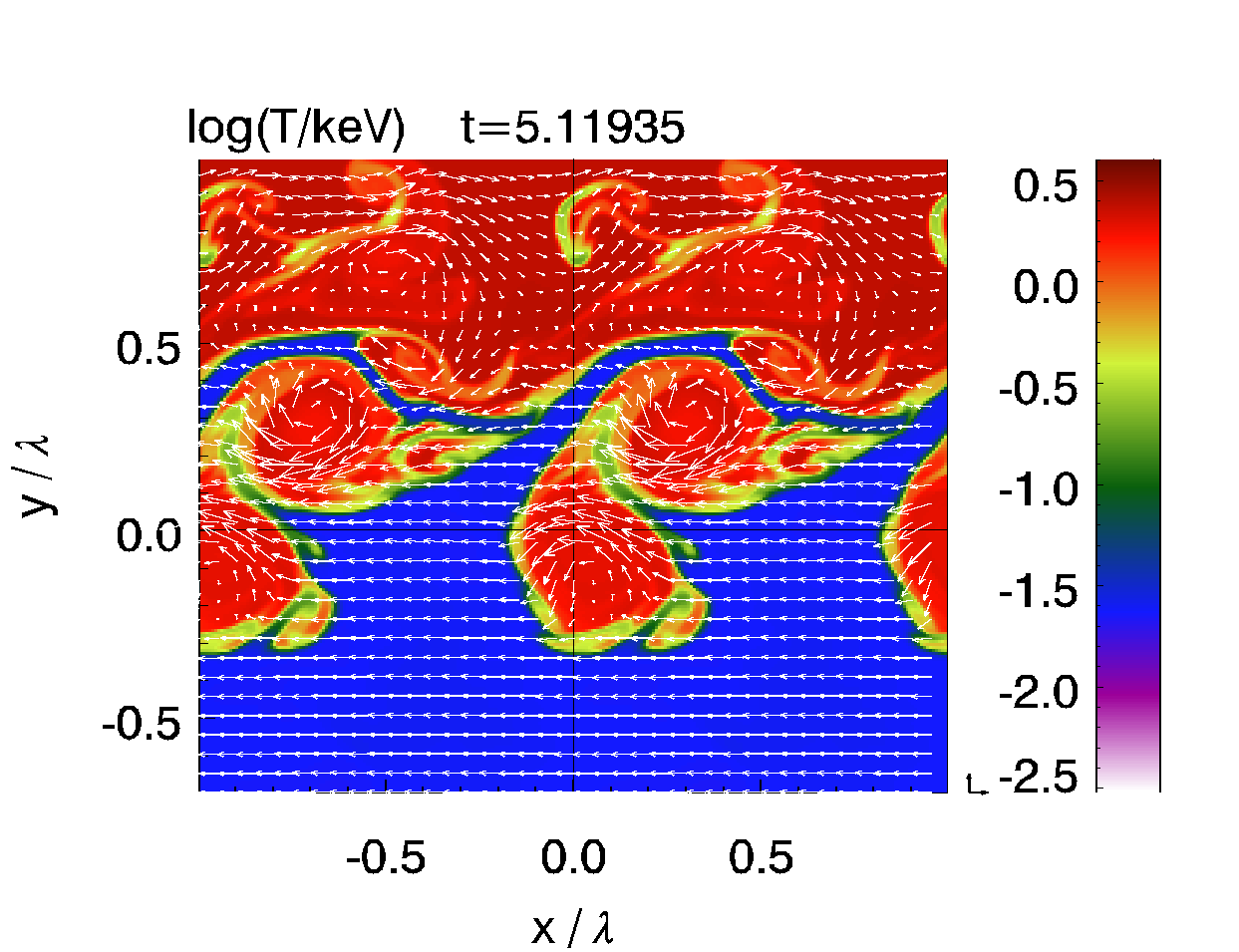}
\includegraphics[trim=200 0     0 160,clip,height=3.3cm]{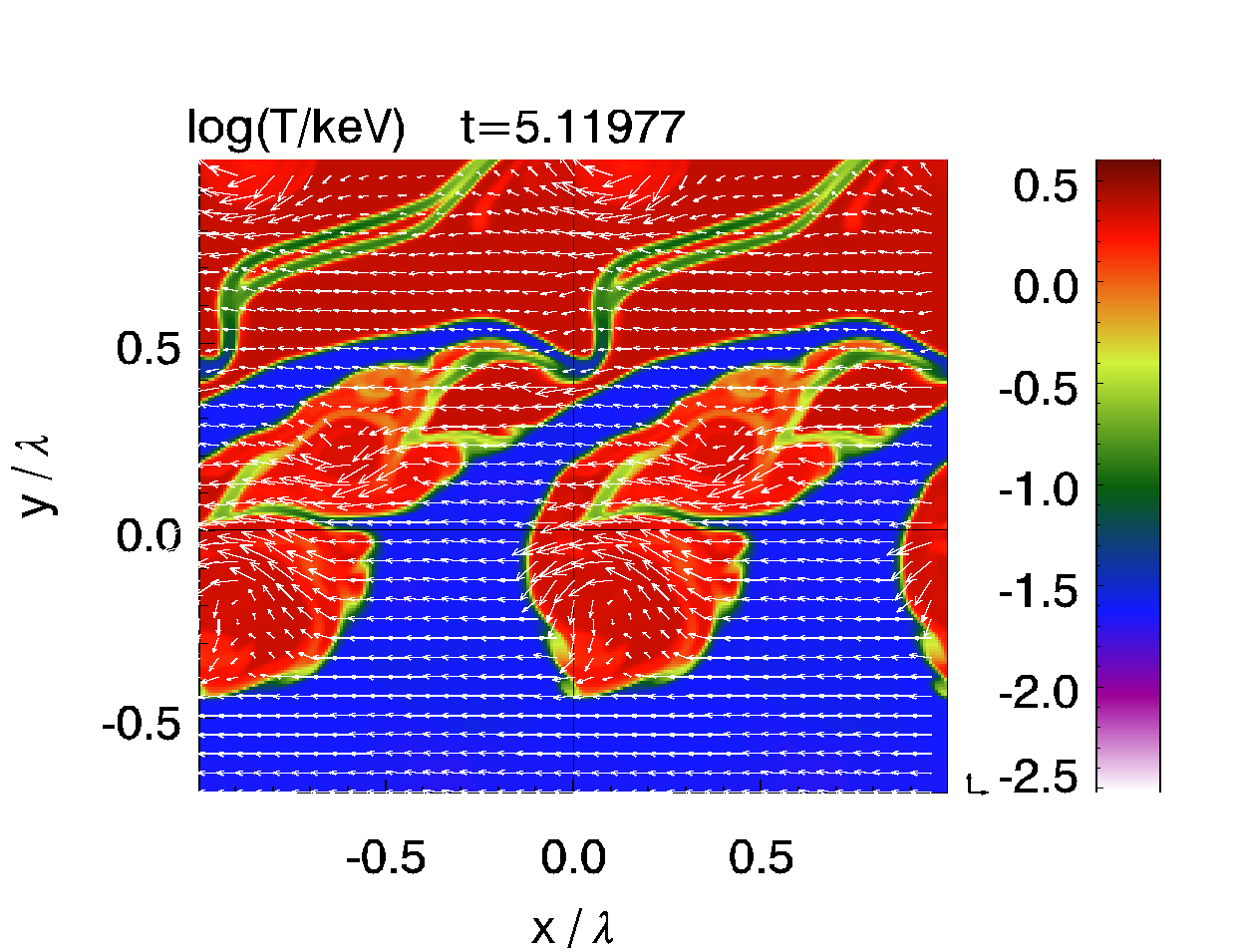}
\caption{Snapshots for simulations with constant kinematic viscosity, density ratio of 100, shear flow Mach 0.5, Reynolds number $10^4$. The colour codes $\log(T/\KeV)$,  velocity vectors are overlaid. The left column has an initially discontinuous interface, in the right column the interface is smoothed over 1/100 of the perturbation length. Rows from top to bottom are for timesteps 0.5, 1, 2, 3, 4, 5 $\tau\KHinvisc$.  The motion of the cold gas resembles "growing mountains". The hot gas is flowing over this distorted surface, developing turbulences. Hardly any turbulence is injected in the dense cold gas.}
\label{fig:rolls_nu_D100_M05_Re10000}
\end{figure}

At even higher density contrasts, the dynamics of the KHI change also at low viscosities. The inertia of the denser layer is enormous, and the flow patterns in the cool dense gas hardly change over many inviscid growth times. At such high density contrasts, the shear velocity is highly supersonic with respect to the cold layer while still subsonic in the hot layer.  The cold gas keeps its motion parallel to the interface and the initial perturbation velocity perpendicular to it. Thus, the cold gas causes the interface between both layers to resemble ``growing mountains", over which the wind of hot  gas is flowing (Fig.~\ref{fig:rolls_nu_D100_M05_Re10000}). The vertical extent of the mixing layer (height of KH rolls, or distortions) closely follows the linear widening expected from the initial perturbation velocity. Almost all turbulence is induced in the hot gas that flows around the ``mountains" of cold gas. Vortices on the lee side of the ``mountains" form on a timescale $\sim \lambda/c$, where $c$ is the sound speed of the hot gas. With increasing density contrast the KHI timescale can exceed the vortex formation timescale, and the KHI will not dominate the gas flow anymore. Consequently, the maximum or minimum $v_y$, or any velocity patterns, are not a tracer of the KHI anymore. 

Interestingly, after about 3 $\tau\KHinvisc$, enough momentum has been transferred from the huge reservoir in the cold layer to the hot gas between the "cool mountains", such that this hot gas between the cool mountains moves along with the cool gas. Thus results in a \textit{new} shear flow interface at the level of the  "top" of the mountains, and a second generation of KHI evolves. Slightly smoothing the initial interface brings out this effect more clearly (right column of Fig.~\ref{fig:rolls_nu_D100_M05_Re10000}). Without the initial smoothing, numerous secondary instabilities occur. However, the overall dynamics are similar with and without the interface smoothing.

\subsubsection{High viscosity}

\begin{figure*}
\begin{center}
\hspace{1.5cm} $\Reyn=10$ \hfill $\Reyn=100$ \hfill  $\Reyn=1000$ \hfill $\Reyn=10^4$ \hfill\phantom{x}\newline
\includegraphics[trim=0     0 290 160,clip,height=3.9cm]{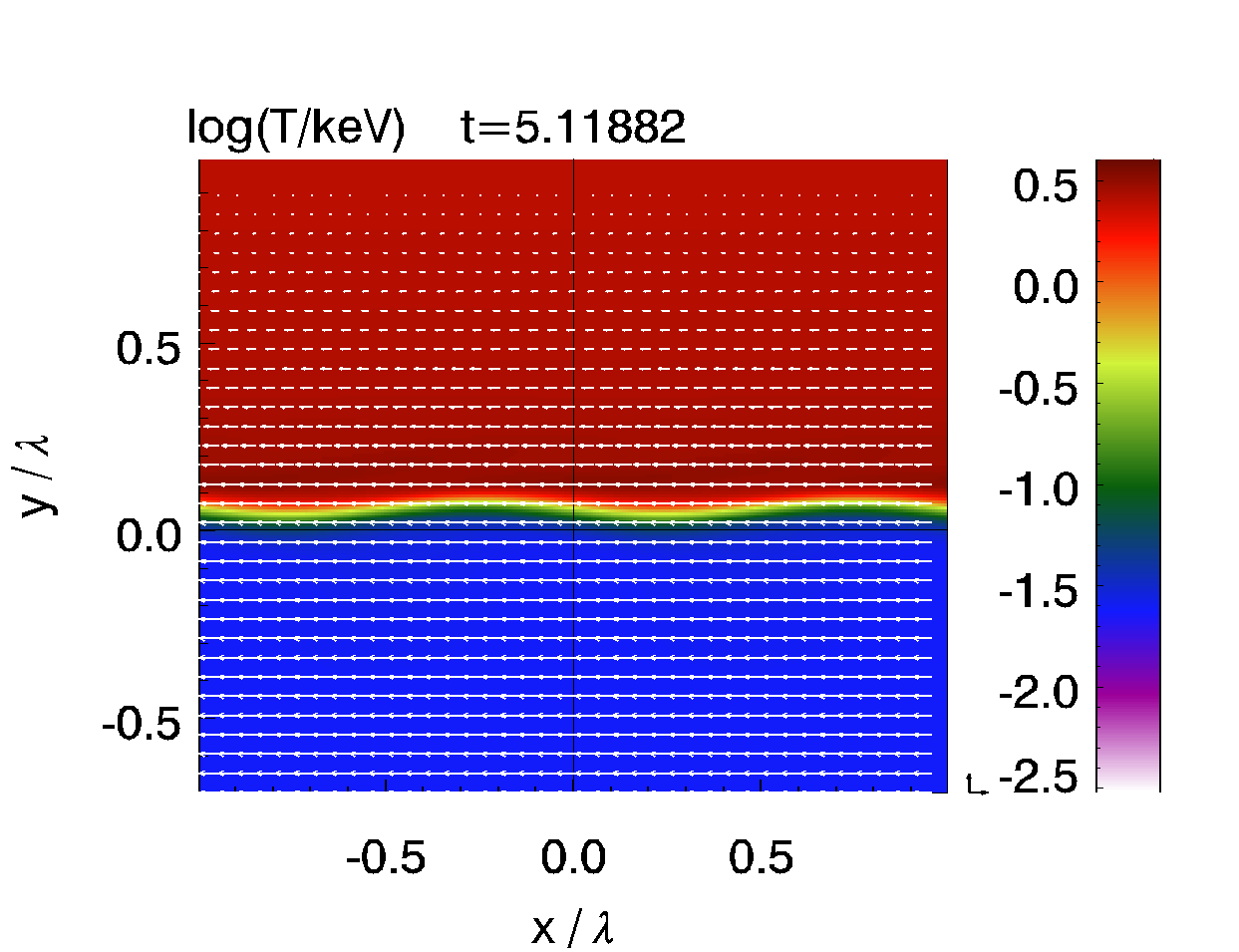}
\includegraphics[trim=190 0 290 160,clip,height=3.9cm]{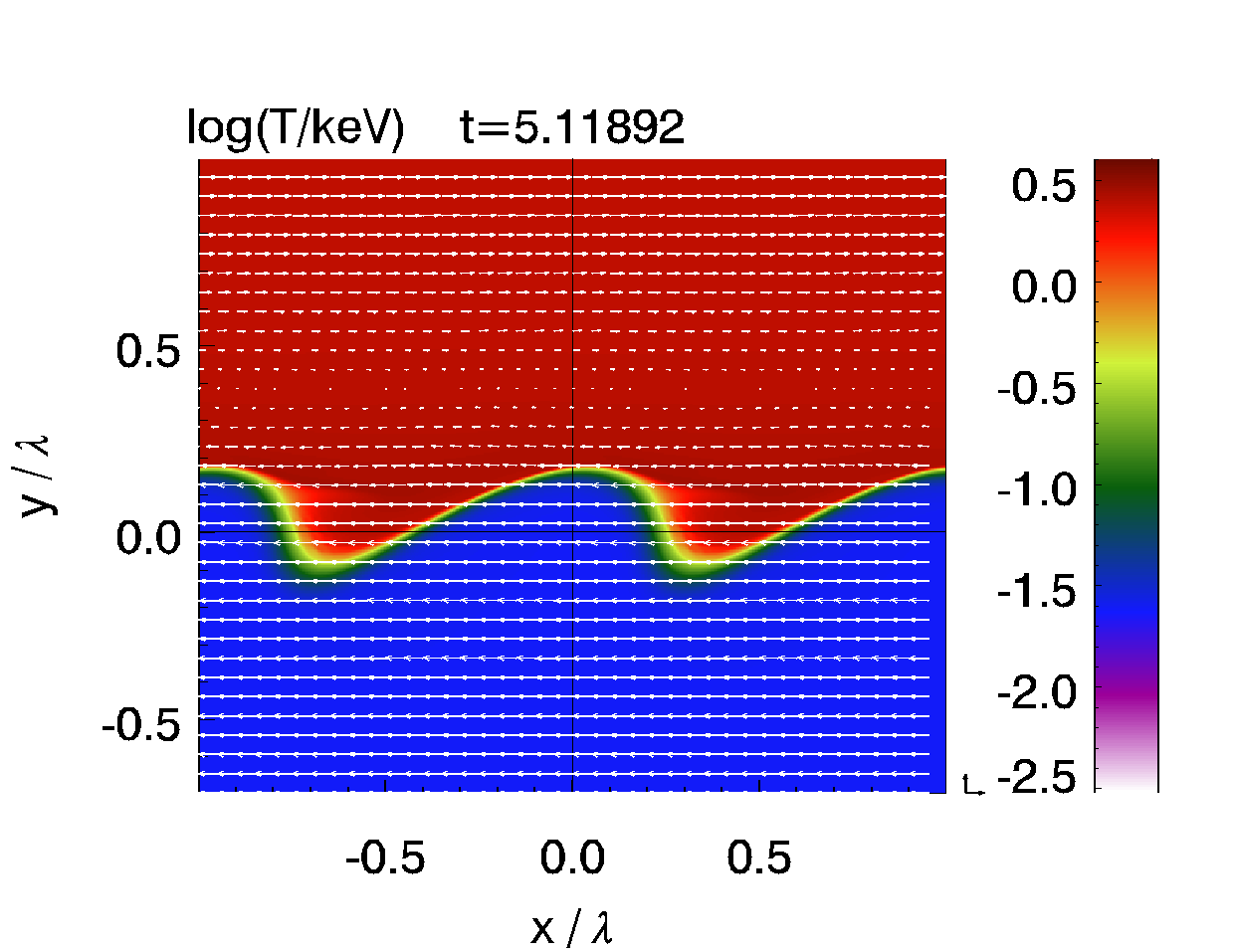}
\includegraphics[trim=190 0 290 160,clip,height=3.9cm]{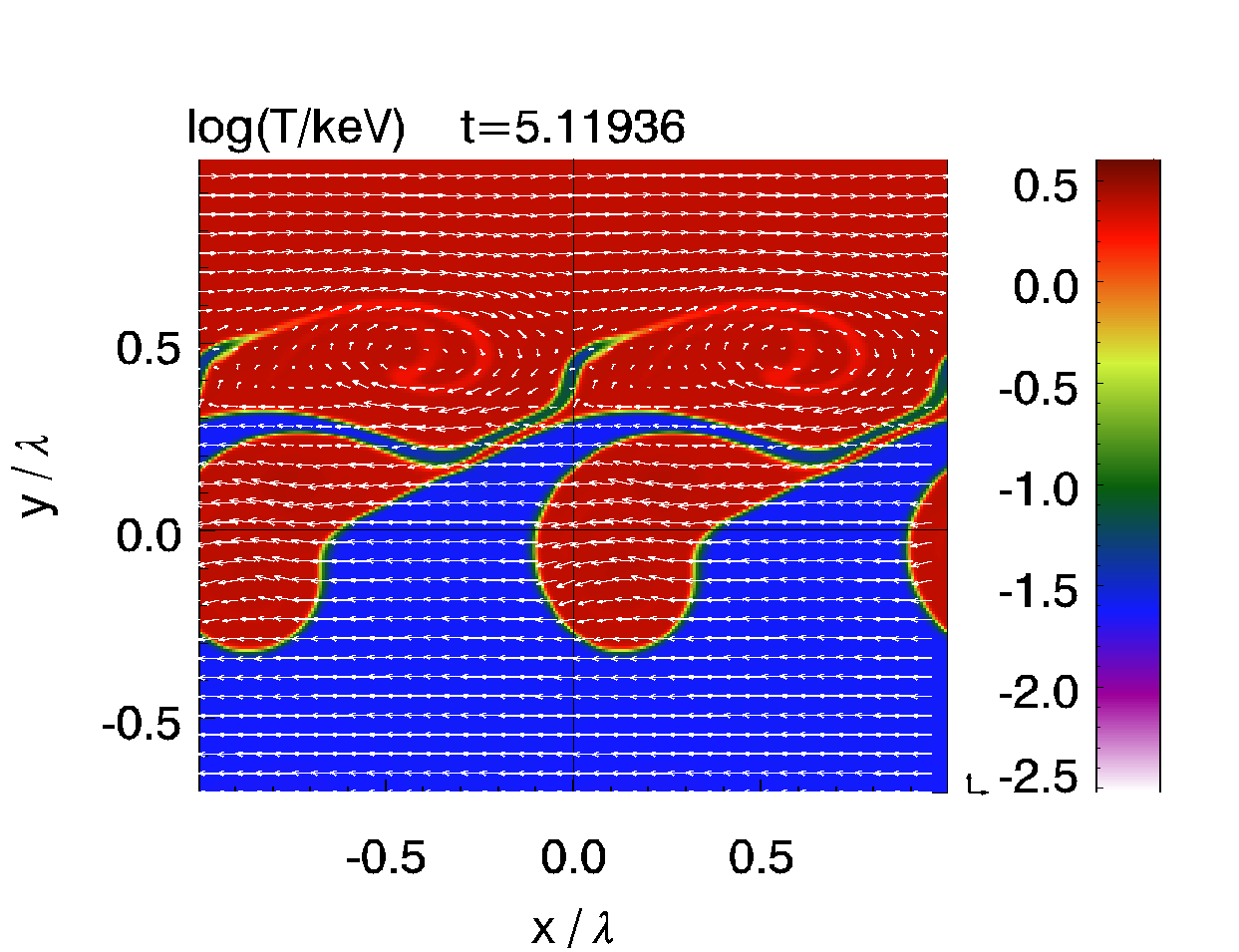}
\includegraphics[trim=190 0     0 160,clip,height=3.9cm]{Figs/Nu_D100/Templog_Re10000_0200}
\caption{Snapshots for simulations with constant kinematic viscosity, a high density ratio of 100, shear flow of Mach 0.5, different $\Reyn$ as given at the top. The colour codes $\log(T/\KeV)$,  velocity vectors are overlaid. We show the timestep 5 $\tau\KHinvisc$. Note the reversed flow direction in the hot layer near the interface for $\Reyn \le 100$. For $\Reyn \le 100$ we smoothed the initial interface over 1\% of the perturbation length scale to avoid excessive viscous heating at the shear flow discontinuity (see Eqn.~\ref{eq:smooth}).}
\label{fig:rolls_nu_D100_M05}
\end{center}
\end{figure*}

\begin{figure*}
\begin{center}
\hspace{1.5cm} $\Reyn=10$ \hfill $\Reyn=100$ \hfill  $\Reyn=1000$ \hfill $\Reyn=10^4$ \hfill\phantom{x}
\newline
\includegraphics[trim=0     0 290 160,clip,height=3.9cm]{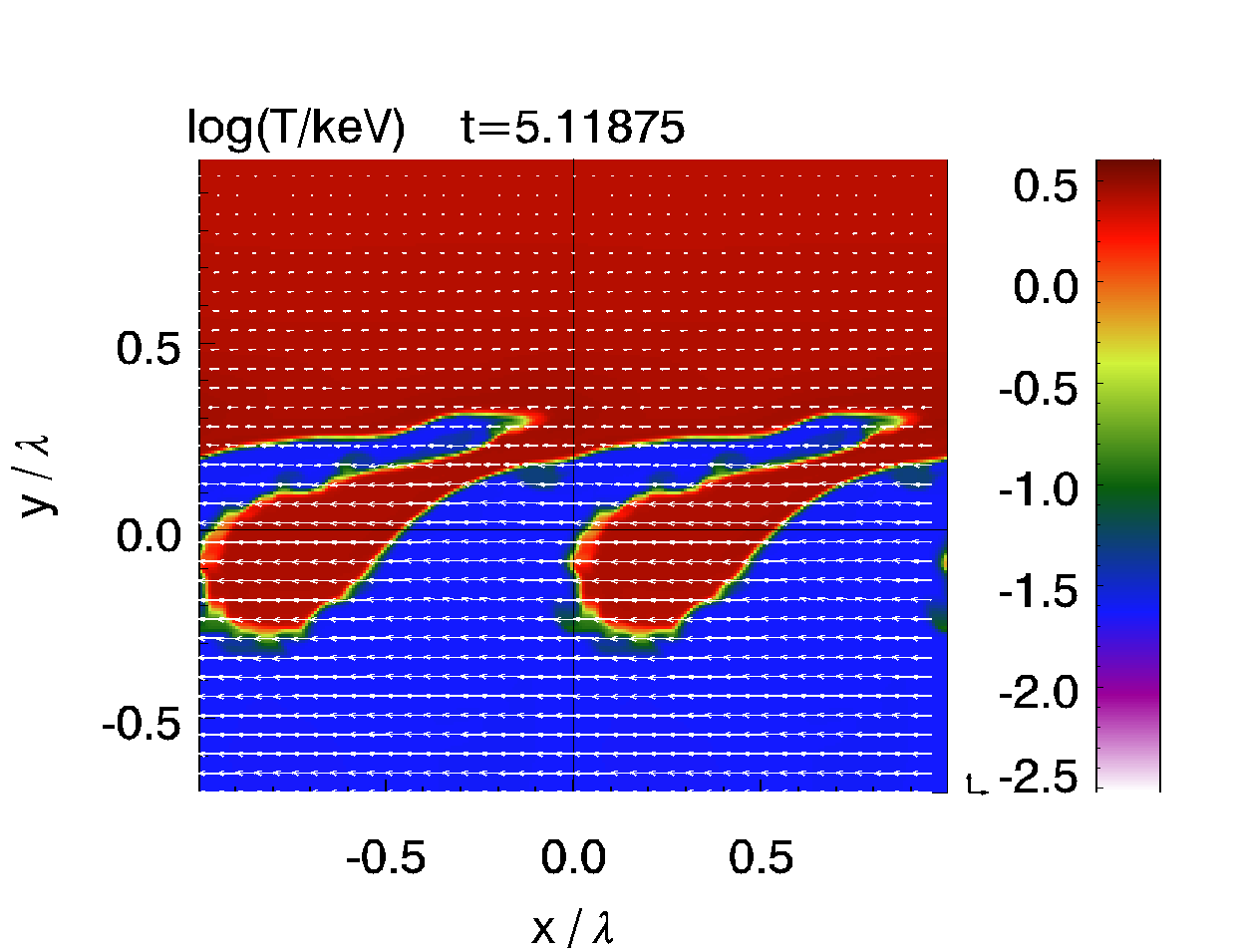}
\includegraphics[trim=190 0 290 160,clip,height=3.9cm]{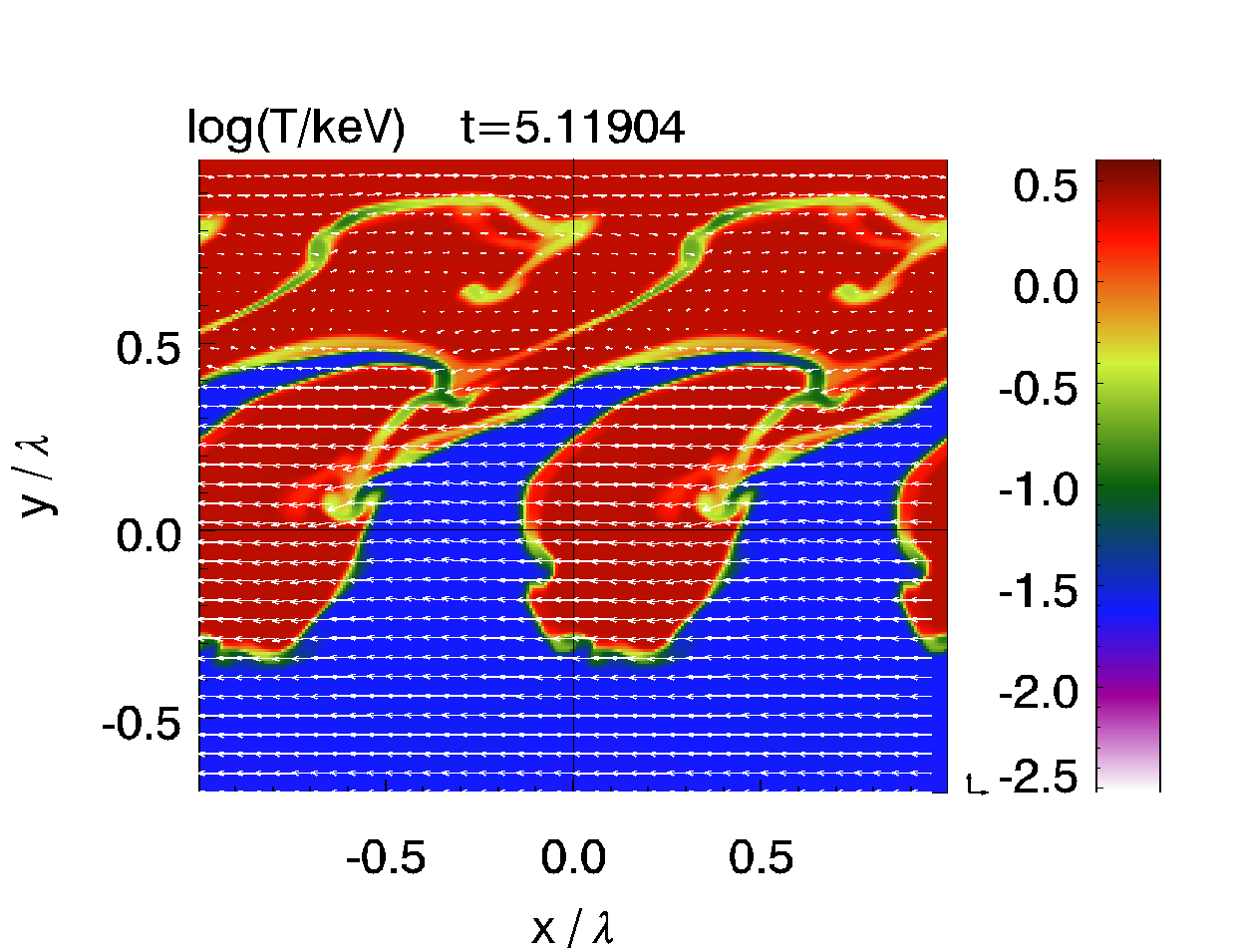}
\includegraphics[trim=190 0 290 160,clip,height=3.9cm]{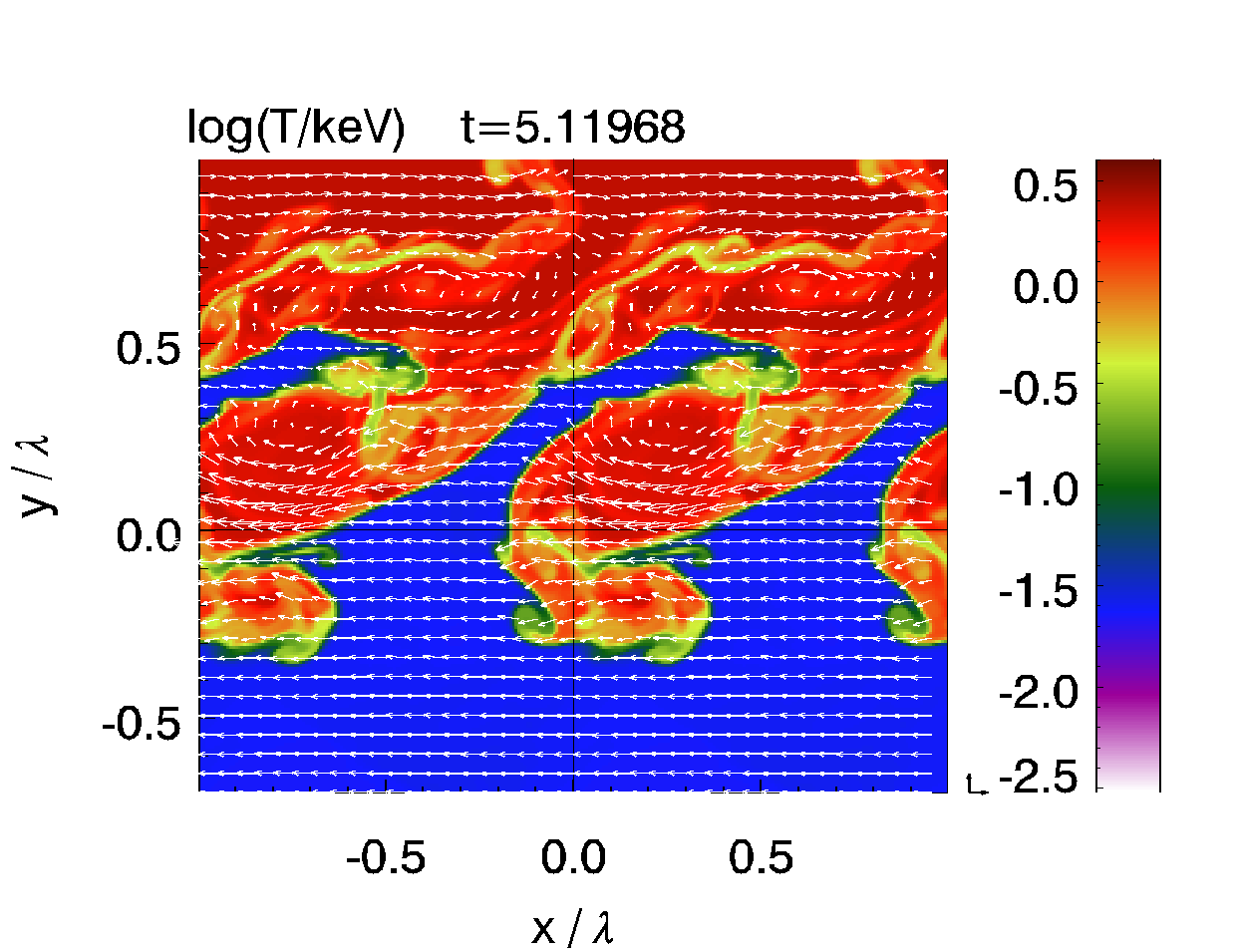}
\includegraphics[trim=190 0     0 160,clip,height=3.9cm]{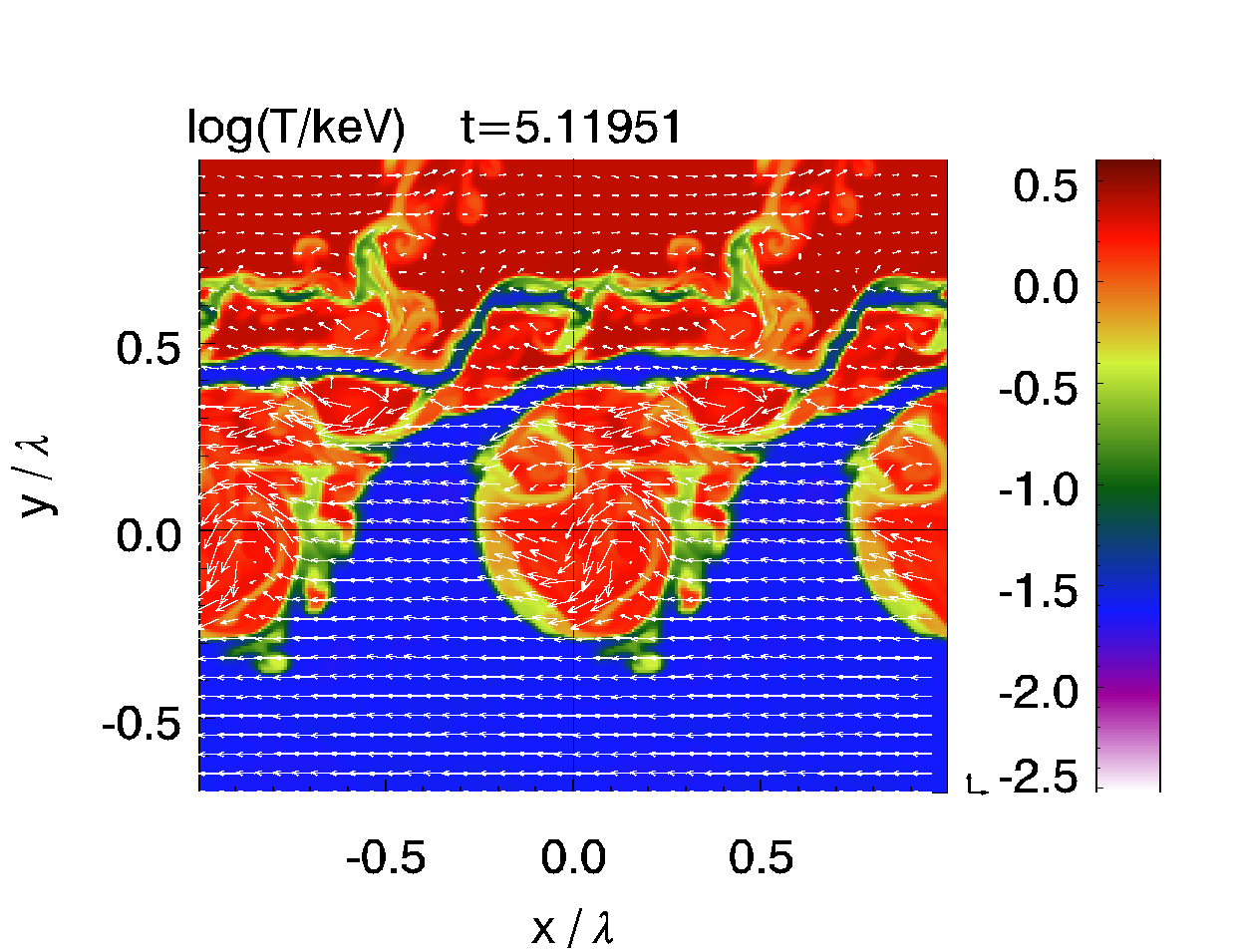}
\caption{Same as Fig.~\ref{fig:rolls_nu_D100_M05} (density ratio of 100, shear flow of Mach 0.5), but with Spitzer viscosity. No smoothing of the initial interface is applied. Note the reversed flow direction in the hot layer near the interface for $\Reyn \le 100$.}
\label{fig:rolls_Sp_D100_M05}
\end{center}
\end{figure*}

In case of a constant kinematic viscosity, any instability or turbulence is suppressed for $\Reyn \le 100$ (see Fig.~\ref{fig:rolls_nu_D100_M05}). For  a Spitzer-like viscosity only the hot gas provides viscosity, and a lower Reynolds number of $\Reyn \le 10$ is needed (Fig.~\ref{fig:rolls_Sp_D100_M05}). At Re=10, the interface resembles frozen KH rolls, but the instability is not evolving anymore. Instead, the hot gas has reversed its direction of motion near the original interface already.

\subsection{Supersonic shear flow} \label{sec:supersonic}

\begin{figure*}
\begin{center}
\hspace{1.5cm} $\Reyn=30$ \hfill $\Reyn=100$ \hfill  $\Reyn=300$ \hfill $\Reyn=1000$ \hfill\phantom{x}
\newline
\includegraphics[trim=0     155 260 160,clip,height=1.9cm]{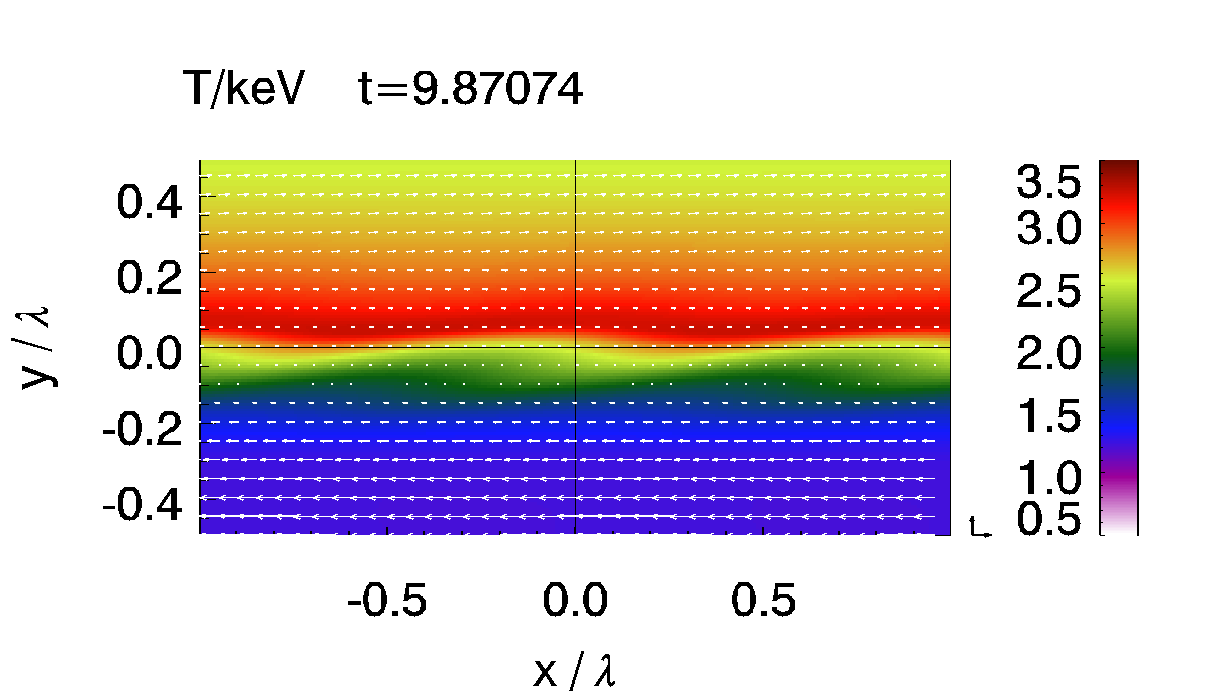}
\includegraphics[trim=200 155 260 160,clip,height=1.9cm]{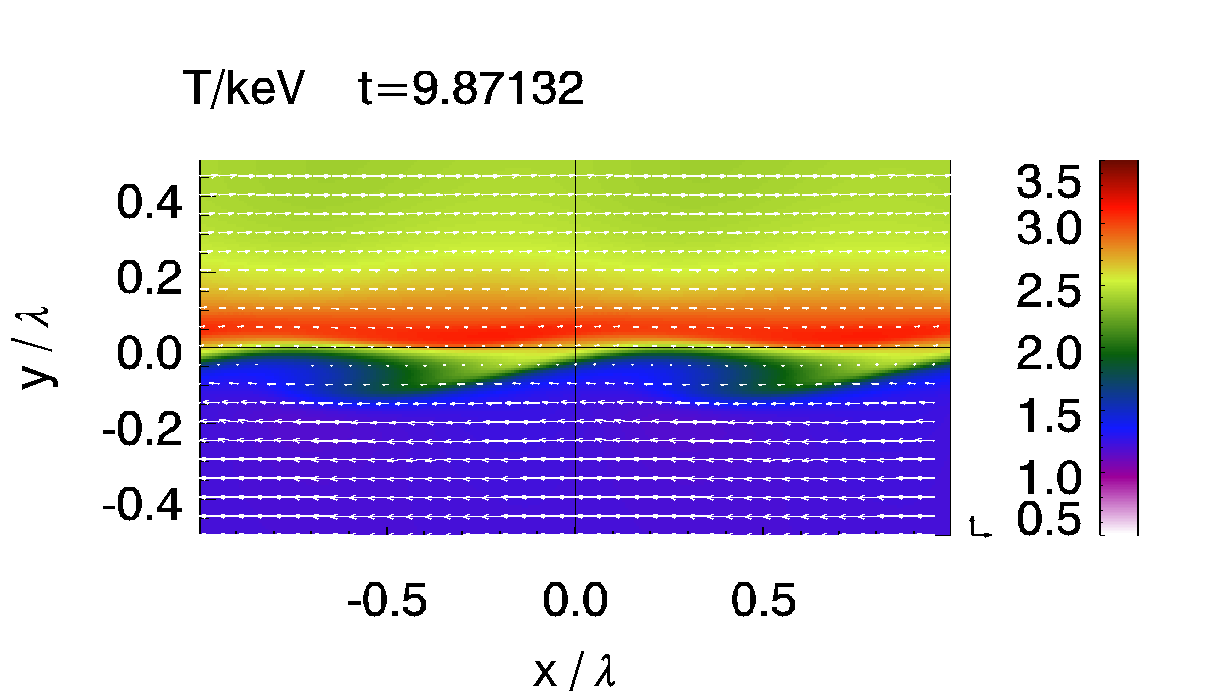}
\includegraphics[trim=200 155 260 160,clip,height=1.9cm]{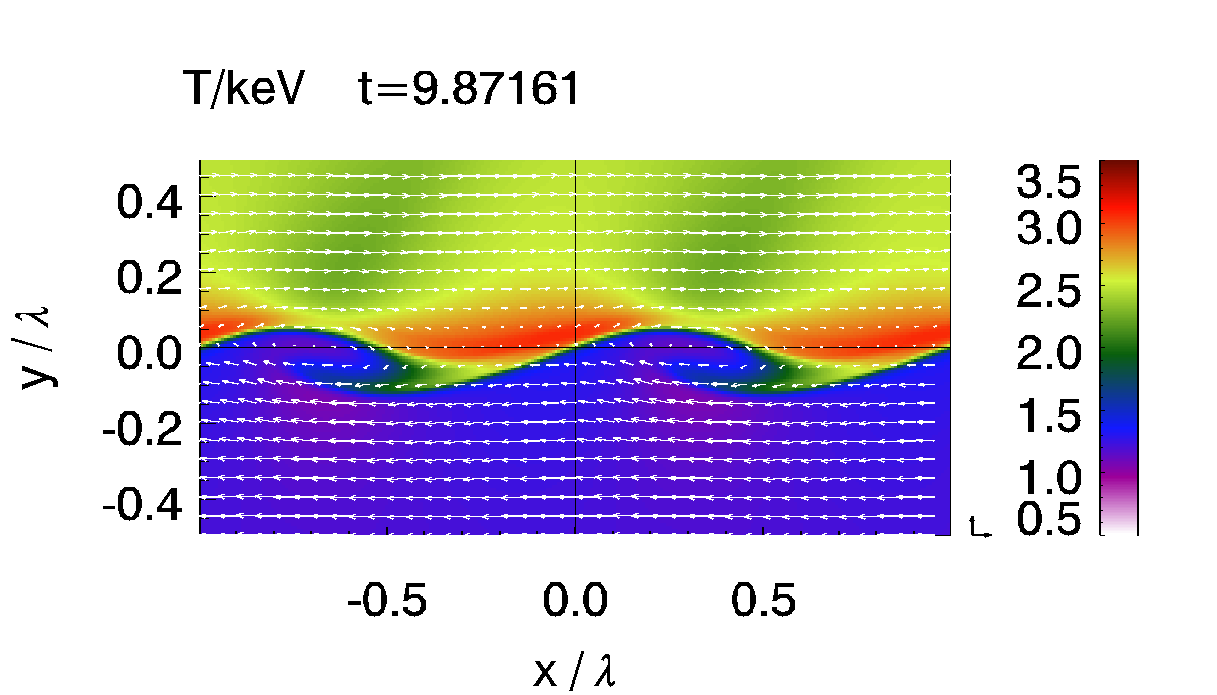}
\includegraphics[trim=200 155     0 160,clip,height=1.9cm]{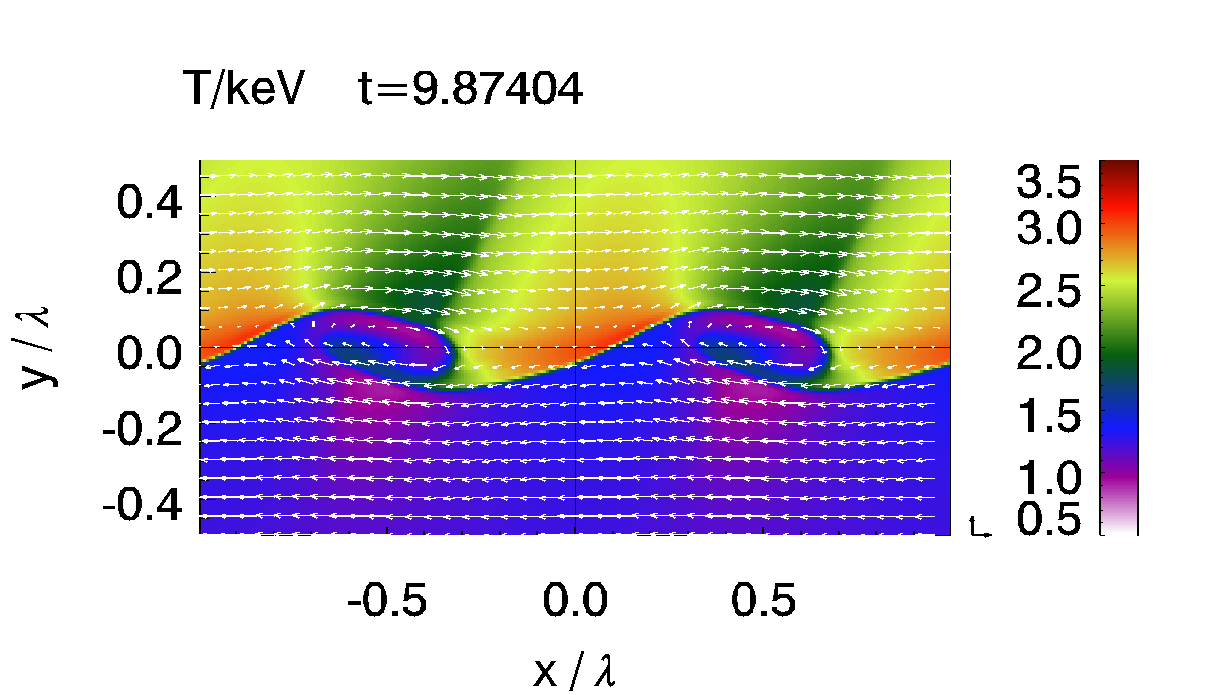}
\includegraphics[trim=    0 155 260 160,clip,height=1.9cm]{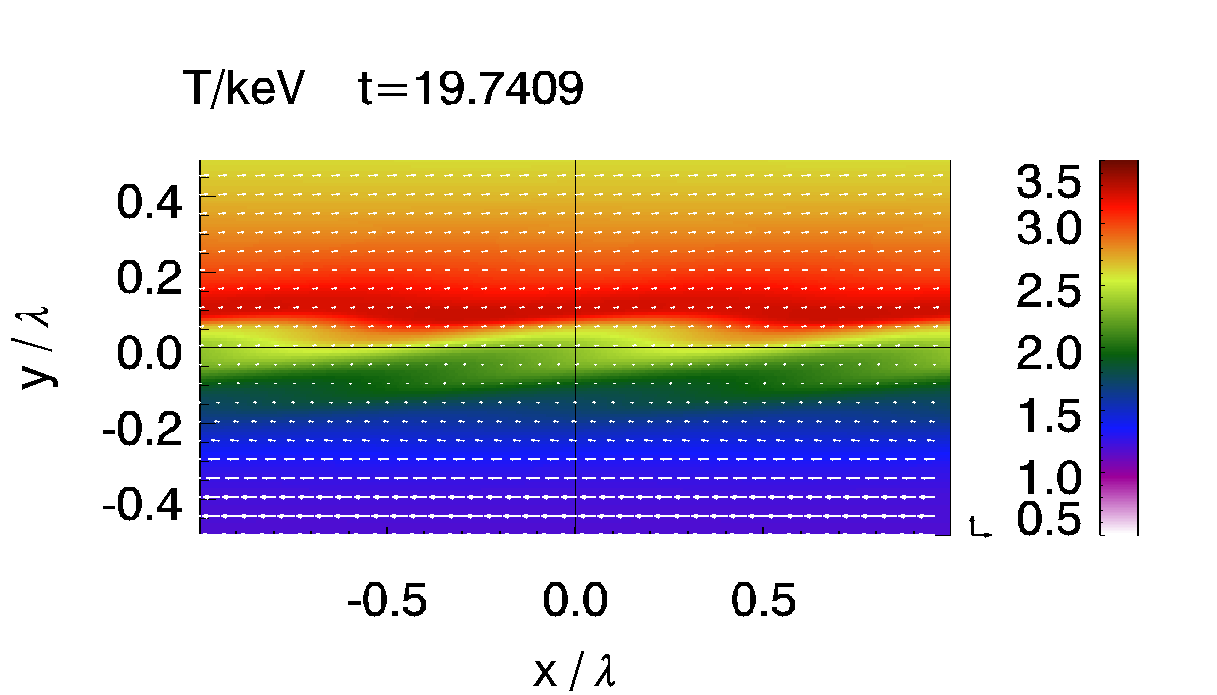}
\includegraphics[trim=200 155 260 160,clip,height=1.9cm]{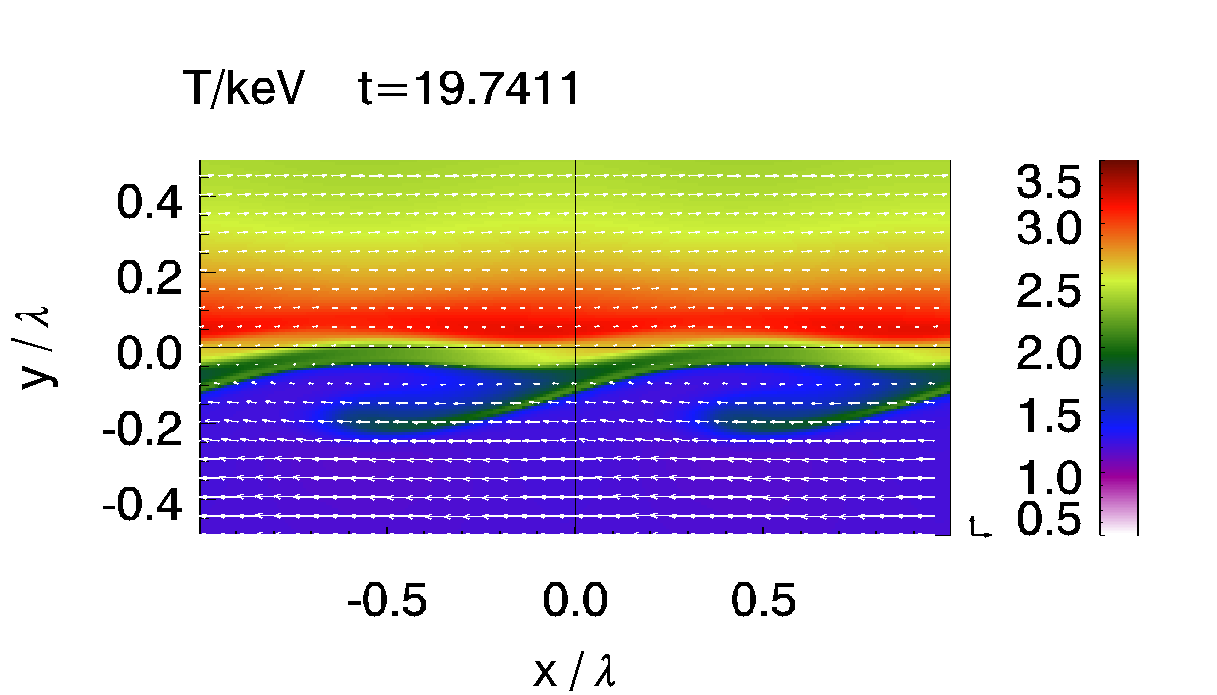}
\includegraphics[trim=200 155 260 160,clip,height=1.9cm]{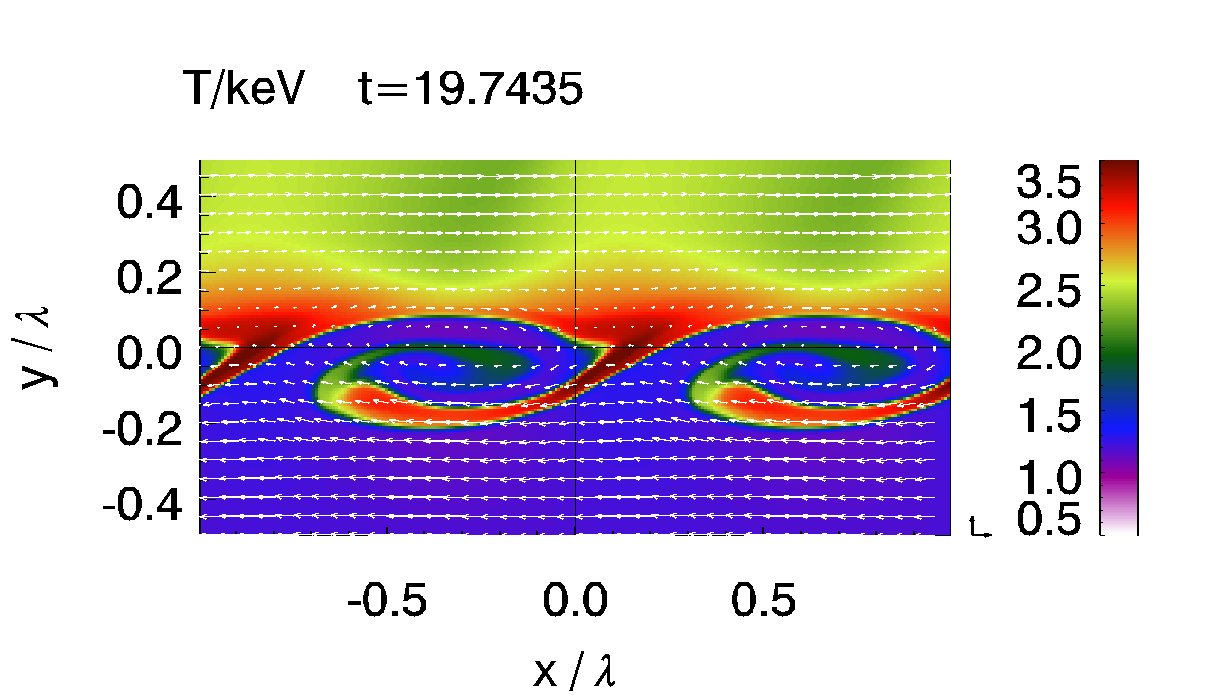}
\includegraphics[trim=200 155     0 160,clip,height=1.9cm]{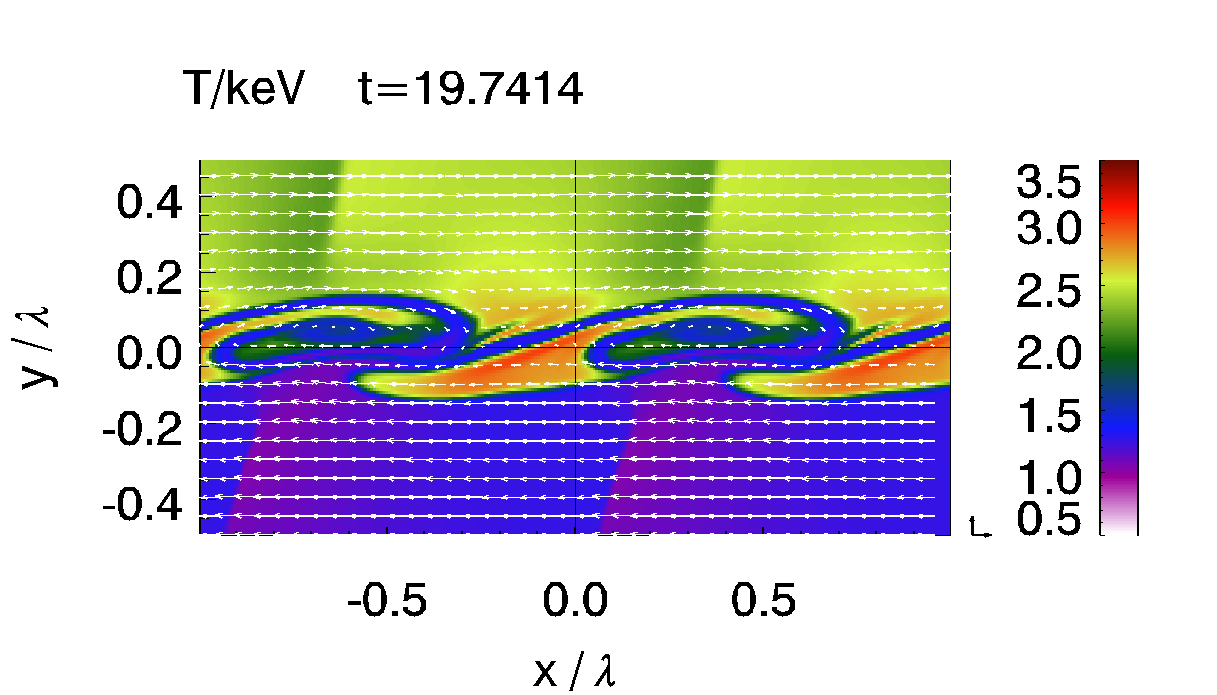}
\includegraphics[trim=    0 0 260 160,clip,height=2.65cm]{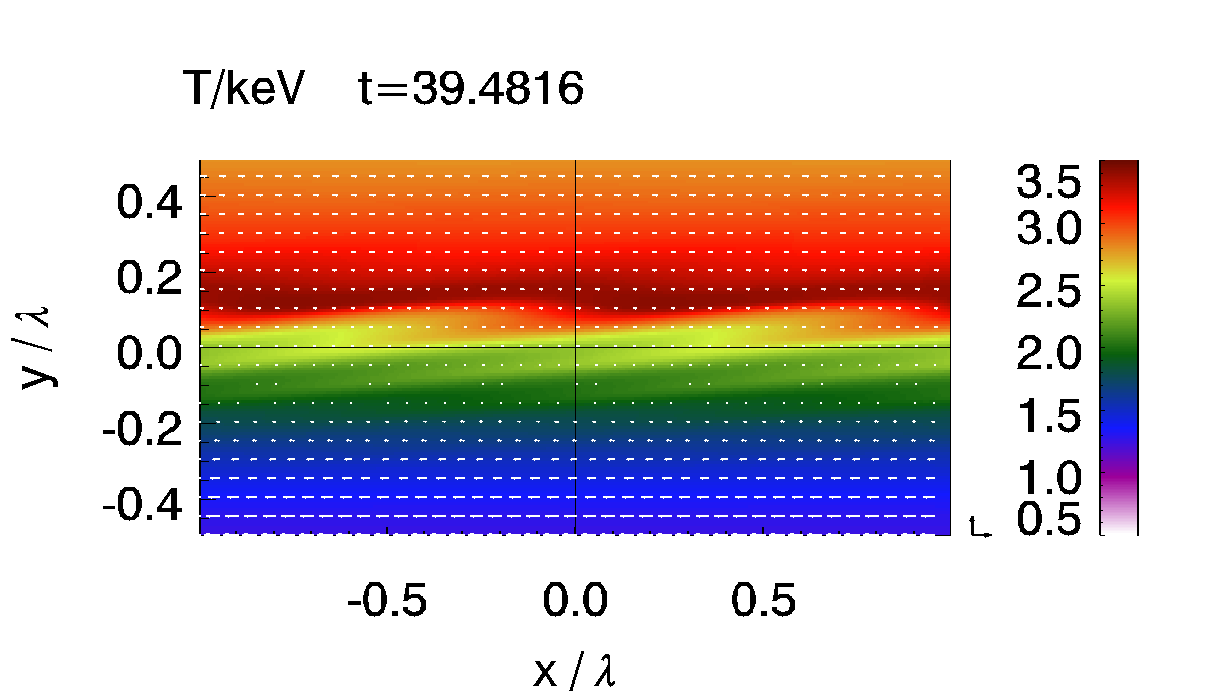}
\includegraphics[trim=200 0 260 160,clip,height=2.65cm]{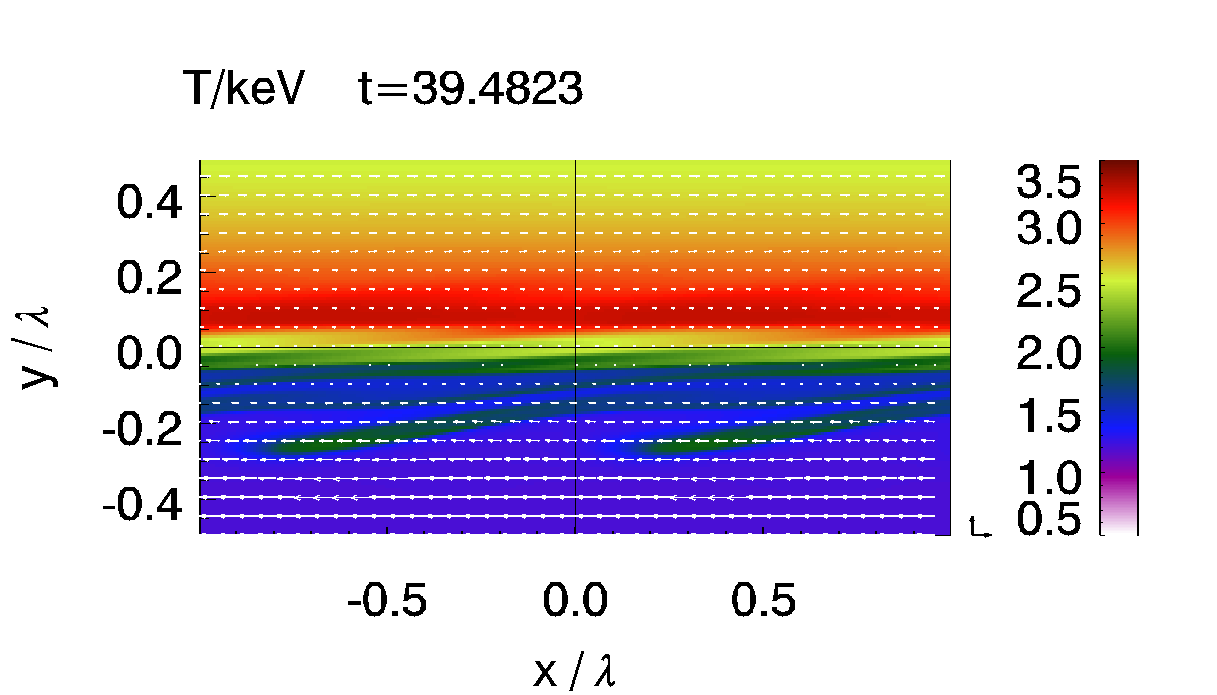}
\includegraphics[trim=200 0 260 160,clip,height=2.65cm]{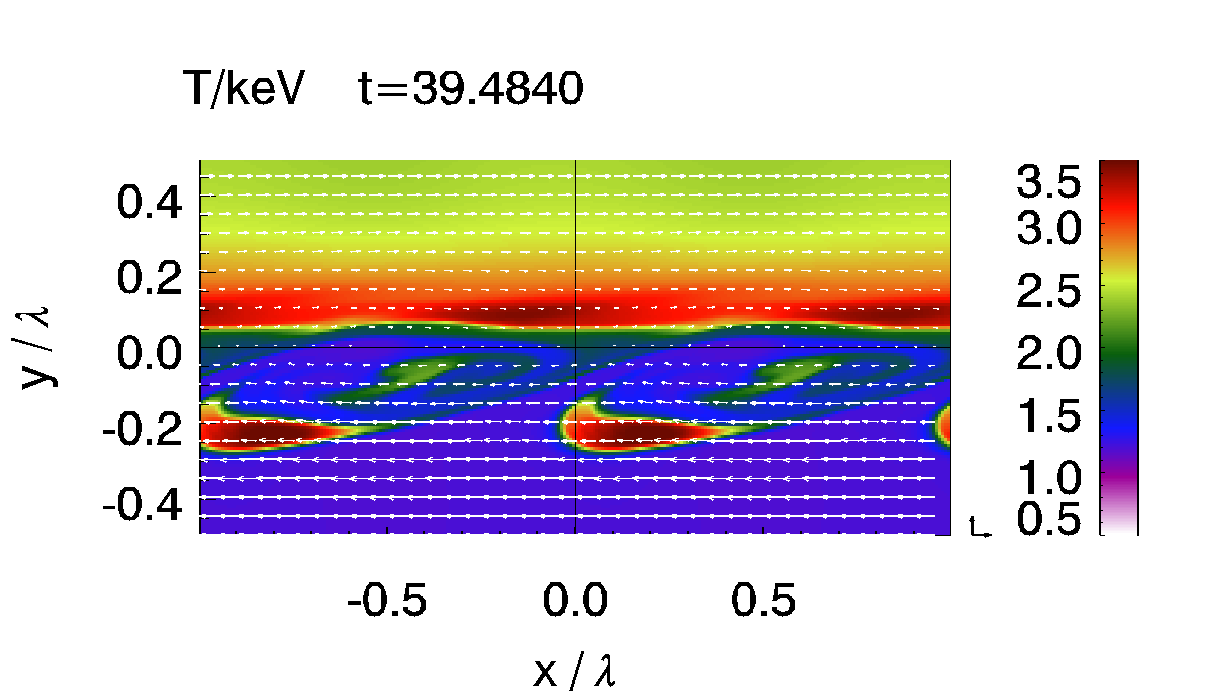}
\includegraphics[trim=200 0     0 160,clip,height=2.65cm]{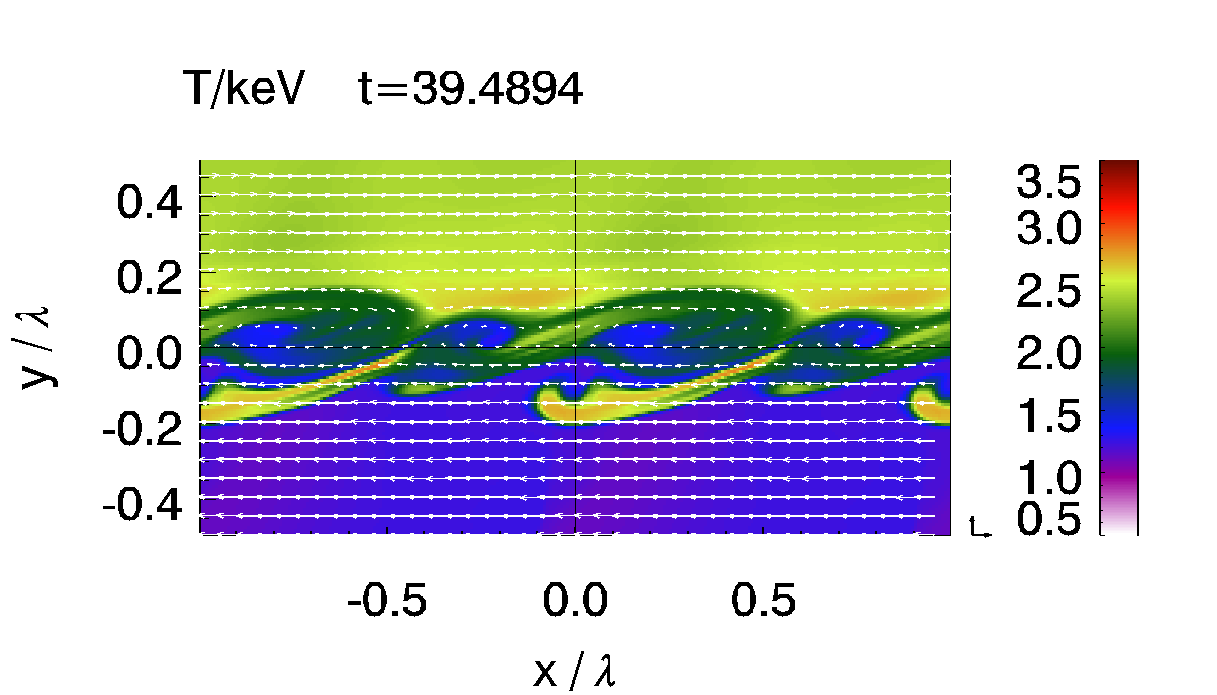}
\caption{Snapshots for simulations with Spitzer viscosity, density ratio of 2, shear flow of Mach 1.5 (relative to hot layer), different $\Reyn$ as given at the top. The colour codes $T/\KeV$,  velocity vectors are overlaid.
We show the timesteps 10, 20, 40 in units of $\tau\KHinvisc$. 
We note that the comparison to the KHI growth time derived in Eqn.~\ref{eq:tau_invisc} are only marginally meaningful, because this growth time assumes an incompressible gas, which is not true for supersonic flows.}
\label{fig:rolls_Sp_D2_M1.5}
\end{center}
\end{figure*}

\begin{figure*}
\begin{center}
\hspace{1.5cm} $\Reyn=30$ \hfill $\Reyn=100$ \hfill  $\Reyn=300$ \hfill $\Reyn=1000$ \hfill\phantom{x}
\newline
\includegraphics[trim=0     155 260 160,clip,height=1.9cm]{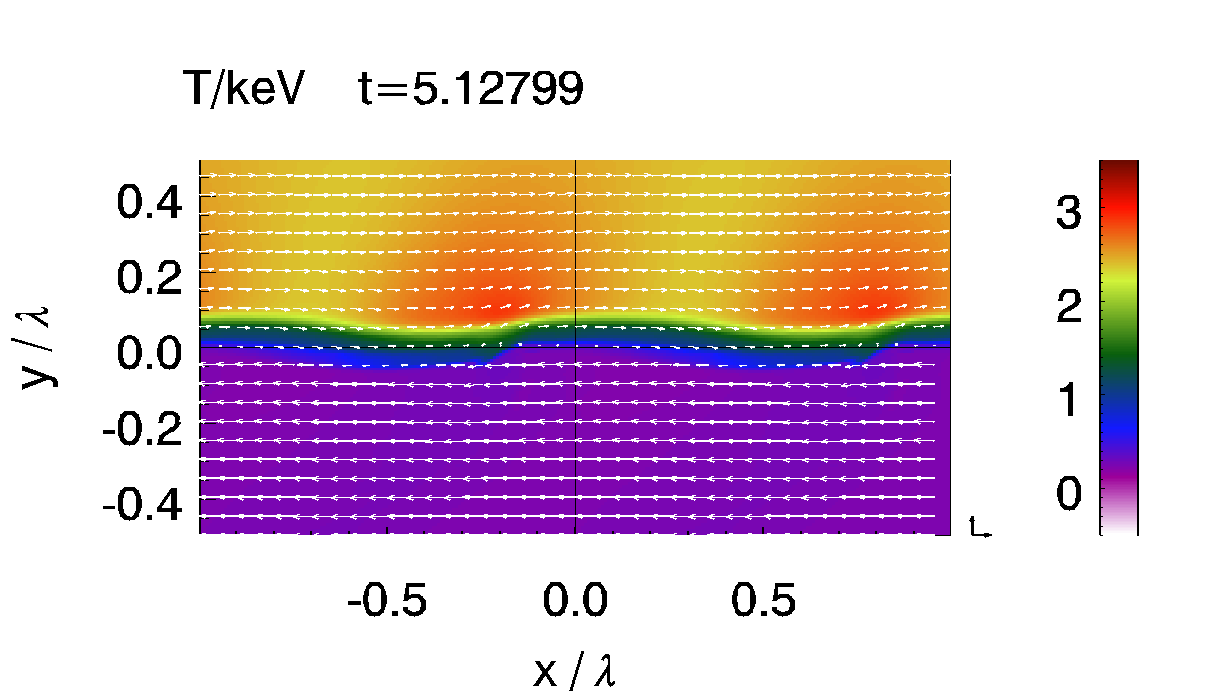}
\includegraphics[trim=200 155 260 160,clip,height=1.9cm]{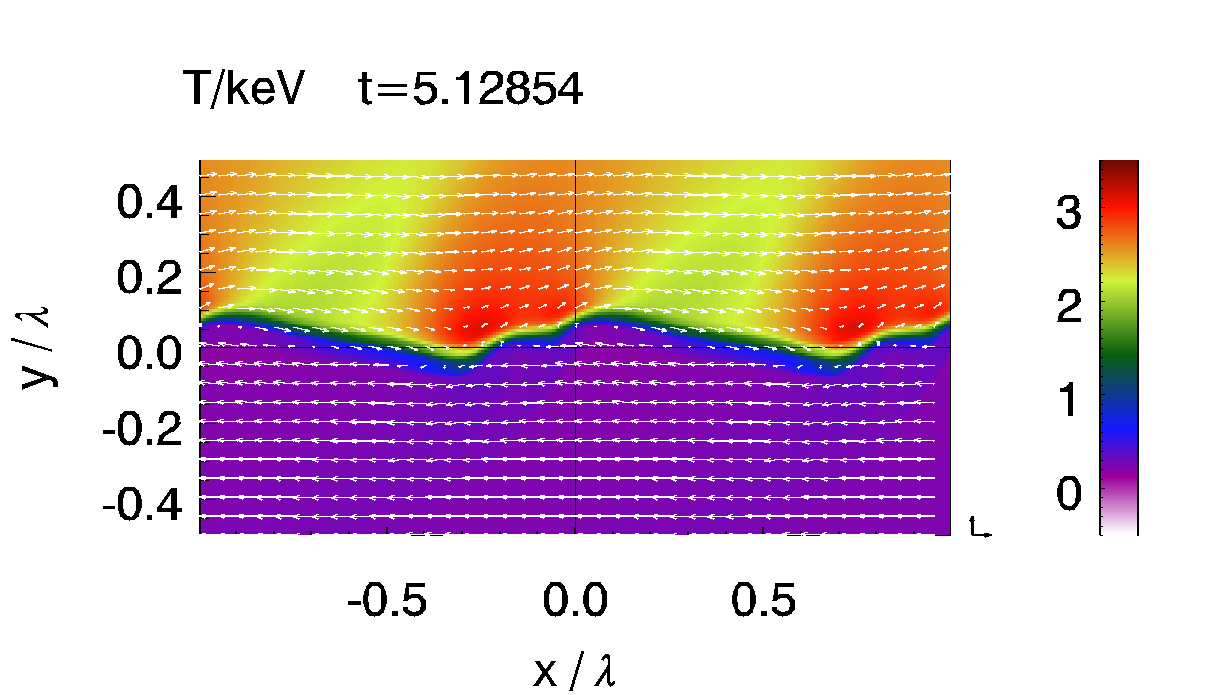}
\includegraphics[trim=200 155 260 160,clip,height=1.9cm]{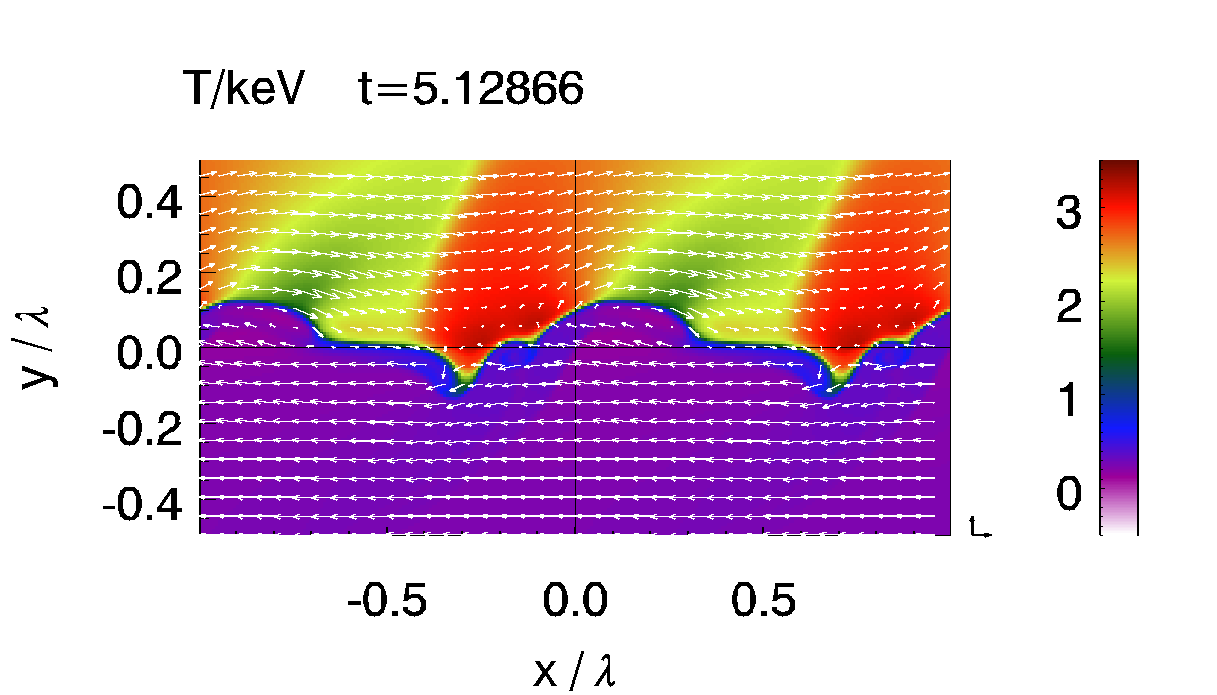}
\includegraphics[trim=200 155     0 160,clip,height=1.9cm]{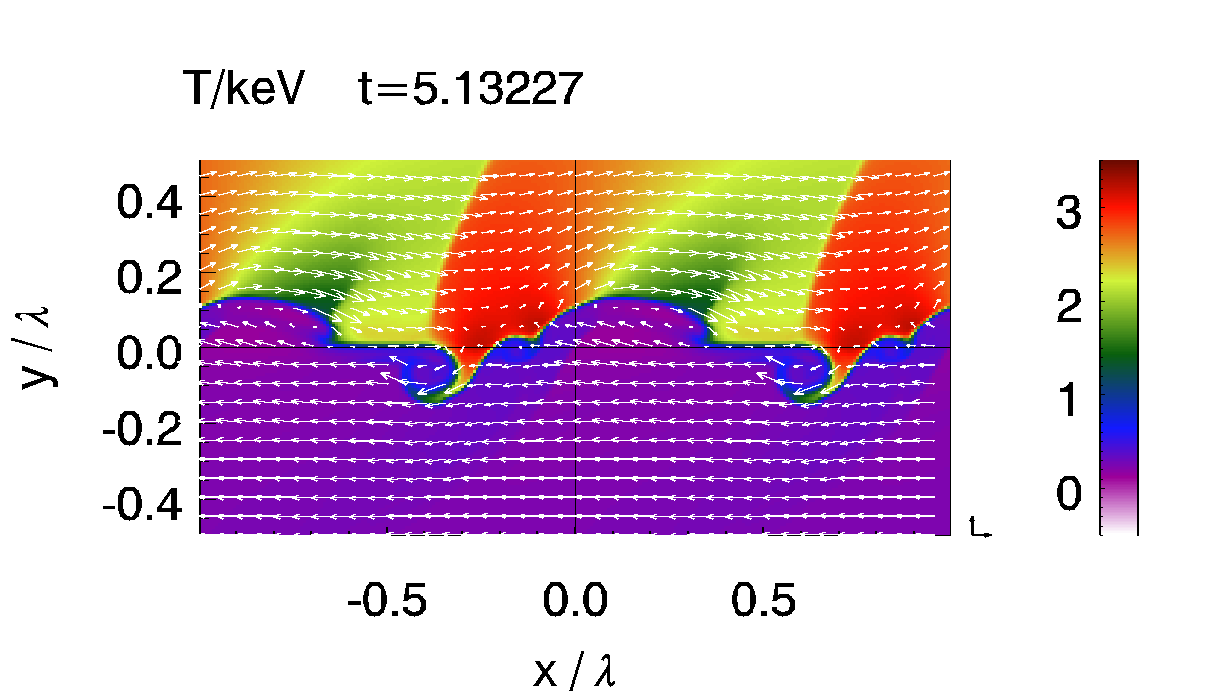}
\includegraphics[trim=0     155 260  160,clip,height=1.9cm]{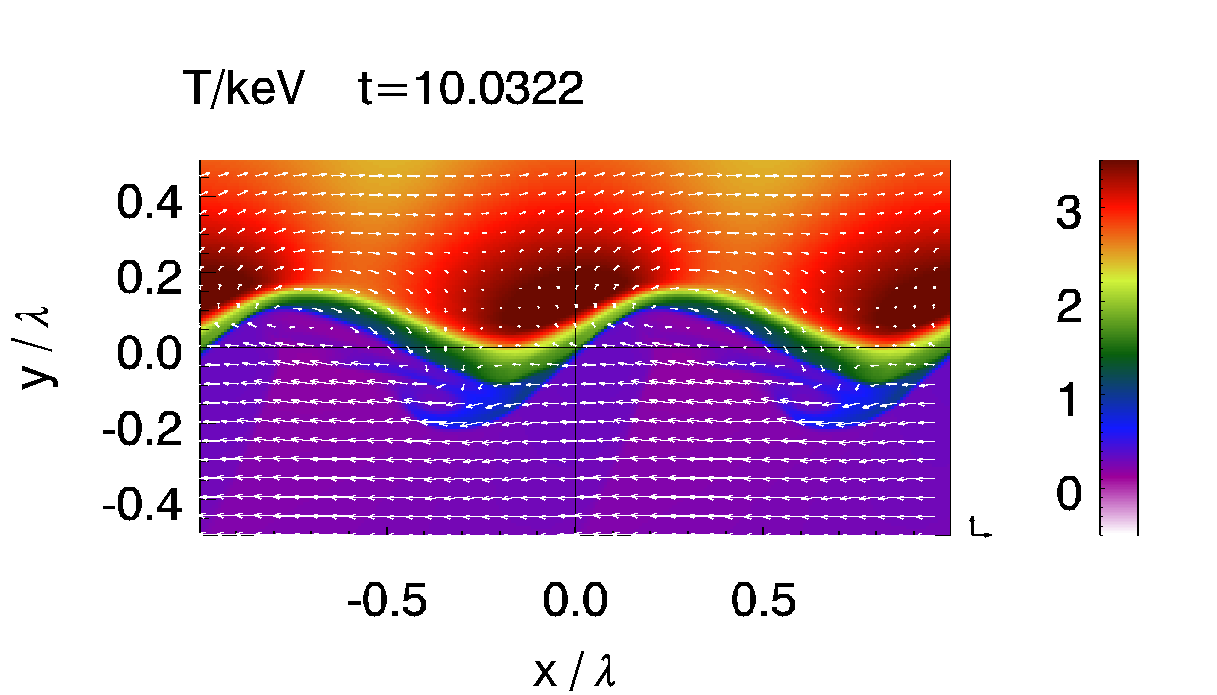}
\includegraphics[trim=200 155 260 160,clip,height=1.9cm]{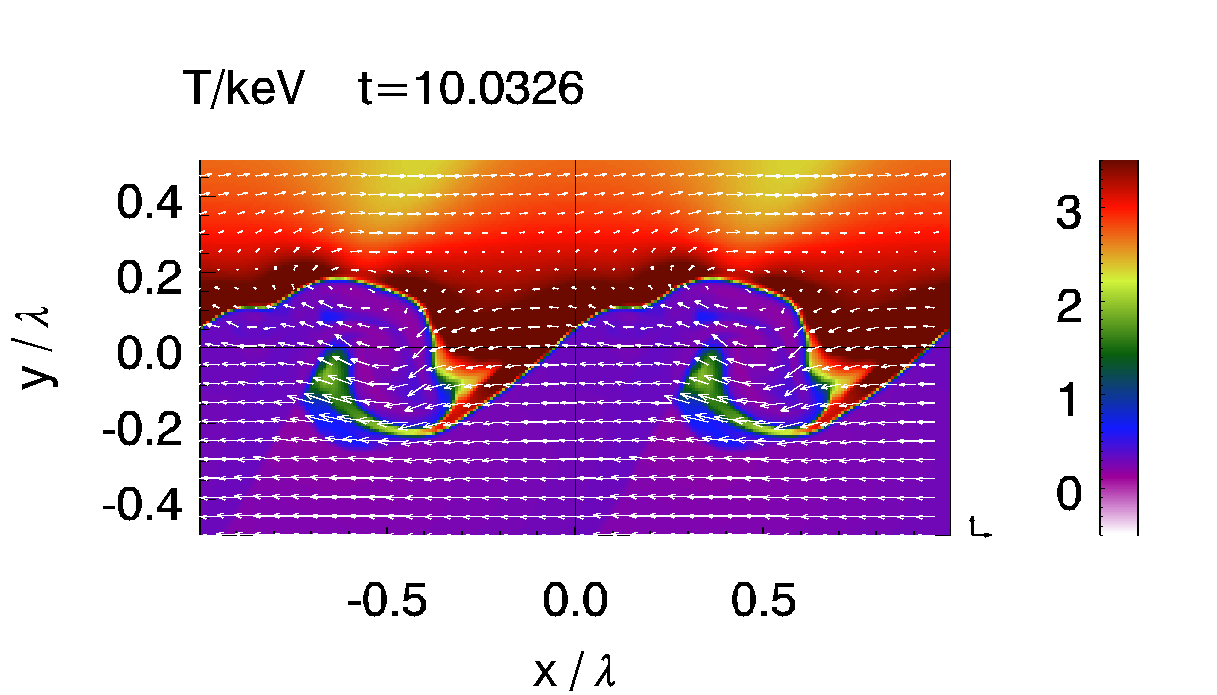}
\includegraphics[trim=200 155 260 160,clip,height=1.9cm]{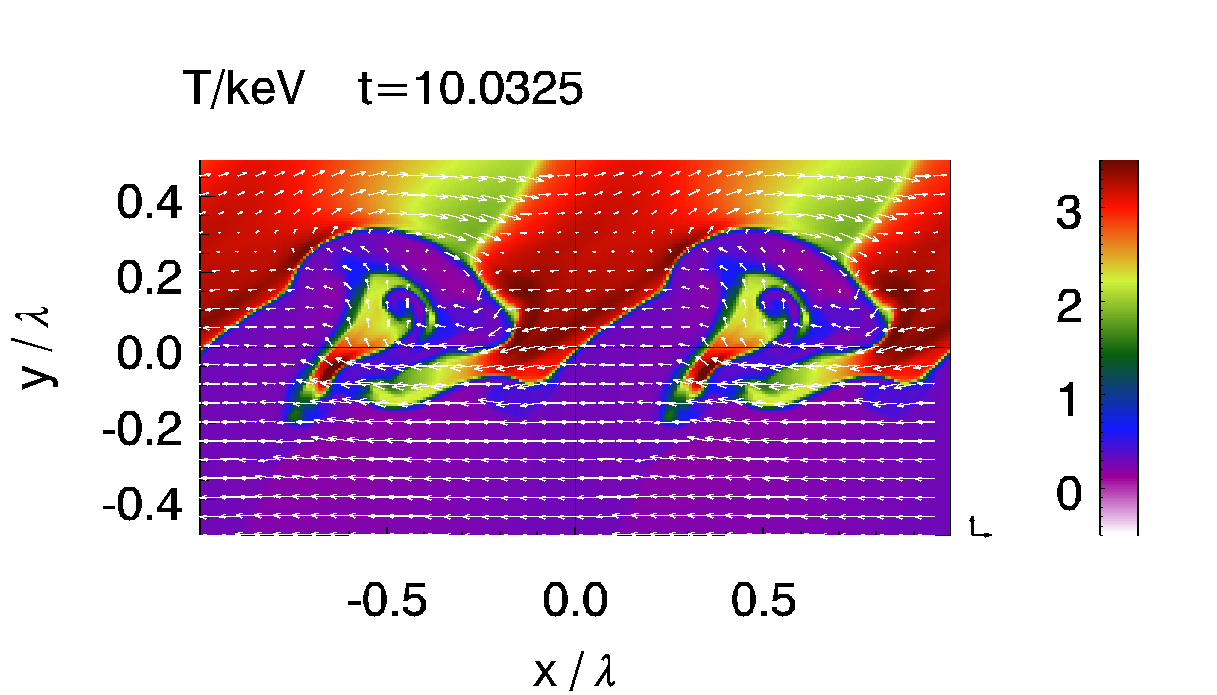}
\includegraphics[trim=200 155 0     160,clip,height=1.9cm]{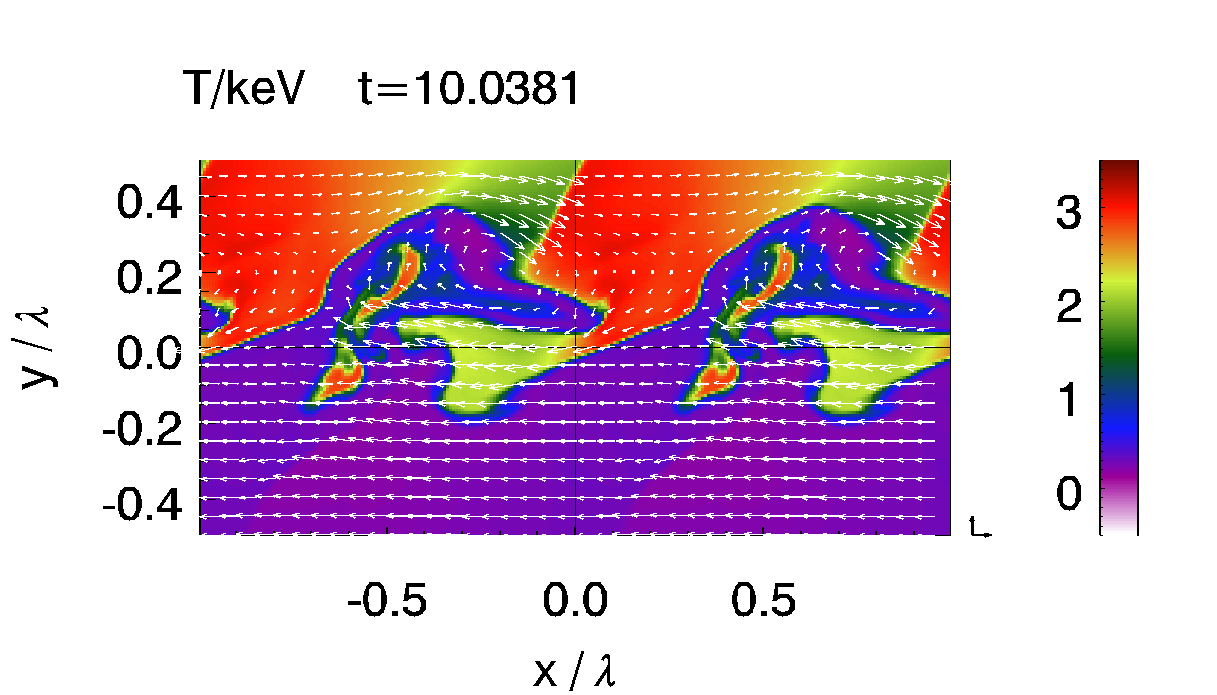}
\includegraphics[trim=0     0 260 160,clip,height=2.65cm]{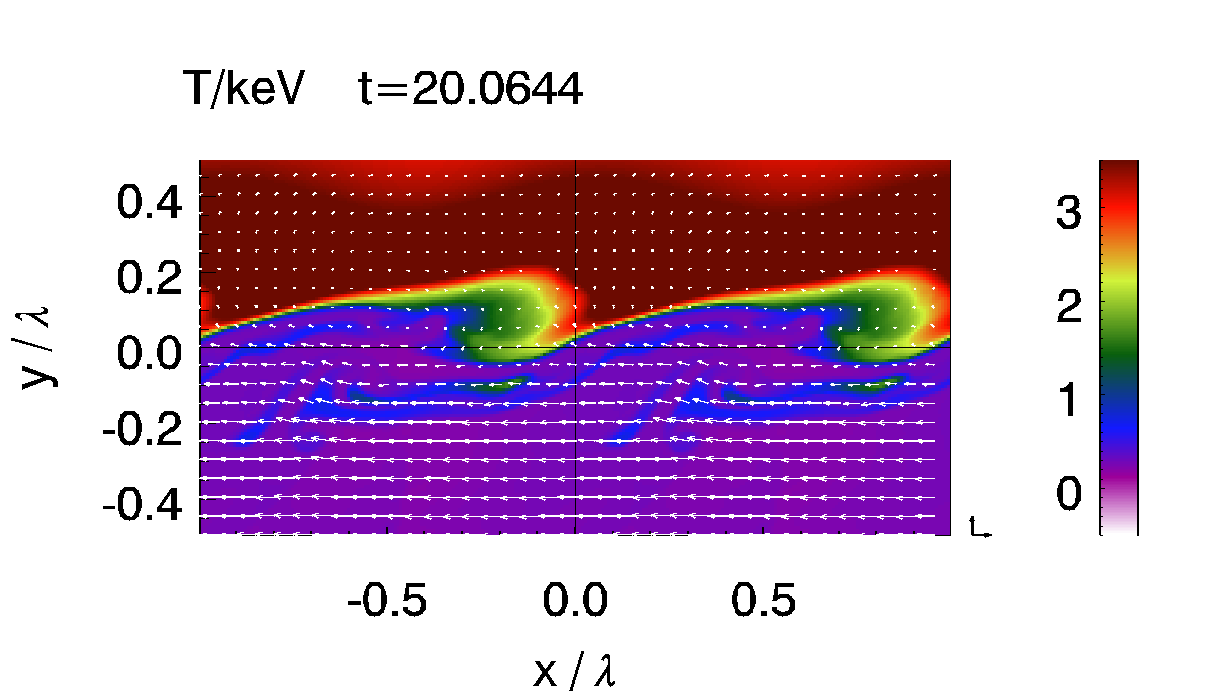}
\includegraphics[trim=200 0 260 160,clip,height=2.65cm]{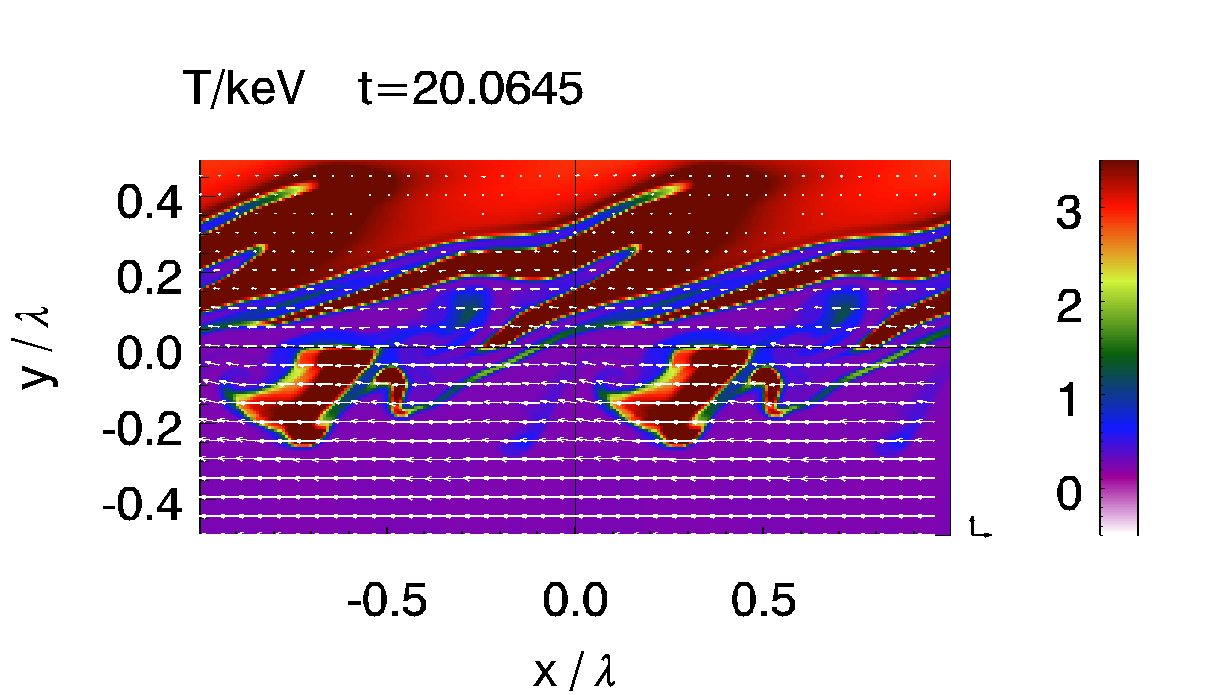}
\includegraphics[trim=200 0 260 160,clip,height=2.65cm]{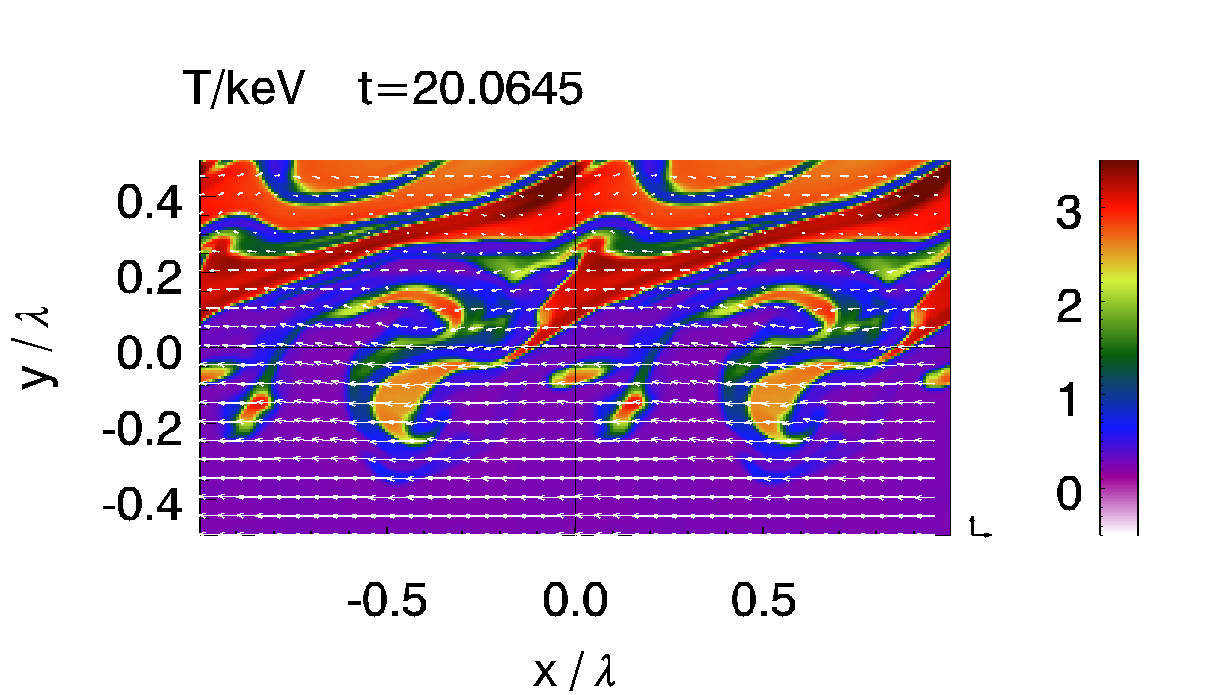}
\includegraphics[trim=200 0 0     160,clip,height=2.65cm]{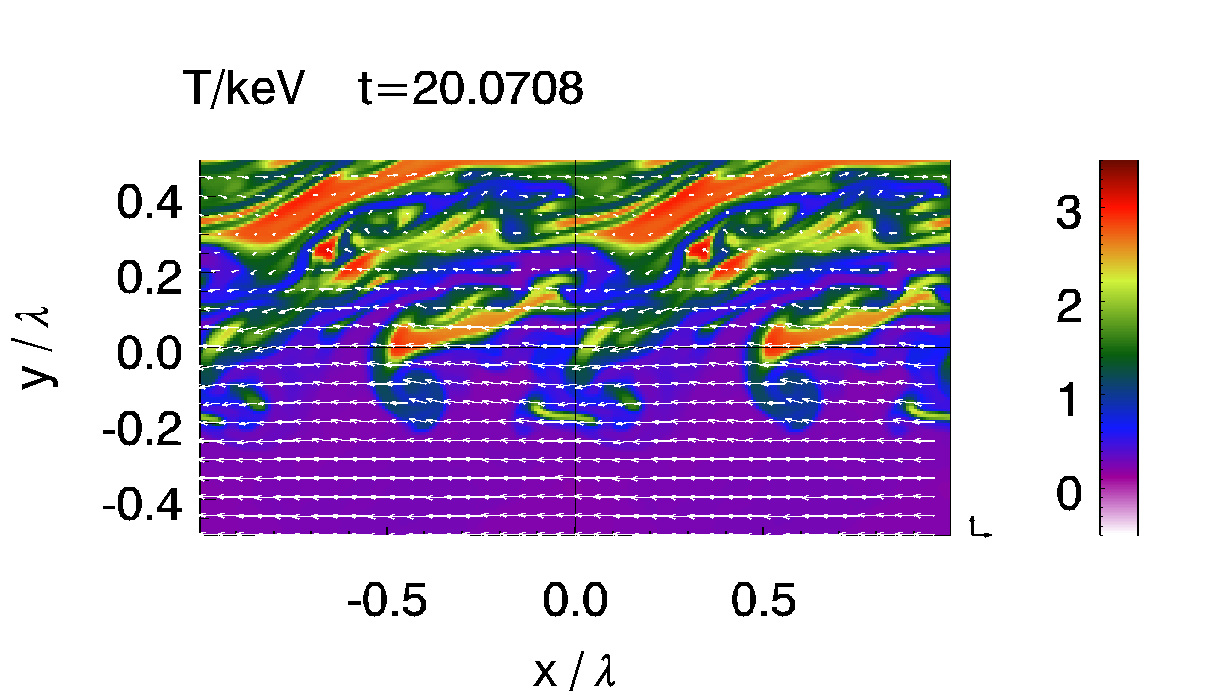}
\caption{Snapshots for simulations with Spitzer viscosity, density ratio of 10, shear flow of Mach 1.5 (relative to hot layer), different $\Reyn$ as given at the top. The colour codes $T/\KeV$,  velocity vectors are overlaid.
We show the timesteps 5, 10, 20 in units of $\tau\KHinvisc$.}
\label{fig:rolls_Sp_D10_M1.5}
\end{center}
\end{figure*}

In this section we investigate shear velocities that are mildly supersonic also with respect to the hot layer.  Figures~\ref{fig:rolls_Sp_D2_M1.5} and \ref{fig:rolls_Sp_D10_M1.5} show series of snapshots for a shear flow of Mach number 1.5 relative to the hot layer and for different Re and for density contrasts 2 and 10, respectively.  The "hills" cool gas now cause bow shocks in the hot gas ahead of them. Due to the larger amount of  momentum available for dissipation the viscous heating is more prominent at  higher shear velocity and at higher viscosity. For all supersonic runs we smoothed the initial interface over 1\% of the perturbation length scale to suppress secondary instabilities and to avoid excessive viscous heating (see Eqn.~\ref{eq:smooth}).

We  applied the more interesting case of a Spitzer viscosity in Figs.~\ref{fig:rolls_Sp_D2_M1.5} and \ref{fig:rolls_Sp_D10_M1.5}. At high Reynolds number, the KHI still evolves similar to the subsonic case. However, at such high velocities the compressibility of the gas affects the growth rate of the KHI and the classic estimate in Eqn.~\ref{eq:tau_invisc} is not valid anymore (see review by \citealt{Gerwin1968} and references therein). One consequence is that in the 2D case a shear flow above a certain critical Mach number stabilises the KHI. The case studied here is still below this limit. The viscous suppression of the KHI proceeds very similar to the subsonic case. At low density contrast (Fig.~\ref{fig:rolls_Sp_D2_M1.5}) the KHI is suppressed below $\Reyn \lesssim 100$, at a higher density contrast of 10 the critical Re is $\sim 30$ (Fig.~\ref{fig:rolls_Sp_D10_M1.5}).  The instability again enters the hybrid state where some vorticity is inserted in the cold layer.

\section{Discussion} \label{sec:discussion}

We investigated the long-term evolution of the viscous KHI for the case of a constant kinematic viscosity and a strongly  temperature-dependent Spitzer-like viscosity. We showed that viscosity suppresses the KHI below a critical Reynolds number and derived  the dependence of this critical Reynolds number on density contrast, shear flow velocity and constant or temperature-dependent viscosity.

At high Reynolds numbers ($\gtrsim 10^4$) and moderate density contrasts ($\le 10$), our simulations reproduce the results of earlier inviscid KHI simulations and of the analytical estimate. Most of those simulations, however, concentrate on the early phase of the KHI and use the numerically derived growth rate as a test for the code performance. 
\citet{Junk2010} presented  viscous KHI simulations in order to determine the intrinsic viscosity of SPH codes, and compared to viscous grid simulations with the FLASH code. They find a significantly weaker impact of viscosity than reported here. However, their analysis focusses on the onset of the instability only, i.e., long before the interface starts rolling up.  

Our simulations were run in two dimensions, where modes with wavevectors inclined to the shear direction cannot exist.  These can be the fastest growing modes for transonic shear speeds and the only unstable modes for supersonic shear (see App.~\ref{app:compressibility}). However, the effect of viscosity is not expected to be significantly different in three dimensions because the viscous spreading of the shear flow discontinuity proceeds in the same manner. The reduction of the local shear velocity should affect  all modes. The presence of the third dimension allows for more complex morphologies of the KH rolls. This effect along with the influence of a realistic perturbation spectrum are relevant for the detailed structure of the mixing layer, which will be the focus of a subsequent paper.

\subsection{Plausible Reynolds numbers in the ICM} 
\label{sec:sensibleRe}

In the hydrodynamic paradigm, transport processes are related to the mean free path $\lambda\Mfp$ in the medium. For example, the dynamic viscosity is (\citealt{Sarazin1988})
\begin{equation}
\mu \approx \frac{1}{3}\rho c \lambda\Mfp \label{eq:mfp_mu},
\end{equation}
where $c$ the sound speed. In the hydrodynamic paradigm $\lambda\Mfp$ must be much smaller than scales of interest. This condition put limits on reasonable viscosities, or Reynolds numbers. For the KHI, we could demand that the mean free path is much smaller than the perturbation wavelength $\lambda$, say,
\begin{equation}
\lambda\Mfp < \lambda/10. 
\end{equation}
With Eqns.~\ref{eq:mfp_mu} and \ref{eq:ourRe} this condition corresponds to 
\begin{equation}
\Reyn > 15\; \frac{\Mach}{0.5},
\end{equation}
where $\Mach=U/c$  is the Mach number of the shear velocity $U$. Thus, our derived critical Reynolds numbers are in the hydrodynamic regime where the required viscosity still implies a reasonable mean free path. 

A second issue to consider is a possible saturation of viscosity, or more precisely, momentum transport. If the scale length of the velocity gradient is smaller than the mean free path, momentum transport as treated here becomes supersonic. As this is unrealistic, it would saturate to a maximum momentum flux (\citealt{Sarazin1988}).
 Thus, for any given KHI setup with shear velocity $U$, density $\rho$, perturbation wavelength $\lambda$ and initial width of the interface $w$, we can rewrite the condition $w>\lambda\Mfp$ by using Eqn.~\ref{eq:mfp_mu}  to
\begin{equation}
\Reyn \; \frac{w}{\lambda}> 3 \, \Mach, \label{eq:Resat}
\end{equation}
thus expressing the minimal Reynolds number where viscosity can still act unsaturated. We note that $\Reyn \,w/\lambda = wU\rho/ \mu = \Reyn_w$, the Reynolds number related to the initial width of the interface. We verified that an initial interface width of up to $w=0.03\lambda$ influences the growth of the KHI only weakly. From $w=0.1\lambda$ on the initial smoothing of the interface alone drastically slows down the KHI. Demanding $w \le 0.03\lambda$, Eqn.~\ref{eq:Resat} requires $\Reyn > 100 \,\Mach$ to remain in the unsaturated viscosity domain. At smaller $\Reyn$, our simulations progressively overestimate the effect of viscosity. Consequently, our results for the constant kinematic viscosity are unaffected by the issue of saturation, because the KHI is already suppressed for $\Reyn \sim 100$ to 200. In the case of a Spitzer-like viscosity Reynolds numbers of $\sim 30$ are required to suppress the KHI even without taking saturation effects into account. This critical $\Reyn$ is only slightly below the saturation limit of Eqn.~\ref{eq:Resat}, hence we conclude that also a Spitzer-like viscosity has a significant effect on the KHI. Given that the real behaviour of the ICM is even more complex than the question of saturated or unsaturated momentum transport, we refrain from investigating the details of a saturated momentum transport here. 

We note that the nature of the effective viscosity in the ICM is still unclear, and the potentially large mean free path from Coulomb collisions may not even be a concern. For example, \citet{Guo2012b} discuss in some detail the possible origin and amplitude of viscosity and conclude that Spitzer-like amplitude plausible.

\subsection{Application to specific shear layers} 
For any given shear layer, the condition for the growth of the KHI, $\Reyn > \Reyn\Crit$, can be translated into a critical wavelength above which the KHI can grow:
\begin{eqnarray}
\lambda >\lambda\Crit &= & 30\Kpc\; \; \frac{\Reyn\Crit}{30} f_{\mu}\; \left(\frac{U}{400\Kms}\right)^{-1} \nonumber \\
&&  \left(\frac{n_e}{10^{-3}\ccm}\right)^{-1}  \;\left( \frac{kT\ICM}{2.4\KeV}  \right)^{5/2}.
\label{eq:lambdacrit}
\end{eqnarray}
As a typical value, we used the more conservative $\Reyn\Crit$ for Spitzer-like viscosity here. We apply this relation to observed shear layers in the ICM in the following subsections. 

\subsubsection{KHI at sloshing CFs} \label{sec:obs_cfs}

\begin{table*}
\begin{minipage}{\textwidth}
\caption{Full Spitzer viscosity should suppress the KHI for perturbations  $<\lambda\Crit$ (Eqn.~\ref{eq:lambdacrit}).  This table lists $\lambda\Crit$ at the sloshing CFs in different clusters, along with assumed values for temperature $T\ICM$ and electron density $n_e$ on the hotter side, and shear velocity $U$ at each CF. Furthermore, we list for each CF its distance to the cluster/group centre $r\CF$, and the scale of observed KHIs. Where no observed scale is given, the CF is not obviously distorted and a dedicated investigation is needed to determine upper limits on instability length scales.}
\label{tab:examples}
\centering
\begin{tabular}{lcccccc}
\hline\hline
object & $T/$ & $n_e/ $ & $U/ $ & $\lambda\Crit/$ & $r\CF/$ & observed $\lambda\KHI/$ \\
 & keV & $10^{-3}\ccm$ & $\Kms$ & kpc & kpc & kpc \\
\hline
N7618/U12491\footnote{\citealt{Roediger2012n7618}} &  1.2 & 2.5 & 200 & 4 & 20 & 15  \\
Virgo\footnote{\citealt{Roediger2011}} northern CF & 2.5 & 2 & 300 & 22 & 90  &  \\
A496\footnote{\citealt{Roediger2012a496}} northern CF & 4.2 & 8 & 400 & 15 & 60 & 20 \\
A2142\footnote{\citealt{Markevitch2000}} south-east CF & 8 & 10 & 400 & 60 & 70 &  \\
A2142 north-west CF & 8 & 2 & 600 & 200  & 360 & \\
\hline
\end{tabular}
\end{minipage}
\end{table*}

In \citet{Roediger2012n7618} we applied a relation like Eqn.~\ref{eq:lambdacrit} to sloshing cold fronts observed in several clusters and estimated whether or not these particular shear interfaces are expected to be KH unstable. In this work we used the too low critical Reynolds number based on the dispersion relation of \citet{Junk2010}. Hence we update the estimates for $\lambda\Crit$ at these sloshing CFs with our improved, higher $\Reyn\Crit$ in Table~\ref{tab:examples}. At full Spitzer viscosity, the KHI should be suppressed in the hottest cluster in this list, A2142, but KHIs could occur in cooler systems. Indeed, A496 and NGC7618/UGC12491 show characteristically distorted CFs.

The above estimate includes \textit{solely} the effect of viscosity. Sloshing CFs, however are special interfaces, and  viscous suppression is not the only effect on KHIs here. The gas flow patterns at sloshing CFs lead to an enhanced temperature on the hotter side of the CFs by about a factor of 1.3 compared to the azimuthal average (see simulations, e.g., \citealt{Roediger2012fastslosh,ZuHone2010}), which boosts the effect of viscosity by a factor of $\sim 2$ compared to the estimates given in Table~\ref{tab:examples}. Additionally, modes above but close to the critical length scale are slowed down. Furthermore,  gravity slows down or suppresses long modes even further. Finally, a sloshing CF itself, i.e., the shear interface, is a dynamic phenomenon and not a stationary interface. Thus, KHIs originating from a given perturbation have only a finite amount of time to grow into recognisable patterns. For example, at the northern CF in the Virgo cluster gravity suppresses KHI modes longer than 50 kpc. Our Virgo-specific simulations (\citealt{Roediger2013virgovisc}) demonstrate that, given the additional complexities of sloshing CFs, already 10\% of the Spitzer viscosity suppresses the KHI along the northern CF. Thus, in the context of sloshing cold fronts, the critical length scales given in Table~\ref{tab:examples} are lower limits only.

\subsubsection{KHI at gas-stripped elliptical galaxies and subcluster merger cores} \label{sec:obs_ellgal}
%
\begin{table*}
\begin{minipage}{\textwidth}
\caption{Full Spitzer viscosity should suppress the KHI for perturbations  $<\lambda\Crit$ (Eqn.~\ref{eq:lambdacrit}).  This table lists the critical wavelength $\lambda\Crit$ for several gas stripped elliptical galaxies and remnant merger cores, along with the assumed ambient ICM temperature $T\ICM$, ambient electron density $n_e$, galaxy infall velocity $v\Gal$. We assume a shear velocity at the side of the galaxy/merger core of $0.5\times v\Gal$. 
 Furthermore, we list for each galaxy/merger core the radius of its current ISM atmosphere  $r\Gas$, and the scale of observed KHIs, $\lambda\KHI$.}
\label{tab:rps}
\centering
\begin{tabular}{lcccccc}
\hline\hline
object & $T\ICM/$ & $n_e/ $ & $v\Gal/ $ & $\lambda\Crit/$ & $r\Gas/$ & observed $\lambda\KHI/$ \\
 & keV & $10^{-3}\ccm$ & $\Kms$ & kpc & kpc & kpc \\
\hline
NGC 4552 (M89)\footnote{\citealt{Machacek2006a}} & 2.5 &  0.3  & 1700 & 50  & 3  &  3 \\
NGC 4472 (M49)\footnote{\citealt{Kraft2011}}           & 2 to 2.5 & 0.1 to 0.2 & 950 to 1500 & 50 to 270 &  20 & 10 \\
M86 \footnote{\citealt{Randall2008}}                           & 2.5 & 0.2 & 1700 & 76 & 25  &  \\
NGC 1404\footnote{\citealt{Machacek2005}}                    & 1.5 & 1 & 600      & 12  & 8 &  \\
Bullet Cluster\footnote{\citealt{Owers2009hifid,Markevitch2002,Springel2007}} & 18 & 3 & 1350 & 440 & 20 &  \\
A3667\footnote{\citealt{Owers2009hifid,Vikhlinin2002}} & 8 & 0.8 & 700 &  420  & 300 & 200 \\
\hline
\end{tabular}
\end{minipage}
\end{table*}

Galaxies moving through the ICM experience a ram pressure that can remove part of their interstellar medium (ISM), leading to a tail of stripped gas trailing behind the galaxy. Recent observations reveal the complex structure of such stripping tails for both spiral galaxies (\citealt{Owers2012}, \citet{Sun2010a}, 
Zhang et al., submitted)  and elliptical galaxies (e.g., for the Virgo ellipticals  M89, \citet{Machacek2006a}, Kraft et al., in prep.,  NGC 4472, \citealt{Kraft2011}, and M86, \citealt{Randall2008}, and the Fornax elliptical NGC 1404, \citealt{Machacek2005}). \citet{Churazov2004} demonstrated that already a tiny intrinsic width of the interface between the galaxy gas and the ICM at the upstream side of such galaxies can suppress local KHIs. Consequently, KHIs appear only at the sides of the galaxy or along their tails, as confirmed by numerical simulations (e.g., \citealt{Iapichino2008}). These KHIs at the sides and the tail are a major agent to mix the cold stripped gas with the ambient ICM. 

We applied Eqn.~\ref{eq:lambdacrit} to several galaxies and calculated expected critical wavelengths for full Spitzer viscosity. The result along with assumptions regarding shear flow and gas densities in the ICM are listed in Table~\ref{tab:rps}. For the Virgo ellipticals NGC 4552, NGC 4472, and M86, the critical perturbation length exceeds the current radius of the galaxies' ISM atmospheres by a factor of several, whereas for the Fornax elliptical NGC 1404 the radius of the gaseous halo and the critical length are comparable. Thus, at full Spitzer viscosity, none of the galaxies should experience KHIs at its sides, and the tails should start mixing with the ICM only several ISM radii downstream. If viscosity is suppressed at a comparable level in all cases, NGC 1404 is most likely to experience KHIs. 

The observations of these galaxies draw a mixed picture. M86 has a spectacular cold and 150 kpc long  tail, suggesting a significant suppression of mixing. NGC 4472 and NGC 4552 show distorted upstream edges resembling KHIs on scales smaller than the critical length scale. NGC 1404 has a smooth upstream edge, but a filamentary and warm tail.
 
The shear flows at gas stripped elliptical galaxies are in the intermediate density contrast regime $2<D_{\rho}<10$. Depending on the nature of viscosity, the flow patterns may be particularly interesting.  If viscosity is Spitzer-like, the galaxy stripping can occur in the hybrid KHI regime where viscous damping suppresses the mixing of the cool gas stripped from the galaxy, but the cool stripped tail could be internally turbulent.

The cores of subclusters falling through their host clusters undergo a  scenario very similar to the gas stripped galaxies.  We include in Table~\ref{tab:rps} the Bullet cluster and A3667. In the former case, the small remnant merger core is surrounded by hot, shocked gas, and we expect KHIs only on scales much larger than the merger core. No obvious distortions in its upstream cold front are observed. In contrast, the CF in A3667 shows kinks on scales of $\sim 200\Kpc$ (see, e.g., Fig.~3 in \citealt{Owers2009hifid}, also \citealt{Mazzotta2002}). This is roughly consistent with our expected critical perturbation length of $\sim 300\Kpc$.

\subsubsection{High velocity clouds}
High velocity clouds (HVC) falling into the halos of the Milky Way or other galaxies are  to some degree a scaled-down version of elliptical galaxies falling into clusters. While the HVCs themselves have a cold core of atomic gas, the galactic halo gas is an ionised plasma like the ICM. Indeed, at full Spitzer viscosity, the infall of HVCs leads to a similar low Reynolds number as for the galaxy stripping:
\begin{eqnarray}
\Reyn &=&  7  \; f_{\mu}^{-1}\; 
\left(\frac{V}{100\Kms}\right)\; 
\left(\frac{R}{0.1\Kpc}\right) \times \nonumber\\
&& \left(\frac{n_e}{10^{-4}\ccm}\right) \; \left( \frac{kT\ICM}{0.1\KeV}  \right)^{-5/2}. \label{eq:ReICM_HVC}
\label{eq:Re_hvc}
\end{eqnarray}
To our knowledge, HVCs have been studied only in the high Reynolds number regime (e.g., \citealt{Heitsch2009}, \citealt{Pittard2010}).
The density contrasts between the cloud surface and the halo gas can be high, $\gtrsim 100$, but we would expect a full Spitzer viscosity to have a stabilising effect on instabilities at the cloud surface. The mixing between the stripped gas and the ambient halo gas is of particular interest as it can be traced in various ion lines (see \citealt{Kwak2011} and refs.~therein).  As discussed by \citet{Kwak2010} and \citet{Kwak2011}, the mixing and the interpretation of the observational data is complex due to additional effects like radiative cooling and non-equilibrium ionisation. Our results indicate that even if the viscosity may not fully suppress KHIs at scales of the cloud radius, it could be relevant for the final mixing levels in the cloud tails, i.e., whether stripped and ambient gas truly mix or whether cold clumps can remain. In addition to viscosity, also thermal conduction (\citealt{Vieser2007}) and magnetic draping could be relevant for the evolution of HVCs.

\subsubsection{AGN cavities}

In inviscid simulations, buoyantly rising AGN inflated bubbles are disrupted by RTIs and KHIs. 
The simulations of \citet{Reynolds2005} and \citet{Guo2012b} demonstrated that a viscosity {below} full Spitzer viscosity can stabilise both buoyantly rising cavities as well as recently inflated cavities. The density contrast between the gas exterior and interior of the simulated cavities  is around 100. The AGN inflated cavities are thought to be filled with a relativistic plasma whose viscosity is unconstrained either. Both of the above simulations use a constant dynamic viscosity as a first approximation in studying the stability of the cavities. \citet{Reynolds2005} find viscosity stabilises the buoyantly rising cavities for $\Reyn\lesssim 250$ ($\sim 1/4$ {Spitzer for their values}), where their Reynolds number refers to the size of their cavities. The resolved KHI that disrupt the cavities in their inviscid simulations are a factor of a few smaller than the bubble size, hence the Reynolds number for these dominant KHI is of the order of 100 when they are suppressed. This agrees with our prediction. 

\citet{Guo2012b} simulated the impact of viscosity on the stability of the Fermi bubbles observed in our Galaxy. To this end, they simulate the inflation of such bubbles in a viscous galaxy halo gas by an AGN jet. Already for viscosities of 0.1 to 1\% of the Spitzer level, viscosity can prevent KHIs at the boundaries of the inflated bubbles. {The lower level of viscosity compared to the Spitzer value is sufficient here because the bubble inflation causes shock heating of the ambient gas, where the higher temperature also boosts the viscosity.} Given that their simulations cover the highly dynamic cavity inflation phase, also the critical Reynolds number will change with time. Using canonical values for their simulation of shear velocity of $1000\Kms$ at the bubble boundary, a density of $10^{-29}\gccm$ inside the bubble,  $10^{-28}\gccm$  outside the bubble, a dynamic viscosity of  1 g cm$^{-1} \Sec^{-1}$, and perturbation length scale of 2 kpc, results in  Reynolds numbers of 6 and 60 in the hot and cold layer, respectively. At this viscosity the KHI at the bubble boundary is significantly suppressed in Guo et al.'s simulations. Given the high density contrast and the dynamic context, this is in rough agreement with our estimate. 

{\citet{Dong2009} investigate the interplay of anisotropic viscosity and magnetic field strength on rising cavities. They show that the evolution of the bubbles depends significantly on the field geometry. E.g., horizontal magnetic fields lead to bubbles that are stabilised only along the direction of the field lines. They thus would appear coherent when seen along a line-of-sight perpendicular to the field lines, but disrupted otherwise. With toroidal fields the bubbles transform into a stable ring. The situation of tangled magnetic fields has not been studied.}

\subsubsection{Turbulence in the ICM}

In hot clusters full Spitzer viscosity implies that important scales such as Kolmogorov scale $\eta \sim \frac{L}{Re^{3/4}}$ or Taylor scale $\lambda_T \sim \frac{3L}{Re^{1/2}}$ may be resolvable with the current generation of X-ray observatories. Here $L$ and $Re$ are the integral scale the corresponding Reynolds number. For example, in the Coma cluster core $T\approx 8.5$ keV, $n_e \approx 4~10^{-3}~{\rm cm^{-3}}$. Using Spitzer viscosity,  the Mach number $\sim 0.25$ and driving scale $L\sim 500$ kpc, yields $\eta\sim 20$ kpc and $\lambda_T\sim 170$ kpc.   The scales above $\sim 30$ kpc can be probed via the surface brightness analysis \citep[e.g.,][]{Churazov2012}. The scales predicted by the Eqn.~\ref{eq:lambdacrit} (for the velocity of 400 km/s) also falls into this range. In fact, numerically it is close to the Taylor scale $\lambda$. It will therefore be possible to search for structural changes in the density/velocity perturbation spectrum around these scales.  It is interesting that in the recent simulations of the Coma cluster \citep{Gaspari2013} with the effective viscosity at the level comparable to the Spitzer value, the Mach number of $\sim 0.25$ and the Reynolds number $\sim 10^2$, clear steepening of the density fluctuations power spectrum is indeed seen at scales of few tens of kpc.

\section{Summary} \label{sec:summary}
%
We investigated the long-term evolution of the viscous KHI for the case of a constant kinematic viscosity and a strongly  temperature-dependent Spitzer-like viscosity. We considered density ratios between the shear flow layers from 1 to 100. We expressed our results in terms of the Reynolds number that relates to the perturbation scale, i.e., as defined in Eqn.~\ref{eq:ourRe}. 

We showed that a constant kinematic viscosity suppresses the KHI for Reynolds numbers $\le 100$, and already for $\Reyn \le 200$ for density contrasts $\lesssim 2$. This agrees well with our  analytic estimate of the critical Reynolds number at low density contrasts.  We note that the long-term evolution of the boundary layer over 10 or 20 KHI growth times, i.e., the spinning of the KHI rolls, is affected already for Reynolds numbers of $\sim 1000$.  We derive an empirical relation between the viscous KHI growth time and $\Reyn$ (see Eqn.~\ref{eq:tau_Re_fit} combined with Eqns.~\ref{eq:ReCritFit} and \ref{eq:Re0Fit}). 

The strong temperature dependence of the Spitzer viscosity causes a significant difference of Reynolds numbers between the hot and the cold layer in a shear flow. The ratio of Reynolds numbers scales  as $D_{\rho}^{7/2}$. Consequently, only the viscosity in the hotter layer can suppress the instability, and Reynolds numbers below $\sim 30$ or 10 are required for density contrast of $2$ or $> 10$, respectively. In fact, at intermediate density contrasts around 10 the KHI enters a hybrid state where it does not mix both fluids, but induces turbulence in the cold layer. At lower density contrasts the evolution becomes more symmetric between both layers. 

At higher density contrast (at 100), the inertia of the cold dense layer is so large that turbulence is not induced in the cold layer even for high Reynolds numbers. 

{We apply our results to potential mixing layers in the ICM in galaxy clusters, i.e., sloshing cold fronts, gas stripped galaxies, AGN cavities, and turbulence. The difference between ongoing or suppressed mixing is observable with current X-ray observatories. There are several observations that indicate a viscosity significantly below the Spitzer value, but not all observations fit this picture straightforwardly.  It may well be that additional ICM properties such as magnetic fields or anisotropic transport processes even on macroscopic scales are required to explain all observations consistently.}

\section*{Acknowledgments}
%
E.R.~acknowledges the support of the Priority Programme 
ÓPhysics of the ISMÓ of the DFG (German Re- 
search Foundation), the supercomputing grants NIC 
 5027 and  6006 at the John-Neumann Institut at the 
Forschungszentrum J\"ulich, the hospitality of CfA during a Visiting Scientist Fellowship.
We thank Fulai Guo and Robi Banerjee for helpful discussions.

%
\bibliographystyle{mn2e}
\bibliography{library}

\begin{thebibliography}{}

\bibitem[\protect\citeauthoryear{Amsden \& Harlow}{Amsden \&
  Harlow}{1964}]{Amsden1964}
Amsden A.~A.,  Harlow F.~H.,  1964, Physics of Fluids, 7, 327

\bibitem[\protect\citeauthoryear{Batchelor}{Batchelor}{2000}]{BatchelorHydro}
Batchelor G.~K.,  2000, {An Introduction to Fluid Dynamics}, cambridge edn.
Cambridge University Press, Cambridge

\bibitem[\protect\citeauthoryear{Bonafede, Feretti, Murgia, Govoni, Giovannini,
  Dallacasa, Dolag \& Taylor}{Bonafede et~al.}{2010}]{Bonafede2010}
Bonafede A.,  Feretti L.,  Murgia M.,  Govoni F.,  Giovannini G.,  Dallacasa
  D.,  Dolag K.,    Taylor G.~B.,  2010, A\&A, 513, A30

\bibitem[\protect\citeauthoryear{Chandrasekhar}{Chandrasekhar}{1961}]{Chandrasekhar1961}
Chandrasekhar S.,  1961, {Hydrodynamic and hydromagnetic stability}.
Clarendon, Oxford

\bibitem[\protect\citeauthoryear{Churazov \& Inogamov}{Churazov \&
  Inogamov}{2004}]{Churazov2004}
Churazov E.,  Inogamov N.,  2004, MNRAS, 350, L52

\bibitem[\protect\citeauthoryear{Churazov, Vikhlinin, Zhuravleva, Schekochihin,
  Parrish, Sunyaev, Forman, B\"{o}hringer \& Randall}{Churazov
  et~al.}{2012}]{Churazov2012}
Churazov E.,  Vikhlinin A.,  Zhuravleva I.,  Schekochihin A.,  Parrish I.,
  Sunyaev R.,  Forman W.,  B\"{o}hringer H.,    Randall S.,  2012, MNRAS, 421,
  1123

\bibitem[\protect\citeauthoryear{Dong \& Stone}{Dong \& Stone}{2009}]{Dong2009}
Dong R.,  Stone J.~M.,  2009, ApJ, 704, 1309

\bibitem[\protect\citeauthoryear{Drazin \& Reid}{Drazin \& Reid}{2004}]{Drazin}
Drazin P.~G.,  Reid W.~H.,  2004, {Hydrodynamic Stability}, second edn.
Cambridge University Press, Cambridge

\bibitem[\protect\citeauthoryear{Dubey, Antypas, Ganapathy, Reid, Riley,
  Sheeler, Siegel \& Weide}{Dubey et~al.}{2009}]{Dubey2009}
Dubey A.,  Antypas K.,  Ganapathy M.~K.,  Reid L.~B.,  Riley K.,  Sheeler D.,
  Siegel A.,    Weide K.,  2009, Parallel Computing, 35, 512

\bibitem[\protect\citeauthoryear{Dupke, {White III} \& Bregman}{Dupke
  et~al.}{2007}]{Dupke2007}
Dupke R.,  {White III} R.~E.,    Bregman J.~N.,  2007, ApJ, 671, 181

\bibitem[\protect\citeauthoryear{Dursi \& Pfrommer}{Dursi \&
  Pfrommer}{2008}]{Dursi2008}
Dursi L.~J.,  Pfrommer C.,  2008, ApJ, 677, 993

\bibitem[\protect\citeauthoryear{Esch}{Esch}{1957}]{Esch1957}
Esch R.~E.,  1957, Journal of Fluid Mechanics, 3, 289

\bibitem[\protect\citeauthoryear{Ferrari, Govoni, Schindler, Bykov \&
  Rephaeli}{Ferrari et~al.}{2008}]{Ferrari2008}
Ferrari C.,  Govoni F.,  Schindler S.,  Bykov A.~M.,    Rephaeli Y.,  2008,
  Space Sci. Rev., 134, 93

\bibitem[\protect\citeauthoryear{Gaspari \& Churazov}{Gaspari \&
  Churazov}{2013}]{Gaspari2013}
Gaspari M.,  Churazov E.,  2013, eprint arXiv:1307.4397

\bibitem[\protect\citeauthoryear{Gerwin}{Gerwin}{1968}]{Gerwin1968}
Gerwin R.,  1968, Reviews of Modern Physics, 40, 652

\bibitem[\protect\citeauthoryear{Guo, Mathews, Dobler \& Oh}{Guo
  et~al.}{2012}]{Guo2012b}
Guo F.,  Mathews W.~G.,  Dobler G.,    Oh S.~P.,  2012, ApJ, 756, 182

\bibitem[\protect\citeauthoryear{Heitsch \& Putman}{Heitsch \&
  Putman}{2009}]{Heitsch2009}
Heitsch F.,  Putman M.~E.,  2009, ApJ, 698, 1485

\bibitem[\protect\citeauthoryear{Iapichino, Adamek, Schmidt \&
  Niemeyer}{Iapichino et~al.}{2008}]{Iapichino2008}
Iapichino L.,  Adamek J.,  Schmidt W.,    Niemeyer J.~C.,  2008, MNRAS, 388,
  1079

\bibitem[\protect\citeauthoryear{Junk, Walch, Heitsch, Burkert, Wetzstein,
  Schartmann \& Price}{Junk et~al.}{2010}]{Junk2010}
Junk V.,  Walch S.,  Heitsch F.,  Burkert A.,  Wetzstein M.,  Schartmann M.,
  Price D.,  2010, MNRAS, 407, 1933

\bibitem[\protect\citeauthoryear{Kaiser, Pavlovski, Pope \& Fangohr}{Kaiser
  et~al.}{2005}]{Kaiser2005}
Kaiser C.~R.,  Pavlovski G.,  Pope E. C.~D.,    Fangohr H.,  2005, MNRAS, 359,
  493

\bibitem[\protect\citeauthoryear{Kraft, Forman, Jones, Nulsen, Hardcastle,
  Raychaudhury, Evans, Sivakoff \& Sarazin}{Kraft et~al.}{2011}]{Kraft2011}
Kraft R.~P.,  Forman W.~R.,  Jones C.,  Nulsen P. E.~J.,  Hardcastle M.~J.,
  Raychaudhury S.,  Evans D.~A.,  Sivakoff G.~R.,    Sarazin C.~L.,  2011, ApJ,
  727, 41

\bibitem[\protect\citeauthoryear{Kunz, Bogdanovi\'{c}, Reynolds \& Stone}{Kunz
  et~al.}{2012}]{Kunz2012}
Kunz M.~W.,  Bogdanovi\'{c} T.,  Reynolds C.~S.,    Stone J.~M.,  2012, ApJ,
  754, 122

\bibitem[\protect\citeauthoryear{Kwak, Henley \& Shelton}{Kwak
  et~al.}{2011}]{Kwak2011}
Kwak K.,  Henley D.~B.,    Shelton R.~L.,  2011, ApJ, 739, 30

\bibitem[\protect\citeauthoryear{Kwak \& Shelton}{Kwak \&
  Shelton}{2010}]{Kwak2010}
Kwak K.,  Shelton R.~L.,  2010, ApJ, 719, 523

\bibitem[\protect\citeauthoryear{Lamb}{Lamb}{1932}]{Lamb}
Lamb H.,  1932, {Hydrodynamics}, sixth edn.
Cambridge University Press, Cambridge

\bibitem[\protect\citeauthoryear{Landau \& Lifschitz}{Landau \&
  Lifschitz}{1991}]{Landau_hydro}
Landau L.~D.,  Lifschitz E.~M.,  1991, {Lehrbuch der theoretischen Physik}.
Akademie Verlag, Berlin

\bibitem[\protect\citeauthoryear{Lyutikov}{Lyutikov}{2006}]{Lyutikov2006}
Lyutikov M.,  2006, MNRAS, 373, 73

\bibitem[\protect\citeauthoryear{Machacek, Dosaj, Forman, Jones, Markevitch,
  Vikhlinin, Warmflash \& Kraft}{Machacek et~al.}{2005}]{Machacek2005}
Machacek M.~E.,  Dosaj A.,  Forman W.~R.,  Jones C.,  Markevitch M.,  Vikhlinin
  A.,  Warmflash A.,    Kraft R.~P.,  2005, ApJ, 621, 663

\bibitem[\protect\citeauthoryear{Machacek, Jones, Forman \& Nulsen}{Machacek
  et~al.}{2006}]{Machacek2006a}
Machacek M.~E.,  Jones C.,  Forman W.~R.,    Nulsen P. E.~J.,  2006, ApJ, 644,
  155

\bibitem[\protect\citeauthoryear{McNally, Lyra \& Passy}{McNally
  et~al.}{2012}]{McNally2012khi}
McNally C.~P.,  Lyra W.,    Passy J.-C.,  2012, ApJSS, 201, 18

\bibitem[\protect\citeauthoryear{Markevitch, Gonzalez, David, Vikhlinin,
  Murray, Forman, Jones \& Tucker}{Markevitch et~al.}{2002}]{Markevitch2002}
Markevitch M.,  Gonzalez A.~H.,  David L.,  Vikhlinin A.,  Murray S.,  Forman
  W.~R.,  Jones C.,    Tucker W.,  2002, ApJ, 567, L27

\bibitem[\protect\citeauthoryear{Markevitch, Ponman, Nulsen, Bautz, Burke,
  David, Davis, Donnelly, Forman, Jones, Kaastra, Kellogg, Kim, Kolodziejczak,
  Mazzotta, Pagliaro, Patel, {Van Speybroeck}, Vikhlinin, Vrtilek, Wise \&
  Zhao}{Markevitch et~al.}{2000}]{Markevitch2000}
Markevitch M.,  Ponman T.~J.,  Nulsen P. E.~J.,  Bautz M.~W.,  Burke D.~J.,
  David L.~P.,  Davis D.,  Donnelly R.~H.,  Forman W.~R.,  Jones C.,  Kaastra
  J.,  Kellogg E.,  Kim D.,  Kolodziejczak J.,  Mazzotta P.,  Pagliaro A.,
  Patel S.,  {Van Speybroeck} L.,  Vikhlinin A.,  Vrtilek J.,  Wise M.,    Zhao
  P.,  2000, ApJ, 541, 542

\bibitem[\protect\citeauthoryear{Mazzotta, Fusco-Femiano \& Vikhlinin}{Mazzotta
  et~al.}{2002}]{Mazzotta2002}
Mazzotta P.,  Fusco-Femiano R.,    Vikhlinin A.,  2002, ApJ, 569, L31

\bibitem[\protect\citeauthoryear{Narayan \& Medvedev}{Narayan \&
  Medvedev}{2001}]{Narayan2001}
Narayan R.,  Medvedev M.~V.,  2001, ApJ, 562, L129

\bibitem[\protect\citeauthoryear{Nulsen}{Nulsen}{1982}]{Nulsen1982}
Nulsen P. E.~J.,  1982, MNRAS, 198, 1007

\bibitem[\protect\citeauthoryear{Owers, Couch, Nulsen \& Randall}{Owers
  et~al.}{2012}]{Owers2012}
Owers M.~S.,  Couch W.~J.,  Nulsen P. E.~J.,    Randall S.~W.,  2012, ApJ, 750,
  L23

\bibitem[\protect\citeauthoryear{Owers, Nulsen, Couch \& Markevitch}{Owers
  et~al.}{2009}]{Owers2009hifid}
Owers M.~S.,  Nulsen P. E.~J.,  Couch W.~J.,    Markevitch M.,  2009, ApJ, 704,
  1349

\bibitem[\protect\citeauthoryear{Pittard, Hartquist \& Falle}{Pittard
  et~al.}{2010}]{Pittard2010}
Pittard J.~M.,  Hartquist T.~W.,    Falle S. A. E.~G.,  2010, MNRAS, 405, 821

\bibitem[\protect\citeauthoryear{Randall, Nulsen, Forman, Jones, Machacek,
  Murray \& Maughan}{Randall et~al.}{2008}]{Randall2008}
Randall S.~W.,  Nulsen P. E.~J.,  Forman W.~R.,  Jones C.,  Machacek M.~E.,
  Murray S.~S.,    Maughan B.,  2008, ApJ, 688, 208

\bibitem[\protect\citeauthoryear{Reynolds, McKernan, Fabian, Stone \&
  Vernaleo}{Reynolds et~al.}{2005}]{Reynolds2005}
Reynolds C.~S.,  McKernan B.,  Fabian A.~C.,  Stone J.~M.,    Vernaleo J.~C.,
  2005, MNRAS, 357, 242

\bibitem[\protect\citeauthoryear{Roediger \& Br\"{u}ggen}{Roediger \&
  Br\"{u}ggen}{2008}]{Roediger2008visc}
Roediger E.,  Br\"{u}ggen M.,  2008, MNRAS, 388, L89

\bibitem[\protect\citeauthoryear{Roediger, Br\"{u}ggen, Simionescu,
  B\"{o}hringer, Churazov \& Forman}{Roediger et~al.}{2011}]{Roediger2011}
Roediger E.,  Br\"{u}ggen M.,  Simionescu A.,  B\"{o}hringer H.,  Churazov E.,
    Forman W.~R.,  2011, MNRAS, 413, 2057

\bibitem[\protect\citeauthoryear{Roediger, Kraft, Forman, Nulsen \&
  Churazov}{Roediger et~al.}{2013}]{Roediger2013virgovisc}
Roediger E.,  Kraft R.~P.,  Forman W.~R.,  Nulsen P. E.~J.,    Churazov E.,
  2013, ApJ, 764, 60

\bibitem[\protect\citeauthoryear{Roediger, Kraft, Machacek, Forman, Nulsen,
  Jones \& Murray}{Roediger et~al.}{2012}]{Roediger2012n7618}
Roediger E.,  Kraft R.~P.,  Machacek M.~E.,  Forman W.~R.,  Nulsen P. E.~J.,
  Jones C.,    Murray S.~S.,  2012, ApJ, 754, 147

\bibitem[\protect\citeauthoryear{Roediger, Lovisari, Dupke, Ghizzardi,
  Br\"{u}ggen, Kraft \& Machacek}{Roediger et~al.}{2012}]{Roediger2012a496}
Roediger E.,  Lovisari L.,  Dupke R.,  Ghizzardi S.,  Br\"{u}ggen M.,  Kraft
  R.~P.,    Machacek M.~E.,  2012, MNRAS, 420, 3632

\bibitem[\protect\citeauthoryear{Roediger \& ZuHone}{Roediger \&
  ZuHone}{2012}]{Roediger2012fastslosh}
Roediger E.,  ZuHone J.~A.,  2012, MNRAS, 419, 1338

\bibitem[\protect\citeauthoryear{Rosin, Schekochihin, Rincon \& Cowley}{Rosin
  et~al.}{2011}]{Rosin2011}
Rosin M.~S.,  Schekochihin A.~A.,  Rincon F.,    Cowley S.~C.,  2011, MNRAS,
  413, 7

\bibitem[\protect\citeauthoryear{Ruszkowski, Bruggen, Lee \& Shin}{Ruszkowski
  et~al.}{2012}]{Ruszkowski2012}
Ruszkowski M.,  Bruggen M.,  Lee D.,    Shin M.-S.,  2012, eprint
  arXiv:1203.1343

\bibitem[\protect\citeauthoryear{Ruszkowski, En\ss~lin, Br\"{u}ggen, Heinz \&
  Pfrommer}{Ruszkowski et~al.}{2007}]{Ruszkowski2007}
Ruszkowski M.,  En\ss~lin T.~A.,  Br\"{u}ggen M.,  Heinz S.,    Pfrommer C.,
  2007, MNRAS, 378, 662

\bibitem[\protect\citeauthoryear{Sarazin}{Sarazin}{1988}]{Sarazin1988}
Sarazin C.~L.,  1988, {X-ray emission from clusters of galaxies}, cambridge
  edn.
Cambridge University Press, Cambridge

\bibitem[\protect\citeauthoryear{Spitzer}{Spitzer}{1956}]{Spitzer1956}
Spitzer L.,  1956, {Physics of Fully Ionized Gases}.
Interscience Publishers, New York

\bibitem[\protect\citeauthoryear{Springel \& Farrar}{Springel \&
  Farrar}{2007}]{Springel2007}
Springel V.,  Farrar G.~R.,  2007, MNRAS, 380, 911

\bibitem[\protect\citeauthoryear{Sun, Donahue, Roediger, Nulsen, Voit, Sarazin,
  Forman \& Jones}{Sun et~al.}{2010}]{Sun2010a}
Sun M.,  Donahue M.,  Roediger E.,  Nulsen P. E.~J.,  Voit G.~M.,  Sarazin C.,
  Forman W.~R.,    Jones C.,  2010, ApJ, 708, 946

\bibitem[\protect\citeauthoryear{Vieser \& Hensler}{Vieser \&
  Hensler}{2007}]{Vieser2007}
Vieser W.,  Hensler G.,  2007, A\&A, 472, 141

\bibitem[\protect\citeauthoryear{Vikhlinin \& Markevitch}{Vikhlinin \&
  Markevitch}{2002}]{Vikhlinin2002}
Vikhlinin A.~A.,  Markevitch M.,  2002, Astron. Let., 28, 495

\bibitem[\protect\citeauthoryear{Villermaux}{Villermaux}{1998}]{Villermaux1998}
Villermaux E.,  1998, Physics of Fluids, 10, 368

\bibitem[\protect\citeauthoryear{ZuHone, Markevitch \& Johnson}{ZuHone
  et~al.}{2010}]{ZuHone2010}
ZuHone J.~A.,  Markevitch M.,    Johnson R.~E.,  2010, ApJ, 717, 908

\bibitem[\protect\citeauthoryear{ZuHone, Markevitch \& Lee}{ZuHone
  et~al.}{2011}]{ZuHone2011}
ZuHone J.~A.,  Markevitch M.,    Lee D.,  2011, ApJ, 743, 16

\end{thebibliography}

\appendix

\section{Compressibility} \label{app:compressibility}

\newcommand\sh{s_{\rm h}}
\newcommand\scold{s_{\rm c}}

\begin{figure}
\includegraphics[trim=0     30 60 80,clip,angle=0,width=0.43\textwidth]{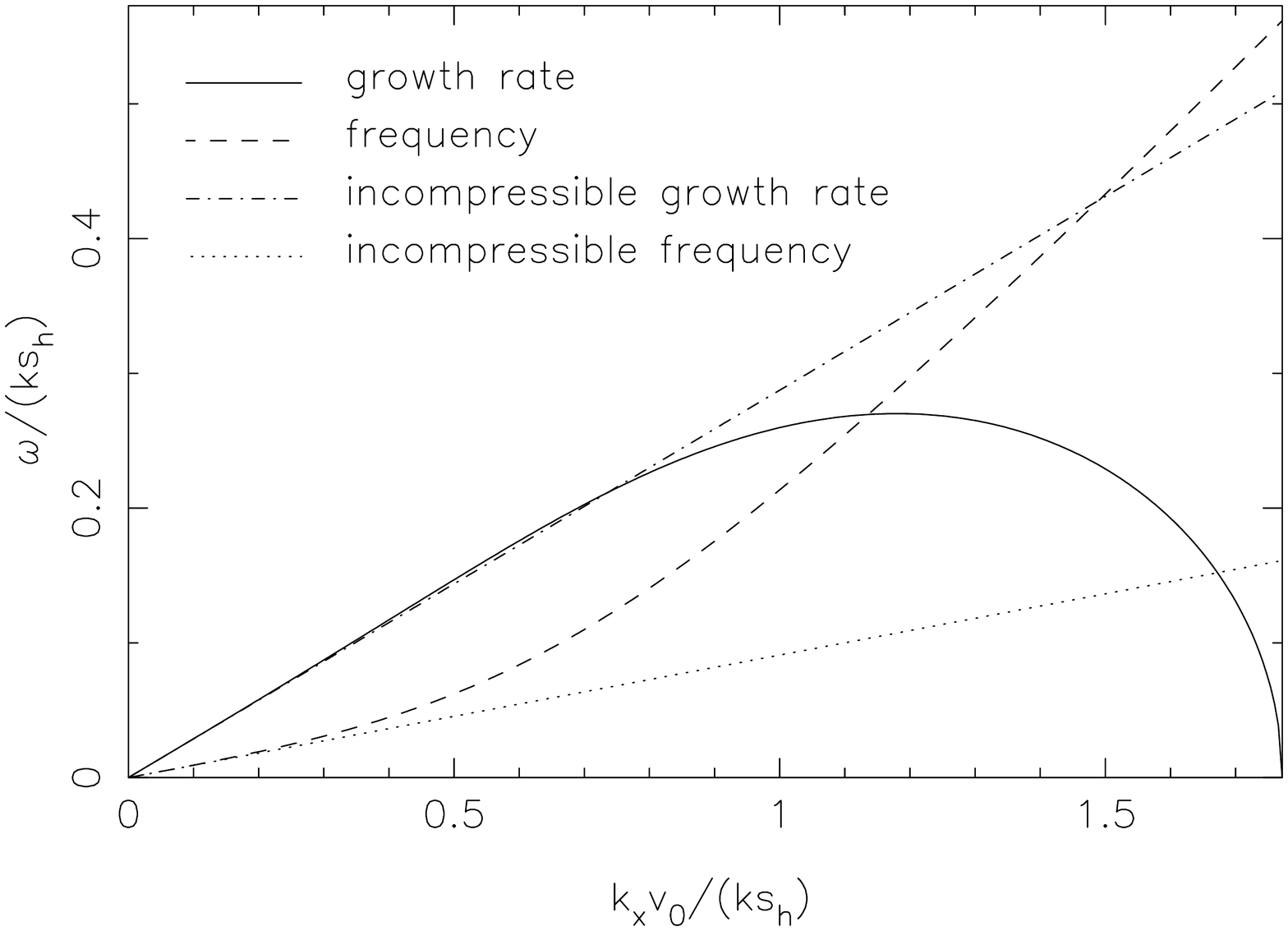}
\caption{Real and imaginary parts of the frequency for unstable KHI
  modes in compressible fluids when $\scold^2/\sh^2 = 0.1$.  As
  indicated in the legend, the curves show the real part (frequency)
  and imaginary part (growth rate) of the scale frequency, $\omega /
  k \sh$, for the growing KHI modes in a shear layer of speed $v_0$.
  The independent variable, $k_x v_0 / (k \sh)$, gives the Mach number
  measured in the hotter fluid of the shear velocity projected onto
  the wavevector of a mode.}
\label{fig:compressible}
\end{figure}

At shear speeds approaching the sound speed, compressibility alters
the behaviour of the KHI.  The dispersion relation for modes in a
shear layer between two compressible fluids may be written \citep{Gerwin1968,
  Nulsen1982}
\begin{equation}
\left({\omega \over k \sh} - {k_x v_0 \over k \sh}\right)^2 \left\{
  \left(\omega \over k \sh\right)^2 - \left(\scold \over \sh\right)^2
\right\} - \left(\omega \over k \sh\right)^2 = 0,
\end{equation}
where $v_0$ is the shear speed, $\sh$ is the sound speed in the
hotter, lower density phase and $\scold$ is the sound speed in the
cooler phase.  For gases of interest here, the density ratio is
generally related to these by $D_\rho = \sh^2 / \scold^2$ (although
this is modified, for example, when the shearing layers are in
different states).  The wavevector is confined to the shear layer, but
it need not be parallel to the shear direction.  Its magnitude is $k$,
while $k_x$ is its component in the shear direction, so that $k_x v_0
/ (k \sh)$ is the Mach number in the hot phase of the shear velocity
projected onto the wavevector.  The complex frequency, $\omega$, here
is measured in a frame at rest with respect to the cooler phase.  This
quartic equation for $\omega$ always has two real roots.  The
remaining two roots are a complex conjugate pair, so that one
corresponds to a growing mode, if
\begin{equation}
{k_x v_0 \over k \sh} < \left\{1 + \left(\scold \over \sh\right)^{2/3}
\right\}^{3/2}. 
\end{equation}

Figure~\ref{fig:compressible} shows the real and imaginary parts of
the scaled frequency, $\omega / (k \sh)$, plotted against the
effective Mach number, $k_x v_0 / (k \sh)$, for the case when
$\scold^2 / \sh^2 = 0.1$, corresponding to a density ratio of 10 (cf.\
Fig.~\ref{fig:rolls_Sp_D10_M1.5}).  The full line shows the growth rate, i.e.\ the imaginary
part of $\omega/(k\sh)$, while the dash-dot line shows the same thing
for the incompressible case.  The dashed line shows the real part of
$\omega/(k\sh)$ and the dotted line shows the same thing for the incompressible case.  As expected, the incompressible approximation is
good for low Mach numbers, but fails as the effective Mach number approaches unity.  For effective Mach numbers $k_x v_0 / (k \sh)
\gtrsim 1.77$ in this case there are no growing modes.  However, for larger Mach numbers, inclined modes with $k_x < k$ can still grow.
Note also that for a Mach number of 1.5 (Sect.~\ref{sec:supersonic}), the growth rate
of the parallel mode with $k_x = k$ is slower than for inclined modes with $k_x v_0 / (k \sh) \simeq 1.2$.  Such modes are excluded in our
2D simulations, so we may have underestimated the true growth rate in this one case.  It is the only case we have simulated where inclined modes may make an appreciable difference to the outcome.

\section{Code tests} \label{sec:visctests}

We tested the viscosity implementation on two setups with analytic solutions. 

\subsection{Viscous flow between plates}
\begin{figure}
\begin{center}
\includegraphics[width=0.4\textwidth]{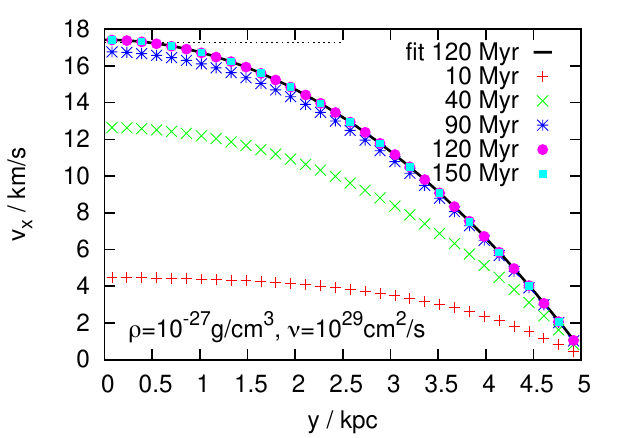}
\caption{Viscous flow through a 2D pipe -- profile of velocity component parallel to the pipe $v_x$ as a function of position perpendicular to the pipe, $y$. The centre of the pipe is at $y=0$, the outer boundaries are at $y=\pm 5\Kpc$. Symbols of different colour denote different timesteps. As the flow started from zero velocity, some time is needed to reach the predicted steady-state.  The black line is the fit of the analytic prediction (Eqn.~\ref{eq:pipeflow}) to the simulation result, leaving the viscosity as a free parameter. We recover the viscosity within 1 or 2 percent.}
\label{fig:plates}
\end{center}
\end{figure}

We set up the classic viscous flow between two plates, i.e., through a 2D pipe. Initially, the fluid has a homogeneous density and zero velocity. A constant pressure gradient is applied in $x$-direction. Boundary conditions are no-slip, i.e., $v_x$=0, at the $y$-boundaries, and open in $x$-direction. The pressure gradient accelerates the fluid. Viscous forces lead to a parabolic profile in $v_x(y)$,
\begin{eqnarray}
v_x(y) &=& v\Max \left(y^2 - d^2  \right) \;\;\textrm{with} \label{eq:pipeflow} \\
v\Max &=& \frac{1}{2\nu\rho} \frac{\partial p}{\partial x},
\end{eqnarray}
where $d$ is half the distance between the plates, $\nu$  the kinematic viscosity, $\rho$ the gas density, and $\partial p/\partial x$ the pressure gradient.
Our chosen parameters (pipe diameter $10\Kpc$, maximum flow velocity $17.3\Kms$, kinematic viscosity $\nu=10^{29}$ cm$^2/$s) correspond to a Reynolds number of 0.5. We ran the same test for a 30 times higher $\Reyn$.
The pipe is sufficiently long such that at its centre in $x$-direction a steady-state is reached before boundary effects reach the centre. Figure~\ref{fig:plates} shows velocity profiles  $v_x(y)$ at different timesteps. We measure the actual viscosity in the simulation by fitting Eqn.~\ref{eq:pipeflow} to the final timestep, leaving the viscosity as a free parameter. We recover the intended viscosity within 1 or 2 percent.

%

\subsection{Viscous smoothing of a shear flow discontinuity}  \label{eq:smoothshearjump}
\begin{figure}
\begin{center}
\includegraphics[width=0.48\textwidth]{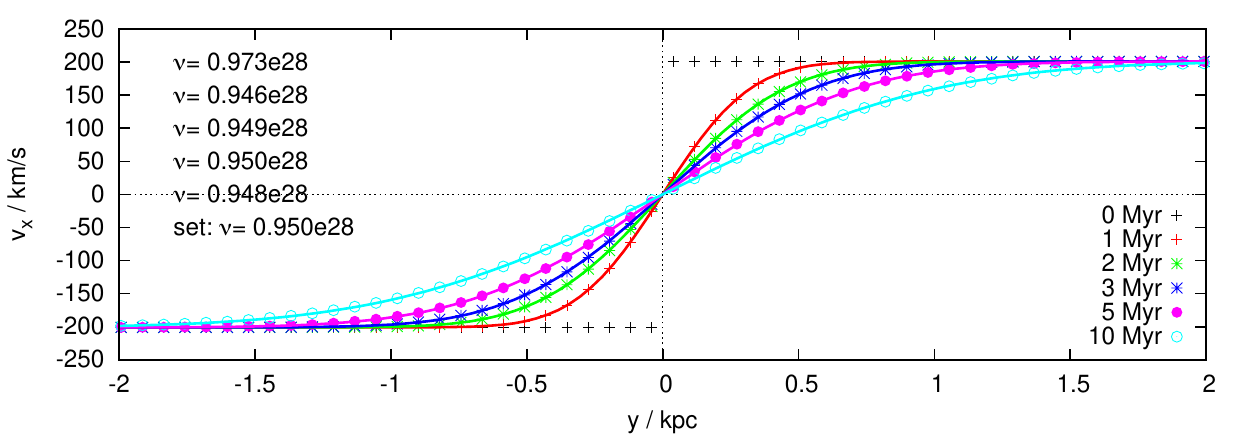}
\caption{Viscous smoothing of shear flow discontinuity. Initially, $v_x$ has a discontinuity in $y$-direction, which is progressively smoothed out  by viscosity. Symbols denote simulation results, colours and symbol styles code different times. Lines of matching colour present fits of the analytic prediction (Eqn.~\ref{eqn:smoothstep}) to the simulation results. The viscosity is left as a free parameter. The fit results and the originally set viscosity are listed on the left.}
\label{fig:diffusion}
\end{center}
\end{figure}

We set up the same shear box like in our KHI tests, but induce no perturbation.  Due to momentum diffusion, viscosity spreads the initial discontinuity in horizontal velocity $v_x(y)$, according to
\begin{equation}
v_x(y) = \Erf\left(\frac{y}{2\sqrt{\nu t}}\right).  \label{eqn:smoothstep}
\end{equation}
This behaviour is shown in Fig.~\ref{fig:diffusion}. Symbols denote simulation results. For each timestep, we fit the simulation results with the expected analytical function (lines in Fig.~\ref{fig:diffusion}), leaving the viscosity as a free parameter. We recover the intended viscosity within 1 percent.

%

\section{Analysis methods} \label{sec:appx_analysis}
%
\subsection{Height of KH rolls} \label{sec:appx_rollheight}
We measure the height of the KH rolls, or the width of the KH mixing interface, by using the fluid tracer $F$. Initially, $F$ is set to 1 below the initial interface, and to 0 above the interface. In Fig.~\ref{fig:rolls_nu_Re1000_D1_M05} we mark the "edges" of the upper and lower fluid by the two thin black lines: for each $x$ we find  the maximum $y\Up(x)$ for gas with $F>0.9$ and the minimum $y\Down(x)$ of gas with $F<0.1$. We define the upwards height of the KH rolls as $h\Up=\max(y\Up(x))$, and the downwards height as $h\Down=\min(y\Down(x))$. In case of equal densities as shown in Figs.~\ref{fig:rolls_nu_Re1000_D1_M05} and  \ref{fig:rolls_nu_D1_M05}, the KHI evolves symmetrically and  $h\Up = h\Down$. For unequal densities the KHI evolves asymmetrical, see Figs.~\ref{fig:thick_vely_nu_M05} and \ref{fig:thick_vely_Sp_M05}.

\subsection{Measuring KH growth times}
We measure the growth time of the KHI by tracking the evolution of the maximum and minimum velocity in $y$-direction, $v_y{}\Max$ and $v_y{}\Min$. We note that these are not the velocity of the interface itself, but reflect also the spin-up of the KH rolls as demonstrated in the left column in Fig.~\ref{fig:rolls_nu_Re1000_D1_M05}. This simple method has the potential risk of contamination by noise-seeded, unintentional secondary KHIs. Many previous works used Fourier filtering to bypass this problem. We find that the presence of at least a small amount of viscosity avoids secondary instabilities in most of our simulations. For example, in Fig.~\ref{fig:rolls_nu_Re1000_D1_M05} even at a Reynolds number of 1000 only the intended instability exists. At still higher Re, we slightly smooth the initial interface (see Eqn.~\ref{eq:smooth}), which mostly avoids the secondary instabilities. Only in the equal density case does the KHI evolve symmetrically. For all other cases we follow the minimum and maximum $v_y$ separately. 

For sufficiently high $\Reyn$, $v_y{}\Max$ and $v_y{}\Min$ show an initial exponential increase which reflects the growth of the KHI (except for very high density contrasts, see Sect.~\ref{sec:highcontrast}).  We fit this initial increase with an exponential function
\begin{equation}
v\Max{}_{/}{}\Min (t) = v_0 \exp(t/\tau\KHvisc)
\end{equation}
with free amplitude $v_0$ and growth time $\tau\KHvisc$. The latter is the derived viscous growth time. Examples for the fits are shown in Fig.~\ref{fig:thick_vely_nu_M05_D1}. In all other corresponding plots, we only state the derived growth times for both directions, but do not plot the fitted function to avoid confusion.  At low $\Reyn$ we can perform a similar fit during the corresponding early evolution time. This results, however, in negative growth times, i.e., a suppressed instability.

\section{Impact of perturbation strength and region} \label{sec:testperturb}
\begin{figure}
\begin{center}
\includegraphics[width=0.4\textwidth]{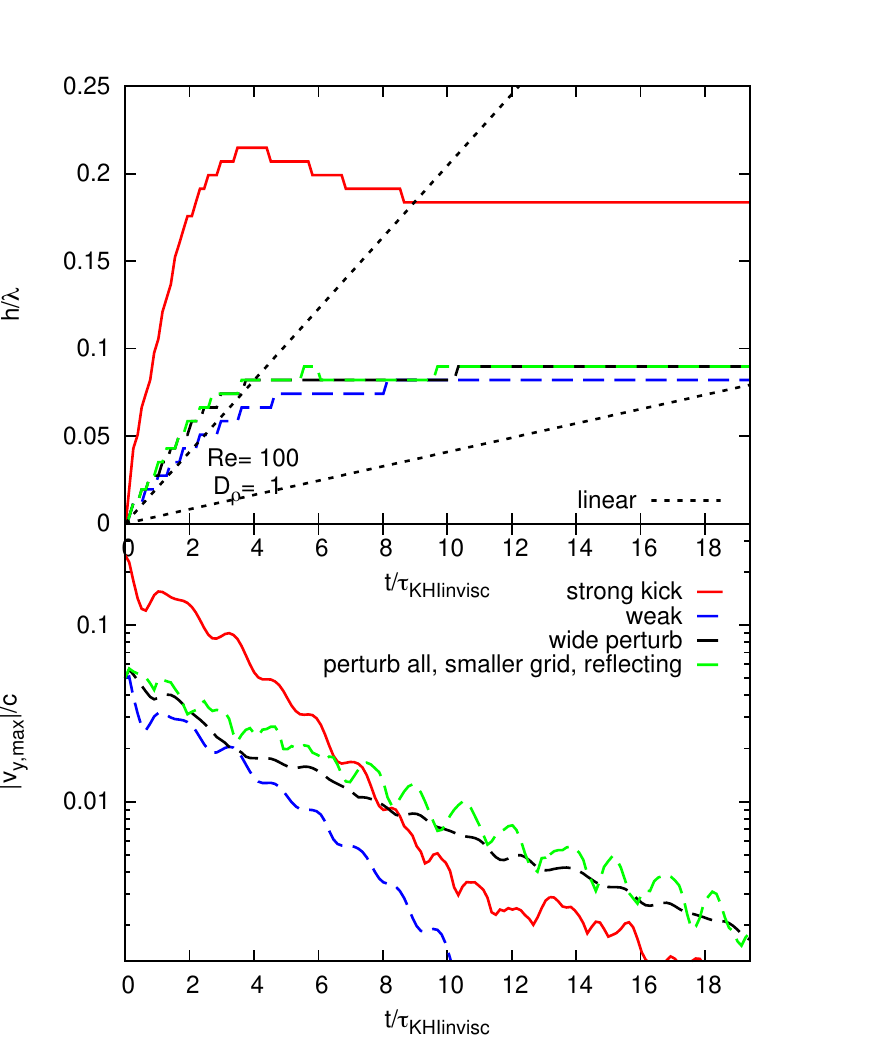}
\caption{Impact of perturbation strength and region width. A higher initial perturbation does not help the instability to grow, neither does a wider perturbation layer around the interface. These simulations are for a constant kinematic viscosity, a density contrast of 1, a shear velocity of Mach 0.5, and a Reynolds number of 100.}
\label{fig:testperturb}
\end{center}
\end{figure}

We verify that our simulations do not over-predict the viscous damping of the KHI due to an insufficient perturbation. Figure~\ref{fig:testperturb} demonstrates that neither a perturbation of much higher velocity amplitude nor a wider perturbation region can revive the KHI. We tested a perturbation velocity of half the shear velocity, and a perturbation scale width of 10 kpc instead of the standard 3 kpc. We also tested perturbing throughout the simulation grid, using reflecting boundaries at $\pm 10\Kpc$ above and below the interface in this case. The viscous damping of the KHI is a robust result.
%

\section{Resolution test} \label{sec:resolution}

\begin{figure}
\includegraphics[trim=0 150 0 150,clip,height=1.8cm]{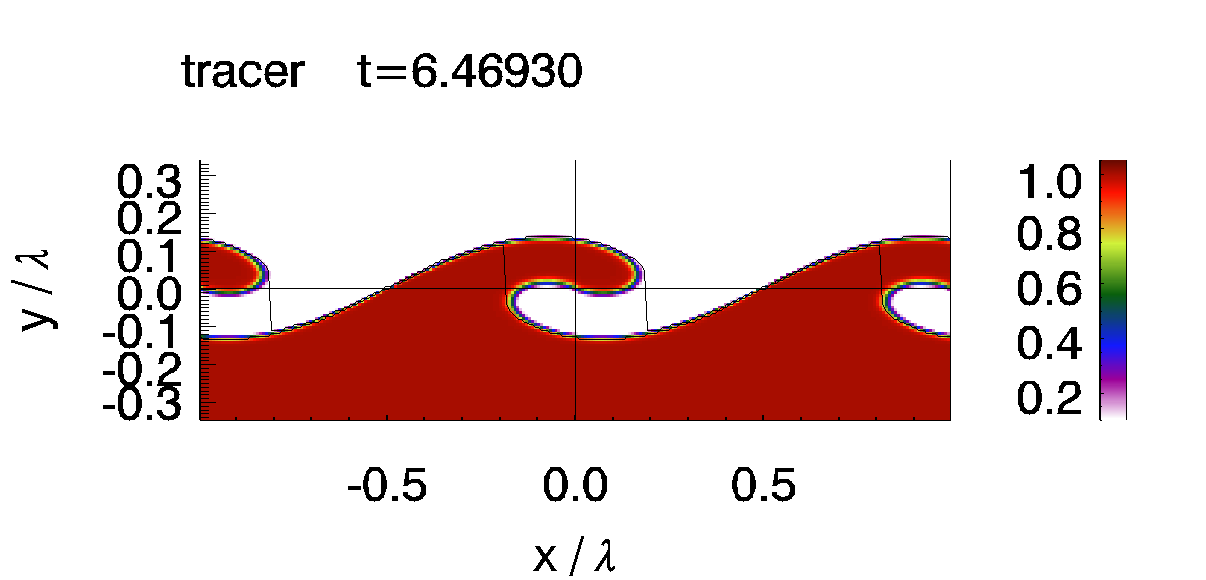}
\includegraphics[trim=0 150 250 150,clip,height=1.8cm]{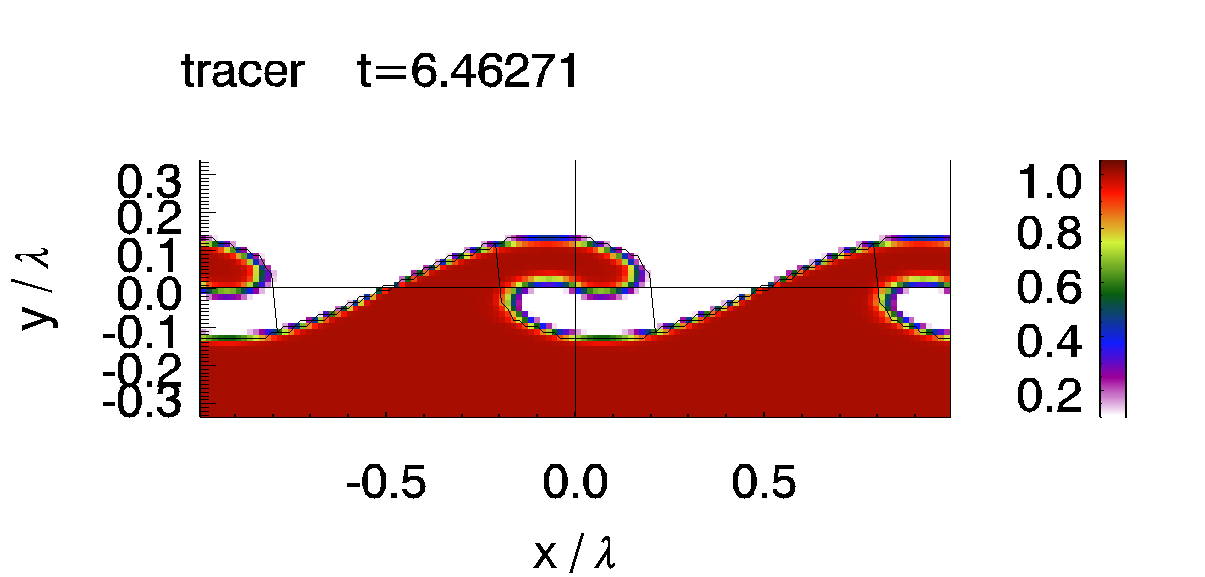}
\includegraphics[trim=0 150 250 150,clip,height=1.8cm]{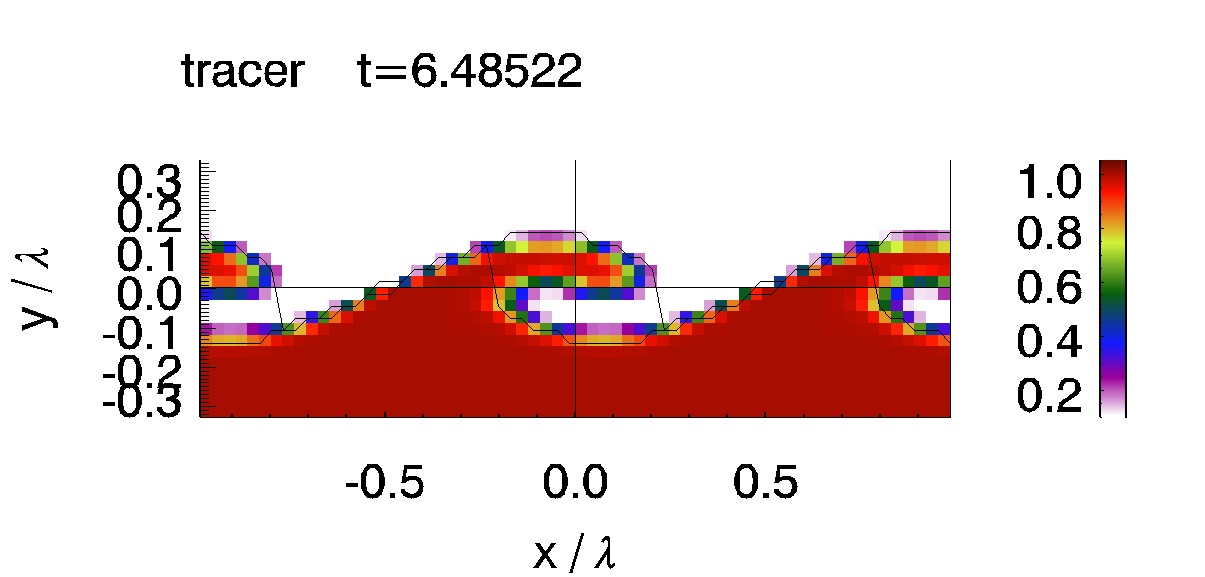}
\includegraphics[trim=0 0 250 150,clip,height=2.8cm]{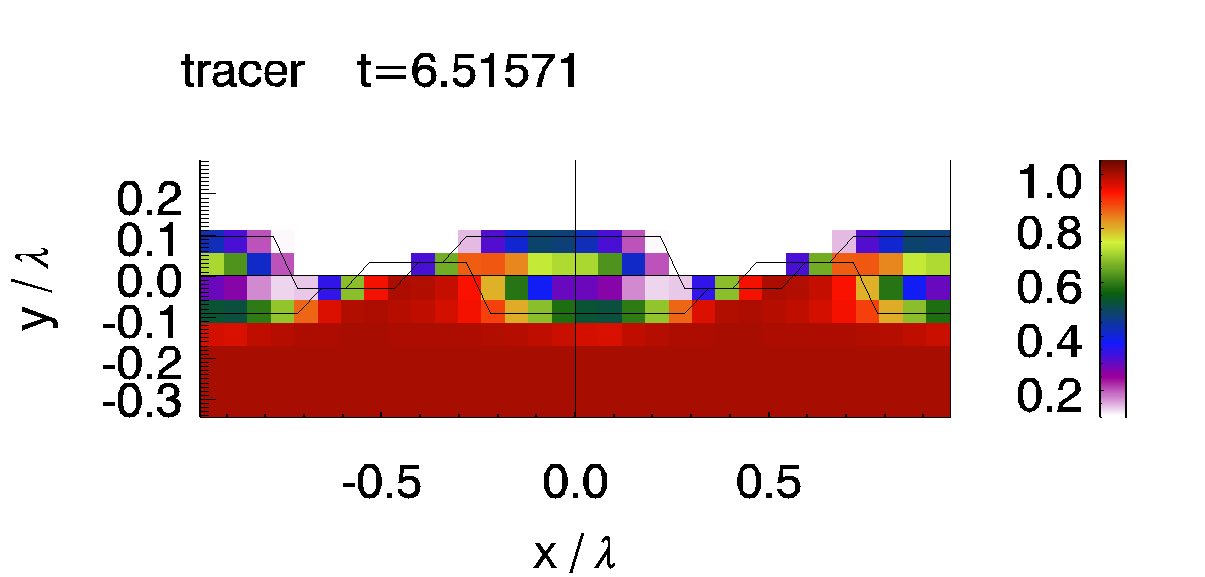}
\includegraphics[width=0.45\textwidth]{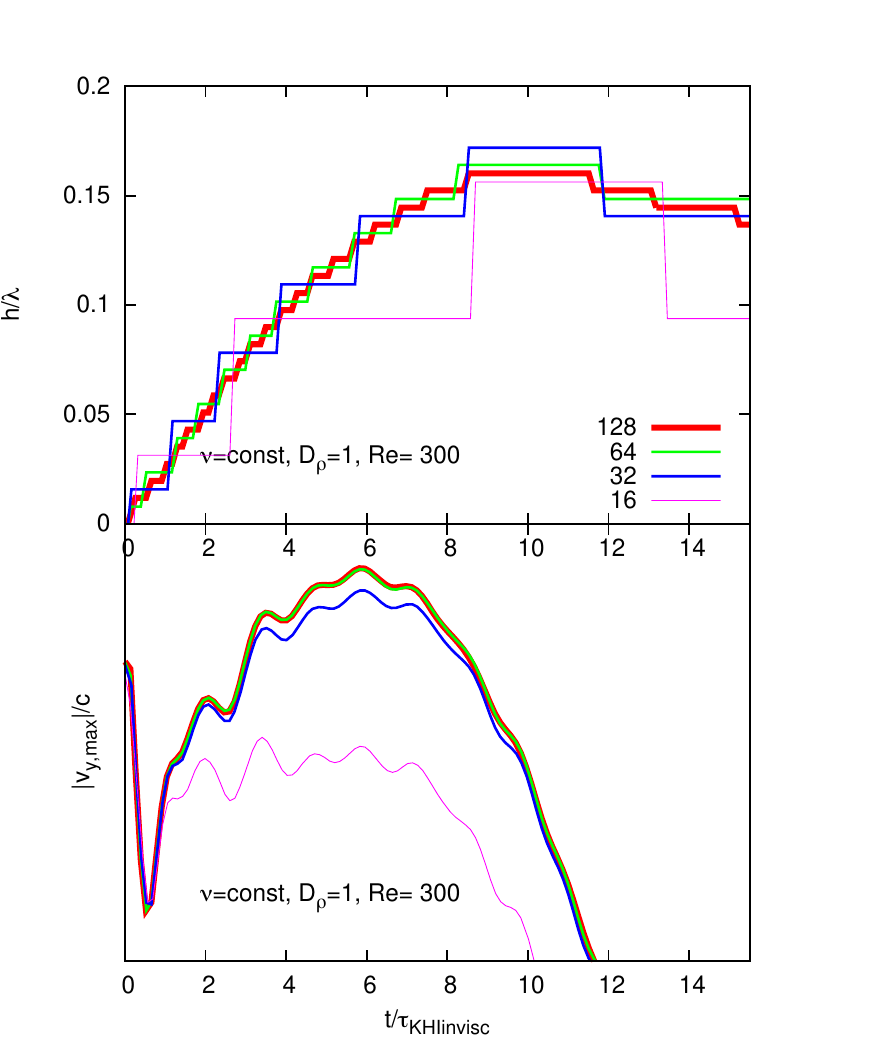}
\caption{Resolution test: comparison of tracer slices for constant kinematic viscosity, density ratio 1, Reynolds number 300, shear flow Mach 0.5, timestep $50\Myr=6.5\tau\KHinvisc$. Top to bottom is 128, 64, 32, 16 cells per perturbation wavelength. The bottom line plot compares the height of the KHI rolls and the maximum vertical velocity.}
\label{fig:resolution_Nu_Re300_D1_Ma05}
\end{figure}

In realistic contexts like gas stripping from galaxies or clouds, or cluster and galaxy mergers, the KHI is just one of several processes and is resolved with only tens of grid cells per wavelength or less, but not $>100$ grid cells as in idealised KHI tests. Therefore we test not only the convergence of our results, but also the ability of the FLASH code to capture the KHI with low resolution.

We compare tracer slices, the height of the KHI rolls and $v_y{}\Max$ for different resolutions in Fig.~\ref{fig:resolution_Nu_Re300_D1_Ma05}. The height of the KHI rolls is captured correctly for a resolution as low as 16 grid cells per perturbation length. We note that in this low resolution the height of the KH rolls above and below the original interface is only $\pm 2$ grid cells. The morphology of the KH rolls is captured well down to resolutions of 32 cells per perturbation wavelength. The same is true for the evolution of $v_y{}\Max$. The peak $v_y{}\Max$ is about 10\% smaller than for the higher resolutions, but the initial increase and later decline are reproduced accurately. With only 16 grid cells per perturbation length, however, the scales on which the peak velocities occur are not resolved, and the growth rate would be underestimated. This is not surprising given that the thickness of the mixing layer is covered by only 4 grid cells.

The PPM hydro scheme spreads discontinuities over 2-3 grid cells as evident from the tracer slices in Fig.~\ref{fig:resolution_Nu_Re300_D1_Ma05}. Consequently, the internal structure of the KH rolls requires resolutions of the \textit{width} of the KH layer exceeding several grid cells. For example, the intermediate tracer values in the KH rolls at the lowest resolution are due only to numerical diffusion. The low resolution results resemble the higher resolution results well except for this caveat.

\bsp

\label{lastpage}

\end{document}